%% file: english.tex
\documentclass[a4paper,UKenglish,cleveref, autoref, thm-restate,numberwithinsect]{lipics-v2021}
\usepackage{xparse}
\usepackage{stmaryrd}
\usepackage{csquotes}
\usepackage{babel}
\usepackage{tikz}

\input{preambule}

\usetikzlibrary{shapes.geometric}
\usetikzlibrary{positioning, automata, arrows}

\makeatletter
\newcommand{\myref}[1]{%
  \hyperref[{#1}]{%
    \cref{#1}%
    \if\vcenter\getrefbykeydefault{#1}{name}{}\vcenter
    \else 
      ~(\nameref*{#1})%
    \fi
  }%
}

\makeatother

\title{Keyboards as a New Model of Computation} 

\author{Yoan GÉRAN}{ENS Paris-Saclay, France}{yoan.geran@ens-paris-saclay.fr}{}{}
\author{Bastien LABOUREIX}{ENS Paris-Saclay, France}{bastien.laboureix@ens-paris-saclay.fr}{}{}
\author{Corto MASCLE}{ENS Paris-Saclay, France}{corto.mascle@ens-paris-saclay.fr}{}{}
\author{Valentin D. RICHARD}{ENS Paris-Saclay, France}{valentin.richard@ens-paris-saclay.fr}{https://orcid.org/0000-0002-9236-8106}{}

\authorrunning{Y. Géran, B. Laboureix, C. Mascle and V.D. Richard} 

\Copyright{Yoan Géran, Bastien Laboureix, Corto Mascle and Valentin Richard} 

\ccsdesc[500]{Theory of computation}

\keywords{formal languages, models of computation, automata theory}

\category{} 

\relatedversion{Here is a French version of the paper: \url{https://arxiv.org/abs/2102.10182}} 

\acknowledgements{We want to thank Pierre Béaur, Lucas Buéri, Jean-Baptiste Daval, Paul Gastin, Colin Geniet, Valentin Maestracci and Clément Théron for their helpful advice and feedback.}

\nolinenumbers 

\hideLIPIcs  

\EventEditors{John Q. Open and Joan R. Access}
\EventNoEds{2}
\EventLongTitle{42nd Conference on Very Important Topics (CVIT 2016)}
\EventShortTitle{CVIT 2016}
\EventAcronym{CVIT}
\EventYear{2016}
\EventDate{December 24--27, 2016}
\EventLocation{Little Whinging, United Kingdom}
\EventLogo{}
\SeriesVolume{42}
\ArticleNo{23}

\begin{document}

\maketitle

\begin{abstract}
 We introduce a new formalisation of language computation, called keyboards. We consider a set of atomic operations (writing a letter, erasing a letter, going to the right or to the left) and we define a keyboard as a set of finite sequences of such operations, called keys. The generated language is the set of words obtained by applying some non-empty sequence of those keys. Unlike classical models of computation, every key can be applied anytime. We define various classes of languages based on different sets of atomic operations, and compare their expressive powers. We also compare them to rational, context-free and context-sensitive languages. We obtain a strict hierarchy of classes, whose expressiveness is orthogonal to the one of the aforementioned classical models. We also study closure properties of those classes, as well as fundamental complexity problems on keyboards.
\end{abstract}

\input{introduction}
\input{preliminaries}

\input{definitions}
\input{general_properties}
\input{key_behaviour}
\input{characterizations}
\input{comparisons}
\input{complexity}
\input{closure}

\input{conclusion}

\bibliography{english}


\newpage

\appendix 
\section{Proofs of general properties}

\subsection*{Proof of \protect\myref{lem-localite}}

    \lemLocalite*

\begin{proof}
    Let $t = \sigma_1 \ldots \sigma_n$. We set $\config{u_0}{v_0} =
    \config{u}{v}$ and  $\config{u_i}{v_i} = \config{u_0}{v_0} \cdot \sigma_1 \cdots \sigma_i$ for all $0 \leq i \leq n$
    (note that for all $0 < i \leq n$, $\config{u_i}{v_i} =
    \config{u_{i - 1}}{v_{i - 1}} \cdot \sigma_i$).

    Further, for all $i$ we use the notations $w_p(u, i) = u[1, \taille{u} - i]$ and
    $w_s(v, i) = v[i+ 1, \taille{v}]$.
    

    We will now show by induction on $i$ that for all
    $0 \leq i \leq n$, $w_p(u, i)$ is a prefix of $u_i$ and
    $w_s(v, i)$ a suffix of $v_i$. 
    This clearly holds for $i = 0$ ($u_0 = u$ and $v_0 = v$).

    Now suppose the property holds for some $i \in \intervE[0,n-1]$. Then $w_p(u, i)$ is a prefix of
    $u_i$ and $w_s(v, i)$ a suffix of $v_i$. 
    Note that
    $w_p(u, i + 1)$ is a prefix of $w_p(u, i)$ (and thus of $u_i$) and $w_s(v, i + 1)$
    a suffix of $w_p(v, i)$ (and thus of $v_i$).

    We distinguish cases according to the nature of $\sigma_{i + 1}$.
    
    \begin{itemize}
        \item If $\sigma_{i + 1} = a \in A$, then $u_{i + 1} = u_ia$ and
              $v_{i + 1} = v_i$, hence the property holds.
        \item If $\sigma_{i + 1} = \retour$, we have $v_{i + 1} = v_i$. In order to show that $w_p(u, i + 1)$
        is a prefix of $u_{i + 1}$, we consider two subcases
              \begin{itemize}
                    \item If $u_i$ is empty , then $u_{i+1}=u_i$.
                    \item Otherwise, $u_i$ is of the form $w_p(u, i)u'c$ with $u\in A^*$ and $c \in A$, and
                          thus $u_{i + 1} = w_p(u, i)u'$.
              \end{itemize}
              In both cases, $w_p(u, i + 1)$ is indeed a prefix of $u_{i + 1}$.
        \item If $\sigma_{i + 1} = \gauche$, we again consider two subcases.
              \begin{itemize}
                  \item If $u_i$ is empty, then $u_{i + 1} = u_i$ and $v_{i + 1} = v_i$.
                    \item Otherwise, $u_i$ is of the form $w_p(u, i)u'c$ for some $c \in A$,
                          thus $u_{i + 1} = w_p(u, i)u'$ and $v_{i + 1} = c v_i$.
              \end{itemize}
              We obtain the result in both cases.
        \item If $\sigma_{i + 1} = \droite$, we proceed similarly.
              \begin{itemize}
                  \item If $v_i$ is empty, then $u_{i + 1} = u_i$ and $v_{i + 1} = v_i$.
                    \item Otherwise, $v_i$ is of the form $cv'w_s(v, i)$ for some $c \in A$
                          hence $u_{i + 1}= u_i c$ and $v_{i + 1} = v'w_s(v, i)$.
              \end{itemize}
              We obtain the result in both cases.
    \end{itemize}
    The first part of the lemma is proven as it corresponds to the case $i=n$. 
    
    A nearly identical induction gives the second part of the lemma.
\end{proof}

\subsection*{Proof of \protect\myref{lem-efficience_loin_bords}}

    \lemEfficienceLoinBords*
    
\begin{proof}
Let us proceed by induction on $n$. If $n=0$, then the property trivially holds.

Now suppose the property holds for some $n$, we prove it for $n+1$. Let
$t = \sigma_1 \ldots \sigma_{n + 1}$ be a key, and $\config{u}{v}$ a configuration such that
$n + 1 \leq \min{\taille{u}, \taille{v}}$. 
Let $t'$ be the key $\sigma_1 \ldots \sigma_n$. There exists
$\config{u_n}{v_n}$ such that $\config{u}{v} \actEff{t'} \config{u_n}{v_n}$.
Further, by \myref{lem-localite} we have that $u_n$ and $v_n$ are non-empty. Thus $u_n$ is of the form $u'a$ and $v_n$ of the form $bv'$
for some $a, b \in A$. We have
\begin{align*}
    \config{u_n}{v_n} &\actEff{c} \config{u_n c}{v_n} \text{ if $c \in A$}.\\
    \config{u_n}{v_n} &\actEff{\retour} \config{u'}{v_n}.\\
    \config{u_n}{v_n} &\actEff{\gauche} \config{u'}{av_n}.\\
    \config{u_n}{v_n} &\actEff{\droite} \config{u_nb}{v'}.
\end{align*}
Thus, no matter the nature of $\sigma_{n + 1}$, there exists a configuration
$\config{u_{n + 1}}{v_{n + 1}}$ such that $\config{u_n}{v_n} \actEff{\sigma_{n + 1}}
\config{u_{n + 1}}{v_{n + 1}}$ and therefore
$\config{u}{v} \actEff{t} \config{u_{n + 1}}{v_{n + 1}}$.

The lemma is proven.
\end{proof}

\subsection*{Proof of \protect\myref{lem-encadrement_taille}}

\lemEncadrementTaille*

\begin{proof}
Let us proceed by induction on $n$. The property clearly holds for $n = 0$.

    Suppose the property holds for some $n$, we show it for $n+1$. Let
    $t = \sigma_1 \ldots \sigma_{n + 1}$ be a key, let $t'$ be the key $\sigma_1 \ldots \sigma_n$,
    let $\config{u_n}{v_n} = \config{u}{v} \cdot t'$ and $\config{u_{n + 1}}{v_{n + 1}} = \config{u}{v} \cdot t$.
    We set
    \begin{align*}
        M_n       &= \size{uv} - \size{t'}_{\retour} + \sum_{x \in A} \size{t'}_x &&
        N_n       = \size{uv} + \sum_{x \in A} \size{t'}_x\\
        M_{n + 1} &= \size{uv} - \size{t}_{\retour} + \sum_{x \in A} \size{t}_x &&
        N_{n + 1} = \size{uv} + \sum_{x \in A} \size{t}_x
    \end{align*}
    We have $M_n \leq \size{u_nv_n} \leq N_n$ by induction hypothesis. 
    We then consider cases based on the nature of $\sigma_{n+1}$
    \begin{itemize}
        \item If $\sigma_{n + 1} = a \in A$ then $M_{n + 1} = M_n + 1$,
                $N_{n + 1} = N_n + 1$
                and $\size{u_{n + 1}v_{n + 1}} = \size{u_n v_n} +1$.
        \item If $\sigma_{n + 1} \in \set{\droite,\gauche}$
              then $M_{n + 1} = M_n$, $N_{n + 1} = N_n$
              and $\size{u_{n + 1}v_{n + 1}} = \size{u_n v_n}$.
      \item If $\sigma_{n + 1} = \retour$ then $M_{n + 1} = M_n - 1$,
            and $N_{n + 1} = N_n$. If $u_n$ is empty then $\size{u_{n + 1}v_{n + 1}}=\size{u_n v_n}$, otherwise $\size{u_{n + 1}v_{n + 1}}=\size{u_nv_n} - 1$.
    \end{itemize}
    In all cases we have $M_{n + 1} \leq \size{u_{n + 1}v_{n + 1}} \leq N_{n + 1}$, proving the first part of the lemma. Similar inductions prove the second part, i.e., $\abs{\enlarger \size{u_n}-\size{u}} \leq n$ and
    $\abs{\enlarger \size{v_n}-\size{v}} \leq n$.
\end{proof}

\subsection*{Proof of \protect\myref{lem-egalite_taille_loin_bords}}

\lemEgaliteTailleLoinBords*

\begin{proof}
We proceed by induction on $n$. The lemma holds trivially for $n = 0$.

Suppose the property holds for some $n$. Let
$t = \sigma_1 \ldots \sigma_{n + 1}$ be a key
and $\config{u}{v}$ a configuration such that
$\size{u} \geq n + 1$. Let $t' = \sigma_1 \ldots \sigma_n$,
$\config{u_n}{v_n} = \config{u}{v} \cdot t'$ and $\config{u_{n + 1}}{v_{n + 1}} = \config{u}{v} \cdot t$.

By \myref{lem-localite}, $u_n$ is non-empty. Thus, $u_n$ is of the form $u'a$, with $a,b \in A$.
We then have
\[
\left\{
\begin{aligned}
    u_{n + 1} &= u_n c && \; v_{n + 1} = v_n  \enspace &&\text{ if $\sigma_{n + 1} = c \in A$}\\
    u_{n + 1} &= u'    && \; v_{n + 1} = v_n  \enspace &&\text{ if $\sigma_{n + 1} = \retour$}\\
    u_{n + 1} &= u'    && \; v_{n + 1} = av_n \enspace &&\text{ if $\sigma_{n + 1} = \gauche$}\\
    u_{n + 1} &= u'    && \; v_{n + 1} = v_n \enspace &&\text{ if $\sigma_{n + 1} = \droite$ with $v=\epsilon$}\\
    u_{n + 1} &= u_n b && \; v_{n + 1} = v'   \enspace &&\text{ if $\sigma_{n + 1} = \droite$ with $v = bv'$ for some $b \in A$}
\end{aligned}
\right.
\]
By induction hypothesis applied to $t'$ and $\config{u_n}{v_n}$, we have
\[
    \size{u_nv_n} = \size{uv} - \size{t'}_{\retour} + \sum_{x \in A} \size{t'}_x
\]
which allows us to obtain the following equality:
\[
    \size{u_{n + 1}v_{n + 1}} = \size{uv} - \size{t}_{\retour} + \sum_{x \in A} \size{t}_x.
\]
which proves the lemma.
\end{proof}

\subsection*{Proof of \protect\myref{prop-croissance_bornee}}

\propCroissanceBornee*

\begin{proof}
We start by proving the result for automatic keyboards, and then extend it to the general case.

Let $K$ be an automatic keyboard, $L$ its language, we prove that for all $w \in L$ of size greater than $\normeInf{K}$, there exists $w'$ such that $w' \neq w$ and
    \[
        \size{w} - \normeInf{K} \leq \size{w'} < \size{w},
    \]
proving the proposition.

If for all $w \in L$, $\size{w} \leq \normeInf{K}$, then the property is trivially true.

    Otherwise, let $w \in L$ be such that $\size{w} > \normeInf{K}$. By \myref{lem-encadrement_taille} we know that any execution yielding $w$ is of length superior or equal to $2$, as otherwise we would have:
    \[
    \size{w} \leq \size{\epsilon} + \sum_{a \in A} \size{t}_a \leq \normeInf{K}.
    \]
    Hence there exist $t_0 \ldots t_{n + 1}$ ($n \in \nats$)
     and $u,v \in A^*$ such that $\config{\epsilon}{\epsilon} \cdot t_0 \cdots t_{n + 1} = \config{u}{v}$ and $w = uv$. For all $i \in\intervE[0, n + 1]$,
    let
    \[
        \left\{
        \begin{aligned}
            \config{u_i}{v_i} &= \config{\epsilon}{\epsilon} \cdot t_1 \cdots t_i\\
            w_i               &= u_iv_i
        \end{aligned}.
        \right.
    \]
    All $w_i$ are in $L$ ($K$ being automatic) and by \myref{lem-encadrement_taille}
    we know that for all $i \in \intervE[0, n]$,
    $\abs{\enlarger \size{w_{i + 1}} - \size{w_i}} \leq \normeInf{K}$.
    Further, $0 \leq \size{w_0} \leq \normeInf{K}$ and $\size{w_n} = \size{w} > \normeInf{K}$, hence there exists
    $k \in \intervE[0, n - 1]$ such that $w_k \neq w$ and
    \[
        \size{w} - \normeInf{K} \leq \size{w_k} < \size{w}.
    \]
    which demonstrates the lemma for automatic keyboards.
    
    Now let $K = (T,F)$ be a keyboard of language $L$. We show that for all $w \in L$ of size greater than $5\normeInf{K}$,
there exists $w' \in L$ such that $\size{w}-5 \normeInf{K} \leq \size{w'} < \size{w}$.
    
    If for all $w \in L$, $\taille{w} \leq 5 \normeInf{K}$, then the result holds trivially.

Otherwise, let $w$ be such that $\taille{w} > 5 \normeInf{K}$. By \myref{lem-encadrement_taille}, we cannot obtain
$w$ by only executing one key. By seeing $T$ as an automatic keyboard, we have $\normeInf{T} \leq \normeInf{K}$ and there exist $w' \in \langage{T}$, $f \in F$ and two configurations
$\config{u}{v}$ and $\config{u'}{v'}$ such that
\[
    w = uv, w' = u'v' \text{ and } \config{u'}{v'} \cdot f = \config{u}{v}.
\]
By \myref{lem-encadrement_taille}, we have $\size{w'} > 4\normeInf{K}$.
Then the first part of the proof ensures the existence of a word $w'' \in \langage{T}$ such that
\[
    \size{w'} - 3\normeInf{K} \leq \size{w''} < \size{w'}- 2\normeInf{K}.
\]
In particular, there exists $\tau \in T^*$ such that
$\config{\epsilon}{\epsilon} \cdot \tau = \config{u''}{v''}$ with $u''v'' = w''$.
Let $\config{x}{y} = \config{u''}{v''} \cdot \tau \cdot f$. We have
$xy \in L$ and \myref{lem-encadrement_taille} gives us the inequalities
\[
    \size{w'} - 4\normeInf{K} \leq \size{xy} < \size{w'}- \normeInf{K}
\]
and therefore
\[
    \size{w} - 5\normeInf{K} \leq \size{xy} < \size{w}.
\]

This proves the lemma.

\end{proof}

\subsection*{Proof of \myref{lem-efficience_correct}}

\label{proof-lem-efficience-correct}

\lemEfficienceCorrect*

\begin{proof}
    We use an induction on $n$. The result is clear for $n = 0$. Let $n > 0$, we have 
    \[
        \config{u}{v} \actEff{\sigma_1 \ldots \sigma_n} \config{u_n}{v_n}
        \actEff{\sigma_{n + 1}} \config{u_{n + 1}}{v_{n + 1}}.
    \]
    By induction hypothesis, 
    \[
        \config{xu}{vy} \actEff{\sigma_1 \ldots \sigma_n} \config{xu_n}{v_ny}.
    \]
    We separate cases depending on $\sigma_{n + 1}$.
    \begin{itemize}
        \item If $\sigma_{n + 1} = a \in A$, 
              then $\config{u_{n + 1}}{v_{n + 1}} = \config{u_na}{v_n}$ and
              $\config{xu_n}{v_ny} \actEff{a} \config{xu_na}{v_ny}$.
        \item If $\sigma_{n + 1} = \retour$, then, 
        $\config{u_{n + 1}}{v_{n + 1}} = \config{u'}{v_n}$ and
            $\config{xu_n}{v_ny} \actEff{\retour} \config{xu'}{v_ny}$.
        \item If $\sigma_{n + 1} = \gauche$, then, as the execution is effective, $u_n$ is not empty, hence of the form $u'a$ and thus 
        $\config{u_{n + 1}}{v_{n + 1}} = \config{u'}{av_n}$ and
            $\config{xu_n}{v_ny} \actEff{\gauche} \config{xu'}{av_ny}$.
        \item If $\sigma_{n + 1} = \droite$, then, as the execution is effective, $v_n$ is not empty, hence of the form $av'$ and thus
        $\config{u_{n + 1}}{v_{n + 1}} = \config{u_na}{v'}$ and
        $\config{xu_n}{v_ny} \actEff{\droite} \config{xu_na}{v'y}$.
    \end{itemize}
    The induction is proven.
\end{proof}

\subsection*{Proof of \myref{lem-efficience_meme_taille}}

\label{proof-lem-efficience_meme_taille}

\lemEfficienceMemeTaille*

\begin{proof}
    For all $i \in \intervE[1; n]$, we set
    \[
    \left\{
    \begin{aligned}
        \config{u_i}{v_i} &= \config{u}{v} \cdot \sigma_1 \ldots \sigma_i\\
        \config{x_i}{y_i} &= \config{x}{y} \cdot \sigma_1 \ldots \sigma_i
    \end{aligned}
    \right.
    \]
    An induction on $i$ shows that for all
    $i \in \intervalleEntier{1}{n}$, we have $\size{u_i} = \size{x_i}$ and 
    $\size{v_i} = \size{y_i}$, yielding the result
    (whether a non effective action happens depends only on the lengths of the $x_i$ and $y_i$, which are the same as the $u_i$ and $v_i$).
\end{proof}

\section{Proofs of properties on keys}

\subsection*{Proof of \myref{prop-ecriture}}

\label{proof-prop-ecriture}

\propEcriture*

\begin{proof}
    For all $i \in \intervE[1; n]$, we set
    \[
    \begin{aligned}
        \config{u_i}{v_i}         &= \config{u}{v} \cdot \sigma_1 \ldots \sigma_i       \qquad&
            \config{x_i'}{y_n'}  &= \fc{u}{v} \cdot \sigma_1 \ldots \sigma_i\\
        \config{u_i'}{v_i'}      &= \config{u}{v} \cdot f_t(\sigma_1 \ldots \sigma_i)  \qquad &
            \config{x_i}{y_i}    &= \fc{u}{v} \cdot f_t(\sigma_1 \ldots \sigma_i)
    \end{aligned}
    \]
    An induction on $i$ then shows that
    $i \in \intervalleEntier{1}{n}$,
    \[
        \size{u_i} = \size{x_i} = \size{u_i'} = \size{x_i'} 
        \quad \text{ and } \quad 
        \size{v_i} = \size{y_i} = \size{v_i'} = \size{y_i'}
    \]
    and for all $a \in A$,
    \[
    \begin{aligned}
        \config{u_i}{v_i}_j = a 
        &\text{ iff } \config{x_i}{y_i}_j = \cEnt{k} \text{ and } \config{u}{v}_k = a  &&\text{ or }
            \config{x_i}{y_i}_j = \tEnt{k} \text{ and } t_k = a\\ 
        &\text{ iff } \config{u_i'}{v_i'}_j = a &&\text{ or } 
            \config{u_i'}{v_i'}_j = \tEnt{k} \text{ and } t_k = a\\ 
        &\text{ iff } \config{x_i'}{y_i'}_j = \cEnt{k} \text{ and } \config{u}{v}_k = a &&\text{ or } \config{x_i'}{y_i'}_j = a.
    \end{aligned}
    \]
\end{proof}

\subsection*{Proof of \protect\myref{prop-alternative_ecrire_a}}

\label{proof-prop-alternative_ecrire_a}

\propAlternativeEcrireA*
\begin{proof}
    We set $\config{x}{y} = \fc{u}{v} \cdot f_t(t)$. By \myref{prop-ecriture}, 
    $\config{u'}{v'}_i = a$ if and only if $\config{u}{v}_i = a$ or
    $\config{x}{y}_i = \cEnt{k}$ and $t_k = a$.
    
    As $\config{u}{v}_i$ does not contain any $a$, then $\config{x}{y}_i = \cEnt{k}$ 
    and $t_k = a$, i.e., $t$ writes an $a$ from $\config{u}{v}$.
\end{proof}

\subsection*{Proof of \protect\myref{prop-ajoute_a}}

\label{proof-prop-ajoute_a}

\propAjouteA*
\begin{proof}
    We set $\config{u'}{v'} = \config{u}{v} \cdot t$ and 
    $\config{x}{y} = \fc{u}{v} \cdot f(t)$. By
    \myref{prop-ecriture}, $\config{u'}{v'}_i = a$ if and only if 
    $\config{x}{y}_i = \cEnt{k}$ and $\config{u}{v}_k = a$ or 
    $\config{x}{y}_i = \tEnt{k}$ and $t_k = a$.
    
    Further, $\size{\config{u'}{v'}}_a > \config{u}{v}_a$, thus there exists
    $i$ such that $\config{x}{y}_i = k$ and $t_k = a$, i.e., $t$ writes an $a$ from $\config{u}{v}$.
\end{proof}

\subsection*{Proof of \myref{prop-ecriture_efficiente}}

\label{proof-prop-ecriture_efficiente}

\propEcritureEfficiente*

\begin{proof}
    Let $n = \size{u}, m = \size{v}, p = \size{x}$ and $q = \size{y}$
    and let $a \in A$.
    We define 
    \[
        \config{u'}{v'} = \config{a^n}{a^m} \cdot f_t(t)
        \quad \text{and} \quad 
        \config{x'}{y'} = \config{a^p}{a^q} \cdot f_t(t).
    \]
    By \myref{cor-touche_aveugle} and \myref{prop-touche_aveugle}, 
    $t$ writes its $k$\ieme symbol at position $i$ from 
    $\config{u}{v}$ (resp. from $\config{x}{y}$) 
    if and only if $\config{u'}{v'}_i = \tEnt{k}$ (resp. 
    $\config{x'}{y'}_i = \tEnt{k}$).
    Let $N = \max{n, p}$ and $M = \max{m, q}$. By
    \myref{lem-efficience_meme_taille}, $t$ acts effectively on $\config{a^n}{a^m}$ and on  $\config{a^p}{a^q}$ , thus by \myref{lem-efficience_correct},
    \[
        \config{a^N}{a^M} \cdot f_t(t) = \config{a^{N - n}u'}{v'a^{M - m}}
        \quad \text{and} \quad
        \config{a^N}{a^M} \cdot f_t(t) = \config{a^{N - p}x'}{y'a^{M - q}}.
    \]
    In particular, we have that $\config{a^N}{a^M}_i = \tEnt{k}$ 
    if and only if $\config{u'}{v'}_i = \tEnt{k}$ and that
    $\config{a^N}{a^M}_i = \tEnt{k}$ if and only if
    $\config{x'}{y'}_i = \tEnt{k}$. 
    The result is proven.
\end{proof}

\section{Proofs of properties of \REK{}}

\subsection*{Proof of \protect\myref{lem-forme_normale_touche_rek}}

    \lemFormeNormaleREK*

\begin{proof}
Suppose $k$ is of the form $k_1 a \retour k_2$
with $k_1, k_2 \in S^*$ and $a \in A$. Let $u$
be a configuration, and let $v = u \cdot k_1$. We get
\begin{align*}
    u \cdot k &=
    \bigl((v \cdot a) \cdot \retour\bigr) \cdot k_2\\
    &= (va \cdot \retour) \cdot k_2\\
    &= v \cdot k_2\\
    &= (u \cdot k_1) \cdot k_2
    = u \cdot k_1k_2.
\end{align*}
Thus, if $k$ contains a letter followed by a backspace, we can erase both from $k$ to obtain an equivalent key.

By iterating the removal of such patterns as many times as possible, we end up with a key $k'$, equivalent to $k$ and of the form $\retour^mw$ for some $m \in \nats$ and $w \in A^*$.

Detecting those patterns can be done in time linear in the size of the key. Further, the size of the key decreases at each step, and thus we only have to iterate at most $\size{t}$ times. As a result turning a key into normal form requires at most quadratic time.
\end{proof}

\subsection*{Proof of \protect\myref{thm-rek_in_rat}}
\label{proof-rek_in_rat}

\thmREKInRat*

We will first show this result for $\RK$. Let $K$ be a keyboard of $\RK$. In order to construct an NFA recognizing $L(K)$,
we decompose executions of $K$ into blocks, each block being associated with the writing of a prefix of the final word. Those blocks allow us to see the execution in a ``monotonous'' way. This decomposition is given by \myref{decompositionsansFLeche}. The following technical lemmas provide a method to construct 
an automaton recognizing $L(K)$ in polynomial time. This automaton is defined in \myref{defautoRK}.

The following definitions will be useful. We assume all keyboards to be in normal form, i.e., to be subsets of $\retour^*A^*$ (see \myref{lem-forme_normale_touche_rek}).

 \begin{definition}[Action on natural numbers]
    Let $K$ be a keyboard of $\RK$. 
    We extend the $\cdot$ operator to natural numbers by setting, for all $n \in \Nat$ and and $t = \retour^kw \in K$
    \[
        n \cdot t = \max{0, n - k} + \taille{w}.
    \]
    We extend this definition to $K^*$ by setting, for all $n\in \nats$, $t\in K$ and $\tau \in K^*$,
    $n \cdot \epsilon = n$ and $n \cdot (\tau t) = (n \cdot \tau) \cdot t$.
\end{definition}

\begin{remark}
    This definition allows us to abstract away the different letters in an execution, to focus on the evolution of the size of the word:
    For all $u \in A^*$ and $\tau \in K^*$,
    we have $\taille{u \cdot \tau} = \taille{u} \cdot \tau$.
\end{remark}
\begin{remark}
    We extend the definitions of $\act{\tau}$, $\actEff{\tau}$ and $\cdotEff$ to natural integers similarly.
\end{remark}

Let us start by a quick study of the behaviour of the keys of $\RK$ keyboards.
Let $u_0,v \in A^*$, say applying a sequence of keys $\tau$ turns $u_0$ into $vu$, with $u$ some extra letters. 
Then the action of $\tau$ on $u_0$ decomposes as follows:

\begin{enumerate}
    \item At the start $u_0$ is of the form $v_0x_0$ with 
          $v_0$ a common prefix of $v$ and $u_0$ (and $v_0$ will not be affected throughout the execution).
    \item Then we would like to have a key erasing $x_0$ and writing letters following $v_0$ in $v$, but this key does not necessarily exist.
    First, some sequence of keys $\tau_0$ modifies $x_0$ (without affecting $v_0$), turning it into a word $x_0'$ of different size
    such that there exists a key $t_0 = \retour^{\size{x'_0}} v_1 x_1$ erasing $\size{x_0'}$ and writing $v_1x_1$ such that $v_0v_1$ is a prefix of $v$
    (and $v_0 v_1$ will not be affected in the rest of the execution).
    \item We apply $t_0$ to obtain $v_0v_1x_1$.
    \item And so on until we obtain $v$ (plus some extra suffix $u$)
\end{enumerate}

The following lemma formalizes this idea. Note that in this lemma we only consider effective executions.

\begin{lemma}
\label{decompositionSansFlecheacteff}%
Let $K$ be a keyboard of $\RK$ and $u_0, v \in A^*$. Let $\ell \in \nats$.
There exist 
$\tau \in K^*$ and $u \in A^*$ such that $\size{u} = \ell$ and $u_0 \actEff{\tau} vu$ if 
and only if there exist an integer $k > 0$, $v_0, \ldots , v_k \in A^+, x_0, \ldots, x_k \in A^*$, $0 \leq s_1, \ldots, s_k \leq \normeInf{K}$ and $\tau_0, \ldots, \tau_k \in K^*$ such that:

\begin{enumerate}
    \item $u_0 = v_0x_0$ and $v = v_0 \cdots v_k$.
    
    \item For all $1 \leq i \leq k$, $\retour^{s_i} v_ix_i \in K$.
    
    \item By setting $s_{k+1} = \ell$, for all $0 \leq i \leq k$, $\size{x_i} \actEff{\tau_i} s_{i+1}$.
\end{enumerate}
\end{lemma}

\begin{proof}
    The $\retour^{s_i} v_i x_i$ are the keys writing parts of $v$, and the $\tau_i$ are the sequences of keys turning $x_i$ into $x_i'$
    of length $s_{i + 1}$ (allowing us to apply 
    $\retour^{s_{i + 1}}v_{i + 1}x_{i + 1}$).
    
    \begin{description}
        \item[$\impliedby$] Let $\tau = \tau_0 t_1 \tau_1 \cdots t_k\tau_k$, one easily checks that $u_0 \actEff{\tau} vu$ for some $u$ such that $\size{u} = \ell$.
        
        \item[$\implies$] 
        Let $\ell \in \nats$, suppose there exists $u \in A^*$ and $\tau \in K^*$ such that $\size{u} = \ell$ and $u_0 \actEff{\tau} vu$.
        $\tau$ is of the form $t_1 \ldots t_n$, with $t_i = \retour^{r_i}w_i$.
        
        For all $i \in \intervE[1; n]$, let 
        \begin{gather*}
            u_i = u_0 \cdot (t_1 \ldots t_i)\\
            y_i = u_0 \cdot (t_1 \ldots t_{i - 1}) \retour^{r_i}
        \end{gather*}
        
        Thus $u_i$ is the word obtained after applying the $i$ first keys of $\tau$, and $y_i$ is the word obtained after applying the $i-1$ first keys and the erasing part of $t_i$.
        In particular, $y_i$ is a prefix
        of $u_{i - 1}$ and $u_i = y_iw_i$.
        
        We also set
        \[
            j = \min\ensC{i}{\forall i' > i, \text{$v$ is a prefix of $y_{i'}$}}
        \]
        with $j = 0$ if $v$ is a prefix of all $y_i$. Intuitively,
        $j$ is the index after which $v$ is written and will no longer be affected. We proceed by induction on $v$.
        
        If $j = 0$ (which includes the case $v = \epsilon$),
        then $v$ is a prefix of all $y_i$ and thus all $u_i$.
        By setting $v_0=v$ and $x_0$ such that $u_0 = v_0x_0$, we have $x_0 \actEff{\tau} u$. 
        We then obtain the result by setting $k = 0$ and $\tau_0 = \tau$.
        
        If $j > 0$, then $u_j = v z$ (if $j < n$ then $v$ is a prefix of $y_{j+1}$ 
        and thus of $u_j$, and if $j = n$ then $u_j = vu$) with $z \actEff{t_{j+1} \ldots t_{n}} u$.
        Hence $t_j$ is in fact the last key affecting the $v$ prefix, and $t_{j + 1} \ldots t_{n}$ turns
        $u_j = vz$ into $vu$ without touching $v$. 
        
        There exists $u'$ of length $r_j$ such that $u_{j - 1} = y_ju'$. As $j$ is minimal, 
        $v$ is not a prefix of $y_j$. Further, as they are both prefixes of $u_j$, $y_j$ is a strict prefix of $v$.
        
        Then we can apply the induction hypothesis to $v' = y_j$.
        We have $u_0 \actEff{t_1...t_{j-1}} v' u'$, thus there exist 
        $k' \in \nats$, $v_0 \in A^*$, $v_1, \ldots, v_{k'} \in A^+, x_0, \ldots, x_{k'} \in A^*$, 
        $0 \leq s_1, \ldots, s_{k'} \leq \normeInf{K}$ and $\tau_0,\tau_1, \ldots, \tau_{k'} \in K^*$ such that:
        \begin{enumerate}
            \item $u_0 = v_0x_0$ and $v' = v_0\cdots v_{k'}$.
            \item For all $1 \leq i \leq k'$, $\retour^{s_i} v_ix_i \in K$.
            \item By setting $s_{k'+1} = \size{u'} = r_j$, for all $0 \leq i \leq k$, $\size{x_i} \actEff{\tau_i} s_{i+1}$.
        \end{enumerate}
        
        We hereby obtain the first part of the wanted decomposition. We then set:
        \begin{gather*}
            k = k' + 1\\
            \tau_k = t_{j+1} \ldots t_n\\
            v_k \text{ such that } v = v'v_k\\
            s_k = \size{u'} = r_j\\
            s_{k + 1} = \ell\\
            x_k = z 
        \end{gather*}
        allowing us to satisfy the three conditions of the lemma.
    \end{description}
\end{proof}

We now slightly refine this lemma. We want to write some word $v$ from $\epsilon$. 
The idea is that we can assume that every key of the execution acts effectively except possibly for the first one: 

If we write $v$ by applying $t_1\ldots t_n$ and some $t_j$ (with $j > 1$) 
does not act effectively on $\epsilon \cdot t_1 \cdots t_{j - 1}$, 
then that means $t_j$ erases all that was written before, so applying $t_j \cdots t_n$ instead of $t_1 \cdots t_n$ would yield the same result.

\begin{lemma}[Decomposition of a $\RK$ execution]
\label{decompositionsansFLeche}%
Let $K$ be a keyboard of $\RK$ and let $v \in A^*$. There exists
$\tau \in K^+$ such that $\epsilon \act{\tau} v$ if and only if
there exists an integer $k > 0$, $v_1, \ldots , v_k \in A^*$, 
$x_1, \ldots, x_k \in A^*$, $0 \leq s_1, \ldots, s_k \leq \normeInf{K}$ and $\tau_1, \ldots, \tau_k \in K^*$ such that:

\begin{enumerate}
    \item $v = v_1\cdots v_k$.
    \item For all $1 \leq i \leq k$, $\retour^{s_i} v_ix_i \in K$.
    \item By setting $s_{k+1} = 0$, for all $1 \leq i \leq k$, $\size{x_i} \actEff{\tau_i} s_{i+1}$.
\end{enumerate}
In particular, we have $(\epsilon \cdot t_1) \actEff{\tau'} v$, with
$\tau' = t_2 \tau_2 \ldots t_n \tau_n$ and $t_1 = \retour^{s_1} v_1 x_1$.
\end{lemma}

\begin{proof}
\begin{description}
    \item[$\impliedby$] We simply observe that $\tau = \retour^{s_1} v_1 x_1 \tau_1 \cdots \retour^{s_k} v_k x_k \tau_k$ yields $v$ when applied on $\epsilon$.
    
    \item[$\implies$] $\tau$ is of the form $t_1\ldots t_n$ with $t_i = \retour^{r_i} w_i$.
    Let $u_0 = \epsilon$ and for all $i \in \intervE[1; n]$,
    \begin{gather*}
        u_i = \epsilon \cdot (t_1 \ldots t_i)\\
        y_i = \epsilon \cdot (t_1 \ldots t_{i - 1}) \retour^{r_i}
    \end{gather*}
    Now let $j$ be the maximal index such that $y_j = \epsilon$. 
    Then we get $\epsilon \cdot (t_j \cdots t_n) = v$. Moreover,
    for all $i > j$, $y_i \neq \epsilon$, which implies that
    the action of the $t_i$ for $i > j$ is effective. 
    As a result, $(\epsilon \cdot t_j) \actEff{t_{j + 1} \ldots t_n} v$.
    
    We finally get the result by applying \myref{decompositionSansFlecheacteff} with $\ell = 0$.
\end{description}
\end{proof}

The following lemma isolates a part of the previous one for clarity.

\begin{lemma}
\label{oneWastefulKey}
Let $v \in A^*$, let $K \subseteq \retour^*A^*$ be a keyboard. There exists 
$\tau \in K^+$ such that $\epsilon \act{\tau} v$ if and only if there exists 
$t \in K$ and $\tau' \in K^*$ such that $(\epsilon \cdot t) \actEff{\tau'} v$.
\end{lemma}

\begin{proof}
This is a direct consequence of \myref{decompositionsansFLeche}.
\end{proof}

In order to produce a word $v$, we have to find a decomposition $v = v_1 \ldots v_k$, with some keys $\retour^{r_i}v_ix_i$
and for all $i$ a sequence turning
$x_i$ into some $x_i'$ of size $r_{i + 1}$.

This allows us to construct from $K$ an automaton
recognizing its language. The idea behind the construction is to memorize in the states the number of extra letters we have to erase and to have two types of transitions.

\begin{itemize}
    \item We can go from $n_1$ to $n_2$ reading $v$ if there exists
          a key $\retour^{n_1} vw$ with $\size{w} = n_2$ (we erase
          the $n_1$ extra letters, write $v$ and get a configuration with $n_2$ letters to erase.
    
    \item We can go from $n_1$ to $n_2$ with an $\epsilon$-transition
          if there exists $\tau$ such that $n_1 \actEff{\tau} n_2$ (we transform
          the extra $x$ of size $n_1$ into an extra $x'$ of size $n_2$).
\end{itemize}

The automaton then allows us to simulate an execution from its decomposition.
If we reach state $i$ after reading some word $v$, it means there exists
a sequence of keys $\tau$ leading to configuration $vx$ with $x$ of length $i$. 
The only accepting state is then $0$. This automaton will be formally defined in \myref{defautoRK}.

We give in \myref{fig-exemple-automate-rek} the automaton obtained by applying
this construction to the keyboard $\set{\touche{\retour abc}, \touche{\retour^4 bb}}$. We simplified it by removing state 3 which turns out not to be co-accessible.


\begin{figure}[htbp]
    \centering
    \begin{tikzpicture}[shorten >=1pt,node distance=2.5cm,on grid,auto] 
        \node[state] (i)   {}; 
        \node[state] (q_1) [below=of i] {$1$}; 
        \node[state, accepting] (q_0) [left=of q_1] {$0$}; 
        \node[state] (q_2) [right=of q_1] {$2$}; 
        \node[state] (q_4) [below=of q_1] {$4$}; 
        \draw[->,thick, >=latex] (0, 1.3) to (i);
        \path[->,thick, >=latex, color3, bend right = 20, above left = 0.5cm and 0.1cm] (i) edge node {$abc$} (q_0);
        \path[->,thick, >=latex, color2, dashed, bend left = 5] (i) edge node {$bb$} (q_0);
        
        \path[->,thick, >=latex, color3, bend right = 10, left = 0.5cm] (i) edge node {$ab$} (q_1);
        \path[->,thick, >=latex, color2, dashed, bend left = 10, right = 0.5cm] (i) edge node {$b$} (q_1);
        
        \path[->,thick, >=latex, color3] (i) edge node {$a$} (q_2);
        
        \path[->,thick, >=latex, color3] (q_1) edge node {$abc$} (q_0);
        \path[->,thick, >=latex, color3, out=350,in=320,looseness=4,below right = 0.1cm and 0.3cm] (q_1) edge node {$ab$} (q_1);
        \path[->,thick, >=latex, color3] (q_1) edge node {$a$} (q_2);
        
        \path[->,thick, >=latex, color2, dashed] (q_4) edge node {$bb$} (q_0);
        \path[->,thick, >=latex, color2, dashed] (q_4) edge node {$b$} (q_1);
        
        \path[->,thick, >=latex, color1, bend left = 20, dotted] (q_2) edge node {$\epsilon$} (q_4);
        \path[->,thick, >=latex, color1, dotted] (q_4) edge node {$\epsilon$} (q_2);
    \end{tikzpicture}
    \caption{Automaton of $\set{\touche{\retour abc}, \touche{\retour^4 bb}}$}
    \label{fig-exemple-automate-rek}
\end{figure}

Red transitions (solid arrows) denote the action of $\touche{\retour abc}$ and blue ones (dashed arrows) 
the action of $\touche{\retour^4 bb}$. A black transition (dotted arrows) from $i$ to $j$ means that there is a sequence of keys turning effectively the $i$ extra letters (that should be erased)
into $j$ letters.

However, we want to be able to construct this automaton (efficiently), thus we still have to be able to decide, for all 
$0 \leq n_1, n_2 \leq \normeInf{K}$, if there is an execution $\tau$ turning $n_1$ into $n_2$ effectively
(and thus turning the $x_i$ into $x_i'$).
We are now interested in this transformation.
We show that we can decide in polynomial time whether there exists 
a sequence of keys transforming effectively a word of length
$n_1$ into one of length $n_2$. 
Note that as the $x_i$ and $x_i'$ from the decomposition are respectively written and erased by a single key, their lengths are at most $\normeInf{K}$, thus it is enough for us to focus on integers lower or equal to $\normeInf{K}$.

We start by giving necessary and sufficient conditions
for the existence of a sequence of keys turning a word of size $n_1$ into one of size $n_2$ effectively.
For that we consider several cases.

In our informal explanations we will sometimes confuse the $x_i$ and $x'_i$ with their lengths (as we are interested in their lengths and not their content).

\begin{lemma}
\label{Frobenius}%
Let $K$ be a keyboard of $\RK$. We set $E = \ensC{\abs{w} - r }{\retour^r w \in K}$
and $p = \PGCD{E}$. Let $x$ and $x'$ be two integers such that $0 \leq x, x' \leq \normeInf{K}$.
If there exists $\retour^{r_1}w_1, \retour^{r_2}w_2 \in K$ such that
\begin{itemize}
    \item $\size{w_1} - r_1 < 0 < \size{w_2} - r_2$ ;
    \item $r_2 \leq x$ and $\size{w_1} \leq x'$;
    \item $p$ divides $x' - x$,
\end{itemize}
then there exists a sequence of keys $\tau = t_1 \ldots t_n \in K^*$ such that
$x \actEff{\tau} x'$.
\end{lemma}

\begin{proof}
    The elements of $E$ are the numbers of letters written by each key, minus the number of letters it erases. We split $E$ between positive integers (corresponding to keys which write more letters than they erase), and negative ones, (keys which erase more than they write). We are not interested in the keys which write as much as they erase for this proof. We set
    \[
        E_+ = \ensC{i}{i \in E \text{ and } i > 0} \text{ and }
        E_- = \ensC{i}{i \in E \text{ and } i > 0}
    \]
    and $p^+$ and $p^-$ their respective gcds, $p^+ = \PGCD{E_+}$
    and $p^- = \PGCD{E_-}$. We have $\PGCD{p^+, p^-} = p$, thus, by Bézout's identity, there exist $\alpha, \beta \in \N$ such
    that $\alpha p^+ - \beta p^- = p$.
    
    We then consider $M$ a multiple of $p^+$ and $p^-$ greater than $\normeInf{K}$
    and
    \[
        D = x' + (M^2 + M)(r_1 - \size{w_1}) - x - M(\size{w_2} - r_2).
    \] 
    $D$ is non-negative because $(M^2 + M)(r_1 - \size{w_1}) \geq M^2 + M \geq M\norme{K} + \norme{K}
    \geq M(\size{w_2} - r_2) + x$. Further, $p$ divides $D$, hence there exist 
    $c \in \N$ such that $D = cp$.
    
    We define $A = c\alpha p^+ + M^2$ and $B = c \beta p^- + M^2$. They are 
    respectively divisible by $p^+$ and $p^-$ and we further have
    $A - B = cp = D$.
    
    We have $A \geq M^2 \geq \normeInf{K}^2 \geq \max{E_+}^2$ and similarly 
    $B \geq \max{E_-}^2$. A result first proven by Schur \cite{alfonsin2005diophantine} (and then proven again many times) states that given a finite set $S$ of positive integers, any number greater than $\max(S)^2$ and divisible by $\PGCD{S}$ can be written as a linear combination of elements of $S$ with natural numbers as coefficients (the bound given by Schur is actually better, but we do not need it here).  
    There exist non-negative coefficients $\suite[i]{a_i}[E_+]$ and 
    $\suite[i]{b_i}[E_-]$ such that
    \[
        \sum_{i \in E_+} ia_i = A \text{ and }
        \sum_{i \in E_-} ib_i = B.
    \]
    We can now construct the sequence of keys allowing us to go from $x$ to $x'$ effectively.
    
    \begin{itemize}
        \item We start from $x$. We apply $M$ times the key $\retour^{r_2}w_2$.
        The execution is effective ($r_2 \leq x$ and the key is positive) and leads to $x + M(\size{w_2} - r_2)$.
        
        \item For all $i \in E$, let $t_i = \retour^{r_i} w_i$ be a key
        such that $i = \size{w_i} - r_i$. For all $i \in E_+$, we apply $a_i$ 
        times $t_i$ (in an arbitrary order). The execution is effective
        as we start with more than $M$ letters and we only apply positive keys. We obtain configuration $x + M(\size{w_2} - r_2) + A$.
        
        \item For all $i \in E_-$, we apply $b_i$ times $t_i$
        (in an arbitrary order). This leads to configuration
        \[
            x + M(\size{w_2} - r_2) + A - B = x + M(\size{w_2} - r_2) + D
                                            = x' + (M^2 + M)(r_1 - \size{w_1}).
        \]
        The final configuration is greater than $M$ and we applied only negative keys, thus the execution is again effective.
        
        \item Finally, we apply $M + M^2$ times the key $\retour^{r_1}w_1$, yielding the final configuration $x'$. 
        As $x' \geq r_1$ and as this key is negative, the execution is effective.
    \end{itemize}
    
    We indeed obtain a sequence of keys $\tau \in K^*$ such that
    $x \actEff{\tau} x'$.
\end{proof}

The previous lemma presented a necessary condition for the existence of an execution from $x$ to $x'$, the two following ones will give necessary and sufficient conditions in other cases. 
Another lemma will then use those three to give a necessary and sufficient condition in the general case.
To begin with, if every negative key writes more than $x'$ letters, 
then we can only use positive keys (as we then have no way of getting back to $x'$ from a larger configuration) and in particular $x \leq x'$.

\begin{lemma}
\label{nogoodnegative}%
Let $K$ be a keyboard of $\RK$. Let $x$ and $x'$ be two integers such that
$0 \leq x, x' \leq \normeInf{K}$. 
Suppose that for all
$\retour^{r} w \in K$ such that $\size{w} < r$, $\size{w} > x'$. 
Then the following conditions are equivalent.
\begin{itemize}
    \item There exists $\tau \in K^*$ such that $x \actEff{\tau} x'$.
    \item There exists $\tau = t_1 \ldots t_n \in K^*$ such that $x \actEff{\tau} x'$
          and for all $i$, $x \leq x \cdot (t_1 \ldots t_i) \leq x'$.
\end{itemize}
\end{lemma}

\begin{proof}
\begin{description}
    \item[$\impliedby$] Clear as $\tau$ is already constructed.
    \item[$\implies$] Suppose there exists $\tau = t_1 \ldots t_n$ such that
    $x \actEff{\tau} x'$. We proceed by contradiction and assume there exists
    $i \in \intervE[1; n]$ such that $t_i$ is of the form $\retour^r w$
    with $\size{w} < r$ (and thus $\size{w} > x'$, by the lemma's hypothesis). 
    Let $i$ be the maximal index satisfying that property.
    
    As $t_{i + 1}, \ldots, t_n$ write at least as many letters as they erase,
    we have $x \cdot (t_1 \ldots t_i) \leq x' < \size{w}$. However, $x \cdot (t_1 \ldots t_i)$
    is the word obtained after applying $t_i$, thus
    $x \cdot (t_1 \ldots t_i) \geq \size{w}$, yielding a contradiction.
    
    As a consequence, all keys in $\tau$ add at least as many letters as they erase. The sequence $x \cdot (t_1 \ldots t_i)$ is therefore nondecreasing,
    giving us the inequalities:
    \[
        \forall i \in \intervE[1; n], x \leq x \cdot (t_1 \ldots t_i) \leq x'.
    \]
\end{description}
\end{proof}

We now present the symmetric case, in which we cannot apply any positive key. The lemma and proof are similar to the previous ones.

\begin{lemma}
\label{nogoodpositive}%
Let $K$ be a keyboard of $\RK$. Let $x$ and $x'$ be two integers such that
$0 \leq x, x' \leq \normeInf{K}$. 
Suppose that for all
$\retour^{r} w \in K$ such that $\size{w} > r$, $r > x$. 
Then the following conditions are equivalent.
\begin{itemize}
    \item There exists $\tau \in K^*$ such that $x \actEff{\tau} x'$.
    \item There exists $\tau = t_1 \ldots t_n \in K^*$ such that $x \actEff{\tau} x'$
          and for all $i$, $x \geq x \cdot (t_1 \ldots t_i) \geq x'$.
\end{itemize}
\end{lemma}

\begin{proof}
\begin{description}
    \item[$\impliedby$] Clear as $\tau$ is already constructed.
    \item[$\implies$] Suppose there exists $\tau = t_1 \ldots t_n$ such that
    $x \actEff{\tau} x'$. We proceed by contradiction and assume there exists
    $i \in \intervE[1; n]$ such that $t_i$ is of the form $\retour^r w$
    with $\size{w} > r$ (and thus $r > x$, by the lemma's hypothesis). 
    Let $i$ be the minimal index satisfying that property.
    
    As $t_{1}, \ldots, t_{i-1}$ erase at least as many letters as they write,
    we have $x \cdot (t_1 \ldots t_{i-1}) \leq x < r$. However, $t_i$ acts effectively on $x \cdot (t_1 \ldots t_{i-1})$, thus
    $x \cdot (t_1 \ldots t_{i-1}) \geq r$, yielding a contradiction.
    
    As a consequence, all keys in $\tau$ erase at least as many letters as they write. The sequence $x \cdot (t_1 \ldots t_i)$ is therefore nonincreasing,
    giving us the inequalities:
    \[
        \forall i \in \intervE[1; n], x \geq x \cdot (t_1 \ldots t_i) \geq x'.
    \]
\end{description}
\end{proof}

Equipped with these results, we can now decide in polynomial time
if we can go from $x$ to $x'$ effectively.

\begin{lemma}
\label{Polynomialeff}%
The following problem is decidable in polynomial time.
\[
\begin{cases}
    \text{\textsc{Input}}:   & 
    \text{$K$ a keyboard of $\RK$ and $0 \leq x, x' \leq \normeInf{K}$} \\
    \text{\textsc{Output}}: & 
    \text{Does there exist $\tau = t_1 \ldots t_n \in K^*$ such that $x \actEff{\tau} x'$ ?}
\end{cases}
\]
\end{lemma}

\begin{proof}
    Let $K$ be a keyboard of $\RK$ and $0 \leq x, x' \leq \normeInf{K}$. We distinguish four cases:
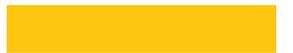
\begin{itemize}
    \item If for all $\retour^r w \in K$ such that $\size{w} - r < 0$ we have $\size{w} > x'$, then we construct a graph whose vertices are integers from $0$ to $\normeInf{K}$ and such that
    there is a transition from $s$ to $s'$ if and only if there exists $t \in K$ such that 
    $s \cdot t = s'$. By \myref{nogoodnegative}, there exists an execution 
    $x \actEff{\tau} x'$ if and only if there is a path from $x$ to $x'$ in that graph.
    \item If for all $\retour^r w \in K$ such that $0 < \size{w} - r$ we have $r > x$, then
    we construct the same graph as in the previous case. By \myref{nogoodpositive},
    there exists $\tau$ such that $x \actEff{\tau} x'$ if and only if there is a path from 
    $x$ to $x'$ in that graph.
    \item If there exists $\retour^{r_1} {w_1}, \retour^{r_2} {w_2} \in K$ such that 
    $\size{w_1} - r_1 < 0 <\size{w_2} - r_2$, $r_2 \leq x$, $\size{w_1} \leq x'$ and 
    $p$ divides $x' - x$, then, by \myref{Frobenius}, there exists $\tau$ such that 
    $x \actEff{\tau} x'$.
    \item Otherwise, $p$ does not divide $x - x'$: a straightforward induction on
    $\tau$ shows that for all $\tau \in K^*$, if $\tau$ acts effectively on $x$, then $p$ divides $(x \cdot \tau) - x$ and thus $x \cdot \tau \neq x'$.
\end{itemize}
\end{proof}

We have all the necessary tools to construct, given a keyboard $K$ of \RK, an automaton recognizing the language $\langage{K}$.

\begin{remark}\label{rem-wordtransitions}
In the following definition we will allow transitions to be labelled with words of $A^*$ (of length at most $\normeInf{K})$ and not only letters. 
This allows for a clearer presentation and one can easily get from this construction an automaton labelled with letters, by decomposing each transition $n_1 \act{a_1 \cdots a_k} n_2$ into a sequence of transitions reading each $a_i$ one by one. 
As we only use words of size at most $\normeInf{K}$, the resulting automaton is still of size polynomial in the one of $K$.
\end{remark}

\begin{definition}
\label{defautoRK}%
Let $K$ be a keyboard of $\RK$, we define $\Auto{K}$ the NFA
whose states are elements of $\set{\Init} \cup \intervE[0; \normeInf{K}]$.

The only initial state is $\Init$, the only final state is $0$.

The set of transitions is 
$\Delta = \Delta_{\Init} \cup \Delta_{\text{t}} \cup \Delta_{\tau}$ with

\begin{align*}
    \Delta_{\Init}         &= \ensC{\Init \act{v} n}
                           {\exists m, w, \retour^m vw \in K \land \size{w} = n}\\
    \Delta_{\text{t}} &= \ensC{n_1 \act{v} n_2}
                           {\exists w, \retour^{n_1} vw \in K \land \size{w} = n_2}\\
    \Delta_{\tau}          &= \ensC{n_1 \act{\epsilon} n_2}
                           {\exists \tau \in K^*, n_1 \actEff{\tau} n_2}
\end{align*}
\end{definition}

\begin{lemma}
\label{AutoEqSansFleche}%
Let $K$ be a keyboard of \RK, $\Auto{K}$ and $K$ recognize the same language.
\end{lemma}

\begin{proof}
Let $v \in A^*$, suppose $v \in \langage{K}$, then by \myref{decompositionsansFLeche}, there exist $k>0$, $v_1, \ldots , v_k \in A^*$, 
$x_1, \ldots, x_k \in A^*$, $0 \leq s_1, \ldots, s_k \leq \normeInf{K}$ and $\tau_1, \ldots, \tau_k \in K^*$ such that:

\begin{enumerate}
    \item $v = v_1 \cdots v_k$.
    \item For all $1 \leq i \leq k$, $\retour^{s_i} v_i x_i \in K$.
    \item By setting $s_{k + 1} = 0$, for all $1 \leq i \leq k$, $\size{x_i} \actEff{\tau_i} s_{i + 1}$.
\end{enumerate}

One can then easily observe that the execution $\Init \act{v_1} \size{x_1} \act{\epsilon} s_2 \act{v_2} \size{x_2} \cdots \act{v_k} \size{x_k} \act{\epsilon} 0$ accepts $v$.

Now suppose $v \in \langage{\Auto{K}}$, there exists a run on $\Auto{K}$ accepting $v$.

First observe that:
\begin{itemize}
    \item for all state $n \in \intervE[0; \normeInf{K}]$ there is a transition 
    $n \act{\epsilon} n$ ;
    
    \item for all transitions $n_1 \act{\epsilon} n_2, n_2 \act{\epsilon} n_3$, there is a transition $n_1 \act{\epsilon} n_3$.
\end{itemize}

As a consequence, we can assume the run of the automaton to be of the form
\[
    \Init \act{v_1} n_1 \act{\epsilon} n'_2 
          \act{v_2} n_2 \act{\epsilon} \cdots 
          \act{v_k} n_k \act{\epsilon} n'_{k + 1} = 0.
\]

By definition of $\Auto{K}$, we can then define:

\begin{itemize}
    \item for all $1 \leq i \leq k$, $\tau_i \in K^*$ such that $n_i \actEff{\tau_i} n'_i$ ;
    \item for all $1 \leq i \leq k$, $x_i \in A^*$ and $n'_1$ such that 
          $\retour^{n'_i}v_ix_i \in K$ and $\size{x_i} = n_i$;
    \item $s_i = n'_i$ for all $1 \leq i \leq k+1$.
\end{itemize} 
We obtain that $v \in \langage{K}$ by applying \myref{decompositionsansFLeche}.
\end{proof}

\begin{theorem}
    All languages recognized by a keyboard of $\RK$ are rational.
\end{theorem}

\begin{proof}
    Let $K$ be a keyboard of $\RK$, by \myref{AutoEqSansFleche}, its language 
    is the one of $\Auto{K}$, $\langage{K}$ is thus rational.
\end{proof}

\begin{lemma} \label{PolyAutoSansfleche}%
    Let $K$ be a keyboard of $\REK$, we can construct $\Auto{K}$ in polynomial time in the size of $K$.
\end{lemma}

\begin{proof}
By \myref{Polynomialeff}, for all $0 \leq n_1, n_2 \leq \normeInf{K}$, 
we can decide in time polynomial in $\normeInf{K}$ whether there exists $\tau \in K^*$ 
such that $n_1 \actEff{\tau} n_2$.

As a consequence (and by \myref{rem-wordtransitions}), the automaton $\Auto{K}$ can be constructed in polynomial time.
\end{proof}

We now extend the previous results to $\REK$.
The idea is that an execution of a keyboard $(T, F)$ of $\REK$, is
an execution of $T$ as an $\RK$ keyboard 
followed by the application of a key from $F$.

\begin{lemma} \label{AutoREK}%
    Let $K = (T,F)$ be a keyboard of $\REK$. We can construct in polynomial time an 
    NFA $\Auto{K}$ recognizing $\langage{K}$.
\end{lemma}

\begin{proof}
Let $K \eqDef (T, F)$ be a keyboard of $\REK$. 
Let $L \eqDef \langage{K}$ and 
$L_T = \langage{T}$ where $T$ is seen as a keyboard of $\RK$.

By \myref{AutoEqSansFleche} and \myref{PolyAutoSansfleche}, 
$L_T$ is rational we can construct in polynomial time an NFA recognizing it. Further, note that 
\[
    L = \bigcup_{f \in F} L_f
\]
where $L_f \eqDef \ensC{w \cdot f}{w \in L_T}$.

Let $f$ be a final key of the form $\retour^k u$. 
Then $L_f = (L_T/A^k)u + u$ where 
\[
    L_T/A^k = \ensC{w}{\exists v \in A^k, wv \in L_T}
\]
is the right quotient of $L_T$ by $A^k$ and is thus rational.

Union, concatenation and quotient by $A^k$ all correspond to polynomial-time operations on NFAs, proving the result.
\end{proof}

A more explicit construction of this automaton is also possible, 
for instance by transforming the automaton associated with $T$ of $\RK$ (see \myref{defautoRK}).
\begin{itemize}
    \item The state $0$ is no longer final, and we add a final state $\Fin$.
    \item For all states $i$ and $t_f = \retour^i w \in F$, we add 
          a transition from $i$ to $\Fin$ labeled by $w$. 
          This simulates the action of, after
          producing $vx$ (with $\size{x} = i$) with $T$, applying $t_f$.
    \item For all $t_f = \retour^r w \in F$, we add a transition
          from $\Init$ to $\Fin$ labeled by $w$.
\end{itemize}

\section{Proofs of properties of \GREK{}}

\subsection*{Proof of \protect\myref{lem-insensibilite-position}}

\lemInsensibilitePosition*

\begin{proof}
    We set $t = \sigma_1 \cdots \sigma_n$.
   Let $\config{x}{y}$ be a configuration. By \myref{prop-touche_aveugle}, we can assume it does not contain any $a$ and by \myref{rem-extension-touche-ecriture}, 
   we can assume $\sigma_1 =a$ and $t$ writes its first symbol from 
    $\config{u}{v}$. We then define for all $1 \leq i \leq n$ 
    \[
        \config{u_i}{v_i} = \config{u}{v} \cdot \sigma_1 \ldots \sigma_i
        \qquad \text{ and } \qquad
        \config{x_i}{y_i} = \config{x}{y} \cdot \sigma_1 \ldots \sigma_i.
    \]
    Let $i$ be the smallest index such that $u_i$ or $x_i$ is empty, and $i = n$
    if such an index does not exist (note that $i > 1$ as $u_1$ and $x_1$
    contain the $a$ written by $\sigma_1$). 
    As $t$ writes its first symbol from $\config{u}{v}$, 
    $\sigma_1 \ldots \sigma_i$ writes an $a$ from $\config{u}{v}$ in some position $j$. 
    
    Further, $\sigma_1 \ldots \sigma_i$ acts efficiently on
    $\config{u}{v}$ and on $\config{x}{y}$, thus by 
    \myref{prop-ecriture_efficiente}, $t$ writes an $a$ from $\config{x}{y}$
    at position $j$. If $i = n$, the result is proven.
    
    Otherwise, as either $u_i$ or $x_i$ is empty, we have $j > 0$ and 
    $y_i[j] = v_i[j] = a$. By \myref{lem-suffix-grek},  $y_n$ contains an $a$, while $\config{x}{y}$ does not by assumption, showing the result by
    \myref{prop-alternative_ecrire_a}.
\end{proof}

\subsection*{Proof of \protect\myref{thm-fundamental-grek}}

\ThmFondamentalGREK*

\begin{proof}
We use an induction on $n$. If $n=0$ then the property is clear.

Let $n > 0$, let $\config{u}{v}$ be a configuration. We set 
$\config{x_{n-1}}{y_{n-1}} = \config{\epsilon}{\epsilon} \cdot \sigma_1 \cdots \sigma_{n-1}$ and $c_n = \config{u}{v} \cdot \sigma_1 \cdots \sigma_n$. 

By induction hypothesis, there exists $u_{n-1}$ and $v_{n-1}$ such that 
\[
    c_{n-1} = \config{u_{n-1}x_{n-1}}{v_{n-1}v}
\]
with $y_{n-1}$ a subword of $v_{n-1}$ and $u_{n-1}$ a prefix of $u$.

We distinguish cases according to $\sigma_n$.
\begin{itemize}
    \item If $\sigma_n = a \in A$ then $x_n = x_{n-1} a$ and $y_n = y_{n-1}$. We set $u_n = u_{n-1}$ and $v_n = v_{n-1}$.
    
    \item If $\sigma_n = \retour$ and $x_{n-1} = \epsilon$, then $x_n = \epsilon$ 
    and $y_n = y_{n-1}$. We set $v_n = v_{n - 1}$ and
    \[
        u_n = \begin{cases}
            u_{n-1}' & \text{if } u_{n-1} \text{ is of the form } u_{n-1}'a \text{ with }a \in A\\
            \epsilon & \text{otherwise}
        \end{cases}.
    \]
    
     \item If $\sigma_n = \retour$ and $x_{n-1} = x_{n-1}'a$ with $a \in A$ then
     $x_n = x_{n - 1}'$ and $y_n = y_{n - 1}$. We set $u_n = u_{n - 1}$ and $v_n = v_{n - 1}$.

    \item If $\sigma_n = \gauche$ and $x_{n-1} = \epsilon$ then          
    $x_n = \epsilon$ and $y_n = y_{n + 1}$. We separate two cases.
    \begin{itemize}
        \item If $u_{n-1} = \epsilon$, we set $u_n = \epsilon$ and $v_n = v_{n-1}$.
        \item If $u_{n-1} = u_{n-1}' a$ with $a \in A$, we set $u_n = u_{n - 1}'$
              and $v_n = a v_{n-1}$.
    \end{itemize}
    
    \item If $\sigma_n = \gauche$ and $x_{n-1} = x_{n-1}' a$ with $a \in A$, then
          $x_n = x_{n - 1}'$ and $y_n = a y_{n - 1}$. We set $u_n = u_{n-1}$ and 
          $v_n = a v_{n-1}$.
\end{itemize}
In all cases, we have $c_n = \config{u_n x_n}{v_n v}$ with $y_n$ a subword of $v_n$ 
and $u_n$ a prefix of $u$.

The theorem is proven.
\end{proof}

\subsection*{Proof of \protect\myref{thm-grek_dans_alg}}

\thmGrekAlg*

\begin{definition}
Let $K = (T,F)$ be a keyboard of $\GREK$. We define the pushdown automaton $\AutoMiroir{K}$ as follows.

\begin{itemize}
    \item Its set of states is $\Pref(T \cup F) \cup \set{\Fin}$.

    \item Its input alphabet is $A$, its stack alphabet is $A \cup \set{\bot}$, $\bot$ being its initial stack symbol.

    \item $\epsilon$ is the only initial state.

    \item $\Fin$ is the only final state. The automaton accepts on an empty stack in state $\Fin$.

    \item The set of transitions is
    $\Delta = \Delta_A \cup \Delta_{\gauche} \cup \Delta_{\retour} \cup \Delta_{\text{loop}} \cup \Delta_{\Fin}$
    with
\begin{align*}
    \Delta_{A} =
        &\set{t \transpile{\epsilon}{\rien}{\depile a} ta \mid ta \in \Pref(T \cup F), a \in A}\\
   \Delta_{\gauche} =
        &\set{t \transpile{a}{\empile a}{\rien} t\gauche \mid t\gauche \in \Pref(T \cup F), a \in A} \cup \\
        &\set{t \transpile{\epsilon}{\empile \bot}{\depile \bot} t\gauche \mid t\gauche \in \Pref(T \cup F)}\\
    \Delta_{\retour} =
        &\set{t \transpile{\epsilon}{\empile a}{\rien} t\retour \mid t\retour \in \Pref(T \cup F), a \in A} \cup \\
        &\set{t \transpile{\epsilon}{\empile \bot}{\depile \bot} t\retour \mid t\retour \in \Pref(T \cup F)}\\
    \Delta_{\text{loop}} =
        &\set{t \transpile{\epsilon}{\rien}{\rien} \epsilon \mid t \in T}\\
    \Delta_{\Fin} =
        &\set{t \transpile{\epsilon}{\rien}{\rien} \Fin \mid t \in F} \cup\\
        &\set{\Fin \transpile{a}{\empile a}{\rien} \Fin \mid a \in A} \cup\\
        &\set{\Fin \transpile{\epsilon}{\empile \bot}{\rien} \Fin}
\end{align*}
\end{itemize}
\end{definition}

\begin{lemma}\label{AutoGREK}%
Let $K = (T,F)$ be a keyboard of $\GREK$, then $\langage{K} = \miroir{\langage{\AutoMiroir{K}}}$.
\end{lemma}

\begin{proof}
Let $t = \sigma_1\cdots \sigma_k \in  T$.

We use the notation $\etatpile{s}{u}$ for the configuration of the automaton $\AutoMiroir{K}$ in state $s$ with $u$ as stack content (from bottom to top).

First we observe that for all configuration $\etatpile{s}{u}$ of $\AutoMiroir{K}$, for all state $s'$ there is at most one transition from $\etatpile{s}{u}$ to a configuration of the form $\etatpile{s'}{u'}$. 

We also observe that the only accepting runs from $\epsilon$ to $\Fin$ are of the form:

\[\epsilon \act{t_1} t_1 \transpile{\epsilon}{\rien}{ \rien} \epsilon \act{t_2} t_2 \transpile{\epsilon}{\rien}{ \rien} \epsilon \cdots t_k \transpile{\epsilon}{\rien}{ \rien} \epsilon \act{t'} t_f \transpile{\epsilon}{\rien}{ \rien} \Fin \act{read} \Fin\]

with $t_1,\ldots, t_k \in T, t_f \in F$, and where $\act{t}$ stands for the run of the form

\[\epsilon \transpile{x_1}{op_1}{op'_1} \sigma_1 \transpile{x_2}{op_2}{op'_2} \sigma_1 \sigma_2 \cdots \transpile{x_k}{op_k}{op'_k} \sigma_1 \cdots \sigma_k\]

and $\Fin \act{read} \Fin$ stands for the run 

\[\Fin \transpile{a_1}{\rien}{\depile a_1} \Fin \transpile{a_2}{\rien}{\depile a_2} \cdots \transpile{a_m}{\rien}{\depile a_m} \Fin \transpile{\epsilon}{\rien}{\depile \bot}.\]

A straightforward case distinction on $\sigma_i$ shows that for all $1 \leq i \leq k$, for all $u,v,u' \in A^*$, $x \in A \cup \set{\epsilon}$ there is a transition reading $x$ from $\etatpile{(\sigma_1\cdots \sigma_{i-1})}{\bot u}$ to  $\etatpile{(\sigma_1\cdots \sigma_{i})}{\bot u'}$ if and only if $\config{u}{v}\cdot \sigma_i = \config{u'}{xv}$.

We easily infer that for all $t_1,\ldots,t_k \in T, t_f \in F$, the run $\epsilon \act{t_1} t_1 \transpile{\epsilon}{\rien}{ \rien} \epsilon \act{t_2} t_2 \transpile{\epsilon}{\rien}{ \rien} \epsilon \cdots t_k \transpile{\epsilon}{\rien}{ \rien} \epsilon \act{t'} t_f \transpile{\epsilon}{\rien}{ \rien} \Fin$ reads $\miroir{v}$ and ends in configuration $\etatpile{\Fin}{\bot u}$, with $\config{u}{v} = \config{\epsilon}{\epsilon} \cdot (t_1 \cdots t_k t_f)$.

As a result, a word $w$ is accepted if and only if there exist $u,v \in A^*$ and $t_1, \ldots t_n \in T, t_f \in F$ such that $w = \miroir{v}\miroir{u}$ and $\config{u}{v} = \config{\epsilon}{\epsilon} \cdot t_1\cdots t_n t_f$.

In other words, a word $w$ is accepted by $\AutoMiroir{K}$ if and only if its mirror is in $\langage{K}$.

We have shown $\langage{\AutoMiroir{K}} = \miroir{\langage{K}}$.
\end{proof}

Given a pushdown automaton, one can construct in polynomial time a pushdown automaton recognizing its mirror language. Thus, as $\AutoMiroir{K}$ can clearly be constructed in polynomial time, we obtain the result.

Note that we use $\epsilon$-transitions in the constructions, however these can be eliminated in polynomial time as well.

\section{Proofs of \FEK{} results}

\subsection*{Proof of \protect\myref{lem-croissance-fek}}

\lemCroissanceFEK*

 \begin{proof}
    We simply have to prove the lemma for $\tau = \sigma \in S$ an atomic operation, and the general case follows by a straightforward induction. 
    
    If $\sigma \in \set{\gauche, \droite}$ then $u'v' = uv$ and thus $uv$ is clearly a subword of $u'v'$.
    
    If $\sigma = a \in A$ then $u'v' = uav$ and again $uv$ is clearly a subword of $u'v'$.
    
    This concludes our proof.
\end{proof}

\subsection*{Proof of \protect\myref{lem-borne-touches-fek}}

\lemBorneTouchesFEK*

 \begin{proof}
    Let  $w \in \langage{K}$, and let $u,v \in A^*$, $\tau = t_1 \cdots t_n \in T^*F$ be such that $w=uv$ and $\config{\epsilon}{\epsilon} \cdot \tau = \config{u}{v}$.
    
    We show that if $n > \size{w}^2+1$ then there exists a shorter execution writing $\config{u}{v}$. 
    
    Suppose $n > \size{w}^2+1$, and for all $0 \leq i \leq n$ let $k_i = \size{\config{\epsilon}{\epsilon} \cdot t_1 \cdots t_i}$ and $\config{u_i}{v_i} = \config{\epsilon}{\epsilon} \cdot t_1 \cdots t_i$. By \myref{lem-croissance-fek}, the sequence $(k_i)$ is nondecreasing. As $n > \size{w}^2+1$ and $k_n = \size{w}$, there exists $0 \leq i \leq n-\size{w}$ such that $k_i = k_{i+1} = \cdots = k_{i+\size{w}}$. Again by \myref{lem-croissance-fek}, we have $u_iv_i = u_{i+1}v_{i+1} = \cdots = u_{i+\size{w}}v_{i+\size{w}}$. If $u_iv_i = w$ then the execution $t_1 \cdots t_i$ writes $w$.
    
    If $u_i v_i \neq w$ then $k_i < \size{w}$. There are $\size{w} > k_i$ configurations $\config{u'}{v'}$ such that $u'v' = u_iv_i$, thus there exist $i \leq j_1 < j_2 \leq i+\size{w}$ such that $\config{u_{j_1}}{v_{j_1}}= \config{u_{j_2}}{v_{j_2}}$.
    
    As a result, the execution $t_1 \cdots t_{j_1} t_{j_2+1} \ldots t_n$ writes $w$.
    
    The lemma is proven.
    \end{proof}

\subsection*{Proof of \protect\myref{thm-fek_contextuel}}

\thmFEKContextuel*

\begin{proof}
    We construct $\Auto{K}$ the linear bounded automaton which, given an input $w$, proceeds as follows:
    
    \begin{itemize}
        \item It divides the tape in three parts: one to memorize the input (of linear size), one to simulate an execution of $K$ (of linear size as well by \myref{lem-croissance-fek}) and one containing a counter (of logarithmic size).
        
        \item It guesses a sequence of keys of $T$ followed by a key of $F$ and computes their effect on the fly. After the application of each key the counter is incremented. 
        
        \item If the counter goes beyond $\size{w}^2+1$ then the automaton rejects.
        
        \item If not, the automaton compares the obtained word to $w$, accepts if they are equal, and rejects otherwise.
    \end{itemize}
    
    This machine guesses a sequence of at most $\size{w}^2+1$ keys and accepts if the word obtained by their actions is the input.
    
    Clearly if a word is accepted by $\Auto{K}$ then it is in $\langage{K}$. 
    Conversely, if a word $w$ is in $\langage{K}$ then by \myref{lem-borne-touches-fek} there exists an execution of length at most $\size{w}^2+1$ accepting it, thus $\Auto{K}$ can  guess this execution and accept $w$.
    
    As a result, $\langage{K} = \langage{\Auto{K}}$.
\end{proof}

\section{Comparisons between classes}

\eknotingrk*

\begin{proof}
Consider the language $L = (a^2)^*(b+b^2)$, which is clearly in $\EK$ via the keyboard $(\set{a^2}, \set{b,b^2})$. We show that there is no $\GRK$ keyboard recognizing it.

Suppose there exists a keyboard $K$ of $\GRK$ recognizing $L$.

As $b^2 \in L$, some sequence of keys $\tau_{b^2} \in K^+$ leads to some $\config{x}{y}$ with $xy = b^2$ from $\config{\epsilon}{\epsilon}$.

We first prove that for all $\tau \in K^*$, $\config{\epsilon}{\epsilon} \cdot \tau$ is of the form $\config{u}{\epsilon}$.

Let $\tau \in K^*$, let $\config{u}{v} = \config{\epsilon}{\epsilon}\cdot \tau$. By \myref{thm-fundamental-grek}, $\config{\epsilon}{\epsilon}\cdot \tau\tau_{b^2}$ is of the form $\config{u'x}{v'v}$ with $y$ a subword of $v'$. As $uv \in L$, $uv$ contains a $b$. If $v \neq \epsilon$, $v$ contains a $b$ and thus $u'xv'v$ contains at least three $b$'s, contradicting $u'xv'v \in L$.

Therefore, the right component of every configuration obtained with $K$ is empty. 

We infer that left arrows $\gauche$ are only ever applied on the configuration $\config{\epsilon}{\epsilon}$, as otherwise we would obtain a nonempty right component which would then stay nonempty throughout the execution by \myref{thm-fundamental-grek}, contradicting the previous statement.

We can thus delete every $\gauche$ from the keys of $K$ to obtain an equivalent keyboard of $\RK$.

In what follows we will assume $K$ to be a $\RK$ keyboard.

By \myref{lem-forme_normale_touche_rek}, we can assume it to be in normal form.

As $b^2 \in L$, there exists $\tau \in K^*$ such that
$\epsilon \cdot \tau = b^2$. Again by \myref{lem-forme_normale_touche_rek} we have a sequence of atomic operations $\retour^kb^2$ avec $k \in \N$ equivalent to $\tau$. 

We distinguish cases depending on $k$.
\begin{itemize}
    \item If $k = 0$, then $\tau \sim b^2$ ; we then have
          $\epsilon \cdot \tau \cdot \tau = b^2 \cdot \tau = b^4 \notin L$.
    \item If $k = 1$, then $\tau \sim \retour b^2$ ; then
          $\epsilon \cdot \tau \cdot \tau = b^2 \cdot \tau = b^3 \notin L$.
    \item If $k > 1$ and $\mathbf{k}$ \textbf{is even} ; as $a^{2k}b \in L$, $a^{2k}b$ is a reachable configuration for $K$. Then
    we have $a^{2k}b \cdot \tau = a^{k + 1}b^2 \notin L$.
    \item If $k > 1$ and $\mathbf{k}$ \textbf{is odd} ; as $a^{2k}b^2 \in L$, $a^{2k}b$ is a reachable configuration for $K$. Then we have
          $a^{2k}b^2 \cdot \tau = a^{k}b^2 \notin L$.
\end{itemize}
In all cases, $K$ recognizes a word outside of $L$, yielding a contradiction.
\end{proof}

\eknotinfk*

\begin{proof}
Consider the language $\set{a}$. It is recognized by the keyboard $(\emptyset, \set{a})$ of $\EK$.

Suppose there exists $K$ a keyboard of $\FK$ recognizing $\set{a}$. Then there exists $t \in K$ exactly containing an $a$. As $K$ does not contain any $\retour$, applying $t$ twice from $\config{\epsilon}{\epsilon}$ yields a configuration with two $a$, and thus a word outside of $\set{a}$. We obtain a contradiction.
\end{proof}

\subsection*{Proof of \myref{lem-rknotinfek}}

\label{proof-rk_not_in_fek}

In all that follows we will use the notations $t_a = \retour a \diese \dollar$ and $t_b = \retour \retour b \diese\dollar\dollar$. 

We define $L_{\diese\dollar}$ as the language recognized by the $\RK$ keyboard $\set{t_a,t_b}$.

\begin{lemma}
\label{lem-forme-lc}
Let $x = x_1 \cdots x_n \in \set{a,b}^+$, we have

\[ \config{\epsilon}{\epsilon}\cdot t_{x_1}\cdots t_{x_n} = x_1 w_{x_1x_2} x_2 w_{x_2x_3} x_3 \cdots w_{x_{n-1}x_n} x_n v_{x_n}\]

with $w_{aa} = w_{bb} = \diese$, $w_{ab} = \epsilon$, $w_{ba} = \diese\dollar$, $v_a = \diese\dollar$ and $v_b = \diese\dollar\dollar$.

In particular, for all $u_1, u_2, u_3 \in A^*$, if $u_1b u_2 a u_3 \in L_{\diese\dollar}$ then $u_2$ contains a $\dollar$. 
\end{lemma}

\begin{proof}
We proceed by induction on $n$. 
For $n = 1$ the property is clear.

Let $n > 1$, let $x = x_1 \cdots x_n \in \set{a,b}^+$, we define $x' = x_1 \cdots x_{n-1}$. Suppose $\config{\epsilon}{\epsilon}\cdot t_{x_1}\cdots t_{x_{n-1}} = x_1 w_{x_1x_2}  \cdots x_{n-1} v_{x_{n-1}}$, we separate four cases:

\begin{itemize}
    \item $x_{n-1} = x_n =a$: then $v_{x_{n-1}} = \diese\dollar$ and thus 
    \[
        \config{\epsilon}{\epsilon}\cdot t_{x_1}\cdots t_{x_n} = x_1 w_{x_1x_2}  \cdots x_{n-1} \diese a \diese\dollar.
    \]
    
    \item $x_{n-1} = x_n =b$: then $v_{x_{n-1}} = \diese\dollar\dollar$ and thus 
    \[
        \config{\epsilon}{\epsilon}\cdot t_{x_1}\cdots t_{x_n} = x_1 w_{x_1x_2}  \cdots x_{n-1} \diese b \diese\dollar\dollar.
    \]
    
    \item $x_{n-1} =a$ and $x_n =b$: then $v_{x_{n-1}} = \diese\dollar$ and thus 
    \[
        \config{\epsilon}{\epsilon}\cdot t_{x_1}\cdots t_{x_n} = x_1 w_{x_1x_2}  \cdots x_{n-1} b \diese\dollar\dollar.
    \]
    
    \item $x_{n-1} =b$ and $x_n =a$: then $v_{x_{n-1}} = \diese\dollar\dollar$ and thus 
    \[
        \config{\epsilon}{\epsilon}\cdot t_{x_1}\cdots t_{x_n} = x_1 w_{x_1x_2}  \cdots x_{n-1} \diese \dollar a \diese\dollar.
    \]
\end{itemize}

The property is verified in all cases.
\end{proof}

\rknotinfek*

\begin{proof}
Suppose there exists a keyboard $K = (T,F)$ of $\FEK$ recognizing $L = L_{\diese\dollar}$.
In this proof we will use the letter $k$ to denote keys of $K$,
in order to avoid confusion with the keys $t_a$ and $t_b$.

As $a \diese \dollar \in L$, there exists $\tau_f \in T^*F$ yielding that word.

Moreover, $w = ({a\diese})^{3\normeInf{K}}a\diese \dollar \in L$ (obtained by applying $t_a$
$3\normeInf{K}$ times). 
Hence there exists an execution of $K$ 
writing that word. By \myref{lem-croissance-fek}, as $\size{k}_a \leq \normeInf{K}$ 
for all $k \in T \cup F$, this execution contains at least three keys
containing an $a$, including at least two in $T$ (as the execution only has one final key).

As $w$ contains a single $\dollar$, there exists a key $k_a$ of $T$ writing an $a$, but writing neither $\dollar$ nor $b$. By applying $k_a$ then
$\tau_f$, we obtain a word of $L$ containing a single $\dollar$ and no $b$. By lemma \myref{lem-forme-lc} this word must be of the form $(a \diese)^{i} a \diese \dollar$.
We infer that $k_a$ contains as many $a$ and $\diese$. Let
$n$ be its number of $a$.

We can similarly show, using the word $({b\diese})^{3\normeInf{K}}b\diese \dollar \dollar$,
that there exists $k_b \in T$ containing as many $b$ and $\diese$ but neither
$\dollar$ nor $a$. Let $m$ be its number of $b$ (and of $\diese$).

Let $\tau = k_a k_b$. We write 
\[
    \config{u}{v} = \config{\epsilon}{\epsilon} \cdot \tau
    \quad \text{ and } \quad
    \config{u'}{v'} = \config{u}{v} \cdot \tau_f
\]
We have that $u'v'$ contains a single $\dollar$. As $u'v' \in L$, $u'v'$ has to be of the form $wa\diese\dollar$ with 
$w$ not containing any $\dollar$,
$\size{w}_a = n$, $\size{w}_b = m$ and $\size{w}_{\diese} = n + m$.

As $wa$ contains at least a $b$, $wa$ it is of the form $u_1bu_2a$, thus by 
\myref{lem-forme-lc}, $u_2$ contains a $\dollar$, yielding a contradiction. As a conclusion, $L \not\in \FEK$. 
\end{proof}

\subsection*{Proof of \myref{lem-fknotingrek}}

Consider the keyboard $K = \set{\touche{a \gauche^2\droite b}}$. 

    By applying $a \gauche^2\droite b$ on $\config{\epsilon}{\epsilon}$ we obtain 
    $\config{ab}{\epsilon}$, and by applying it on a configuration of the form $\config{ub}{v}$ 
    we obtain $\config{ubb}{av}$. Hence after applying it $n$ times on 
    $\config{\epsilon}{\epsilon}$ we get $ab^{n+1}a^n$. The language of $K$ is therefore $L = \ensC{ab^{n+1}a^n}{n \in \N}$.

\begin{lemma} \label{lem-ak}%
    Let $K$ be a keyboard of $\GREK{}$. If $K$ recognizes $L$ then for all $\tau \in T^*$,
    $\config{\epsilon}{\epsilon} \cdot \tau$ is of the form $\config{u}{a^k}$. 
\end{lemma}

\begin{proof}
As $abba \in L$, there exists $\tau \in T^*$ and $t_f \in F$ such that
$\config{\epsilon}{\epsilon} \cdot \tau t_f = \config{x}{y}$ with $xy = abba$.

Let $\config{u}{v}$ be a configuration reachable by a sequence of keys of $\tau$ with $v$ containing a $b$. By  \myref{thm-fundamental-grek},
$\config{u}{v} \cdot \tau t_f$ is of the form $\config{u'x}{v'v}$ with $y$ a subword of $v'$.
As a consequence, $abba$ is a subword of $u'xv'$. With the assumption, we get that $abbab$ is a subword of $u'xv'v$, which contradicts the fact that $u'xv'v \in L$. 
\end{proof}

\fknotingrek*

\begin{proof}
    Let $K = (T, F)$ be a keyboard of $\GREK$, suppose it recognizes $L$.
    Let $\tau \in T^*$ and $t_f \in F$. 
    We set: 
    \[
        \config{x}{y} = \config{\epsilon}{\epsilon} \cdot t_f 
        \quad \text{ and } \quad
        \config{u}{v} = \config{\epsilon}{\epsilon} \cdot \tau.
    \]
    There exists $n \in N$ such that $xy = ab^{n + 1}a^n$.
    By \myref{thm-fundamental-grek}, $\config{\epsilon}{\epsilon} \cdot \tau t_f = \config{u}{v}\cdot t_f$
    is of the form $\config{u'x}{v'v}$ with $y$ a subword of $v'$.
    
    As $ab^{n + 1}a^n$ is a subword of $xv'$ and $u'xv'v \in L$,
    $u'$ is necessarily empty (otherwise $u'$ would contain an $a$ and as $ab$ is a subword of $xv'$, $aab$ would be a subword of $u'xv'v$, contradicting $u'xv'v \in L$). 
    
    By \myref{lem-encadrement_taille}, 
    $\size{v'v} - \size{v} \leq \size{t} \leq \normeInf{K}$,
    i.e., $\size{v'} \leq \normeInf{K}$. 
    Furthermore, again by 
    \myref{lem-encadrement_taille}, $\size{x} \leq \normeInf{K}$.
    
    By \myref{lem-ak}, $v$ is of the form $a^k$. 
    Hence all $b$ in $u'xv'v$ are in $xv'$,  
    thus $u'xv'v$ contains at most $2\normeInf{K}$ $b$.
    
    We have shown that all words in $\langage{K}$ contain at most $2 \normeInf{K}$ $b$, contradicting $\langage{K}= L$ as the number of $b$ of words of $L$ is unbounded.
\end{proof}

\subsection*{Proof of \protect\myref{lem-geknotinfrk}}

Consider the following language over $A = \set{a, b, c}$:

$L = L_1 \cup L_2$ with 
$L_1 = \ensC{w c \miroir{w}}{w \in \set{a,b}^*}$ and
$L_2 = \ensC{w cc \miroir{w}}{w \in \set{a,b}^*}$.

We are going to prove that $L$ is in $\GEK$ but not in $\FRK$.

\begin{lemma} \label{LinGEK}%
    Let $K = (\set{\touche{aa}\gauche, \touche{bb}\gauche}, \set{\touche{c}, \touche{cc}})$. 
    Then $\langage{K} = L$. 
\end{lemma}

\begin{proof}
    We use the notations $t_a = \touche{aa\gauche}$ and $t_b = \touche{bb\gauche}$. A straightforward induction
    shows that $\langage{\set{t_a, t_b}}$ is the language of even palindromes over $\set{a,b}$ and 
    that the configurations reached by application of those keys are of the form $\config{w}{\miroir{w}}$. 
    
    We then easily infer that the language of $K$ is $L$.
\end{proof}

We now show that  $L \notin \FRK$. The intuition is as follows:

Suppose we have a keyboard $K$ of $\FRK$ recognizing $L$. As we want to recognize palindromes, we need to stay close to the center in order to always modify both halves of the word.

If the word is large enough, this means that we have to get far from the edges. In particular, there is a key $t$ writing an $a$ far from the edges (even two $a$, as we have to stay in the language).

We study the behaviour of this key on $b^ncb^n$
and $b^nccb^n$ with large $n$. The cursor being far from the edges, $t$ behaves the same way in both cases. In particular, we make the following remarks:
\begin{itemize}
    \item The maximal distance $d$ between two $a$ will be the same in both words.
    Thus the resulting words have either both two $c$ or both one $c$ in the center.
    
    \item In both cases the key adds a number of letters $\delta$,
    and thus if $b^ncb^n$ is turned into a word of even length, then $b^nccb^n$ is turned into a word of odd length, and vice-versa.
          
    As a result, they have different numbers of $c$ in the center.
\end{itemize}            

As those facts are contradictory, we conclude that there cannot exist such a keyboard.

Now for the formal proof, we start by showing that the maximal distance between two $a$ is the same when we apply the same key to two configurations without $a$ effectively.

\begin{definition}
Let $\config{u}{v}$ be a configuration, let $a\in A$. We define $\distance_a(\config{u}{v})$ as 

$\begin{cases}
    & +\infty \text{ if } \config{u}{v} \text{ contains zero or one a.}\\
    & \max{\set{\size{w} \mid awa \in \Fact{uv}}} \text{ otherwise.}
\end{cases}$

\end{definition}

\begin{lemma} \label{lem-distance_des_a}%
    Let $t \in T$ be a key and $u, v, u', v' \in A^*$ words with no $a$ and of length at least $\size{t}$. Then
    \[
        \distance_a\bigl(\config{u}{v} \cdot t\bigr) =
        \distance_a\bigl(\config{u'}{v'} \cdot t \bigr).
    \]
\end{lemma}
\begin{proof}
    As $u, v, u', v'$ are all of length at least $\size{t}$, $t$ acts effectively on both configurations (by \myref{lem-efficience_loin_bords}). By \myref{prop-ecriture_efficiente}, $t$ writes an $a$ in position $i$ 
    from $\config{u'}{v'}$ if and only if $t$ writes an $a$ in position $i$ 
    from $\config{u}{v}$.
    
    As $\config{u}{v}$ does not contain an $a$, by \myref{prop-ecriture}, 
    $(\config{u}{v} \cdot t)_i = a$ if and only if $t$ writes an $a$ in position $i$
    from $\config{u}{v}$. Similarly, $(\config{u'}{v'} \cdot t)_i = a$ if and only if
    $t$ writes an $a$ in position $i$ from $\config{u'}{v'}$.
    
    Thus $(\config{u}{v} \cdot t)_i = a$ if and only if  $(\config{u'}{v'} \cdot t)_i = a$,
    proving the result.
\end{proof}

We can now prove our result. Suppose
we have $K$ an \FRK{} keyboard recognizing $L$.

We first show that in an execution of $K$ the cursor must stay close to the center of the word.

\begin{lemma}\label{resteaumilieu}%
Let $\config{u}{v}$ a reachable configuration of $K$ and $t \in K$ such that, by setting 
$\config{u'}{v'} = \config{u}{v} \cdot t$, we have $uv \neq u'v'$.
Then $\abs{\size{u}-\size{v}\enlarger} \leq 6\normeInf{K}$
and $\abs{\size{u'}-\size{v'} \enlarger} \leq 8\normeInf{K}$.
\end{lemma}

\begin{proof}
Suppose $\abs{\size{u}-\size{v}\enlarger} > 6\normeInf{K} \geq 6\size{t}$. 
Then, either $\size{u}-\size{v} > 6\size{t}$, or $\size{v} - \size{u} > 6\size{t}$.
We assume to be in the first case, the other one being similar. We thus have
\[
    \size{u} = \frac{\size{u}}{2} + \frac{\size{u}}{2} > 
              \frac{\size{u}}{2} + \frac{\size{v} + 6\size{t}}{2} 
\]
Hence $\size{u} \geq \frac{\size{uv}}{2} + 3\size{t}$.

As $uv \in L$, $uv$ is of the form either $wc\miroir{w}$ or $wcc\miroir{w}$ 
for some $w \in \set{a,b}^*$. 

By \myref{lem-localite}, there exists a common prefix $u_p$ to $u$ and 
$u'$ such that 
\[
    \size{u_p} \geq \size{u} - \size{t} 
               \geq \frac{\size{uv}}{2} + 2\size{t} 
               \geq \frac{\size{uv}}{2} + 2,
\]
The last inequality holds as $t$ is non-empty (as $uv' \neq uv$). 

As a consequence $u_p$ is of the form either $wcy$ (and $uv = wc\miroir{w}$) or $wccy$ (and $uv = wcc\miroir{w}$) 
with $y \in \set{a,b}^+$. As $u'v'$ is in $L$, in both cases
$uv=u'v'$, contradicting our hypothesis.

As a conclusion, we have $\abs{\size{u}-\size{v}\enlarger} \leq 6 \normeInf{K}$.

By \myref{lem-encadrement_taille}, we have 
$\abs{\size{u} - \size{u'}\enlarger} \leq \size{t}$ and 
$\abs{\size{v} - \size{v'}\enlarger} \leq \size{t}$. Consequently,
\[
    \abs{\size{u'}-\size{v'} \enlarger} 
    \leq \abs{\size{u}-\size{v} \enlarger} + \abs{\size{u} - \size{u'} \enlarger} 
                                           + \abs{\size{v} - \size{v'} \enlarger} 
    \leq 6\normeInf{K} + 2 \size{t} \leq 8 \normeInf{K}.
\]
\end{proof}

\begin{lemma}\label{existetouchea}%
If $K \in \FRK$ recognizes $L$, then $K$ contains a key ensuring an $a$ far from the edges.
\end{lemma}

\begin{proof}
There exists $t_1 \cdots t_n \in T^*$ such that $\config{\epsilon}{\epsilon} \cdot t_1 \cdots t_n = \config{u}{v}$ with $uv = a^{6\normeInf{K}}cca^{6\normeInf{K}}$. Let $i$ be the smallest index such that the number of $a$ in $\config{u_i}{v_i} = \config{\epsilon}{\epsilon} \cdot t_1 \cdots t_i$ is maximized. We set 
\[
    \config{u_{i-1}}{v_{i-1}} = \config{\epsilon}{\epsilon} \cdot t_1 \cdots t_{i-1}. 
\]
In particular $\config{u_i}{v_i}$ contains at least $12\normeInf{K}$ $a$.
By \myref{lem-encadrement_taille}, we have
\begin{equation}
    \label{eq:name_todo}
    \size{u_{i-1}v_{i-1}} \geq \size{u_iv_i} - \size{t_i} 
                          \geq 12\normeInf{K} - \normeInf{K}
                          \geq 11 \normeInf{K}.
\end{equation}

Further, by \myref{resteaumilieu}, $\abs{\size{u_{i-1}} - \size{v_{i-1}} \enlarger} \leq 6 \normeInf{K}$, thus by triangle inequality,
\[
    \size{u_{i-1}v_{i-1}} \leq  \size{u_{i-1}} + \bigl(\size{u_{i-1}} + 6 \normeInf{K}\bigr).
\]
Suppose, $\size{u_{i-1}} < \normeInf{K}$, we then have
$\size{u_{i-1}v_{i-1}} \leq 8 \normeInf{K}$, contradicting 
\eqref{eq:name_todo}. Thus
$\size{u_{i-1}} \geq \normeInf{K}$.

We prove similarly that $\size{v_{i-1}} \geq \normeInf{K}$. 
As a consequence, $u_{i-1}$ and $v_{i-1}$ 
are of length at least $\normeInf{K}$, and by \myref{cor-assure_a_loin_bord} $t_i$ ensures an $a$ far from the edges, as
by minimality of $i$, $\config{u_{i-1}}{v_{i-1}}$ contains less $a$
than $\config{u_i}{v_i}$.
\end{proof}

\begin{lemma} \label{LnotinFRK} %
The language $L$ is not recognized by a $\FRK$ keyboard.
\end{lemma}

\begin{proof}
    Suppose there exists $K$ a $\FRK$ keyboard recognizing $L$.
    Then, there exist $\tau$ and $\tau'$ of minimal length leading respectively to configurations $\config{u}{v}$ and $\config{u'}{v'}$
    with $uv = b^{5\normeInf{K}} c b^{5\normeInf{K}}$
    and $u'v' = b^{5\normeInf{K}} cc b^{5\normeInf{K}}$.
    
    Let $\config{x}{y}$ be the penultimate configuration in the execution of $\tau$. By minimality of $\tau$, $xy \neq uv$ and by \myref{resteaumilieu} we then have $\abs{\size{u} - \size{v} \enlarger} \leq 8\normeInf{K}$. 

    We show similarly that $\abs{\size{u'} - \size{v'} \enlarger} \leq 8 \normeInf{K}$.  
    
    In particular, as $\size{uv}, \size{u'v'} > 10 \normeInf{K}$ we have $\size{u}, \size{v}, \size{u'}, \size{v'} \geq \normeInf{K}$.
    
    By \myref{existetouchea}, there exists $t \in T$ such that $t$ ensures an $a$ far from the edges.
    We then set
    \[
        \config{u_a}{v_a} = \config{\epsilon}{\epsilon}\cdot \tau t 
        \quad \text{and} \quad
        \config{u'_a}{v'_a} = \config{\epsilon}{\epsilon} \cdot \tau' t.
    \] 
    By \myref{lem-distance_des_a}, we have 
    $\distance_a(u_av_a) = \distance_a(u_a' v_a')$. We set 
    \[
        d = \distance_a(u_av_a)
        \quad \text{and} \quad
        \delta = \sum_{x \in A} \size{t}_x - \size{t}_{\retour}.
    \]
    We have $\size{u} \geq \size{t}$, $\size{v} \geq \size{t}$ and $t$ ensures an
    $a$ far from the edges, hence $u_av_a$ contains at least an $a$ and thus at least two as $u_av_a \in L$. We therefore have $d \in \Nat$.
    
    Further, as $u_av_a$ and $u_a'v_a'$ are in $L$, then if $d$ is even
    they must both be in $L_2$ and otherwise both in
    $L_1$.
    
    However, if $\delta$ is even then by \myref{lem-egalite_taille_loin_bords},
    \[
    \begin{cases}
        \size{u_av_a} = \size{uv} +\delta = 10\normeInf{K} + 1 + \delta 
                            & \text{is odd}\\
        \size{u_a'v_a'} = \size{u'v'} +\delta = 10\normeInf{K} + 2 + \delta 
                            & \text{is even}
    \end{cases}
    \]
    thus $u_av_a \in L_1$ and $u_a'v_a' \in L_2$,
    contradicting the fact that they must be in the same $L_j$.

    Similarly, if $\delta$ is odd then $u_av_a \in L_2$ and
    $u_a'v_a' \in L_1$, again contradicting the fact that they must be in the same $L_j$.

    We obtained a contradiction. As a result, $L$ is not recognized by an $\FRK$ keyboard.
\end{proof}

\geknotinfrk*

\begin{proof}
We have shown in \myref{LinGEK} that $L$ is in $\GEK$ and in \myref{LnotinFRK} that it is not in $\FRK$, hence the result.
\end{proof}

\section{Complexity}

\subsection*{Proof of \protect\myref{prop-ek_complexity}}

\propEKComplexity*

    \begin{proof}
        The complexities of the membership problem on $\MK$ and the universality problem on $\EK$ arise directly from the complexities of the membership and universality problems on rational expressions.
        
        Let $K \subset A^*$ be a minimal keyboard. Since $\langage{K} = K^+$,
        $K$ is universal if and only if $K^+ = A^*$. Let us show that 
        $K^+= A^*$ if and only if $A \cup \set{\epsilon} \subseteq K$.
        \begin{itemize}
            \item If $A \cup \set{\epsilon} \subseteq K$, then $A^* = (A \cup \set{\epsilon})^+ \subset K^+$.
            \item If $A \cup \set{\epsilon} \not\subseteq K$, then either $\epsilon \notin K$ (and $\epsilon \notin K^+$)
                  or there exists $a \in A$ such that $a \notin K$ (and $a \notin K^+$). In both cases, $A^* \neq K^+$.
        \end{itemize}
        To check the universality of a minimal keyboard, we just have to check 
        (in polynomial time) if $A \cup \set{\epsilon} \subset K$.
    \end{proof}

\subsection*{Proof of \myref{prop-rk-universality}}

\propRKUniversality*

\begin{proof}
We prove that if a $\RK$ keyboard $K$ is not universal then there is a word of length at most $\normeInf{K} +1$ it does not recognize.

Let $K$ be a keyboard of $\RK$, suppose there exists $w \in A^*$ not recognized by $K$. We take $w$ of minimal length. If $\size{w}\leq  \normeInf{K}+1$ then the property holds. 

If $\size{w}> \normeInf{K}+1$ then there exist $a \in A, v \in A^+$ such that $av = w$. As we assumed $w$ to be of minimal length, $v$ is recognized by $K$. By \myref{oneWastefulKey}, there exist $t \in K, \tau \in K^*$ such that $(\epsilon \cdot t) \actEff{\tau} v$. Let $u = \epsilon \cdot t$, $k \in \nats$ be such that $t = \retour^k u$. 

As $\size{a^{k+1}} = k+1 \leq \normeInf{K} +1 < \size{w}$, $a^{k+1}$ is recognized by $K$. Let $\tau'$ be such that $\epsilon \cdot \tau' = a^{k+1}$, then we have $\epsilon \act{\tau'} a^{k+1} \act{t} au \actEff{\tau} av =w$.

This contradicts the fact that $w$ is not recognized by $K$. The property is proven.

\end{proof}

\section{Closure properties}

\subsection*{Proof of \protect\myref{lem-mirrors}}

\mirrors*

\begin{lemma} \label{lem-fek_stable_miroir}%
    $\MK$, $\FEK$ and $\FK$ are stable under mirror.
\end{lemma}

\begin{proof}
For $\MK$ we simply observe that given $K$ an $\MK$ keyboard, we have $(\miroir{K})^+ = \miroir{K^+}$. As a consequence, the $\MK$ keyboard $\miroir{K}$ recognizes the mirror of $\langage{K}$.

For $\FEK$ and $\FK$ we define $\overline{\cdot}$ the word morphism generated by
\begin{align*}
    \overline{\droite} = \gauche \qquad
    \overline{\gauche} = \droite \qquad
    \overline{a} = a \gauche  \text{ if } a \in A
\end{align*}

If $\config{u}{v}$ is a configuration, we write $\miroir{\config{u}{v}} = \config{\miroir{v}}{\miroir{u}}$. We can show by a straightforward induction on $n$ that for all sequence of elementary operations $\sigma_1 \ldots \sigma_n$, 
\[
c_0 \act{\sigma_1...\sigma_n} c_n \iff 
\miroir{c_0} \act{\overline{\sigma_1}...\overline{\sigma_n}} \miroir{c_n}.
\]

Let $L$ be a language of $\FEK$ and $K = (T, F)$ a keyboard recognizing it. 
Let $\overline{K} = (\overline{T}, \overline{F})$ where the morphism is applied on all keys. For all $w \in A^*$, the execution
$\config{\epsilon}{\epsilon} \act{t_1\ldots t_n} w$ can be reversed into an execution
$\config{\epsilon}{\epsilon} \act{\overline{t_1} \ldots \overline{t_n}} \miroir{w}$ 
of $\overline{K}$ and vice versa. 
As a result, $w \in L$ if and only if $\miroir{w} \in \langage{\overline{K}}$,
where $\langage{\overline{K}} = \miroir{L}$. 
Further, 
$\overline{K}$ is an $\FEK$ keyboard, and is automatic if and only if $K$ is automatic as well, showing the result for both $\FK$ and $\FEK$.
\end{proof}

\begin{lemma}
    $\REK$, $\RK$ and $\EK$ are not stable under mirror.
\end{lemma}

\begin{proof}
    Let $L = b^*a$, we have $\miroir{L} = ab^*$. We have $b^*a \in \EK$
    (with the keyboard $(\set{\touche{b}}, \set{\touche{a}})$) and $b^*a \in \RK$,
    with the keyboard $\set{\touche{\retour a}, \touche{\retour b a}}$.
    
    We now show that $\miroir{L} \not \in \REK$. Suppose there exists $K = (T, F)$ a $\REK$ keyboard recognizing $\miroir{L}$. Since $w = ab^{\normeInf{K}} \in \miroir{L}$, there exists $\tau \in T^*, t_f \in F$ yielding $w$. In particular, $t_f$ only wrote some $b$ and is thus equivalent to $\retour^k b^n$ for some $k,n \in \nats$.
    
    By applying $t_f$ from $\epsilon$, we obtain $b^n$, which is impossible as $b^n \notin \miroir{L}$.
    As a result, $\miroir{L} \not \in \REK$.
    
    We then have $L$ in $\EK$ and $\RK$, and $\miroir{L} \not\in \REK$,
    proving the lemma.
\end{proof}

\begin{lemma}
    $\GRK$ is not stable under mirror.
\end{lemma}

\begin{proof}
    Let $L = a^*(b + b^2)$. We showed in the proof of \myref{lem-eknotingrk} that
    $L \not\in \GRK$. We then simply observe that $\miroir{L} = (b + b^2)a^*$
    is recognized by the following $\GRK$ keyboard: 
    \[
        K = \set{
            \touche{\retour^2 a\gauche b},
            \touche{\retour^2 a\gauche bb},
            \touche{\retour^2 b},
            \touche{\retour^2 bb}
        }.
    \]
\end{proof}

\subsection*{Proof of \myref{lem-intersections}}

We give, for each keyboard language class $\mathcal{C}$, two languages of $\mathcal{C}$ whose intersection is not in $\mathcal{C}$.

\begin{lemma}
\label{intersectionEKMK}
The language $L = \set{bw \mid w \in (a+b)^*, \size{w} \text{ odd}, aa \notin \Fact{w}}$ is not in $\EK$.
\end{lemma}

\begin{proof}
Suppose there exists $K = (T,F)$ an $\EK$ keyboard recognizing $L$. Then $L = T^*F$. As all words in $L$ are of even length, so are all keys in $F$. 

Further, $(ba)^{\normeInf{K}}$ is in $L$, thus accepted by $K$, and thus of the form $t_1\cdots t_n f$ with $t_i \in T$ for all $i$, $f \in F$ and $n>0$ (as its length is greater than $\normeInf{K}$). As $\size{f}$ is even, $t_n$ ends with an $a$ (we can assume $t_n \neq \epsilon$). 

Similarly, $bb(ab)^{\normeInf{K}}$ is in $L$, thus of the form $t'_1\cdots t'_m f'$ with $t_i' \in T$ for all $i$, $f' \in F$ and $m>0$. 
As $\size{f'}$ is an even suffix of $(ab)^{\normeInf{K}}$, $f'$ starts with an $a$. As a result, $t_nf'$ contains $aa$ as a factor, and is thus not in $L$, yielding a contradiction.
\end{proof}

\begin{lemma} \label{MKintersection}%
$\MK$ and $\EK$ are not stable by intersection.
\end{lemma}

\begin{proof}
The language $L$ from \myref{intersectionEKMK} is the intersection of $(ab+ba+bb)^*$ and $(ba+b)^*$, which are both in $\MK$. By \myref{intersectionEKMK}, their intersection is not in $\EK$.
\end{proof}

\begin{lemma} \label{REKintersection}%
$\RK$ and $\REK$ are not stable by intersection.
\end{lemma}

\begin{proof}
We consider again the language $L_{\diese\dollar}$ described in the proof of \myref{lem-rknotinfek}. Recall that $L_{\diese\dollar}$ is defined as the language of $\set{t_a,t_b}$ with $t_a = \retour a \diese \dollar$ and $t_b = \retour \retour b \diese\dollar\dollar$. 

We also define $L'$ the language $(a+b+\diese)^+\dollar^2$, recognized by the $\RK$ keyboard 
\[
    \set{\retour^2a\dollar^2,\retour^2b\dollar^2, \retour^2\diese\dollar^2}.
\]
By \myref{lem-forme-lc}, the intersection of $L_{\diese\dollar}$ and $L'$ is the language $((a\diese)^*a + \epsilon)(b \diese)^+\dollar^2$. We now prove that this language is not in $\REK$. 

Suppose there is a keyboard $K = (T,F)$ of $\REK$ recognizing $L = L_{\diese\dollar} \cap L'$. We assume all keys of $K$ to be in normal form (see \myref{lem-forme_normale_touche_rek}). As $L$ contains $(a\diese)^{2\normeInf{K}+1}ab \diese \dollar^2$, there is a sequence of keys $\tau \in T^*$ writing $(a\diese)^{\normeInf{K}+1}w$ for some $w$. 

Similarly as $L$ contains $(b\diese)^{(\size{\tau}+1)\normeInf{K}+1}\dollar^2$, there is a sequence of keys $\tau' \in T^*$ writing $(b\diese)^{\size{\tau}\normeInf{K}+1}w'$ for some $w'$. 

By applying $\tau'\tau$ we get, by \myref{lem-localite}, a word starting with $b$ and containing $\normeInf{K}+1$ $a$. Thus by then applying some $f \in F$, again by $\myref{lem-localite}$, we get a word accepted by $K$ starting with $b$ and containing an $a$, which is impossible as there is no such word in $L$. 
\end{proof}

\begin{lemma} \label{GKintersection} %
No class containing $\GK$ is stable by intersection.
\end{lemma}

\begin{proof}
We consider the following languages:
\begin{itemize}
    \item $(a+b)^*c^*$, recognized by $\set{a,b,c\gauche}$
    
    \item $a^*(b+c)^*$, recognized by $\set{a,b\gauche,c\gauche}$
    
    \item $\set{w \in (a+b+c)^* \mid \size{w}_a = \size{w}_b}$, recognized by $\set{ab, ba, c, \gauche}$
    
    \item $\set{w \in (a+b+c)^* \mid \size{w}_b = \size{w}_c}$, recognized by $\set{bc, cb, a, \gauche}$
\end{itemize}

Those languages are all in $\GK$. Their intersection is $L = \set{a^n b^n c^n \mid n \in \nats}$. We now show that $L$ is not in $\FREK$.

Suppose there exists $K = (T,F)$ in $\FREK$ recognizing $L$. 
Let $\tau f \in T^*F$ be the shortest execution of $K$ accepting 
$w = a^{6\normeInf{K}+1}b^{6\normeInf{K}+1}c^{6\normeInf{K}+1}$.

We have $\config{\epsilon}{\epsilon} \cdot \tau f = \config{u}{v}$ with $uv = w$. 
Clearly as $\size{uv} > \normeInf{K}$, $\tau$ is non-empty, thus there exist 
$\tau' \in T^*$ and $t \in T$ such that $\tau = \tau' t$.

We have either $a^{6\normeInf{K}}b^{3\normeInf{K}+1}$ as a prefix of $u$ or $b^{3\normeInf{K}+1}c^{6\normeInf{K}+1}$ as a suffix of $v$. We assume that we are in the first case, as the other one is similar. Let $\config{u'}{v'} = \config{\epsilon}{\epsilon} \cdot \tau'$ and $\config{x}{y} = \config{\epsilon}{\epsilon}\cdot \tau' f$.
By \myref{lem-localite}, $a^{6\normeInf{K}+1}b^{\normeInf{K}+1}$ is a prefix of $u'$, and again by \myref{lem-localite}, $a^{6\normeInf{K}+1}b$ is a prefix of $x$. As $xy$ is accepted by $K$, $xy \in L$, hence 
\[
    xy = a^{6\normeInf{K}+1}b^{6\normeInf{K}+1}c^{6\normeInf{K}+1} = w.
\]
This contradicts the minimality of $\tau f$. As a result, $L$ is not in $\FREK$.
\end{proof}

\intersections*

\begin{proof}
We proved in \myref{MKintersection} that $\MK$ and $\EK$ are not stable by intersection, then in \myref{REKintersection} that neither are $\RK$ and $\REK$. We finally proved it in \myref{GKintersection} for the rest of the classes.
\end{proof}

\subsection*{Proof of \myref{lem-unions}}

\unions*

We decompose the proof into several parts.

To start with, we consider the languages $a^*$ and $b^*$, both in $\MK$, and prove that their union is neither in $\GREK$ nor in $\FEK$.

\begin{lemma}
The language $L = a^* + b^*$ is not in $\GREK$. 
\end{lemma}

\begin{proof}
Suppose there exists a keyboard $K = (T,F)$ of $\GREK$ recognizing $L$. 
Then there exists $\tau_a \in T^*$, $f_a \in F$ such that $\config{\epsilon}{\epsilon} \cdot \tau_a f_a = \config{u_a}{v_a}$ with $u_av_a = a$.

There also exists $\tau_b \in T^*$, $f_b \in F$ such that $\config{\epsilon}{\epsilon} \cdot \tau_b f_b = \config{u_b}{v_b}$ with $u_bv_b = b^{1+\normeInf{K}(\size{\tau_a}+2)}$.

By \myref{lem-localite} applying $\tau_b$ to $\config{\epsilon}{\epsilon}$ yields a configuration with at least ${1+\normeInf{K}(\size{\tau_a} +1)}$ $b$. 
We apply $\tau_b\tau_af_a$ to $\config{\epsilon}{\epsilon}$, by \myref{thm-fundamental-grek} the resulting configuration contains an $a$, and as $\tau_af_a$ can only erase at most $(\size{\tau_a} +1)\normeInf{K}$ letters, it contains a $b$. This is impossible as the resulting word should be in $L$.
\end{proof}

\begin{lemma}
The language $L = a^*+b^*$ is not in $\FEK$.
\end{lemma}

\begin{proof}
Suppose there exists $K = (T,F)$ a keyboard of $\FEK$ recognizing $L$. As $a^{\normeInf{K}+1}$ and $b^{\normeInf{K}+1}$ are both in $L$, there exist $t_a, t_b \in T$ such that $t_a$ writes an $a$ and $t_b$ a $b$ (and those letters are never erased as we do not have $\retour$). Let $f \in F$, $t_at_bf$ writes a word containing both $a$ and $b$, thus not in $L$. 
\end{proof}

\subsection*{Proof that \FRK{} and \FREK{} are not stable under union.}

We define the language $L = L_a \cup L_b $ with $L_a = \set{a^n c a^n \mid n \in \nats}$ and $L_b = \set{b^n c b^n \mid n \in \nats}$ and prove that $L$ is not in $\FREK$. 
Note that $L_a$ and $L_b$ are in $\FRK{}$ as they are recognized by $\set{\retour c, \retour aca \gauche}$ and $\set{\retour c, \retour bcb \gauche}$, respectively.

Suppose we have a keyboard $K = (T,F)$ of $\FREK{}$ recognizing $L$. 

We start by proving that there is an execution of $K$ leading to a configuration with the cursor far from the edges.

\begin{lemma} \label{lem-union-frek-center} %
Let $M\in \N$, there is a sequence of keys $\tau_a \in T^*$ such that $\config{\epsilon}{\epsilon} \cdot \tau_a = \config{u_a}{v_a}$ with $\size{u_a} > M$ and $u_a$ starts with $\normeInf{K}+1$ $a$.

There is also a sequence of keys $\tau_b \in T^*$ such that $\config{\epsilon}{\epsilon} \cdot \tau_b = \config{u_b}{v_b}$ with $\size{u_b} > M$ and $u_b$ starts with $\normeInf{K}+1$ $b$.
\end{lemma}

\begin{proof}
As $w = a^{M + 4\normeInf{K}} c a^{M+4\normeInf{K}}$ is in the language, there exists $\tau \in T^*$ and $f \in F$ such that $\config{\epsilon}{\epsilon} \cdot \tau f = \config{x}{y}$ with $w = xy$. Let $\config{u}{v} = \config{\epsilon}{\epsilon} \cdot \tau$. We can assume $\tau$ to be of minimal length. Clearly $\tau$ is not empty, thus let $\tau' \in T^*, t \in T$ be such that $\tau = \tau' t$. Let $\config{u'}{v'} = \config{\epsilon}{\epsilon} \cdot \tau'$.

Suppose $\size{u} \leq M$, then $\size{x} \leq M + \normeInf{K}$ by \myref{lem-encadrement_taille}, hence $\size{y} \geq M + 7\normeInf{K}+1$. Then $a^{3\normeInf{K}}c a^{M+4\normeInf{K}}$ is a suffix of $y$, and by \myref{lem-localite}, $a^{2\normeInf{K}}c a^{M+4\normeInf{K}}$ is a suffix of $v'$. Again by \myref{lem-localite}, $c a^{M+4\normeInf{K}}$ is a suffix of the word given by $\config{\epsilon}{\epsilon}\cdot \tau' f$. As this word is in $L$, it has to be $w$, contradicting the minimality hypothesis on $\tau$. 

As a result, $\size{u} \geq M$. Another application of \myref{lem-localite} gives that $u$ starts with $\normeInf{K}+1$ $a$. We thus set $u_a = u$, $v_a = v$ and $\tau_a = \tau$.

A similar proof shows the second part of the lemma.
\end{proof}

The next lemma states that the keys of $K$ erase more than they write, and thus the number of letters of the configuration can only increase when the cursor is close to the left end of the word.

\begin{lemma} \label{lem-union-frek-retours-lettres}%
Every key $t \in T$ contains at least as many $\retour$ as letters, i.e., $\size{t}_{\retour} \geq \size{t}_a + \size{t}_b + \size{t}_c$.
\end{lemma}

\begin{proof}
We consider $\tau_a, \tau_b$, $u_a, u_b$ and $v_a, v_b$ as in \myref{lem-union-frek-center}, with $M = 3(\normeInf{K}+4)^2$. Suppose there exists $t \in T$ such that $\size{t}_{\retour} < \size{t}_a + \size{t}_b + \size{t}_c$. Then by applying $t$ $3(\normeInf{K}+4)$ times from $\config{u_a}{v_a}$ we write either $\normeInf{K}+2$ $a$, $\normeInf{K}+2$ $b$ or $\normeInf{K}+2$ $c$.

If applying $t$ $3(\normeInf{K}+4)$ times effectively writes $\normeInf{K}+2$ $b$ or $\normeInf{K}+2$ $c$, then from $\config{u_a}{v_a}$ we get a configuration starting with $\normeInf{K}+1$ $a$ (by \myref{lem-localite}) and containing at least $\normeInf{K}+2$ $b$ or $c$, which after applying some final key $f \in F$ still contains an $a$ and two $b$ or $c$. This is impossible as the resulting word should be in $L$.

Similarly, if applying $t$ $3(\normeInf{K}+4)$ times effectively writes $\normeInf{K}+2$ $a$, then from $\config{u_b}{v_b}$ we get a configuration starting with $\normeInf{K}+1$ $b$ (by \myref{lem-localite}) and containing at least $\normeInf{K}+2$ $a$, which after applying some final key $f \in F$ still contains a $b$ and two $a$. This is impossible as the resulting word should be in $L$.

Thus there is no such key.
\end{proof}

We now show that the number of letters in the configuration cannot increase too much while staying far from the center.

\begin{lemma} \label{lem-union-frek-bord-gauche}%
Let $\tau_0 \in T^*, t_1, \ldots, t_n \in T$. For all $0 \leq i \leq n$ we define $\config{u_i}{v_i} = \config{\epsilon}{\epsilon}\cdot \tau_0t_1\cdots t_i$.

If $\size{u_n} + \size{v_n} \geq \size{u_0} + \size{v_0} + 4\normeInf{K} + 1$ then there exists $0 \leq i \leq n$ such that $\abs{\size{u_i} - \size{v_i}} \leq 8\normeInf{K}+2$.
\end{lemma}

\begin{proof}
Suppose $\size{u_n} + \size{v_n} \geq \size{u_0} + \size{v_0} + 4\normeInf{K} + 1$ and for all $0 \leq i \leq n$, $\abs{\size{u_i} - \size{v_i}} > 8\normeInf{K}+2$.

Then as $\abs{\size{u_{i+1}}-\size{u_{i}}}$ and $\abs{\size{v_{i+1}}-\size{v_{i}}}$ are both at most $\normeInf{K}$, and thus $\size{u_i} - \size{v_i}$ always has the same sign. 
We assume it to be positive, the other case is similar.

Let $f \in F$, for all $i$ let $\config{u^f_i}{v_i^f} = \config{u_i}{v_i}\cdot f$.
We have by \myref{lem-encadrement_taille} $\size{u_i^f} - \size{v_i^f} \geq 6\normeInf{K} + 2$
As $u_i^fv_i^f \in L$, we thus have $u_i^f$ of the form $a^mca^p$ or $b^mcb^p$, with $p \geq 3\normeInf{K}+1$, because $(m+p) - (m-p) = \size{u_i^f} - \size{v_i^f}$. We assume the first case as the other one is similar.

By \myref{lem-localite}, $a^mca^{2\normeInf{K}+1}$ is a prefix of $u_i$, and $a^mca^{\normeInf{K}+1}$ a prefix of $u_{i+1}$, and $a^mca$ a prefix of $u^f_{i+1}$. Therefore we must have $u^f_{i+1}v^f_{i+1} = a^m c a^m$.

As a result, we have $u^f_nv^f_n = u^f_0v^f_0$, and therefore by \myref{lem-encadrement_taille}, $\abs{\size{u_nv_n} - \size{u_0v_0}} \leq 4 \normeInf{K}$, contradicting the hypothesis.

\end{proof}

We showed that the number of letters could only increase with the cursor close to the left end of the word, and that we can only increase the number of $a$ by a bounded number while staying far from the center.

Thus in order to write a very long word $a^Mca^M$, we need to go back and forth between the left edge and the center. We isolate a moment in the execution at which the cursor is far from the center and the left edge, goes to the left edge, and goes back while increasing the number of letters.

We end up with two configurations with different numbers of letters. As we are far from the edge in both, a final key $f$ has the same behaviour (adds and erases as many letters) in both. Thus the word obtained afterwards cannot be equal. However we are far from the center and we have a same suffix $ca^M$, thus the two words have to be equal, hence a contradiction.

\begin{lemma}
No keyboard of $\FREK{}$ recognizes $L$.
\end{lemma}

\begin{proof}

Let $M = 30 \normeInf{K}$.
Let $\tau = t_1 \cdots t_n t_f \in T^*F$ be an execution of $K$ recognizing $a^Mca^M$. For all $0 \leq i \leq n$ we define $\config{u_i}{v_i} = \config{\epsilon}{\epsilon}\cdot t_1\cdots t_i$. Let $f \in F$, we define for all $0 \leq i \leq n$ $\config{u_i^f}{v_i^f} = \config{u_i}{v_i}\cdot f$. 

Let $i$ be the minimal index such that $\size{u_jv_j} \geq 10\normeInf{K}$ for all $j \geq i$ and $t_i \neq \epsilon$. In particular we have $\size{u_iv_i} \leq 11\normeInf{K}$.

By \myref{lem-union-frek-retours-lettres}, $\size{u_{j+1}v_{j+1}} - \size{u_jv_j}$ is negative or null whenever $\size{u_j} \geq \normeInf{K}$ (as $t_j$ contains at least as many $\retour$ as letters and every $\retour$ is applied effectively).

Let $I$ be the set of indices $j>i$ such that $\size{u_j}> \normeInf{K}$ and $ \size{v_j} - \size{u_j} > 2\normeInf{K}$. 

The size of the configuration increases by at least $9 \normeInf{K}$ between the applications of $t_i$ and $t_f$. By \myref{lem-union-frek-retours-lettres}, the size of the configuration can only increase when the left part of the configuration has length at most $\normeInf{K}$.

As a consequence, by \myref{lem-union-frek-bord-gauche}, there exist $j_1 < j_2$ such that $\size{v_{j_1}} - \size{u_{j_1}}, \size{v_{j_2}} - \size{u_{j_2}} \leq 8 \normeInf{K}+2$ and $\size{u_{j_1}v_{j_1}} < \size{u_{j_2}v_{j_2}}$.

We take $j_1$ maximal and $j_2$ minimal for this property. We obtain that for all $j_1 < j < j_3$, $\size{v_{j_3}} - \size{u_{j_3}} > 8 \normeInf{K}+2 > 2 \normeInf{K}$.

Note that as $\size{u_jv_j}> 10\normeInf{K}$ for all $j>i$, we have that there exists $i_1, i_2 \in I$ such that $j_1 < i_1 < i_2 < j_3$. We choose $i_1$ minimal and $i_2$ maximal for that property. We cannot have any $j_1 < i' < i_1$ such that $\size{u_{i'}} \leq \normeInf{K}$ as otherwise we would necessarily have some index of $I$ between $j_1$ and $i'$, contradicting the minimality of $i_1$. Similarly, we cannot have $i_2 < i' < j_2$ such that $\size{u_{i'}} \leq \normeInf{K}$.

We therefore have $\size{u_{i_1}v_{i_1}} \leq \size{u_{j_1}v_{j_1}} < \size{u_{j_2}v_{j_2}} \leq \size{u_{i_2}v_{i_2}}$ (the second inequality holds by construction of $j_1$ and $j_2$).

We obtain that:

\begin{itemize}
    \item $\normeInf{K} \leq \size{u_{i_1}},\size{u_{i_2}}$

    \item $\size{u_{i_1}v_{i_1}} \neq \size{u_{i_2}v_{i_2}}$
    
    \item For all $i_1 < j < i_2$,  $\size{u_{j}} < \size{v_{j}} - 2 \normeInf{K}$ (by definition of $j_1$ and $j_2$).
\end{itemize}

Let $f \in F$, by an argument similar to the proof of \myref{lem-union-frek-bord-gauche}, we have that $u_{i_1}^fv_{i_1}^f = u_{i_2}^fv_{i_2}^f$ as we stay away from the center and thus the part of the word of the form $ca^m$ is unchanged.

However, in both cases we applied the final key $f$ far from the edges, thus by applying \myref{lem-egalite_taille_loin_bords} we get $\size{u_{i_1}v_{i_1}} = \size{u_{i_2}v_{i_2}}$, yielding a contradiction.

As a result there is no $\FREK{}$ keyboard recognizing $L$.
\end{proof}

\subsection*{Proof of \myref{prop-empty-intersection}}

\emptyintersection*

\begin{proof}
We reduce the Post Correspondence Problem.
Let $(u_i,v_i)_{i \in \intervE[1,n]}$ be a PCP instance. For all $i \in \intervE[1, n]$, let $u^{\fraiche}_i$ and $v^{\fraiche}_i$ be $u_i$ and $v_i$ where we added a $\fraiche$ at the right of every letter, i.e., if $u_i = a_1a_2\cdots a_n$ then $u^{\fraiche}_i = a_1\fraiche a_2 \fraiche \cdots a_n\fraiche$.

We set for all $i \in \intervE[1, n]$
$t_i = u^{\fraiche}_i\miroir{v^{\fraiche}_i} \gauche^{2\taille{v_i}}$.
Let

\[
    K_{pal} = \set{aa\gauche \mid a \in A\cup \set{\fraiche}} \cup \set{\epsilon}
\]

$K_{pal}$ recognizes the language of even palindromes over $A\cup \set{\fraiche}$.

Now let

\[
    K = \ensC{t_i}{i \in \intervE[1,n]}.
\]

We show that $\langage{K} \cap \langage{K_{pal}} \neq \emptyset$ if and only if
$(u_i,v_i)_{i \in \intervE[1,n]} \in \PCP$. 
A straightforward induction on $k \in \N$ shows that
for all $i_1, \ldots, i_k \in \intervE[1,n]$, 
\[
    \config{\epsilon}{\epsilon} \cdot (t_{i_1} \ldots t_{i_k}) = 
    \config{u^{\fraiche}_{i_1} \cdots u^{\fraiche}_{i_k}}{\miroir{v^{\fraiche}_{i_k}} \cdots \miroir{v^{\fraiche}_{i_1}}}.
\]

Note that if $k>0$, $u^{\fraiche}_{i_1} \cdots u^{\fraiche}_{i_k}\miroir{v^{\fraiche}_{i_k}} \cdots \miroir{v^{\fraiche}_{i_1}}$ has exactly one $\fraiche \fraiche$ factor, with the last letter of $u^{\fraiche}_{i_k}$ and the first one of $\miroir{v^{\fraiche}_{i_k}}$. As a consequence, it is an even palindrome if and only if $u^{\fraiche}_{i_1} \cdots u^{\fraiche}_{i_k} = \miroir{\miroir{v^{\fraiche}_{i_k}} \cdots \miroir{v^{\fraiche}_{i_1}}} = v^{\fraiche}_{i_1} \cdots v^{\fraiche}_{i_k}$.

All that is left to show is that $u^{\fraiche}_{i_1} \cdots u^{\fraiche}_{i_k} = v^{\fraiche}_{i_1} \cdots v^{\fraiche}_{i_k}$ if and only if $u_{i_1} \cdots u_{i_k} = v_{i_1} \cdots v_{i_k}$. The left to right direction is shown by projecting the words on $A$, the right to left one by observing that $u^{\fraiche}_{i_1} \cdots u^{\fraiche}_{i_k}$ and $v^{\fraiche}_{i_1} \cdots v^{\fraiche}_{i_k}$ are the images of $u_{i_1} \cdots u_{i_k}$ and $v_{i_1} \cdots v_{i_k}$ under the morphism associating $a\fraiche$ to each letter $a \in A$. 
As a result, $\langage{K}$ contains an even palindrome if and only if there exists $k> 0$ and $i_1, \ldots, i_k \in \intervE[1,n]$ such that $u_{i_1}\cdots u_{i_k} = v_{i_1}\cdots v_{i_k}$
\end{proof}

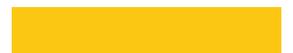
\end{document}


\maketitle

\begin{abstract}
Nous introduisons ici une nouvelle représentation de langages, les \emph{claviers}. On se munit d'un ensemble d'opérations élémentaires (ajout/effacement d'une lettre, déplacement à droite/gauche,...), et on définit un clavier comme un ensemble de suites finies d'opérations élémentaires, appelées touches. Son langage va être l'ensemble des mots obtenus en appliquant une suite quelconque de touches. Contrairement à des modèles de calcul classiques comme les machines de Turing ou les grammaires, toutes les touches peuvent être appliquées à tout moment.

Nous nous intéressons ici à plusieurs problèmes fondamentaux sur ce modèle de calcul. En premier lieu nous définissons différentes classes de claviers en faisant varier l'ensemble des opérations élémentaires autorisées, et nous comparons l'expressivité des classes de langages obtenues entre elles. Nous comparons également ces classes à des classes de langages connues, notamment les langages réguliers, algébriques et contextuels.

Nous étudions également certains problèmes de décision découlant naturellement de l'étude des claviers. Les deux plus importants sont le \emph{problème du mot}, consistant à tester si un clavier peut engendrer un mot donné, et le problème de l'universalité, consistant à tester si un clavier peut engendrer tout mot sur son alphabet.

Nous obtenons que toutes les classes étudiées sont différentes, et nous caractérisons les classes inclues dans les rationnels et les algébriques. Nous déduisons des inclusions de certaines classes dans des classes de langages connues la décidabilité de certains problèmes de décision évoqués plus haut.

Il semble ainsi que l'expressivité des claviers soit orthogonale à celle de nombreux modèles connus. En effet, les claviers sont limités par leur absence de mémoire, mais cette faiblesse peut être compensée par l'ajout d'opérations élémentaires comme l'effacement ou les déplacements, difficiles à simuler dans des modèles comme les automates sans gagner l'expressivité des machines de \bsc{Turing}. 

\end{abstract}


\include{1Heuristique/heuristique}

\include{2Definitions/Definitions}
\include{3Proprietes/proprietes}
\include{4Langages/langages}
\include{5ContreExemples/ce}

\include{6Stabilite/stabilite}
\include{6Conclusion/Conclusion}

\bibliography{french}

\include{Annexes/annexes}

%% file: preambule.tex
\let\phi\varphi

\let\epsilon\varepsilon

\newcommand*\abs[1]{\left\lvert #1\right\rvert}

\newcommand*\taille[1]{\left\lvert #1\right\rvert}

\NewDocumentCommand{\norme}{om}{
    \left\lVert #2\right\rVert \IfValueTF{#1}{_#1}{}
}
\newcommand*\normeInf[1]{\norme[\infty]{#1}}

\let\size\taille

\newcommand*{\retour}{\mathord{\leftarrow}}
\newcommand*{\gauche}{\mathord{\blacktriangleleft}}
\newcommand*{\droite}{\mathord{\blacktriangleright}}
\newcommand*{\entree}{\mathord{\blacksquare}}

\NewDocumentCommand\probleme{smmm}{
    \IfBooleanTF{#1}{
        \text{#2} :
        \begin{cases}
            \text{\textsc{Entrée}}: & \text{#3} \\
            \text{\textsc{Sortie}}: & \text{#4}
        \end{cases}
    }{
        #2 :
        \begin{cases}
            \text{\textsc{Entrée}}: & #4 \\
            \text{\textsc{Sortie}}: & #4
        \end{cases}
    }
}

\NewDocumentCommand\problemeDecision{smmm}{
    \IfBooleanTF{#1}{
        \text{#2} :
        \begin{cases}
            \text{\textsc{Donnée}}:   & \text{#3} \\
            \text{\textsc{Question}}: & \text{#4}
        \end{cases}
    }{
        #2 :
        \begin{cases}
            \text{\textsc{Donnée}}:   & #3 \\
            \text{\textsc{Question}}: & #4
        \end{cases}
    }
}

\let\originalleft\left
\let\originalright\right
\renewcommand{\left}{\mathopen{}\mathclose\bgroup\originalleft}
\renewcommand{\right}{\aftergroup\egroup\originalright}

\let\ensembleNombre\mathbb
\newcommand*\Nat{\ensembleNombre{N}}
\let\nats\Nat
\let\N\Nat
\newcommand{\Z}{\ensembleNombre{Z}}

\newcommand*\ensC[2]{\left\{#1 \mathrel{}\middle|\mathrel{} #2\right\}}

\let\set\ensemble

\newcommand*\intervalle[4]{\left#1 #2 \, ; #3 \right#4}
\newcommand*\intervalleEntier[2]{\intervalle{\llbracket}{#1}{#2}{\rrbracket}}

\newcommand\actimath[1]{\mathcode`#1"8000 \begingroup \lccode`~`#1 \lowercase{\endgroup\def~}}

\newcommand\intervE{%
    \begingroup
    \actimath\;{\mathpunct{}\mathpunct{\mathchar`\;}}%
    \actimath\[{\left\llbracket}%
    \actimath\]{\right\rrbracket\endgroup}%
}

\makeatletter
\NewDocumentCommand\newoperator{smm}{
    \expandafter\NewDocumentCommand
    \csname\expandafter\@gobble\string#2\endcsname{og}{
        \IfBooleanTF{#1}{\operatorname*{#3}}{\operatorname{#3}}%
        \IfNoValueF{##1}{^{##1}}%
        \IfNoValueF{##2}{\left(##2\right)}%
    }
}

\NewDocumentCommand\renewoperator{smm}{
    \expandafter\RenewDocumentCommand
    \csname\expandafter\@gobble\string#2\endcsname{og}{
        \IfBooleanTF{#1}{\operatorname*{#3}}{\operatorname{#3}}%
        \IfNoValueF{##1}{^{##1}}%
        \IfNoValueF{##2}{\left(##2\right)}%
    }
}
\makeatother

\newoperator{\langage}{\mathcal{L}}
\newoperator{\automate}{\mathcal{A}}
\newoperator{\Auto}{\mathcal{A}}
\newoperator{\AutoMiroir}{\mathcal{M}}

\newoperator{\Conf}{\mathcal{C}}

\newoperator{\Supp}{Supp}
\newoperator{\Fact}{Fact}
\newoperator{\Pref}{Pref}
\newoperator{\Suff}{Suff}
\newoperator{\Sumo}{Sub}

\newoperator{\distance}{d}

\renewoperator*{\max}{max}
\renewoperator*{\min}{min}

\let\eqDef\triangleq

\newcommand{\fraiche}{\mathalpha{\blacklozenge}}

\newcommand{\Rat}{\ensuremath{\mathsf{Reg}}}

\newcommand{\Alg}{\ensuremath{\mathsf{CF}}}
\newcommand{\Cont}{\ensuremath{\mathsf{Cont}}}

\newcommand{\MK}{\ensuremath{\mathsf{MK}}}

\newcommand{\EK}{\ensuremath{\mathsf{EK}}}
\newcommand{\RK}{\ensuremath{\mathsf{BK}}}
\newcommand{\REK}{\ensuremath{\mathsf{BEK}}}

\newcommand{\FK}{\ensuremath{\mathsf{AK}}}
\newcommand{\FEK}{\ensuremath{\mathsf{EAK}}}
\newcommand{\FRK}{\ensuremath{\mathsf{BAK}}}

\newcommand{\GK}{\ensuremath{\mathsf{LK}}}
\newcommand{\GRK}{\ensuremath{\mathsf{BLK}}}
\newcommand{\GEK}{\ensuremath{\mathsf{LEK}}}

\newcommand{\FREK}{\ensuremath{\mathsf{BEAK}}}
\newcommand{\GREK}{\ensuremath{\mathsf{BLEK}}}

\newoperator{\T}{\mathcal{T}}

\newcommand*\touche[1]{\mathtt{#1}}

\let\config\conf
\let\miroir\widetilde

\NewDocumentCommand{\suite}{O{n}mO{\Nat}} {
    \left(#2\right)_{#1 \in #3}
}

\newcommand{\act}[1]{\xrightarrow{#1}}
\newcommand{\actEff}[1]{\xrightarrow{#1}_{\text{e}}}

\newcommand*{\cdotEff}{\odot}

\newoperator{\PGCD}{GCD}

\newoperator{\profondeur}{Prof}

\newcommand*{\enlarger}{\vphantom{\frac{a}{b}}}

\NewDocumentCommand{\prooflink}{om}{%
\raggedleft
    \hyperref[#2]{Voir preuve en \ref{#2}}%
    \IfNoValueF{#1}{\strut \newline#1}
\justify
}

\newcommand{\Fin}{\text{Fin}}
\newcommand{\Init}{\text{Init}}

\newcommand{\PCP}{\text{PCP}}

\newcommand{\empile}{\mathord{\downarrow}}
\newcommand{\depile}{\mathord{\uparrow}}
\newcommand{\rien}{-}
\newcommand{\etatpile}[2]{#1^{#2}}

\newcommand*{\transpile}[3]{\xrightarrow[#2, #3]{#1}}

\usepackage{xspace}

\newcommand{\ieme}{\textsuperscript{th}\xspace}
\newcommand{\ier}{\textsuperscript{st}\xspace}

\newcommand*{\cEnt}[1]{\overline{#1}}
\newcommand*{\tEnt}[1]{\widehat{#1}}

\newcommand{\fc}[2]{f_c\left(#1,#2\right)}
\newcommand{\dollar}{\mathalpha{\blacklozenge}}
\newcommand{\diese}{\mathalpha{\Diamond}}

\definecolor{color1}{RGB}{1, 22, 39}    
\definecolor{color2}{RGB}{29, 61, 191} 
\definecolor{color3}{RGB}{231, 29, 54}  
\definecolor{color4}{RGB}{81, 120, 18} 

%% file: introduction.tex
\section{Introduction}

We present a new formalisation of languages, called keyboards. A keyboard $K$ is a finite set of keys, which are finite sequences of atomic operations, such as writing or erasing a letter, going one position to the right or to the left... The language of $K$ is the set of words obtained by applying a sequence of its keys on an initially empty writing space.

This idea of studying the set generated by a set of algebraic operations is far from new: many works exist on the sets generated by a subset of elements of an algebraic structure, for instance in the context of semigroup and group theory~\cite{BEAUDRY198884,JONES1976}, of matrix monoids~\cite{BabaiS1984,paterson1970unsolvability} or the theory of codes~\cite{BerstelPR2010}. There is however, to the best of the author's knowledge, no previous work on a model resembling the one presented here.

The atomic operations we use in this paper are the base of other models of computation, such as forgetting automata and erasing automata~\cite{JancarMrazPlatek1993,JancarMP1992,vonBraunmBV1979}. The use of those operations was originally to simulate some analysis strategies in linguistics. As a first study of the model, we chose the actions of the operations (backspace and arrows) to behave like an actual keyboard in a text editor.

We can define various classes of languages based on the set of atomic operations we consider, and compare their expressive powers between them, and to well-known classes of languages. We obtain a strict hierarchy of classes, with a wide range of expressiveness and difficulty of comprehension. The expressiveness of keyboards seems to be overall orthogonal to the ones of classical models of computation, which we explain by two key differences with the latter.

First, keyboards are blind and memoryless, in that they do not have states and cannot read the tape at any point. Second, because of this weakness, we can allow operations such as moving in the word or erasing letters without blowing up their expressive power too much.

The main interests of keyboards are: 1. to obtain many deep and complex mathematical questions from a deceptively simple model, and 2. that their expressiveness is very different from the ones of classical models. A language that is simple to express with a keyboard may be more complicated for automata, and vice versa. This paper is meant as a first step in the study and comprehension of keyboards and their languages. 

The paper is organised as follows.
In Sections 2 and 3 we establish notations and basic definitions. Section 4 and 5 are dedicated to building properties and tools necessary to the study of keyboards. In Section 6 we dive into the specific properties of each keyboard class, and prove the inclusions of some of them in regular, context-free and context-sensitive languages. Then in Section 7, we study the inclusions between those classes, in particular showing that they are all different. Some complexity results are given in Section 8. Finally, in Section 9 we show that keyboard classes are not stable by union or intersection, and that some (but not all) of them are stable by mirror.

All proofs can be found in the appendices.

%% file: preliminaries.tex
\section{Preliminaries}

Given a finite alphabet $A$, we note $A^*$ the set of finite words over $A$ and $A^+$ for the set of non-empty ones. 
Given a word $w = a_1 \cdots a_n \in A^*$, we write $\size{w}$ for its length and, for all $a\in A$, $\size{w}_a$ the number of occurrences of $a$ in $w$. For all $1 \leq i, j \leq n$ we use the notation $w[i]$ for the $i$\ieme{} letter of $w$ (i.e. $a_i$) and $w[i,j]$ for its factor $a_i\cdots a_j$ (and $\epsilon$ if $j<i$).
We denote the mirror of $w$ by $\Tilde{w} = a_n \cdots a_1$.

We write $\Pref{w}$ for the set of prefixes of $w$, $\Suff{w}$ for its set of suffixes, $\Fact{w}$ for its set of factors and $\Sumo{w}$ for its set of subwords.

We represent a finite automaton $A$ as a tuple $(Q, \Delta, \text{Init}, \text{Fin})$ with $Q$ a finite set of states, $\Delta : Q \times A \to 2^Q$ a transition function, and $\text{Init}, \text{Fin} \subseteq Q$ sets of initial and final states.

We represent a pushdown automaton on $A$ as a tuple $(Q, \Gamma, \bot, \Delta, \Init, \Fin)$ with
\begin{itemize}
    \item $Q$ a finite set of states ;
    \item $\Gamma$ a finite stack alphabet ;
    \item $\bot \in \Gamma$ an initial stack symbol ;
    \item $\Delta \colon Q \times A \times (\Gamma \cup \set{-})^2 \to 2^Q$ a transition function ;
    \item $\Init$ and $\Fin$ sets of initial an final states.
\end{itemize}
We accept a word on final states with an empty stack. We write transitions as follows:
\[
    s_1 \transpile{a}{\text{op}_1}{\text{op}_2} s_2
\]
with
\begin{itemize}
    \item $\text{op}_1 = \depile \gamma$ if we pop $\gamma \in \Gamma$, and $\text{op}_1= \rien$ if we do not pop anything.
    
    \item $\text{op}_2 = \empile \gamma$ if we push $\gamma \in \Gamma$ on the stack, and $\text{op}_2= \rien$ if we do not push anything.
\end{itemize}
We will use $\epsilon$-transitions in both finite and pushdown automata to simplify some proofs.

For more details and properties of those models, we refer the reader to \cite{DBLP:books/lib/HopcroftU69}.

%% file: definitions.tex
\section{Definitions}

We fix a finite alphabet $A$ and the following special symbols, taken out of $A$:
\[
    \text{The backspace}: \retour \qquad
    \text{The left arrow}: \gauche \qquad
    \text{The right arrow}:  \droite
\]
The set of all symbols is $S \eqDef A \cup \set{\retour, \gauche, \droite}.$
An element of $S$ is called an \emph{atomic operation}.

\begin{definition}
    A \emph{configuration} is a pair of words $(u, v) \in A^* \times A^*$. We will use 
    $\Conf{A}$ to denote the set of configurations over $A$, and $\config{u}{v}$ to 
    denote the configuration $(u,v)$. 
    
    We define the notation $\config{u}{v}_i$ as the letter at position $i$ in the configuration with respect to the cursor: 
    $\config{u}{v}_i = \miroir{u}[-i]$ if $i < 0$ and $v[i]$ if $i > 0$.
\end{definition}

\begin{definition}
    The \emph{action} of an atomic operation $\sigma \in S$ on a configuration $\config{u}{v}$ is written $\config{u}{v} \cdot \sigma$ and is defined as follows:
    \begin{gather*}
        \config{u}{v}        \cdot a        = \config{ua}{v} \text{ if } a \in A.\\
        \begin{aligned}
            \config{\epsilon}{v} \cdot \retour  = \config{\epsilon}{v} \quad &\text{ and } \quad
            \config{u'a}{v}      \cdot \retour  = \config{u'}{v} \\
            \config{\epsilon}{v} \cdot \gauche  = \config{\epsilon}{v} \quad &\text{ and } \quad
            \config{u'a}{v}      \cdot \gauche  = \config{u'}{av}\\
            \config{u}{\epsilon} \cdot \droite  = \config{u}{\epsilon} \quad &\text{ and } \quad
            \config{u}{av'}      \cdot \droite  = \config{ua}{v'}
        \end{aligned}
    \end{gather*}  
We will sometimes write $\config{u}{v} \act{\sigma} \config{u'}{v'}$ for $\config{u'}{v'} = \config{u}{v} \cdot \sigma$.
\end{definition}

\begin{example}
    By applying the following sequence of atomic operations $\retour$, $a$, $\droite$, $\droite$, $b$ to the configuration $\config{c}{d}$, we obtain the following rewriting derivation:
    \begin{align*}
        \config{c}{d} \act{\retour} \config{\epsilon}{d}
                      \act{a}       \config{a}{d}
                      \act{\droite} \config{ad}{\epsilon}
                      \act{\droite} \config{ad}{\epsilon}
                      \act{b}       \config{adb}{\epsilon}.
    \end{align*}
\end{example}

\begin{definition}
    We define other semantics for atomic operations, called
    \emph{effective semantics}. The difference with the previous ones is that we forbid application of atomic operations without effect (such as backspace when the left word of the configuration is empty). Formally, given $u,v \in A^*, a \in A$ we have:
    \begin{align*}
        \config{u}{v}   \actEff{a}        & \config{ua}{v}  &
        \config{u'a}{v} \actEff{\retour}  & \config{u'}{v}   \\
        \config{u'a}{v} \actEff{\gauche}  & \config{u'}{av} &
        \config{u}{av'} \actEff{\droite}  & \config{ua}{v'}
    \end{align*}
    We also define the operator $\cdotEff$ by
    $\config{u}{v} \cdotEff \sigma = \config{u'}{v'}$ if and only if 
    $\config{u}{v} \actEff{\sigma} \config{u'}{v'}$.
\end{definition}

\begin{definition}
    A \emph{key} is a sequence of atomic operations, seen as a word on $S$. We will use $\T(S)$ to denote the set of keys on $S$ (variables $k,t,...$), or $\T$ if there is no ambiguity.
\end{definition}

\begin{definition}
    The action of a key over a configuration is defined inductively as follows:
    \[
    \left\{
    \begin{aligned}
         \config{u}{v} \cdot \epsilon &= \config{u}{v}\\
         \config{u}{v} \cdot (\sigma t)      &= (\config{u}{v} \cdot \sigma) \cdot t
    \end{aligned}
    \right.
    \]
    We extend the notation $\config{u}{v} \act{t} \config{u'}{v'}$ to keys. We define $\config{u}{v} \actEff{t} \config{u'}{v'}$ and $\config{u}{v} \cdotEff t$ analogously.

\end{definition}

\begin{remark}
    We will also consider sequences of keys $\tau = t_1 \ldots t_n$. The action of $\tau$ is obtained by composing the actions of the $t_i$, hence applying $\tau$ has the same effect as applying sequentially $t_1, \ldots, t_n$. Note that $\tau$ is seen as a word on $\T(S)$ (and not on $S$), thus $\size{\tau}$ is $n$. 
\end{remark}

\begin{definition}
    The length of a key $t$, written $\size{t}$, is its length as a word on $S$. Further, given $\sigma \in S$, we note $\size{t}_\sigma$ the number of occurrences of $\sigma$ in $t$.
    The size of a configuration $\config{u}{v}$ is defined as $\size{\config{u}{v}} = \size{u}+ \size{v}$.
\end{definition}

\begin{definition}
    Two keys $t$ and $t'$ are equivalent, denoted $t \sim t'$, if for all $u,v \in A^*$, $\config{u}{v} \cdot t = \config{u}{v} \cdot t'$.
\end{definition}

\begin{example}
\label{ex-config_eq}
    $\epsilon$ is equivalent to $a \retour$ for all $a \in A$, but not to $\droite \gauche$, as we have $\config{a}{\epsilon} \cdot \droite \gauche = \config{\epsilon}{a}$ whereas $\config{a}{\epsilon} \cdot \epsilon = \config{a}{\epsilon}$. 
\end{example}

\myref{ex-config_eq} illustrates how $\gauche$, $\droite$ and
$\retour$ act differently if one side of the configuration
is empty. We will see that these \enquote{edge effects} add some expressiveness compared to the effective semantics, but make proofs more complex.

\begin{definition}[Automatic Keyboard]
    An automatic keyboard is a finite subset of $\T(S)$.
\end{definition}

\begin{definition}
An \emph{execution} of an automatic keyboard $K$ on a configuration $c_0 \in \Conf$ is a \textbf{non-empty} finite sequence $\rho = (t_1, c_1), \ldots, (t_{n+1}, c_{n + 1})  \in (K \times C)^{n+1}$ ($n \in \nats$) such that 
\[
    \forall i \in \intervE[1; n + 1], c_{i - 1} \act{t_i} c_i.
\]
    By default, we take as initial configuration $c_0 = \config{\epsilon}{\epsilon}$. We usually write $c_0 \act{\tau} c_{n+1}$ to mean the execution $(\tau[1], c_0 \cdot \tau[1]),..., (\tau[n+1], c_0 \cdot \tau[1,n+1])$.
\end{definition}


\begin{definition}
A word $w \in A^*$ is \emph{recognized} by an automatic keyboard $K$ if there exist $u,v \in A^*$ and an execution $\config{\epsilon}{\epsilon} \act{\tau} \config{u}{v}$ such that $w = uv$.
The language $\langage{K}$ of $K$ is the set of words recognized by $K$.
\end{definition}

We now define keyboards as automatic keyboards to which we added some final 
keys \enquote{with entry}, which mark the end of the execution.

\begin{definition}[Keyboard (with entry)]
    A keyboard $K$ on $S$ is a pair $(T, F)$ of finite sets $T,F \subset \T{S}$.
    We call the elements of $F$ the \emph{final keys} of $K$ and the elements of $T$ its \emph{transient} keys.
\end{definition}

\begin{definition}[Accepting execution of a keyboard]
    Let $K = (T, F)$ be a keyboard and $c_0 = \config{u_0}{v_0}$ an initial configuration. 
    An \emph{accepting execution} of $K$ on $c_0$ is a finite sequence $\rho =  (t_1, c_1), \ldots, (t_{n+1}, c_{n + 1}) \in  (T \times C)^n \cdot (F \times C)$ ($n \in \nats$)
    such that 
    \[
    \forall i \in \intervE[1; n + 1], c_{i - 1} \act{t_i} c_i.
    \]
    By default, an accepting execution is on the empty configuration
    $\config{\epsilon}{\epsilon}$.
\end{definition}

\begin{definition}
A word $w \in A^*$ is \emph{recognized} by a keyboard $K$ if there exist $u,v \in A^*$ and an execution $\config{\epsilon}{\epsilon} \act{\tau} \config{u}{v}$ such that $w = uv$.
The language $\langage{K}$ of $K$ is the set of words recognized by $K$.
\end{definition}

\begin{example}
    The keyboard with one transient key $\touche{aa}$ and one final key 
    $\touche{a}$, recognizes sequences of $a$ of odd length.
\end{example}

\begin{remark}
Let $K_a$ be an automatic keyboard, then $\langage{K_a}$ is recognized by the keyboard with entry $(K_a, K_a)$. In all that follows we will thus see automatic keyboards as a subclass of keyboards.
\end{remark}

\begin{definition}[Size of a keyboard]
    The size of a keyboard $K = (T, F)$ is defined as
    \[
    \normeInf{K} = \max \ensC{\size{t}}{t \in T \cup F}.
    \]
    We may also use another measure $\size{K}$ of the size of $K$ for complexity purposes:
    \[
    \taille{K} = \sum_{t \in T \cup F} \left(\taille{t} + 1\right).
    \]
\end{definition}

\begin{definition}[Minimal keyboard]
    A minimal keyboard $K$ is an automatic keyboard without any operation besides writing letters. It can therefore be seen as a finite subset of $A^*$. We will note $\MK$ the class of minimal keyboards.
\end{definition}

\begin{remark}
    We construct our keyboard classes through the sets of special operations we allow. Class names are obtained by adding
    $\mathsf{B}$ (for $\retour$), $\mathsf{E}$ (for the entry, noted $\entree$),
    $\mathsf{L}$ (for $\gauche$) and $\mathsf{A}$  (for $\gauche$
    and $\droite$) to $\mathsf{K}$. We obtain these classes.
\[
    \begin{aligned}
    \MK   & : \set{}                 &\qquad
    \GK   & : \set{\gauche}          &\qquad
    \FK   & : \set{\gauche, \droite}\\
    \EK   & : \set{\entree}          &\qquad
    \GEK  & : \set{\gauche, \entree} &\qquad
    \FEK  & : \set{\gauche, \droite, \entree}\\
    \RK   & : \set{\retour}          &\qquad
    \GRK  & : \set{\gauche, \retour} &\qquad
    \FRK  & : \set{\gauche, \droite, \retour}\\
    \REK  & : \set{\retour, \entree}         &\qquad
    \GREK & : \set{\gauche, \retour, \entree} &\qquad
    \FREK & : \set{\gauche, \droite, \retour, \entree} \\
    \end{aligned}
\]
\end{remark}

\begin{remark}
    We do not consider classes with $\droite$ without $\gauche$ because, without the $\gauche$ operator, we can only reach configurations of the form $\config{u}{\epsilon}$ and thus $\droite$ never has any effect. 
\end{remark}

\begin{remark}
    We use the class names above to designate both keyboard classes and language classes.
    For instance, we will write that $L$ is in $\FK$ if there exists a keyboard $K \in \FK$ such that $L = \langage{K}$.
\end{remark}

%% file: general_properties.tex
\section{General properties}
    In this section, we establish some properties on keyboard. Although most of them are quite intuitive, we take the time to be as formal as possible in order to build solid bases for the study of keyboards.
    
    Our first lemma states that applying a key can only affect a bounded part of the word around the cursor.
    
\begin{restatable}[Locality]{lemma}{lemLocalite}%
\label{lem-localite}%
    Let $t = \sigma_1 \ldots \sigma_n$ be a key. If $\config{u}{v} \act{t} \config{u'}{v'}$, then $u[1, \size{u} - n]$ is a prefix of $u'$ and $v[n + 1, \size{v}]$ is a suffix of $v'$.
    
    Furthermore, $u'[1, \size{u'} - n]$ is a prefix of $u$ and $v'[n + 1, \size{v'}]$ is a suffix of $v$.
\end{restatable}

Then we formalize the fact that if the cursor is far enough from the extremities of the word then we do not have edge effects.

\begin{restatable}[Effectiveness far from the edges]{lemma}{lemEfficienceLoinBords}%
\label{lem-efficience_loin_bords}%
    Let $t = \sigma_1 \ldots \sigma_n$ be a key, $\config{u}{v}$ a configuration and $\config{u_n}{v_n} = \config{u}{v} \cdot t$. If $n \leq \min{\size{u}, \size{v}}$, then
    $\config{u}{v} \actEff{t} \config{u_n}{v_n}$, meaning that all the arrows and backspaces are applied effectively.
\end{restatable}

The two next lemmas bound the variation in length of the configuration when applying a key.

\begin{restatable}[Bounds on the lengths]{lemma}{lemEncadrementTaille}%
\label{lem-encadrement_taille}%
    Let $t = \sigma_1 \ldots \sigma_n$ be a key, $\config{u}{v}$
    a configuration and $\config{u_n}{v_n} = \config{u}{v} \cdot t$.  Then 
    \[
        \size{uv} - \size{t}_{\retour} + \sum_{x \in A} \size{t}_x
        ~~ \leq ~~ \size{u_nv_n} ~~ \leq ~~
        \size{uv} + \sum_{x \in A} \size{t}_x.
    \]
    In particular $\abs{\enlarger \size{u_nv_n}-\size{uv}} \leq n$.
    Moreover $\abs{\enlarger \size{u_n}-\size{u}} \leq n$ and
    $\abs{\enlarger \size{v_n}-\size{v}} \leq n$.
\end{restatable}

\begin{restatable}[Length evolution without left edge effects]{lemma}{lemEgaliteTailleLoinBords}%
\label{lem-egalite_taille_loin_bords}%
Let $t = \sigma_1 \ldots \sigma_n$ be a key, $\config{u}{v}$ a configuration such that $\size{u} \geq n$. Let $\config{u_n}{v_n} = \config{u}{v} \cdot t$, then
\[
    \size{u_n v_n} = \size{uv} - \size{t}_{\retour} + \sum_{x \in A} \size{t}_x.
\]
\end{restatable}

Then, we obtain the following lemma that can be used to show that some languages 
are not recognized by a keyboard.

\begin{restatable}{lemma}{propCroissanceBornee}%
\label{prop-croissance_bornee}%
    Let $K$ be a keyboard with language $L$. Let $\suite{\ell_n}$
    be the sequence obtained by sorting the lengths of the words in $L$ by increasing order.
    Then $\suite{\ell_{n + 1} - \ell_n}$ is bounded by $3\normeInf{K}$.
\end{restatable}

\begin{example}
    The languages $\ensC{a^{n^2}}{n \in \Nat}$ and $\ensC{a^p}{p \text{ prime}}$ 
    are not recognized by a keyboard.
\end{example}

The two following lemmas will be useful when studying effective executions.

\begin{restatable}{lemma}{lemEfficienceCorrect}%
\label{lem-efficience_correct}%
    Let $t = \sigma_1 \ldots \sigma_n$ be a key such that 
    $\config{u}{v} \actEff{t} \config{u_n}{v_n}$. Then, for 
    all words $x, y$, $\config{xu}{vy} \actEff{t} \config{xu_n}{v_ny}$.
\end{restatable}

\begin{restatable}{lemma}{lemEfficienceMemeTaille}\label{lem-efficience_meme_taille}%
    Let $t = \sigma_1 \ldots \sigma_n$ be a key, $\config{u}{v}$ and 
    $\config{x}{y}$ configurations such that $\size{u} = \size{x}$ and
    $\size{v} = \size{y}$. Then $t$ acts efficiently from $\config{u}{v}$
    if and only if it acts efficiently from $\config{x}{y}$.
\end{restatable}

%% file: key_behaviour.tex
\section{Key behaviour}

This section aims at providing tools to describe the behaviour of a key. How can we formally express the intuitive fact that the $i$\ieme{} symbol of $c \cdot t$ 
was written by $t$ or that the $i$\ieme{} symbol of $c$ was moved by $t$? 
We are going to distinguish letters from $t$ and $c$ in order to keep track of where $t$ writes its letters and how the letters of $c$ were affected. 

\begin{definition}[Tracking function]
    Let $\Z_t$ and $\Z_c$ be two duplicates of $\Z$.
    We denote by $\cEnt{k}$ the elements of $\Z_c$ and by $\tEnt{k}$
    the elements of $\Z_t$.
    
    We define the tracking functions, one for keys
    $f_t \colon S^* \to \left(S \cup \Z_t\right)^*$, defined as follows:
    $f_t(\sigma_1 \ldots \sigma_n) = \sigma'_1 \ldots \sigma'_n$ where
    \[  
        \sigma'_{i} = \begin{cases}
            \tEnt{i}      & \text{if $\sigma_i \in A$}\\
            \sigma_i      & \text{otherwise}
        \end{cases}
    \]
    and one for configurations $f_c \colon \Conf{A} \to \Z^*_c \times \Z^*_c$ defined by 
    \[
        \fc{a_1 \ldots a_k}{b_1 \ldots b_j}
        = \config{\cEnt{-k} \ldots \cEnt{-1}}{\cEnt{1} \ldots \cEnt{j}}.
    \]
\end{definition}

By applying $f_t(t)$ to $f_c(c)$, we can keep track of which letters of the configuration and of the key were written, erased, or displaced, and where.
We need two copies of $\Z$ to differentiate between the symbols of $f_t(t)$ (added by the key) and $f_c(c)$ (already in the configuration).

\begin{definition}
    Let $\config{u}{v}$ be a configuration and $t$ a key. 
    We note $\config{u'}{v'} = \config{u}{v} \cdot t$ and 
    $\config{x}{y} = \fc{u}{v} \cdot f_t(t)$.
    We say that $t$ writes its $k$\ieme{} symbol at position $i$ 
    from $\config{u}{v}$ if $\config{x}{y}_i = \tEnt{k}$.
\end{definition}

\begin{remark} \label{rem-extension-touche-ecriture}%
    Let $t = \sigma_1 \ldots \sigma_n$ be a key,  $\config{u}{v}$ a 
    configuration and $1 \leq j < k \leq n$ integers. Then $t$ writes 
    its $k$\ieme{} symbol at position $i$ from $\config{u}{v}$ if and only if $\sigma_{j+1} \ldots \sigma_n$ writes its $(k - j)$\ieme{} symbol at position $i$ 
    from $\config{u}{v} \cdot \sigma_1 \ldots \sigma_j$.
    In particular, $t$ writes its $k$\ieme{} symbol at position $i$ from 
    $\config{u}{v}$ if and only if $\sigma_k \ldots \sigma_n$ writes its $1$\ier{} symbol from 
    $\config{u}{v} \cdot \sigma_1 \ldots \sigma_{k - 1}$.
\end{remark}

We defined an intuitive notion of writing the $k$\ieme{} symbol of $t$. 
In particular, if $t$ writes its $k$\ieme{} symbol in $i$\ieme{} 
position from $\config{u}{v}$, then $\config{u'}{v'}_i = t_k$, as stated below.

\begin{restatable}{proposition}{propEcriture} \label{prop-ecriture}%
    Let $t = \sigma_1 \ldots \sigma_n$ be a key, $\config{u}{v}$ a 
    configuration. We note 
    \[
    \begin{aligned}
        \config{u_n}{v_n}        &= \config{u}{v} \cdot t       \qquad&
            \config{x_n'}{y_n'}  &= \fc{u}{v} \cdot t\\
        \config{u_n'}{v_n'}      &= \config{u}{v} \cdot f_t(t)  \qquad &
            \config{x_n}{y_n}    &= \fc{u}{v} \cdot f_t(t)
    \end{aligned}
    \]
    Then $\size{u_n} = \size{x_n} = \size{u_n'} = \size{x_n'}$
    and $\size{v_n} = \size{y_n} = \size{v_n'} = \size{y_n'}$.
    And for all $a \in A$,
    \[
    \begin{aligned}
        \config{u_n}{v_n}_j = a 
        &\text{ iff  } \config{x_n}{y_n}_j = \cEnt{k} \text{ and } \config{u}{v}_k = a  &&\text{ or }
            \config{x_n}{y_n}_j = \tEnt{k} \text{ and } t_k = a\\ 
        &\text{ iff  } \config{u_n'}{v_n'}_j = a &&\text{ or } 
            \config{u_n'}{v_n'}_j = \tEnt{k} \text{ and } t_k = a\\ 
        &\text{ iff  } \config{x_n'}{y_n'}_j = \cEnt{k} \text{ and } \config{u}{v}_k = a &&\text{ or } \config{x_n'}{y_n'}_j = a\\
        &\text{ (iff $a$ already in configuration}                     && \text{ or $a$ added by $t$).}
    \end{aligned}
    \]
\end{restatable}

Tracking functions are a convenient formalism to show some results on keyboards.
Besides, they permit to take multiples points of view.

\begin{corollary} \label{cor-touche_aveugle}%
    Let $t$ be a key and $\config{u}{v}$ a configuration. 
    Then $t$ writes its $k$\ieme{} symbol at position $i$ from $\config{u}{v}$
    if and only if $\left(\config{u}{v} \cdot f_t(t)\right)_i = \tEnt{k}$.
\end{corollary}

\begin{definition}
    Let $t$ be a key, $\config{u}{v}$ a configuration. We say that $t$ writes an $a$ in $i$\ieme{} position from $\config{u}{v}$ if there exists $k$ such that $t_k = a$ and $t$ writes its $k$\ieme{} symbol
    in position $i$ from $\config{u}{v}$.
    We say that $t$ writes an $a$ from $\config{u}{v}$ if $t$ writes an $a$ in some position from $\config{u}{v}$.
\end{definition}

Then, we obtain some results, which are direct consequences of \myref{prop-ecriture}.

\begin{proposition} \label{prop-ecriture-contient}%
    If $t$ writes its $k$\ieme{} symbol in $i$\ieme{} position from $\config{u}{v}$ then  
    $\left(\config{u}{v} \cdot t\right)_i = t_k$. In particular, if $t$ writes an $a$ in
    $i$\ieme{} position from $\config{u}{v}$ then  
    $\left(\config{u}{v} \cdot t\right)_i = a$ and if $t$ writes an $a$ from 
    $\config{u}{v}$ then $\config{u}{v} \cdot t$ contains an $a$.
\end{proposition}

\begin{proposition} \label{prop-touche_aveugle}%
    Let $t$ be a key and $\config{u}{v}$, $\config{u'}{v'}$
    two configurations  such that $\size{u} = \size{u'}$ and
    $\size{v} = \size{v'}$. Then $t$ writes its $k$\ieme{} symbol 
    in $i$\ieme{} position from $\config{u}{v}$ if and only if
    $t$ writes its $k$\ieme{} symbol in $i$\ieme{} position from 
    $\config{u'}{v'}$.
    In particular, $t$ writes an $a$ in $j$\ieme{} position from $\config{u}{v}$ if and only if $t$ writes an $a$ in $j$\ieme{} position from $\config{x}{y}$, and 
    $t$ writes an $a$ from $\config{u}{v}$ if and only if $t$ writes an $a$ 
    from $\config{x}{y}$.
\end{proposition}

This proposition makes explicit the fact that keys cannot read the content of
a configuration. This leads to the following characterization.

\begin{restatable}{proposition}{propAlternativeEcrireA} \label{prop-alternative_ecrire_a}%
    Let $t$ be a key, $a \in A$ and $\config{u}{v}$ a configuration containing no $a$. Let 
    $\config{u'}{v'} = \config{u}{v} \cdot t$. $t$ writes an $a$ in position $i$ from 
    $\config{u}{v}$ if and only if $\config{u'}{v'}_i = a$.
    In particular, $t$ writes an $a$ from $\config{u}{v}$ if and only if 
    $\config{u'}{v'}$ contains an $a$.
\end{restatable}

Clearly, if the number of $a$'s in a configuration increases after applying a key then this key writes an $a$.

\begin{restatable}{proposition}{propAjouteA} \label{prop-ajoute_a}
    Let $t$ be a key and $\config{u}{v}$ a configuration.
    If $\size{\config{u}{v}}_a < \size{\config{u}{v} \cdot t}_a$,
    then $t$ writes an $a$ from $\config{u}{v}$.
\end{restatable}

Note that if a key behaves differently from two configurations, then there must be some edge effects. In what follows we focus on effective executions.

\begin{restatable}{proposition}{propEcritureEfficiente} \label{prop-ecriture_efficiente}%
    Let $t$ be a key and $\config{u}{v}$ and $\config{x}{y}$ two
    configurations on which $t$ acts effectively.
    Then $t$ writes its $k$\ieme{} symbol in $i$\ieme{} position from $\config{u}{v}$ if and only if $t$ writes its $k$\ieme{} symbol in $i$\ieme{} position from $\config{x}{y}$.
    Therefore, $t$ writes an $a$ in position $i$ from $\config{u}{v}$ if and only if $t$ writes an $a$ in position $i$ from $\config{x}{y}$.
\end{restatable}

In other words, a key always behaves the same way far from the edges of the configuration.

\begin{definition}
    Let $t$ be a key. We say that $t$ ensures an $a$ in position $i$ far from the edges if
    there exists a configuration $\config{u}{v}$ such that $t$ acts effectively
    on $\config{u}{v}$ and $t$ writes an $a$ in position $i$ from 
    $\config{u}{v}$.
\end{definition}

Then, we immediately obtain the following propositions.

\begin{proposition}
     Let $t$ be a key and $u, v \in A^*$ such that 
    $\size{u} \geq \size{t}$ and $\size{v} > \size{t}$. Then $t$ ensures an $a$ far from the edges
    if and only if $t$ writes an $a$ from $\config{u}{v}$ .
\end{proposition}

\begin{proposition} \label{cor-assure_a_loin_bord}%
    Let $t$ be a key and $u, v \in A^*$ such that 
    $\size{u} \geq \size{t}$ and $\size{v} \geq \size{t}$. 
    If $\size{\config{u}{v}}_a < \size{\config{u}{v} \cdot t}_a$,
    then $t$ ensures an $a$ far from the edges. 
\end{proposition}

\begin{proposition} \label{cor-assure_a_contient}%
    Let $t$ be a key which ensures an $a$ far from the edges
    and $\config{u}{v}$ such that $\size{u} \geq \size{t}$ and 
    $\size{v} \geq \size{t}$. Then
    $\config{u}{v} \cdot t$ contains an $a$.
\end{proposition}

%% file: characterizations.tex
\section{Characterisation of the classes}

\subsection{Languages of \REK{} (without the arrows)}

To begin, we study keyboards that do not contains any arrows. 

\subsubsection{\MK{} and \EK{}}
    $\MK$ and $\EK$ are quite easy to understand. Indeed, since a key of a
    minimal keyboard $K$ is just a word on $A$, $K \subset A^*$.
    
    \begin{remark}
    \label{rm-MKform}
        Let $K = \set{w_1, \ldots, w_n}$ be a minimal keyboard.
        Then $\langage{K} = (w_1 + \cdots + w_n)^+$.
    \end{remark}
    
    $\EK$ languages are rather similar.
    
    \begin{restatable}{lemma}{caracterisationLangageEK}%
\label{lem-caracterisation_langage_ek}%
        Let $K = (T, F)$ be a $\EK$ keyboard. Then $\langage{K} = T^*F$ and this
        regular expression can be computed in $O(\size{K})$.
\end{restatable}
    
    Thus, we can build in linear time a regular expression that recognizes $\langage{K}$.

\subsubsection{\RK{} and \REK{}}
    As a $\REK$ keyboard has no $\gauche$ operation, the right component of
    a configuration in an execution of $K \in \REK$ (starting from 
    $\config{\epsilon}{\epsilon}$) is always empty. Thus, in this part we will sometimes
    denote $u$ for the configuration $\config{u}{\epsilon}$.

Some of the expressiveness of $\REK$ comes from edge effects.
For instance, finite languages are recognized by $\RK$ keyboards.

\begin{example}
    Let $L$ be a finite language and $M = \max \ensC{\size{w}}{w \in L}$. 
    Then $L$ is recognized by $K = \ensC{\retour^M w}{w \in L}$.
\end{example}

These edge effects could make $\REK$ languages quite complex.

\begin{restatable}[Normal form]{lemma}{lemFormeNormaleREK}%
\label{lem-forme_normale_touche_rek}%
    Let $t \in S^*$ be a key from \REK{}. Then there exist $m \in \nats$ and $w \in A^*$ such that $t \sim \retour^m w$. Further, $m$ and $w$ can be computed from $t$ in polynomial time.
\end{restatable}

Using this normal form, we understand that the action of a $\REK$ key always consists
in deleting a bounded number of letters at the end of the word,
then adding a bounded number of letters. This reminds us of stacks. Following this intuition, we can easily
encode the behavior of a $\REK$ keyboard into a pushdown automaton.

However, $\REK$ is even more narrow since all languages of $\REK$ are regular.

\begin{restatable}{theorem}{thmREKInRat}%
\label{thm-rek_in_rat}%
    Let $K$ be a $\REK$ keyboard. Then, $\langage(K)$ is regular and we can build
    an NFA $\automate{K}$ recognizing $\langage{K}$ in polynomial time. 
\end{restatable}

\begin{remark}
There are several ways to prove this result. One of them is to apply the pushdown automaton construction from the proof of \myref{thm-grek_dans_alg} (presented later in the paper) in the particular case of \REK{}. The language of the \REK{} keyboard is then essentially the stack language of this automaton. A slight adaptation of the classical proof that the stack language of a pushdown automaton is rational then yields the result.

We choose to include another proof in this work, as it is elementary, not much longer than the one aforementioned, and seems more elegant to the authors.
\end{remark}

\subsection{Languages of \GREK{} (without the right arrow)}
In this section, we allow the use of $\gauche$. With this symbol, we have the possibility to move
into the word, and then erase or write letters. It opens a new complexity level.

Moreover, we provide a non-regular language of $\GREK$, hence showing that it is more expressive than $\REK$ (see \myref{thm-rek_in_rat}).

\begin{example}
    \label{ex:palindrome}
    Let $K = \set{aa\gauche, bb\gauche}$. Then, $\langage{K} = \ensC{u\miroir{u}}{u \in (a + b)^+}$,
    that is, $K$ recognizes the non-empty palindromes of even length.
\end{example}

Thus, we can represent context-free non-regular languages with
$\GREK$ (one can observe that the keyboard of \myref{ex:palindrome} is actually even in $\GK$).

However, a basic observation helps us to understand the behaviour of a key of $\GREK$:
as we do not have the symbol $\droite$, we cannot go back to the right and all
the letters to the right of the cursor are written forever. The following lemma can be proven easily by induction.

\begin{lemma}
\label{lem-suffix-grek}%
    Let $t = \sigma_1 \ldots \sigma_n$ be a sequence of atomic operations, and
    $\config{u}{v}$ a configuration. Then, $\config{u}{v} \cdot t$ is of the form
    $\config{u'}{v'v}$.
\end{lemma}

Then, we can make some assertions about a key observing its result over a configuration.
\begin{restatable}[Independence from position]{lemma}{lemInsensibilitePosition}%
\label{lem-insensibilite-position}%
Let $t$ be a key of $\GREK$ and $\config{u}{v}$ a configuration. If $t$ writes an $a$
from $\config{u}{v}$, then for all configurations $\config{u'}{v'}$,  $t$ writes an $a$ from $\config{u'}{v'}$.
\end{restatable}

Moreover, we can refine \myref{lem-suffix-grek}.

\begin{restatable}[$\GREK$ fundamental]{theorem}{ThmFondamentalGREK}
\label{thm-fundamental-grek}
    Let $t = \sigma_1 \ldots \sigma_n$ be a sequence of atomic operations, and
    $\config{u}{v}$ a configuration.
    We set $\config{x_n}{y_n} = \config{\epsilon}{\epsilon} \cdot t$.
    Then $\config{u}{v} \cdot t$ is of the form $\config{u_n x_n}{v_n v}$ with $y_n$ a subword of $v_n$ and $u_n$ a prefix of $u$.
\end{restatable}

These observations help us to better understand $\GREK$. A key observation is that we can see the left part of the configuration as a stack, which can be modified, and the right part as the fixed one, just as the prefix of a word that has been read by an automaton. We can then recognize (the mirror of) a \GREK{} language with a pushdown automaton which guesses a sequence of keys, maintains the left part of the configuration in the stack and reads the right part of the configuration.

\begin{restatable}{theorem}{thmGrekAlg}%
\label{thm-grek_dans_alg}%
Let $K$ be a $\GREK$ keyboard. Then, $\langage(K)$ is context-free and we can build
    a non-deterministic pushdown automaton $\automate{K}$ recognizing $\langage{K}$ in polynomial time. 
\end{restatable}

\subsection{Languages of \FEK{} (without backspace)}
    A third interesting class is $\FEK$, where the backspace is not allowed.
    Thus, the size of a configuration does not decrease along an execution.
    The execution of such a keyboard can therefore be easily simulated on
    a linear bounded automaton, as stated in \myref{thm-fek_contextuel}.
    
    In all the proofs of this section we will say that \emph{$t$ writes an $a$} when $t$ contains an $a$ (as we have no $\retour$ if $t$ contains an $a$ then this $a$ will not be erased when applying $t$).  
    
    \begin{restatable}{lemma}{lemCroissanceFEK}
    \label{lem-croissance-fek}%
    Let $K = (T,F)$ be a $\FEK$ keyboard.
    Let $u, v \in A^*$, let $\tau \in (T \cup F)^*$ and let $\config{u'}{v'} = \config{u}{v} \cdot \tau$. Then $uv$ is a subword of $u'v'$.
    In particular $\size{uv}\leq \size{u'v'}$.
    \end{restatable}

    \begin{restatable}{lemma}{lemBorneTouchesFEK}%
    \label{lem-borne-touches-fek}%
    Let $K = (T,F)$ be a $\FEK$ keyboard, let $w \in \langage{K}$.
    There exists an execution $\tau = t_1 \cdots t_n \in T^*F$ such that $\config{\epsilon}{\epsilon} \cdot \tau = \config{u}{v}$ with $uv=w$ and $n \leq \size{w}^2+1$. 
    \end{restatable}

    \begin{restatable}{theorem}{thmFEKContextuel} \label{thm-fek_contextuel}
    For all keyboards $K = (T,F)$ of $\FEK$ we can construct a linear bounded automaton $\Auto{K}$ of polynomial size recognizing $\langage{K}$.
    \end{restatable}

%% file: comparisons.tex
\section{Comparison of the keyboard classes}
    The characterisations that we provide give us some information about each class independently. 
    We now compare the 
    subclasses of \FREK{} in order to find out which inclusions hold between them.
    One of these inclusions, between $\FRK$ and $\FREK$, is
    especially interesting since it shows that keyboards with entry are strictly more powerful than automatic ones.
    
    We decompose our results into the following propositions. Those establish that a class is included in another if and only if that same inclusion holds between their sets of operators, except possibly for the inclusion of \EK{} and \REK{} in \FRK{}, which we do not prove or disprove.

    To start with, we show that a class containing the left arrow cannot be included in a class lacking it. This is a direct consequence of \myref{ex:palindrome} and \myref{thm-rek_in_rat} as we have a language of \GK{} which is not rational, and thus not in \REK{}.
    
\begin{proposition}
    $\GK \not \subseteq \REK$. 
\end{proposition}

We continue with the two next propositions, showing that a class containing the entry cannot be included in a class excluding it, except possibly for \FRK{}.

\begin{restatable}[$\EK \nsubseteq \GRK$]{proposition}{eknotingrk}%
\label{lem-eknotingrk}%
    $\EK$ is not included in $\GRK$.
\end{restatable}

\begin{restatable}[$\EK \nsubseteq \FK$]{proposition}{eknotinfk}%
\label{lem-eknotinfk}%
    $\EK$ is not included in $\FK$.
\end{restatable}

Then we prove that a class with $\retour$ cannot be included in a class without $\retour$. 

\begin{restatable}[$\RK \nsubseteq \FEK$]{proposition}{rknotinfek}%
\label{lem-rknotinfek}%
    $\RK$ is not included in $\FEK$.
\end{restatable}

The next proposition states that a class containing $\droite$ cannot be included in a class lacking it. 

\begin{restatable}[$\FK \nsubseteq \GREK$]{proposition}{fknotingrek}%
\label{lem-fknotingrek}%
    $\FK$ is not included in $\GREK$.
\end{restatable}

And finally we show that, except possibly for \EK{} and \REK{}, a class with entry cannot be included in a class without entry.

\begin{restatable}[$\GEK \nsubseteq \FRK$]{proposition}{geknotinfrk}%
\label{lem-geknotinfrk}%
    $\GEK$ is not included in $\FRK$.
\end{restatable}

The inclusion Hasse diagram of all subclasses of \FREK{} and traditional language classes is displayed in 
\myref{fig:hierarchy}.

\begin{figure}
\centering
    \begin{tikzpicture}
        \node (Alg)   at (0, 1.5) {$\Alg$};
        \node (GREK) at (0, 0) {$\GREK$};
        \node (Rat) at (-2.25, 0) {$\Rat$};
        \node (REK)  at (-2.25, -1.5) {$\REK$};
        \node (GEK)  at (0, -1.5) {$\GEK$};
        \node (GRK)  at (2.25, -1.5) {$\GRK$};
        \node (EK)  at  (-2.25, -3) {$\EK$};
        \node (RK)  at  (0, -3) {$\RK$};
        \node (GK)  at  (2.25, -3) {$\GK$};
        \node (MK)  at (0, -4.5) {$\MK$};
        \node (FK)  at  (4.5, -1.5) {$\FK$};
        \node (FEK)  at (2.25, 0) {$\FEK$};
        \node (FRK)  at (4.5, 0) {$\FRK$};
        \node (FREK)  at (2.25, 1.5) {$\FREK$};
        \node (Cont)  at (4.5, 1.5) {$\Cont$};
        \draw[thick, >=latex] (MK) -- (EK) -- (REK) -- (GREK) -- (GRK) -- (GK) --
        (MK) -- (RK);
        \draw[thick, >=latex] (GEK) -- (GK) -- (GRK) -- (RK) -- (REK) --
        (EK) -- (GEK) -- (GREK) -- (Alg);
        \draw[thick, >=latex] (REK) -- (Rat);
        \draw[thick, >=latex] (GK) -- (FK) -- (FRK) -- (FREK) -- (GREK);
        \draw[thick, >=latex] (GRK) -- (FRK) -- (FREK) -- (FEK);
        \draw[thick, >=latex] (GEK) -- (FEK) -- (FK);
        \draw[thick, >=latex] (FEK) -- (Cont);
        \draw[thick, >=latex] (Rat) -- (Alg);
        \draw[preaction={draw=white,line width=6pt}, thick, >=latex] (REK) -- (RK) -- (GRK);
        \draw[preaction={draw=white,line width=6pt}, thick, >=latex] (GREK) -- (GRK) -- (FRK);
        \draw[preaction={draw=white,line width=6pt}, thick, >=latex] (FREK) -- (FRK);
    \end{tikzpicture}
    \caption{Hierarchy of language classes}
    \label{fig:hierarchy}
\end{figure}

%% file: complexity.tex
\section{Complexity results}

In this section we establish some complexity upper bounds on the membership and universality problems for various keyboard classes.
The following three propositions are direct consequences of the known complexity bounds of the models which we translated keyboards into (in \myref{rm-MKform}, \myref{lem-caracterisation_langage_ek}, \myref{thm-rek_in_rat} and \myref{thm-grek_dans_alg}).

\begin{restatable}{proposition}{propEKComplexity}%
        \label{prop-ek_complexity}%
        The membership problem on $\MK$ and $\EK$ is in \textsc{PTIME}.
        The universality problem is in \textsc{PTIME} on $\MK$ and \textsc{PSPACE} on $\EK$.
\end{restatable}

The problem for $\EK$ seems simple: it amounts to deciding, given two finite sets of words $T$ and $F$, if $T^*F$ is universal. This problem may relate to the factor universality problem, proven to be PSPACE-complete~\cite{MikaS2021}. 

\begin{proposition}
    The membership problem over $\REK$ is in \textsc{PTIME}, and the universality problem over $\REK$ in \textsc{PSPACE}.
\end{proposition}

\begin{proposition}
    The membership problem over $\GREK$ is in \textsc{PTIME}.
\end{proposition}
    
For $\RK$ keyboards, we prove that there exists a word not accepted by a given $\RK$ keyboard if and only if there exists one of polynomial length. 

\begin{restatable}{proposition}{propRKUniversality}%
\label{prop-rk-universality}%
The universality problem for $\RK$ keyboards is in \textsc{coNP}.
\end{restatable}

For $\FEK$ keyboards, we know by \myref{lem-borne-touches-fek} that every word $w$ recognized by a $\FEK$ keyboard can be written with an execution of length polynomial in $\size{w}$, hence this proposition:

\begin{proposition}
    The membership problem for $\FEK$ keyboards is in \textsc{NP}.
\end{proposition}

%% file: closure.tex
\section{Closure properties}

In this section we study closure properties of keyboard classes. We selected three operators, the union and the intersection, as they are the most natural closure operators, and the mirror, under which some classes are stable.

\begin{restatable}[Mirror]{proposition}{mirrors}%
\label{lem-mirrors}%
        $\MK$, $\FK$ and $\FEK$ are stable by mirror.
    $\EK$, $\RK$, $\REK$ and $\GRK$ are not stable by mirror.
\end{restatable}
\begin{restatable}[Intersection]{proposition}{intersections}%
\label{lem-intersections}%
        None of the keyboard language classes are stable by intersection.
\end{restatable}
\begin{restatable}[Union]{proposition}{unions}%
\label{lem-unions}%
        None of the keyboard language classes are stable by union.
\end{restatable}

We end this section with an undecidability result, showing that intersecting keyboards can lead to highly complex languages. This shows another link with context-free languages as the emptiness of the intersection of two context-free languages is undecidable as well.

\begin{restatable}[Intersection emptiness problem]{proposition}{emptyintersection}%
\label{prop-empty-intersection}%
    The following problem is undecidable:   
    
    \textbf{Input:} $K_1, K_2$ two \GK{} keyboards.
    
    \textbf{Output:} Is $\langage{K_1} \cap \langage{K_2}$ empty?
\end{restatable}

%% file: conclusion.tex
\section{Conclusion}

A natural question when it comes to models of computation is what we can do without any memory or any information on the current state of the system. We initiated a line of research aiming at studying such \enquote{blind} models. The one we considered here, keyboards, proved to be mathematically complex and interestingly orthogonal in expressiveness to several of the most classical models.
We have established a number of properties of keyboards, as well as a vocabulary facilitating their study. We separated almost all classes and compared their expressiveness, thereby uncovering the lattice of their power.

\paragraph*{Future work} As keyboards are a completely new model, there are many open problems we are working on or intend to address. 
We conjecture that $\EK$ is not included in $\FRK$, but we do not have a proof. 
We also conjecture that not all $\FK$ languages are algebraic (the language generated by $\{a \droite \droite, b \gauche \gauche\}$ is a candidate as a counter-example) and that $\FREK$ does not contain all rational languages ($a^* + b^*$ being a potential counter-example).
We plan on extending the set of operations to add, for instance, a right erasing operator, symmetric to $\retour$. 
It could also be interesting to study the semantics in which we forbid non-effective operations. 
Finally, we could equip the model with states and transitions labelled with keys. We would then have more control over which keys are applied at which times, thus increasing the expressiveness of the model and facilitating its study.

%% file: 1Heuristique/heuristique.tex
\addchap{Introduction}

On s'intéresse ici à la modélisation mathématique d'un clavier défectueux, où
les différentes touches ne produisent pas forcément l'effet escompté. Par
exemple, la touche $\touche{g}$ peut écrire le mot « rapide » et la touche
$\touche{h}$ efface les trois caractères à gauche du curseur. Dans cette
situation, lorsque l'utilisateur veut écrire le mot « rap », il lui suffit
d'appuyer sur $\touche{g}$ puis sur $\touche{h}$.

Mais peut-on tout écrire ? Peut-on écrire un mot donné ? Tout dépend de notre
clavier ! Nous cherchons donc un algorithme de calcul de la suite de touche à
taper pour obtenir un mot donné. Cela paraît relativement simple à première vue
lorsque les seules opérations effectuées par les touches sont l'écriture de
lettres. Nous avons ainsi ajouté le retour arrière et les flèches
directionnelles, qui compliquent significativement le problème.

Nous nous sommes intéressés à la notion de langage de clavier,
à savoir l'ensemble des mots qui peuvent être écrits à partir d'un ensemble
fini de touches défectueuses. Les différentes classes de clavier (selon que
l'on autorise ou non le retour arrière ou les flèches directionnelles) forment
une hiérarchie de langages et les ajouts d'opérations augmentent
strictement l'expressivité des claviers.

Par exemple, nous montrons dans ce papier que les claviers sans flèche directionnelle ne peuvent exprimer que des langages rationnels. Les claviers
où la flèche droite est interdite (mais la flèche gauche est
autorisée) sont quant à eux tous algébriques.

Il semble ainsi que l’expressivité des claviers soit orthogonale à celle de
nombreux modèles connus. En effet, les claviers sont limités par leur absence
de mémoire et leur incapacité à lire, mais cette faiblesse peut être compensée
par l’ajout d’opérations élémentaires comme l’effacement ou les déplacements,
difficiles à simuler dans des modèles classiques comme les automates.

Le modèle existant le plus proche à notre connaissance est celui des automates oublieux (\textit{forgetting automata}), ainsi que définis par Jan\v car, Mr\'az et Pl\'atek \cite{JancarMrazPlatek1993}. Ceux-ci autorisent des opérations similaires à celles que nous considérons pour les claviers, mais s'en distinguent car les claviers ne peuvent pas lire et ne possèdent pas d'états. 

Nous commencerons par définir la sémantique des claviers via la notion
d'opérations élémentaires et de configurations. Nous introduirons également
deux types de clavier : manuels et automatiques. Nous donnerons ensuite
quelques propriétés fondamentales des claviers, dans le cas général ou pour des
classes plus restreintes. Enfin, nous parlerons de la hiérarchie des langages
de clavier et montrerons qu'elle est stricte.

%% file: 2Definitions/Definitions.tex
\chapter{Définitions et notations}

\section{Mots, langages et automates --- Rappels}
    Dans cette section, nous fixons les notations utilisées
    dans le papier. Soit $A$ un alphabet fini fixé.

\begin{Definition}[Mot]
    Un mot $w = a_1 \ldots a_n$ est une suite finie sur $A$. On note $w_i$ ou 
    $w[i]$ la $i$\ieme lettre de $w$. On note $\size{w} \eqDef N$ la longueur
    de $w$ et $\size{w}_a$ le nombre d'occurrences de la lettre $a$ dans $w$.
    
    On désigne par $\Tilde{w}$ le miroir de $w$. Étant donnés deux indices 
    $1 \leq i \leq j \leq n$, on note $w[i,j]$ le mot $w_i\ldots w_j$. Si 
    $i>j$ alors $w[i,j]$ désigne le mot vide.
\end{Definition}

\begin{Definition}
    Pour tout $k \in \nats$, on note $A^k$ l'ensemble des mots de longueur $k$,
    \[
        A^{\leq k} = \bigcup_{0 \leq i \leq k} A^i \text{ et }
        A^* = \bigcup_{k \in \nats} A^k.
    \]
\end{Definition}

\begin{Definition}[Distance maximale entre deux $a$]
    On définit $\distance_a(u)$ comme le nombre maximal de lettres entre
    deux $a$ dans $u$. Formellement,
    \[
        \distance_a(u) = \max\ensC{\size{w}}{awa \in \Fact{u}}.
    \]
    avec la convention $\max{\emptyset} = - \infty$.
\end{Definition}

On note $\Pref{w}$ l'ensemble des préfixes de $w$, $\Suff{w}$ l'ensemble de ses suffixes
et $\Fact{w}$ l'ensemble de ses facteurs. Un sous-mot de $w$ est un mot de la forme
$w_{i_1}\cdots w_{i_k}$ avec $i_1, \ldots, i_k$ une suite strictement croissante
d'éléments de $\set{1, \ldots, \size{w}}$, et on note $\Sumo{w}$ l'ensemble des sous-mots de $w$.

On représentera un automate fini sur $A$ par un quadruplet $(Q, \Delta, \text{Init}, \text{Fin})$ avec $Q$ un ensemble fini d'états, $\Delta : Q \times A \to 2^Q$ une fonction de transition, $\text{Init}$ et $\text{Fin}$ des ensembles d'états initiaux et finals. On étend $\Delta$ à $Q \times A^*$ en posant, pour tout $q \in Q$, $w \in A^*$ et $a\in A$, $\Delta(q, \epsilon) = \set{q}$ et $\Delta(q,wa) = \Delta(\Delta(q,w),a)$. Dans la suite on pourra autoriser les $\epsilon$-transitions pour simplifier les preuves.

On représentera un automate à pile sur $A$ par un tuple $(Q, \Gamma, \bot, \Delta, \Init, \Fin)$ avec
\begin{itemize}
    \item $Q$ un ensemble fini d'états ;
    \item $\Gamma$ un alphabet de pile fini ;
    \item $\bot \in \Gamma$ un symbole de fond de pile ;
    \item $\Delta \colon Q \times A \times (\Gamma \cup \set{-})^2 \to 2^Q$ une fonction de transition ;
    \item $\Init$ et $\Fin$ des ensembles d'états initiaux et finals.
\end{itemize}
L'acceptation se fait par état final et pile vide et les transitions de $\Delta$ sont sous la forme
\[
    s_1 \transpile{a}{\text{op}_1}{\text{op}_2} s_2
\]
avec
\begin{itemize}
    \item $\text{op}_1 = \depile \gamma$ si on dépile un symbole $\gamma \in \Gamma$, et $\text{op}_1= \rien$ si on ne dépile rien.
    
    \item $\text{op}_2 = \empile \gamma$ si on empile un symbole $\gamma \in \Gamma$, et $\text{op}_2= \rien$ si on n'empile rien.

\end{itemize}

\section{Définitions des claviers}

Notre objectif est
de modéliser l'écriture d'un mot à partir d'une liste
d'actions écrivant des symboles. 
Les caractères de $A$ sont les lettres que nous pourrons
écrire. Nous considérons également les symboles spéciaux
suivants :
\begin{itemize}
    \item le retour arrière $\retour$ ;
    \item La flèche gauche $\gauche$ ;
    \item La flèche droite $\droite$.
\end{itemize}
Nous notons alors $S$ l'ensemble de tous les symboles possibles ;
\[
    S \eqDef A \cup \set{\retour, \gauche, \droite}.
\]

\begin{Definition}[Configuration]
    Nous appelons configuration un élément $(u,v)$ de
    $A^* \times A^*$. Nous notons $\Conf(A)$ l'ensemble
    des configurations sur $A$, et $\Conf$ en l'absence
    d'ambiguïté.
\end{Definition}

\begin{Remarque}
    Intuitivement, être dans la configuration $(u,v)$ correspond à avoir $uv$
    écrit sur la zone de texte et avoir le curseur placé entre $u$ et $v$.
    Nous noterons les configurations $\config{u}{v}$ (où $|$ représente
    le curseur).
\end{Remarque}

Il peut être intéressant de regarder les lettres d'une configuration du
point de vue du curseur, ce qui nous mène à cette définition.

\begin{Definition}
    Soit $\config{u}{v}$ une configuration. On définit $\config{u}{v}_i$ par
    $\config{u}{v}_i = \miroir{u}_i$ si $i < 0$ et $v_i$ si $i > 0$.
\end{Definition}

\subsection{Opérations élémentaires}

\begin{Definition}[Opération élémentaire]
    Nous appelons opération élémentaire de $S$ un élément de $S$.
\end{Definition}

\begin{Definition}[Action]
    L'action $\config{u}{v} \cdot s$ d'une opération élémentaire $s$ sur une configuration
    $\config{u}{v}$ est définie de la manière suivante.
    \begin{gather*}
        \config{u}{v}        \cdot a        = \config{ua}{v} \text{ si } a \in A.\\
        \begin{aligned}
            \config{\epsilon}{v} \cdot \retour  = \config{\epsilon}{v} \quad &\text{ et } \quad
            \config{u'a}{v}      \cdot \retour  = \config{u'}{v} \\
            \config{\epsilon}{v} \cdot \gauche  = \config{\epsilon}{v} \quad &\text{ et } \quad
            \config{u'a}{v}      \cdot \gauche  = \config{u'}{av}\\
            \config{u}{\epsilon} \cdot \droite  = \config{u}{\epsilon} \quad &\text{ et } \quad
            \config{u}{av'}      \cdot \droite  = \config{ua}{v'}
        \end{aligned}
    \end{gather*}
    Soit $s \in S$. Nous noterons $\config{u}{v} \act{s} \config{u'}{v'}$
    pour $\config{u}{v} \cdot s = \config{u'}{v'}$.
\end{Definition}

\begin{Remarque}
    Une opération élémentaire correspond à une action de base. Ainsi,
    une opération élémentaire de $A$ écrit une lettre de $A$, $\gauche$ et $\droite$ permettent de
    se déplacer, et $\retour$ permet de supprimer une lettre.
\end{Remarque}

\begin{Remarque}
    Les touches $\retour$ et $\gauche$ ne font rien dans une configuration de la
    forme $\config{\epsilon}{v}$ (quand le curseur est sur le bord gauche du mot).
    De même, la touche $\droite$ ne fait rien dans une configuration de la
    forme $\config{u}{\epsilon}$ (quand le curseur est sur le bord droit du mot).
\end{Remarque}

\begin{Exemple}
    La suite d'opérations élémentaires $\retour$, $a$, $\droite$, $\droite$, $b$,
    permet de passer de la configuration $\config{c}{d}$ à la
    configuration $\config{adb}{\epsilon}$.
    \begin{align*}
        \config{c}{d} & \act{\retour} \config{\epsilon}{d}\\
                      & \act{a}       \config{a}{d}\\
                      & \act{\droite} \config{ad}{\epsilon}\\
                      & \act{\droite} \config{ad}{\epsilon}\\
                      & \act{b}       \config{adb}{\epsilon}.
    \end{align*}
    Cette même suite d'opérations élémentaires permet de passer de $\config{c}{de}$ à
    $\config{adeb}{\epsilon}$, avec cette fois-ci deux lettres entre $a$ et $b$.
\end{Exemple}

\begin{Remarque}
    Dans la suite, nous confondrons allègrement « action » et
    « opération élémentaire ».
\end{Remarque}

\begin{Definition}[Sémantique efficiente]
    Une sémantique alternative des opérations élémentaire, que nous appellerons
    \emph{sémantique efficiente}, consiste à interdire l'application d'un retour
    arrière ou d'une flèche gauche sur le bord gauche d'un mot, ou d'une flèche
    droite sur le bord droit.

    Formellement, on définit cette fois les opérations élémentaires comme des
    relations de réécriture sur les configurations. Pour tous
    $u,v \in A^*, a \in A$ on a :
    \begin{align*}
        \config{u}{v}   \actEff{a}        & \config{ua}{v}  \\
        \config{u'a}{v} \actEff{\retour}  & \config{u'}{v}   \\
        \config{u'a}{v} \actEff{\gauche}  & \config{u'}{av} \\
        \config{u}{av'} \actEff{\droite}  & \config{ua}{v'}
    \end{align*}
    On définit alors $\cdotEff$ par
    $\config{u}{v} \cdotEff s = \config{u'}{v'}$ si et seulement
    si $\config{u}{v} \actEff{s} \config{u'}{v'}$.
\end{Definition}

\subsection{Touches}

\begin{Definition}[Touche]
    Une touche est un élément de $S^*$ ; c'est un mot sur les opérations élémentaires
    (et donc nous pouvons également voir les touches comme des suites d'actions
    élémentaires).

    Nous notons $\T{S}$ l'ensemble des touches sur $S$. Lorsque $S$
    est donné par le contexte, nous noterons simplement $\T$ l'ensemble
    des touches.
\end{Definition}

\begin{Definition}[Action d'une touche]
    L'action d'une touche de $\T{S}$ sur une configuration est définie
    inductivement par
    \[
    \left\{
    \begin{aligned}
         \config{u}{v} \cdot \epsilon &= \config{u}{v}\\
         \config{u}{v} \cdot st       &= (\config{u}{v} \cdot s) \cdot t
    \end{aligned}
    \right.
    \]
    Nous étendons également l'écriture $\config{u}{v} \act{t} \config{u'}{v'}$
    pour $\config{u}{v} \cdot t = \config{u'}{v'}$ aux touches.
\end{Definition}

\begin{Exemple}
    L'action de la touche $\retour mi$ permet de passer
    de la configuration $\config{a}{ne}$ à la configuration
    $\config{mi}{ne}$.
\end{Exemple}

\begin{Remarque}
    Appliquer une touche revient à appliquer l'action de chaque touche
    élémentaire la composant.
\end{Remarque}

\begin{Remarque}
    Nous définissons de même l'action d'une touche pour la sémantique
    efficiente par $\config{u}{v} \cdotEff{\epsilon} = \config{u}{v}$
    et $\config{u}{v} \cdotEff st =\config{u_2}{v_2}$ si et seulement
    s'il existe $\config{u_1}{v_1}$ telle que
    \[
        \config{u}{v} \cdotEff s = \config{u_1}{v_1} \land
        \config{u_1}{v_1} \cdotEff t = \config{u_2}{v_2}.
    \]
    Nous utiliserons également la notation
    $\config{u}{v} \actEff{t} \config{u_2}{v_2}$.
\end{Remarque}

\begin{Remarque}
    Dans la suite, nous considérerons parfois des suites de touches
    $\tau = t_1 \ldots t_n \in \T{S}^*$. L'action de $\tau$,
    en sémantique classique ou en sémantique efficiente, doit alors
    être vue comme l'action de la touche correspond à la concaténation des $t_i$. Cela correspond à
    appliquer chaque touche l'une après l'autre.
\end{Remarque}

\begin{Definition}[Taille d'une touche et nombre d'occurrences]
    Soit $t$ une touche. La taille de $t$, notée $\taille{t}$
    est sa longueur en tant que mot.

    De plus, pour $s \in S$, on note $\taille{t}_s$ le nombre
    d'occurrences du symbole $s$ dans $t$.

    De même, la taille d'une configuration $\config{u}{v}$ est
    définie par $\size{\config{u}{v}} = \size{u} + \size{v}$.
\end{Definition}

\begin{Definition}[Équivalence contextuelle de touches]
    Soit $t_1, t_2 \in S^*$ deux touches quelconques. On dit que $t_1$
    et $t_2$ sont équivalentes (contextuellement), et on note $t_1 \sim t_2$,
    si, pour toute configuration $\config{u}{v}$, on a
    \[
        \config{u}{v} \cdot t_1 = \config{u}{v} \cdot t_2.
    \]

\end{Definition}

\begin{Exemple}
    Ainsi $\touche{a \retour b} \sim \touche{b}$. Mais on ne peut
    pas dire que $\touche{\gauche \droite} \sim \epsilon$,
    car ces touches ne donnent pas la même configuration à partir de
    $\config{\epsilon}{a}$.

\end{Exemple}

\subsection{Claviers automatiques}

\begin{Definition}[Clavier automatique]
    Soit $S$ un ensemble d'opérations élémentaires. Un clavier automatique sur
    $S$ est un sous-ensemble fini de $\T{S}$.
\end{Definition}

\begin{Definition}[Exécution d'un clavier automatique]
    Soit $K$ un clavier automatique et $c_0 = \config{u_0}{v_0}$ une configuration dite
    initiale. Une exécution de $K$ sur $c_0$ est une suite finie
    $\rho = \left( (t_1, c_1), \ldots, (t_{n + 1}, c_{n + 1}) \right) \in (K \times \Conf)^{n+1}$ avec $n \in \nats$
    telle que
    \[
        \forall i \in \intervE[1; n + 1], c_{i - 1} \act{t_i} c_i.
    \]
    Si cela n'est pas précisé, une exécution porte sur la configuration initiale vide
    $\config{\epsilon}{\epsilon}$.
\end{Definition}

\begin{Definition}[Reconnaissance par un clavier automatique]
    Soit $K$ un clavier automatique et $w \in A^*$. On dit que
    $w$ est reconnu par $K$ s'il existe une exécution
    de $K$ de dernière configuration $\config{u}{v}$
    telle que $uv = w$.
\end{Definition}

\begin{Exemple}\label{a2_plus}
    Soit $K_1 = \set{\touche{aa}}$ et soit $n \in \Nat^*$. $K_1$ reconnaît
    le mot $a^{2n}$ avec l'exécution
    \[
        \config{\epsilon}{\epsilon} \act{\touche{aa}}
        \config{a^2}{\epsilon} \act{\touche{aa}}
        \config{a^4}{\epsilon} \ldots \act{\touche{aa}}
        \config{a^{2n}}{\epsilon}.
    \]
\end{Exemple}

\begin{Remarque}
    Attention, on demande qu'une exécution soit non vide ; il faut donc appuyer
    sur au moins une touche du clavier. En particulier, $\epsilon$ n'est pas
    toujours reconnu par un clavier automatique.
\end{Remarque}

\begin{Definition}[Langage d'un clavier automatique]
    Soit $K$ un clavier automatique. On appelle langage de $K$, noté
    $\langage{K}$, l'ensemble des mots $w$ qui sont
    reconnus par $K$.
\end{Definition}

\begin{Exemple}
    Ainsi, $\langage{K_1} = (aa)^+$ (avec $K_1$ vu en \myref{a2_plus}).
\end{Exemple}

\subsection{Claviers manuels}
    À première vue, les claviers automatiques paraissent un peu limités.
    Il semble par exemple impossible d'obtenir un clavier de langage
    $(aa)^*a$. En effet, nous aimerions écrire deux $a$ (une touche
    $\touche{aa}$) et une touche $\touche{a}$ qui serait utilisée à
    la fin... Toutefois, les claviers n'ont pas de notion de fin de saisie.

\begin{Definition}[Clavier manuel]
    Un clavier manuel sur $S$ est un couple d'ensembles finis de touches
    $(T, F) \subset \T{S}^* \times \T{S}^*$.
    $F$ est appelé ensemble des touches finales du clavier et $T$
    ensemble des touches transientes du clavier.
\end{Definition}

\begin{Definition}[Exécution acceptante d'un clavier manuel]
    Soit $K = (T, F)$ un clavier manuel et $c_0 = \config{u_0}{v_0}$ une configuration
    dite initiale. Une exécution \emph{acceptante} de $K$ sur $c_0$ est une suite
    finie $\rho = ((t_1, c_1), \ldots, (t_{n + 1}, c_{n + 1})) \in (K \times \Conf)^{n+1}$ pour $n \in \nats$
    telle que
    \begin{itemize}
        \item pour tout $i \in \intervE[1; n + 1]$, $c_{i - 1} \act{t_i} c_i$ ;
        \item pour tout $i \in \intervE[1; n]$, $t_i \in T$ ;
        \item $t_{n + 1} \in F$.
    \end{itemize}
    Si cela n'est pas précisé, une exécution porte sur la configuration initiale vide
    $\config{\epsilon}{\epsilon}$.

    On parlera d'exécution dans le cas général lorsque $t_{n+1} \in T \cup F$.
\end{Definition}

\begin{Remarque}
    Avec les claviers manuels, nous avons rajouté la notion de fin de saisie
    à travers les touches acceptantes qui sont les touches qui mènent à
    une fin de saisie après avoir effectué leur action.
\end{Remarque}

\begin{Definition}
    Nous définissons la reconnaissance et les langages pour les claviers manuels
    de la même manière que plus tôt, pour les claviers automatiques.
\end{Definition}

\begin{Exemple}\label{ex-a2_stara}%
    Nous reconnaissons facilement le langage $(aa)^*a$ avec le clavier manuel
    \[
        K_2 = \left(\set{\touche{aa}}, \set{\touche{a}}\right).
    \]
\end{Exemple}

\begin{Theoreme}[Simulation]
    Soit $K_a$ un clavier automatique. Alors il existe un clavier
    manuel qui reconnaît le même langage que $K_a$.
\end{Theoreme}
\begin{Preuve}
    Le clavier $K_m = (K_a, K_a)$ reconnaît le même langage que $K_a$.
\end{Preuve}

Ce théorème nous permet de définir les claviers généraux en se basant uniquement
sur les claviers manuels.

\begin{Remarque}
    Une touche acceptante peut être vue comme une touche se terminant par le symbole
    « Entrée » noté $\entree$. Ceci nous permet de noter les claviers en
    utilisant un seul ensemble. Par exemple, $\set{\touche{aa}, \touche{a\entree}}$
    correspond au clavier $K_2$ vu en \myref{ex-a2_stara}.

    L'entrée peut alors être vue comme un symbole qui n'apparaît qu'à la fin
    des touches et qui permet de finir la saisie. On appelle alors les claviers
    automatiques des claviers « sans entrée », dans le sens où il n'y a pas
    besoin du symbole $\entree$ pour être accepté.
\end{Remarque}

\begin{Definition}[Clavier]
    Nous appelons clavier un clavier manuel. Nous notons $\Clav$
    l'ensemble des claviers.
\end{Definition}

\begin{Definition}[Taille d'un clavier]
    Soit $K = (T, F)$ un clavier. Nous définissons la taille de $K$ notée
    $\taille{K}$ par
    \[
    \taille{K} = \sum_{t \in T} \left(\taille{t} + 1\right) +
    \sum_{t \in F} \left(\taille{t} + 1\right)
    \]
    et nous définissons sa norme infinie par
    \[
    \normeInf{K} = \max_{t \in T \cup F} \taille{t}.
    \]
\end{Definition}

\subsection{Classes de clavier}
    Nous avons défini plusieurs symboles spéciaux qui nous permettent
    d'obtenir des claviers avec (potentiellement) plus d'expressivité.
    Il paraît raisonnable de s'intéresser aux langages que l'on
    peut reconnaître avec des claviers où certains de ces symboles
    ne sont pas disponibles.

\begin{Definition}[Clavier minimal]
    On appelle clavier minimal un clavier ne contenant aucun symbole
    spécial. Il s'agit donc d'un clavier automatique sans $\droite$,
    $\gauche$ et $\retour$. Nous notons $\MK$ l'ensemble des claviers minimaux.
\end{Definition}

\begin{Remarque}
    Nous construisons nos autres classes en partant des claviers
    minimaux auxquels on rajoute certaines possibilités de symboles
    spéciaux. Les noms des classes sont obtenus en rajoutant
    $\mathsf{R}$ (pour $\retour$), $\mathsf{E}$ (pour $\entree$,
    qu'on peut considérer, rappelons-le, comme un symbole),
    $\mathsf{G}$ (pour $\gauche$) et $\mathsf{F}$ (pour $\gauche$
    et $\droite$) à $\mathsf{K}$. Ainsi, en fonction des
    symboles spéciaux autorisés, on a ces différentes classes.

\[
    \begin{aligned}
    \MK   & : \set{}                 &\qquad
    \GK   & : \set{\gauche}          &\qquad
    \FK   & : \set{\gauche, \droite}\\
    \EK   & : \set{\entree}          &\qquad
    \GEK  & : \set{\gauche, \entree} &\qquad
    \FEK  & : \set{\gauche, \droite, \entree}\\
    \RK   & : \set{\retour}          &\qquad
    \GRK  & : \set{\gauche, \retour} &\qquad
    \FRK  & : \set{\gauche, \droite, \retour}\\
    \REK  & : \set{\retour, \entree}         &\qquad
    \GREK & : \set{\gauche, \retour, \entree} &\qquad
    \FREK & : \set{\gauche, \droite, \retour, \entree} \\
    \end{aligned}
\]
\end{Remarque}

\begin{Remarque}
    L'ajout de $\droite$ sans $\gauche$ n'apporte pas d'expressivité car, sans $\gauche$, le curseur reste toujours à droite du mot écrit.
\end{Remarque}

\begin{Remarque}
    Nous dirons souvent qu'un langage de clavier est dans une certaine classe pour indiquer qu'il est reconnu par un clavier de cette classe.
    Ainsi, nous dirons que $L$ est dans $\FK$ s'il existe $K \in \FK$ tel que $L = \langage{K}$.
\end{Remarque}

Nous définissons enfin deux problèmes de décision fondamentaux sur les claviers.

\begin{Definition}
Soit $\mathcal{K}$ une classe de claviers, nous nous intéressons
aux deux problèmes de décision suivants.
\begin{gather*}
    \problemeDecision*{Problème du mot}
                      {$K \in \mathcal{K}$ et $w \in A^*$}
                      {$w \in \langage{K}$ ?} \\
    \vphantom{2}\\
    \problemeDecision*{Problème d'universalité}
                      {$K \in \mathcal{K}$}
                      {$\langage{K} = A^*$ ?}
\end{gather*}
\end{Definition}

%% file: 3Proprietes/proprietes.tex
\chapter{Propriétés des claviers}

Avant de rentrer dans le vif du sujet et de nous attaquer
à l'étude des langages de claviers, nous établissons des
propriétés sur les différents types de claviers. Elles
faciliteront alors nos études ultérieures.

Beaucoup de ces résultats sont sont la formalisation 
d'idées naturelles et sont donc très intuitifs, avec une
preuve par induction.
    \input{3Proprietes/1_generalites}
    \input{3Proprietes/5_touche_6}
    \input{3Proprietes/2_sans_fleches}

    \input{3Proprietes/3_sans_fleche_droite}
    \input{3Proprietes/4_sans_retour}

%% file: 3Proprietes/1_generalites.tex
\section{Généralités}

Commençons par exprimer le fait qu'une touche d'un clavier ne peut effectuer que des transformations locales, et ne peut affecter que les lettres à une distance du curseur inférieure à la longueur de cette touche.

\begin{restatable}[Localité]{Lemme}{lemLocalite}%
\label{lem-localite}%
    Soient $\config{u}{v}$ une configuration et $t = \sigma_1 \ldots \sigma_n$
    une touche. En notant $\config{u_n}{v_n} = \config{u}{v} \cdot t$, on a que
    $u[1, \taille{u} - n]$ et $u_n[1, \taille{u_n} - n]$ sont des préfixes de $u_n$ 
    et de $u$ et que $v[n + 1, \taille{v}]$ et $v_n[n + 1, \taille{v_n}]$
    sont des suffixes de $v_n$ et de $v$.
\end{restatable}
\prooflink{proof-localite}

Ainsi, $t$ ne permet pas de modifier les lettres d'une configuration à une
distance plus grande que $\taille{t}$ du curseur.

Lorsqu'une touche est appliquée avec un curseur suffisamment loin des
extrémités du mot, alors son action ne souffrira pas d'« effets de bord »
tels que l'application d'un retour à l'extrémité gauche.

\begin{restatable}[Efficience loin des bords]{Lemme}{lemEfficienceLoinDesBords}%
\label{lem-efficience_loin_bords}%
    Soient $t = \sigma_1 \ldots \sigma_n$ une touche et $\config{u}{v}$ une
    configuration. Si $n \leq \min{\size{u}, \size{v}}$, alors en notant
    $\config{u_n}{v_n} = \config{u}{v} \cdot t$, on a que
    $\config{u}{v} \actEff{t} \config{u_n}{v_n}$ c'est-à-dire que l'exécution
     efficiente ne bloque pas.
\end{restatable}
\prooflink{proof-efficience_loin_bords}

De plus, une touche ne peut pas ajouter ou retirer un nombre de lettres supérieur à sa longueur.

\begin{Lemme}[Encadrement des tailles]
\label{lem-encadrement_taille}
    Soient $t = \sigma_1 \cdots \sigma_n$ une touche et 
    $\config{u}{v}$ une configuration. On pose $\config{u_n}{v_n} = \config{u}{v} \cdot t$.  Alors
    \[
        \abs{\enlarger \size{u_nv_n}-\size{uv}} \leq n,
        \abs{\enlarger \size{u_n}-\size{u}} \leq n \text{ et }
        \abs{\enlarger \size{v_n}-\size{v}} \leq n.
    \]
\end{Lemme}
\begin{Preuve}
    Chaque opération élémentaire modifie la taille de $u$ et de $v$
    d'au plus un, et une induction sur $n$ donne le résultat.
\end{Preuve}

\begin{Remarque}
    Nous pouvons être un peu plus précis en nous intéressant au nombre
    de retours de $t$.
    Soient $t = \sigma_1 \cdots \sigma_n$ une touche et $\config{u}{v}$
    une configuration. On pose $\config{u_n}{v_n} = \config{u}{v} \cdot t$.  Alors
    \[
        \size{uv} - \size{t}_{\retour} + \sum_{x \in A} \size{t}_x
        \leq \size{u_nv_n} \leq
        \size{uv} + \sum_{x \in A} \size{t}_x.
    \]
\end{Remarque}

On peut même être plus précis et combiner les lemmes précédents pour obtenir que, si le curseur est loin des extrémités du mot, la différence de longueur due à l'application d'une touche est son nombre de lettres moins son nombre de retours.

\begin{restatable}[Égalité des tailles loin des bords]{Lemme}{lemEgaliteTailleLoinDesBords}%
\label{lem-egalite_taille_loin_bords}%
Soient $t = \sigma_1 \cdots \sigma_n$ une touche
et $\config{u}{v}$ une configuration telle que $\size{u} \geq n$ et
$\size{v} \geq n$. En posant $\config{u_n}{v_n} = \config{u}{v} \cdot t$, alors
\[
    \size{u_nv_n} = \size{uv} - \size{t}_{\retour} + \sum_{x \in A} \size{t}_x.
\]
\end{restatable}
\prooflink{proof-egalite_taille_loin_bords}

Ces résultats ont pour conséquence la proposition suivante.

\begin{restatable}{Proposition}{propCroissanceDesAutomatiquesBornee}%
\label{prop-croissance_automatique_bornee}%
    Soit $K$ un clavier automatique de langage $L$. Notons $\suite{\ell_n}$
    les longueurs des mots reconnus par $L$ triés par ordre croissant.
    Alors $\suite{\ell_{n + 1} - \ell_n}$ est bornée par $\normeInf{K}$.
\end{restatable}
\prooflink{proof-croissance_automatique_bornee}

En fait, cette proposition peut s'étendre à tout clavier.

\begin{restatable}{Proposition}{propCroissanceBornee}%
\label{prop-croissance_bornee}%
    Soit $K = (T,F)$ un clavier de langage $L$. Notons $\suite{\ell_n}$
    les longueurs des mots reconnus par $L$ triés par ordre croissant.
    Alors $\suite{\ell_{n + 1} - \ell_n}$ est bornée par $3 \normeInf{K}$.
\end{restatable}
\prooflink{proof-croissance_bornee}

\begin{Remarque} \label{rem-nombes_premiers}%
    Ceci nous permet de montrer qu'on ne peut pas reconnaître
    $\ensC{a^{n^2}}{n \in \Nat}$ ou encore $\ensC{a^p}{p \text{ premier}}$
    avec un clavier.
\end{Remarque}

\begin{restatable}[Monotonie]{Lemme}{lemMonotonie}%
\label{lem-monotonie}%
    Soient $u,v,u',v' \in A^*$ et $t \in S^*$ une touche. Notons
    $\config{x}{y} = \config{u}{v} \cdot t$ et
    $\config{x'}{y'} = \config{u'}{v'} \cdot t$.
    Si $\size{u} \leq \size{u'}$ et $\size{v} \leq \size{v'}$, alors
    $\size{x} \leq \size{x'}$ et $\size{y} \leq \size{y'}$.
\end{restatable}
\prooflink{proof-monotonie}

\begin{Corollaire}
    Soient $t \in T$ et $u,v,u',v' \in A^*$ tels que $\size{u} \leq \size{u'}$
    et $\size{v} \leq \size{v'}$. Si $t$ efface une lettre de $\config{u}{v}$,
    alors $t$ efface une lettre de $\config{u'}{v'}$.
\end{Corollaire}

\begin{Preuve}
    Cela découle du fait que $\retour$ efface une lettre de $\config{u}{v}$
    si et seulement si $\size{u} > 0$.
\end{Preuve}

\begin{restatable}{Lemme}{lemEfficienceCorrect}%
\label{lem-efficience_correct}%
    Soit $t = \sigma_1 \ldots \sigma_n$ une touche telle que 
    $\config{u}{v} \actEff{t} \config{u_n}{v_n}$. Alors pour tous mots 
    $x, y$, 
    \[
        \config{xu}{vy} \actEff{t} \config{xu_n}{v_ny}.
    \]
\end{restatable}
\prooflink{proof-lem-efficience-correct}

\begin{Lemme} \label{lem-efficience_meme_taille}%
    Soient $t = \sigma_1 \ldots \sigma_n$ une touche et 
    $\config{u}{v}$ et $\config{x}{y}$
    des configurations tels que $\size{u} = \size{x}$ et 
    $\size{v} = \size{y}$. Alors $t$ agit de manière efficiente
    sur $\config{u}{v}$ si et seulement si elle agit de 
    manière efficiente sur $\config{x}{y}$.
\end{Lemme}
\begin{Preuve}
    Pour tout $i \in \intervE[1; n]$, on pose
    \[
    \left\{
    \begin{aligned}
        \config{u_i}{v_i} &= \config{u}{v} \cdot \sigma_1 \ldots \sigma_i\\
        \config{x_i}{y_i} &= \config{x}{y} \cdot \sigma_1 \ldots \sigma_i
    \end{aligned}
    \right.
    \]
    Une induction sur $i$ permet alors de montrer que pour tout 
    $i \in \intervalleEntier{1}{n}$, on a $\size{u_i} = \size{x_i}$ et 
    $\size{v_i} = \size{y_i}$, ce qui permet d'établir le résultat
    (si une action non efficiente a lieu, alors $u_i$ ou $v_i$
    est vide, donc $x_i$ ou $y_i$ est vide et la même action non 
    efficiente a lieu sur $\config{x_i}{y_i}$).
\end{Preuve}

Ce dernier lemme suit une idée selon laquelle une touche fait la même chose 
depuis deux configurations de composantes de même taille. Dans la section qui 
suit, nous allons développer cette idée assez intuitive et nous intéresser au 
comportement des touches.

%% file: 3Proprietes/5_touche_6.tex
\section{Les touches}

Le but de cette section est de fournir des outils pour décrire le comportement
d'une touche ; comment dire que le $i$\ieme caractère de $c \cdot t$ 
a été écrit par $t$ ou que le $i$\ieme caractère de $c$ a été déplacé par $t$.  
Nous allons distinguer les lettres de $t$ et de $c$ pour savoir, 
après application de $t$, où $t$ a écrit ses lettres et comment les lettres de $c$ ont été affectées.

\begin{Definition}[Fonction de marquage]
    Posons $\Z_c$ et $\Z_t$ deux copies de $\Z$, dont les éléments sont 
    respectivement notés $\cEnt{k}$ et $\tEnt{k}$.
    On appelle fonction de marquage de touches la fonction 
    $f_t \colon S^* \to \left(S \cup \Z_t\right)^*$ définie par 
    $f_t(\sigma_1 \ldots \sigma_n) = \sigma'_1 \ldots \sigma'_n$ où
    \[  
        \sigma'_{i} = \begin{cases}
            \tEnt{i}      & \text{si $\sigma_i \in A$}\\
            \sigma_i      & \text{sinon}
        \end{cases}
    \]
    et fonction de marquage de configurations la fonction 
    $f_c \colon \Conf{A} \to \Conf{\Z_c}$ définie par 
    \[
        \fc{a_1 \ldots a_k}{b_1 \ldots b_j}
        = \config{\cEnt{-k} \ldots \cEnt{-1}}{\cEnt{1} \ldots \cEnt{j}}.
    \]
\end{Definition}

En appliquant la touche $f_t(t)$ à la configuration $f_c(c)$, on peut alors 
facilement savoir quels symboles de $t$ ont été écrits et quels symboles
de $c$ ont été conservés.

\begin{Definition}
    Soient $\config{u}{v}$ une configuration, $t$ une touche et
    $k \in \Z$. Posons $\config{u'}{v'} = \config{u}{v} \cdot t$ et
    $\config{x}{y} = \fc{u}{v} \cdot f_t(t)$. Alors, on dit que
    \begin{itemize}
        \item $t$ écrit son $k$\ieme symbole en position $i$ 
        depuis $\config{u}{v}$ si $\config{x}{y}_i = \tEnt{k}$ ;
        \item $t$ déplace le $k$\ieme symbole de $\config{u}{v}$ en 
    position $i$ si $\config{x}{y}_i = \cEnt{k}$.
    \end{itemize}
\end{Definition}

\begin{Remarque} \label{rem-extension-touche-ecriture}%
    Soient $t = \sigma_1 \ldots \sigma_n$ une touche,  $\config{u}{v}$ une 
    configuration et $1 < k \leq n$ un entier. Alors $t$ écrit 
    son $k$\ieme symbole en position $i$ depuis $\config{u}{v}$ si et 
    seulement si $\sigma_2 \ldots \sigma_n$ écrit son symbole $k - 1$ en position $i$ 
    depuis $\config{u}{v} \cdot \sigma_1$.
    
    En particulier, une récurrence montre l'équivalence avec $\sigma_k \ldots \sigma_n$ 
    écrit son $1$\ier symbole depuis $\config{u}{v} \cdot \sigma_1 \ldots \sigma_{k - 1}$.
    
    De plus, $t$ écrit son $k$\ieme symbole en position $i$ depuis $\config{u}{v}$
    si et seulement si pour tout $k \leq m \leq n$, on a que $\sigma_1 \ldots \sigma_m$ 
    écrit son $k$\ieme symbole en position $i$ depuis $\config{u}{v}$.
\end{Remarque}

Bien sûr, si $t$ écrit son $k$\ieme symbole en position $i$ depuis 
$\config{u}{v}$, alors $\config{u'}{v'}_i = t_k$, ce qu'on montre,
entre autres choses, dans la proposition qui suit. 

\begin{restatable}{Proposition}{propEcriture} \label{prop-ecriture}%
    Soit $t = \sigma_1 \ldots \sigma_n$ une touche, $\config{u}{v}$ une 
    configuration. On note 
    \[
    \begin{aligned}
        \config{u_n}{v_n}        &= \config{u}{v} \cdot t       \qquad&
            \config{x_n'}{y_n'}  &= \fc{u}{v} \cdot t\\
        \config{u_n'}{v_n'}      &= \config{u}{v} \cdot f_t(t)  \qquad &
            \config{x_n}{y_n}    &= \fc{u}{v} \cdot f_t(t)
    \end{aligned}
    \]
    Alors,
    \[
        \size{u_n} = \size{x_n} = \size{u_n'} = \size{x_n'} 
        \quad \text{et} \quad 
        \size{v_n} = \size{y_n} = \size{v_n'} = \size{y_n'}
    \]
    et de plus, pour tout $a \in A$,
    \[
    \begin{aligned}
        \config{u_n}{v_n}_j = a 
        &\text{ ssi } \config{x_n}{y_n}_j = \cEnt{k} \text{ et } \config{u}{v}_k = a  &&\text{ ou }
            \config{x_n}{y_n}_j = \tEnt{k} \text{ et } t_k = a\\ 
        &\text{ ssi } \config{u_n'}{v_n'}_j = a &&\text{ ou } 
            \config{u_n'}{v_n'}_j = \tEnt{k} \text{ et } t_k = a\\ 
        &\text{ ssi } \config{x_n'}{y_n'}_j = \cEnt{k} \text{ et } \config{u}{v}_k = a &&\text{ ou } \config{x_n'}{y_n'}_j = a\\
        &\text{ (ssi il y avait un $a$}                                         && \text{ ou la touche a écrit un $a$).}
    \end{aligned}
    \]
    Dans chaque cas, on distingue le cas où le $a$ était déjà écrit et celui
    où c'est la touche qui l'a écrit.
\end{restatable}
\prooflink{proof-prop-ecriture}

\begin{Exemple}
    Considérons $u = ab$, $v = abba$ et 
    $t = \touche{\gauche \retour a \droite \droite \droite \retour b a}$.
    On a
    \begin{align*}
        f_t(t) &= 
    \gauche \retour \tEnt{3} \droite \droite \droite \retour \tEnt{8} \tEnt{9}\\
    \fc{u}{v} &= 
    \config{\cEnt{-2}\cEnt{-1}}{\cEnt{1}\cEnt{2}\cEnt{3}\cEnt{4}}.
    \end{align*}
    Donc
    \begin{align*}
        \config{u}{v} \cdot t      &= 
            \config{ababba}{a} &
        \fc{u}{v} \cdot t          &= 
            \config{a\cEnt{-1}\cEnt{1}\cEnt{2}ba}{\cEnt{4}} \\
        \config{u}{v} \cdot f_t(t) &= 
            \config{\tEnt{3}bab\tEnt{8}\tEnt{9}}{a} &
        \fc{u}{v} \cdot f_t(t)     &= 
            \config{\tEnt{3}\cEnt{-1}\cEnt{1}\cEnt{2}\tEnt{8}\tEnt{9}}{\cEnt{4}}
    \end{align*}
    Finalement, après avoir appliqué $t$, nous obtenons une configuration
    où les symboles 3, 8 et 9 de $t$ sont écrits (un $a$, un $b$ et un second 
    $a$), et où les symboles $-1$, $1$, $2$ et $4$ de
    la configuration $\config{u}{v}$ ont été conservés (mais déplacés).
\end{Exemple}

\begin{Remarque}
    Notre but ici est de fournir un cadre général pour démontrer des résultats.
    Les fonctions de marquage semblent alors pertinentes car elles permettant 
    de formaliser des notions très intuitives et de prendre plusieurs points de 
    vue reliées par la \myref{prop-ecriture}.
\end{Remarque}

La \myref{prop-ecriture} explique par exemple qu'il est équivalent de 
regarder le résultat de $f_t(t)$ sur $c$ ou sur $f_c(c)$ lorsqu'on 
s'intéresse à ce qu'a écrit la touche.

\begin{Corollaire} \label{cor-touche_aveugle}%
    Soient $t$ une touche et $\config{u}{v}$ une configuration. Alors, $t$ écrit 
    son $k$\ieme symbole en position $i$ depuis $\config{u}{v}$ si et seulement si $\left(\config{u}{v} \cdot f_t(t)\right)_i = \tEnt{k}$ et $t$ déplace le 
    $k$\ieme symbole de $\config{u}{v}$ en position $i$ si et seulement si 
    $\left(\fc{u}{v} \cdot t\right)_i = \cEnt{k}$.
\end{Corollaire}

Néanmoins, si parler de « l'écriture du $k$\ieme symbole en position $i$ » 
permet de bien formaliser les choses, c'est souvent l'écriture d'une lettre
particulière qui nous intéressera.

\begin{Definition}
    Soit $t$ une touche et $\config{u}{v}$ une configuration. On dit
    que $t$ écrit un $a$ en position $i$ depuis $\config{u}{v}$ s'il
    existe $k$ tel que $t_k = a$ et $t$ écrit son $k$\ieme symbole
    en position $i$ depuis $\config{u}{v}$.
    
    On dit que $t$ écrit un $a$ depuis $\config{u}{v}$ si $t$ écrit
    un $a$ à une certaine position depuis $\config{u}{v}$.
\end{Definition}

La généralisation faite lors de la formalisation nous permet d'obtenir
facilement certains théorèmes comme résultat de la \myref{prop-ecriture}.

\begin{Proposition} \label{prop-ecriture-contient}%
    Si $t$ écrit son $k$\ieme symbole en position $i$ depuis 
    $\config{u}{v}$, alors $\left(\config{u}{v} \cdot t\right)_i = t_k$.
    En particulier, si $t$ écrit un $a$ en position $i$ depuis $\config{u}{v}$, 
    alors  $\left(\config{u}{v} \cdot t\right)_i = a$ et si $t$ écrit un $a$ 
    depuis $\config{u}{v}$, alors $\config{u}{v} \cdot t$ contient un $a$.
\end{Proposition}

\begin{Proposition} \label{prop-touche_aveugle}%
    Soient $t$ une touche et $\config{u}{v}$ et $\config{u'}{v'}$ deux configurations 
    telles que $\size{u} = \size{v}$ et $\size{u'} = \size{v'}$. Alors $t$ écrit son 
    $k$\ieme symbole en $i$-ème position depuis $\config{u}{v}$ si et seulement si 
    $t$ écrit son $k$\ieme symbole en $i$-ème position depuis $\config{u'}{v'}$.
    En particulier, $t$ écrit un $a$ en $i$-ème position depuis $\config{u}{v}$ si et 
    seulement si $t$ écrit un $a$ en $i$-ème position depuis $\config{u'}{v'}$ et
    $t$ écrit un $a$ depuis $\config{u}{v}$ si et seulement si $t$ écrit un $a$ en 
    depuis $\config{u'}{v'}$.
    
    Notons également que $t$ déplace le $k$\ieme symbole de $\config{u}{v}$ en 
    position $i$ si et seulement si $t$ déplace le $k$\ieme symbole de 
    $\config{u'}{v'}$ en position $i$.
\end{Proposition}
\begin{Preuve}
    On a $\fc{u}{v} = \fc{u'}{v'}$, d'où le résultat.
\end{Preuve}

Ceci formalise que les touches sont aveugles au contenu des configurations.
Une manière commode de montrer qu'une touche écrit un $a$ depuis $c$ est alors
de regarder le résultat de $t$ sur une configuration $c'$ similaire à $c$ (de 
même tailles), mais sans $a$.

\begin{Proposition} \label{prop-alternative_ecrire_a}%
    Soient $t$ une touche et $\config{u}{v}$ une configuration sans $a$. 
    On pose $\config{u'}{v'} = \config{u}{v} \cdot t$. Alors $t$ écrit un $a$ 
    en position $i$ depuis $\config{u}{v}$ si et seulement si $\config{u'}{v'}_i = a$.
    En particulier, $t$ écrit un $a$ depuis $\config{u}{v}$ si et seulement si 
    $\config{u'}{v'}$ contient un $a$.
\end{Proposition}
\begin{Preuve}
    Posons $\config{x}{y} = \fc{u}{v} \cdot f_t(t)$. Par la \myref{prop-ecriture}, 
    $\config{u'}{v'}_i = a$ si et seulement si $\config{u}{v}_i = a$ ou
    $\config{x}{y}_i = \cEnt{k}$ et $t_k = a$.
    
    Puisque $\config{u}{v}_i$ ne contient pas de $a$, alors $\config{x}{y}_i = \cEnt{k}$ 
    et $t_k = a$, c'est-à-dire que $t$ écrit un $a$ depuis $\config{u}{v}$.
\end{Preuve}

De plus, si après avoir appliqué une touche, le nombre de $a$ a augmenté,
alors cette touche écrit un $a$.

\begin{Proposition} \label{prop-ajoute_a}%
    Soit $t$ une touche et $\config{u}{v}$ une configuration.
    Si $\size{\config{u}{v}}_a < \size{\config{u}{v} \cdot t}_a$
    alors $t$ écrit un $a$ depuis $\config{u}{v}$.
\end{Proposition}
\begin{Preuve}
    On note $\config{u'}{v'} = \config{u}{v} \cdot t$ et 
    $\config{x}{y} = \fc{u}{v} \cdot f_t(t)$. Par la 
    \myref{prop-ecriture}, $\config{u'}{v'}_i = a$ si et seulement si 
    $\config{x}{y}_i = \cEnt{k}$ et $\config{u}{v}_k = a$ ou 
    $\config{x}{y}_i = \tEnt{k}$ et $t_k = a$.
    
    Or $\size{\config{u'}{v'}}_a > \config{u}{v}_a$, donc il existe
    $i$ tel que $\config{x}{y}_i = \tEnt{k}$ et $t_k = a$, c'est-à-dire
    que $t$ écrit un $a$ depuis $\config{u}{v}$.
\end{Preuve}

Nous remarquons néanmoins qu'une touche n'écrit pas forcément les mêmes 
symboles depuis deux configurations différentes, et ce à cause des effets 
de bords. Mais si l'exécution est efficiente, la touche fait toujours
la même chose.

\begin{restatable}{Proposition}{propEcritureEfficiente}%
\label{prop-ecriture_efficiente}%
Soient $t$ une touche et $\config{u}{v}$ et $\config{x}{y}$ deux
    configurations sur lesquelles $t$ agit de manière efficiente.
    Alors $t$ écrit son $k$\ieme symbole en position $i$ depuis 
    $\config{u}{v}$ si et seulement si $t$ écrit son $k$\ieme
    symbole en position $i$ depuis $\config{x}{y}$. En particulier,
    $t$ écrit un $a$ en position $i$ depuis $\config{u}{v}$ si et 
    seulement si $t$ écrit un $a$ en position $i$ depuis 
    $\config{x}{y}$ et $t$ écrit un $a$ depuis $\config{u}{v}$ si et 
    seulement si $t$ écrit un $a$ depuis $\config{x}{y}$.
\end{restatable}
\prooflink{proof-prop-ecriture_efficiente}

Si l'on est suffisamment loin des bords, une touche fait donc 
toujours la même chose, ce qui motive la définition suivante.

\begin{Definition}
    Soit $t$ une touche. On dit que $t$ assure un $a$ loin des
    bords (en position $i$) s'il existe $\config{u}{v}$ sur 
    laquelle $t$ agit de manière efficiente et tel que $t$ 
    écrit un $a$ depuis (en position $i$) $\config{u}{v}$.
    
    On dit que $t$ assure un $a$ si $t$ écrit un $a$ depuis 
    n'importe quelle configuration. 
\end{Definition}

On obtient alors immédiatement ces différentes propositions.

\begin{Proposition}
    Soient $t$ une touche et $u, v \in A^*$ tels que $\size{u} \geq t$ et 
    $\size{v} \geq t$. Alors, $t$ assure un $a$ loin des bords (en position $i$)
    si et seulement si $t$ écrit un $a$ (en position $i$) depuis $\config{u}{v}$ . 
\end{Proposition}

\begin{Proposition} \label{cor-assure_a_loin_bord}%
    Soit $t$ une touche et $u, v \in A^*$ tels que 
    $\size{u} \geq t$ et $\size{v} \geq t$. 
    Si $\size{\config{u}{v}}_a < \size{\config{u}{v} \cdot t}_a$,
    alors $t$ assure un $a$ loin des bords. 
\end{Proposition}

\begin{Proposition} \label{cor-assure_a_contient}%
    Soit $t$ une touche qui assure un $a$ loin des bords
    et $\config{u}{v}$ tel que $\size{u} \geq \size{t}$ et 
    $\size{v} \geq \size{t}$. Alors
    $\config{u}{v} \cdot t$ contient un $a$.
\end{Proposition}

%% file: 3Proprietes/2_sans_fleches.tex
\section{Les claviers sans flèches}

    Dans cette sous-section, nous considérons des claviers $\REK$, à savoir des claviers où $S = A \cup \set{\retour, \entree}$. En utilisant un tel clavier, la partie
    droite d'une configuration est toujours vide
    (ce que montre le lemme qui suit).
    Ainsi, les configurations obtenues
    seront de la forme
    $\config{u}{\epsilon}$ et il nous arrivera d'utiliser
    la notation $u$ pour une telle configuration.

\begin{restatable}[Caractérisation des configurations \REK]{Lemme}
                  {lemCaracterisationConfigRek}%
\label{lem-caracterisation_config_rek}%
    Soient $K$ un clavier de $\REK$, $u \in A^*$
    et $t_1 \ldots t_n\in K^*$ Alors
    $\config{u}{\epsilon} \cdot t_1 \ldots t_n$ est
    de la forme $\config{v}{\epsilon}$ avec $v \in A^*$.
\end{restatable}
\prooflink{proof-caracterisation_config_rek}

\begin{restatable}[Forme normale d'une touche \REK]{Lemme}
                  {lemFormeNormaleToucheREK}%
\label{lem-forme_normale_touche_rek}%
    Soit $t \in S^*$ une suite d'opérations élémentaires de \REK{}. 
    Alors il existe un unique $k \in \nats$ et un unique $w \in A^*$ 
    tels que $t \sim \retour^k w$. De plus, $k$ et $w$ peuvent être 
    trouvés en temps polynomial en la taille de $t$.
\end{restatable}
\prooflink{proof-forme_normale_touche_rek}

\begin{Corollaire}[Forme normale de \REK] \label{lem-caracterisation_rek}%
    Soit $K$ un clavier de $\REK$. Alors, il existe un clavier $K_0$
    dont les touches sont de la forme $\retour^k w$ avec $w \in A^*$ et tel que
    $\langage{K_0} = \langage{K}$. De plus $K_0$ est constructible en temps polynomial.
\end{Corollaire}

\begin{Remarque}
    Pour la suite, nous supposerons que les claviers $\REK$ considérés sont
    de cette forme. Notons de plus que la construction du clavier équivalent
    peut se faire en temps polynomial, ce qui aura son importance
    lors de résultats de complexité.
\end{Remarque}

%% file: 3Proprietes/3_sans_fleche_droite.tex
\section{Les claviers sans flèche droite}

On considère ici des claviers de $\GREK$, à savoir des claviers n'utilisant pas $\droite$. On se place donc avec $S = A \cup \set{\retour, \gauche, \entree}$.

L'une des grandes idées de $\GREK$ est que l'on ne peut aller à droite et donc revenir sur un mot à droite du pointeur, ce que formalise le théorème suivant.

\begin{restatable}[fondamental de $\GREK$]{Theoreme}
                  {thmFondamentalGREK}%
\label{thm-fondamental_grek}%
    Soient $t = \sigma_1 \cdots \sigma_n$ une suite d'opérations élémentaires et
    $\config{u}{v}$ une configuration.
    Nous posons $\config{x_n}{y_n} = \config{\epsilon}{\epsilon} \cdot t$.
    Alors $\config{u}{v} \cdot t$ est de la forme $\config{u_n x_n}{v_n v}$ avec $y_n$ un sous-mot de $v_n$ et $u_n$ un préfixe de $u$.
\end{restatable}
\prooflink{proof-fondamental_grek}

Ce théorème montre notamment qu si $t$ écrit une lettre depuis la configuration vide,
alors elle l'écrit depuis n'importe quelle configuration. Nous montrons alors que
la réciproque de ce résultat est également vraie et qu'il suffit d'écrire une 
lettre depuis une configuration pour l'écrire depuis n'importe quelle autre
configuration.

\begin{restatable}[Insensibilité à la position]{Lemme}
                  {lemInsensibilitePosition}%
\label{lem-insensibilite-position}%
Soient $t$ une touche de $\GREK$ et $\config{u}{v}$ une configuration. Si
$t$ écrit un $a$ depuis $\config{u}{v}$, alors $t$ écrit un $a$ depuis toute
configuration $\config{x}{y}$.
\end{restatable}
\prooflink{proof-insensibilite-position}

%% file: 3Proprietes/4_sans_retour.tex
\section{Les claviers sans retour}

Dans cette section, nous allons étudier les langages sans retour $\FEK$.
Nous nous plaçons donc avec $S = A \cup \set{\gauche, \droite, \entree}$.

L'absence du retour arrière permet d'obtenir des propriétés très
intéressantes. 

\begin{Theoreme}[du sous-mot] \label{th-sumo}%
Soient $K$ un clavier de $\FEK$, $\tau$ une suite de touches de $K$ et $\config{u}{v}$ une configuration. En posant $\config{u_{\tau}}{v_{\tau}} =
\config{u}{v} \cdot \tau$, on a que $uv$ est un sous-mot de $u_{\tau}v_{\tau}$. En particulier, $\size{uv} \leq \size{u_{\tau}v_{\tau}}$.
\end{Theoreme}

\begin{Preuve}
    L'action d'une suite de touches étant l'action de la concaténation des
    touches de cette suite, il suffit de montrer le résultat pour les touches. Soit donc $t$ une touche, on montre le résultat par induction sur $t$. Si $t$ est vide, le
    résultat est vrai. Sinon, $t = t'\sigma$ et l'hypothèse d'induction
    nous donne que $uv$ est un sous-mot de $u_{t'}v_{t'}$ où
    $\config{u_{t'}}{v_{t'}} = \config{u}{v} \cdot t'$. On distingue
    plusieurs cas.
    \begin{itemize}
        \item Si $\sigma = a \in A$, alors $u_{t} = u_{t'}a$ et $v_{t} = v_{t'}$.
        \item Si $\sigma \in \set{\gauche, \droite}$, alors
              $u_{t}v_{t} = u_{t'}v_{t'}$.
    \end{itemize}
    Dans les deux cas, $uv$ est bien un sous-mot de $u_{t}v_{t}$
    ce qui permet de conclure.
\end{Preuve}

Nous pouvons même être plus précis en nous intéressant à chaque lettre.
Puisqu'il n'y a pas de retour, si une touche contient une lettre, alors
l'action de cette touche permet effectivement d'écrire la lettre.

\begin{Lemme}[Non effacement]\label{lem-non-eff}%
    Soient $t$ une touche, $\config{u}{v}$ une configuration
    et $a \in A$. En posant $\config{u_t}{v_t} = \config{u}{v} \cdot t$, on a
    \[
        \taille{u_tv_t}_a = \taille{uv}_a + \taille{t}_a.
    \]
\end{Lemme}

\begin{Preuve}
    On montre le résultat par induction sur $t$. Si $t$ est vide, le résultat
    est vrai. Sinon, $t = t'
    \sigma$, et l'hypothèse d'induction nous donne
    que $\taille{u_{t'}v_{t'}}_a = \taille{uv}_a + \taille{t'}_a$. On distingue
    plusieurs cas.
    \begin{itemize}
        \item Si $\sigma = a \in A$, $u_t = u_{t'}a$ et $v_t = v_{t'}$.
        \item Si $\sigma \in \set{\gauche, \droite}$, $u_{t}v_{t} = u_{t'}v_{t'}$.
    \end{itemize}
    Dans les deux cas, on a
    \begin{align*}
        \taille{u_tv_t}_a = \taille{u_{t'}v_{t'}}_a + \taille{\sigma}_a
                          = \taille{uv}_a + \taille{t'}_a+ \taille{\sigma}_a
                          = \taille{uv}_a + \taille{t}_a
    \end{align*}
    ce qui permet de conclure notre induction.
\end{Preuve}

Ce résultat nous permet alors de caractériser un peu les langages de claviers
sans retour en nous intéressant aux occurrences des lettres.

\begin{restatable}[Itération de lettre]{Lemme}
                  {lemIterationLettre}%
    \label{lem-iter-lettre}%
    Soient $L \in \FK$, $w \in L$ et $a$ une lettre apparaissant dans $w$.
    Alors il existe une suite $\suite{w_n}$ de mots de $L$ telle que
    $\forall n \in \N, \taille{w_n}_a \geqslant n$.
\end{restatable}

\begin{RecettePreuve}
Des hypothèses on déduit que le clavier contient une touche qui écrit un $a$. Il suffit alors d'itérer cette touche pour écrire autant de $a$ que souhaité, car on est assuré de ne rien effacer !
\end{RecettePreuve}
\prooflink{proof-iter-lettre}

Cette proposition n'est pas vraie telle quelle dans $\FEK$
(la touche écrivant un $a$ pourrait être finale). Cependant,
on peut l'adapter de la manière suivante.

\begin{restatable}[Itération de lettre \FEK{}]{Lemme}
                  {lemIterationLettreFEK}%
    \label{lem-iter-lettre-fek}%
Soit $K = (T, F)$ un clavier de $\FEK$. S'il existe $w \in L$ tel que
$\size{w}_a > \normeInf{K}$, alors pour tout $n \in \nats$,
il existe $w_n \in L$ tel que $\size{w_n}_a \geq n$.
\end{restatable}

\begin{RecettePreuve}
    S'il y a un mot avec plus de $\normeInf{K}$ $a$, alors
    tous les $a$ de ce mot n'ont pas pu être écrits par la touche
    acceptante d'une exécution menant à accepter ce mot.
    Il y a donc une touche transiente qui contient
    un $a$ et qu'on peut itérer avant d'appuyer sur
    une touche acceptante.
\end{RecettePreuve}
\prooflink{proof-iter-lettre-fek}

\begin{Corollaire}
    Soit $L$ un langage de $\FEK$. Il existe $N \in \nats$ tel que
    s'il existe $w \in L$ vérifiant $\size{w}_a \geq N$ alors pour tout $k \geq N$, il existe $w_k \in L$ tel que $\size{w_k}_a \geq k$.
\end{Corollaire}

\begin{Preuve}
Soit $K$ un clavier de $\FEK$ reconnaissant $L$. Le lemme précédent
montre que $N = \normeInf{K}$ convient.
\end{Preuve}

\begin{Lemme}[Distanciation minimale]\label{lem-distanciation_minimale}%
    Soit $L$ un langage de $\FEK$ et $a \in A$. S'il existe $M \in \N$ tel que
    pour tout $w \in L$, $\size{w}_a \leq M$, alors il existe $D \in \N$ tel que
    pour tout $w \in L$, $\distance_a(w) \leq D$.
\end{Lemme}
\begin{Preuve}
    On considère $K$ qui reconnaît $L$ et on prend $D = 2\normeInf{K}$.
    Soit $t$ une touche de $K$ qui contient un $a$. Si $t$ n'est pas acceptante,
    alors le mot obtenu en appuyant $M +1 $ fois sur $t$ contient au moins $M + 1$
    occurrences de $a$ (par le \myref{lem-non-eff}), ce qui est absurde.
    
    Donc toute touche contenant un $a$ est acceptante. Si $w$ écrit par $K$ 
    contient des $a$, ils ont donc tous été écrits par une seule touche $t$
    (la dernière), et le \myref{lem-localite} nous permet de conclure ;
    \[
        \distance_a(w) \leq 2\size{t} \leq 2\normeInf{K}.
    \]
\end{Preuve}

%% file: 4Langages/langages.tex
\chapter{Classification des sous-classes de \FREK}

    Ici, nous caractérisons différentes classes de clavier.
    Ainsi, nous montrerons que $\REK$ est rationnel, 
    $\GREK$ algébrique, et $\FEK$ contextuel.
    \input{4Langages/1_REK}
    \input{4Langages/2_GREK}
    \input{4Langages/3_FEK}
    \input{4Langages/complexite}

%% file: 4Langages/1_REK.tex
\section{Les claviers sans flèches}

\subsection{\MK}
Commençons par étudier les claviers minimaux.

\begin{Remarque} \label{rem-caracterisation_mk}%
    Soit $K \in \MK$. Les touches de $K$ sont des mots de $A^*$
    et on a $\langage{K} = K^+$.
\end{Remarque}

\begin{Exemple}
    Le langage du clavier $\set{\touche{ab}, \touche{bc}}$
    est $(ab + bc)^+$.
\end{Exemple}

\subsection{\EK}
Ici, on travaille avec $S = A \cup \set{\entree}$.

\begin{Remarque} \label{rem-caracterisation_langage_ek}%
    Soit $K = (T,F)$ un clavier de $\EK$. Alors on a
    $\langage{K} = T^*F$. On peut donc construire
    l'expression rationnelle correspondante en
    temps polynomial en $\taille{K}$.
\end{Remarque}

\subsection{\REK{}, la classe rationnelle}

Le but de cette section est de montrer que les langages de $\REK$
sont rationnels. On montre d'abord le résultat pour les claviers de $\RK$.

Soit $K$ un clavier de $\RK$. L'idée est que lorsqu'on exécute une
suite de touches d'un clavier de $\RK$, à tout moment on peut diviser 
le mot obtenu $w$ en deux parties :
\begin{itemize}
    \item un préfixe $v$ de lettres qui ne seront jamais effacées ;
    \item un suffixe $x$ de lettres qui seront effacées à une étape ultérieure de l'exécution.
\end{itemize}

Prenons un exemple : pour écrire le mot $abb$ avec le clavier constitué des touches $\retour^2abcc, \retour^3b, \retour^2cbc$, il nous suffit d'écrire le préfixe $ab$ grâce à $t_1$ et écrire le $b$ restant grâce à $t_2$. Le souci est que si on exécute $t_1$ puis $t_2$, $t_2$ effacera trop de lettres et on obtiendra $ab$ au lieu de $abb$. Il nous faut donc une suite de touches qui n'efface pas le $ab$, mais remplace les deux $c$ par trois lettres quelconques, pour pouvoir appliquer $t_2$. Ceci est possible en exécutant $t_3$, qui remplace $cc$ par $cbc$. La suite de touches $t_1t_3t_2$ permet donc d'écrire $abb$.

Étant donné un clavier $K$ de $\RK$, nous construisons un automate en 
nous basant sur cette idée. Ses états sont les nombres de $0$ à $\normeInf{K}$ et
être dans l'état $i$ signifie qu'il nous faut effacer $i$ lettres avant de lire
la suite du mot. Nous avons alors deux sortes de transitions.
\begin{itemize}
    \item Une transition de $i$ à $j$ étiquetée par le mot $w$ signifie qu'il
          existe une touche qui efface $i$ lettres et écrit $wx$ avec $x$ de 
          taille $j$ ($w$ est une partie du mot qu'on est en train de reconnaître
          et il nous faudra effacer $j$ caractères avant de lire la suite du mot).
    \item Une transition de $i$ à $j$ étiquetée par $\epsilon$ signifie qu'il
          existe une suite de touches nous permettant de passer de $i$ lettres 
          à effacer à $j$ lettres à effacer.
\end{itemize}

Nous présentons en \myref{fig-exemple-automate-rek} l'automate obtenu à partir
du clavier $\set{\touche{\retour abc}, \touche{\retour^4 bb}}$ en utilisant la construction présentée dans la preuve de ce lemme (nous l'avons un peu simplifié en ne gardant que les
états utiles). Son langage est $abc + bb + abb + \bigl[(ab + b)(ab)^*abb\bigr]$. 

\begin{figure}[htbp]
    \centering
    \begin{tikzpicture}[shorten >=1pt,node distance=\distanceAutomatonREK,on grid,auto] 
        \node[state] (i)   {}; 
        \node[state] (q_1) [below=of i] {$1$}; 
        \node[state, accepting] (q_0) [left=of q_1] {$0$}; 
        \node[state] (q_2) [right=of q_1] {$2$}; 
        \node[state] (q_4) [below=of q_1] {$4$}; 
        \draw[->,thick, >=latex] (0, 1.3) to (i);
        \path[->,thick, >=latex, color3, bend right = 20, above left = 0.5cm and 0.1cm] (i) edge node {$abc$} (q_0);
        \path[->,thick, >=latex, color2, dashed, bend left = 5] (i) edge node {$bb$} (q_0);
        
        \path[->,thick, >=latex, color3, bend right = 10, left = 0.5cm] (i) edge node {$ab$} (q_1);
        \path[->,thick, >=latex, color2, dashed, bend left = 10, right = 0.5cm] (i) edge node {$b$} (q_1);
        
        \path[->,thick, >=latex, color3] (i) edge node {$a$} (q_2);
        
        \path[->,thick, >=latex, color3] (q_1) edge node {$abc$} (q_0);
        \path[->,thick, >=latex, color3, out=350,in=320,looseness=4,below right = 0.1cm and 0.3cm] (q_1) edge node {$ab$} (q_1);
        \path[->,thick, >=latex, color3] (q_1) edge node {$a$} (q_2);
        
        \path[->,thick, >=latex, color2, dashed] (q_4) edge node {$bb$} (q_0);
        \path[->,thick, >=latex, color2, dashed] (q_4) edge node {$b$} (q_1);
        
        \path[->,thick, >=latex, color1, bend left = 20, dotted] (q_2) edge node {$\epsilon$} (q_4);
        \path[->,thick, >=latex, color1, dotted] (q_4) edge node {$\epsilon$} (q_2);
    \end{tikzpicture}
    \caption{Automate du clavier $\set{\touche{\retour abc}, \touche{\retour^4 bb}}$}
    \label{fig-exemple-automate-rek}
\end{figure}

Les transitions en rouge (flèches pleines) correspondent à celles qu'il 
est possible de faire avec la touche $\touche{\retour abc}$ et les bleues (tirets) 
à celles qu'on fait avec la touche $\touche{\retour^4 bb}$. Et finalement, les 
transitions noires (pointillés), elles, correspondent à utiliser une suite de touches 
pour passer de $i$ lettres à effacer à $j$ lettres à effacer.

Observons par exemple comment nous pouvons reconnaître le mot $babababb$ avec 
l'automate et le clavier. On le décompose en 
$b \cdot a \cdot b \cdot ab \cdot a \cdot bb$.

\begin{itemize}
    \item On écrit un $b$ avec la touche $\touche{\retour^4 bb}$, ce qui nous mène 
          à l'état $1$ (il faudra effacer le deuxième $b$ avant d'écrire la suite du mot).
    \item La touche $\touche{\retour abc}$ nous permet d'effacer le $b$ et d'écrire $a$,
          ce qui nous mène à l'état $2$ (il faudra effacer le $bc$ restant avant d'écrire
          la suite du mot).
    \item Une suite de touche nous permet de passer de deux lettres à effacer (le $bc$) 
          à quatre lettres à effacer, ce qui nous permet de passer à l'état $4$.
    \item La $\touche{\retour^4 bb}$ nous permet d'effacer les quatre lettres et d'écrire $b$,
          ce qui nous mène à l'état $1$ (il nous faudra effacer le $b$ restant avant d'écrire
          la suite du mot).
    \item La touche $\touche{\retour abc}$ nous permet d'effacer le $b$ et d'écrire $ab$,
          ce qui nous mène à l'état $1$ (il faudra effacer le $c$ restant avant d'écrire la
          suite du mot).
    \item La touche $\touche{\retour abc}$ nous permet d'effacer le $c$ et d'écrire $a$,
          ce qui nous mène à l'état $2$ (il faudra effacer le $bc$ restant avant d'écrire 
          la suite du mot).
    \item Une suite de touche nous permet de passer de deux lettres à effacer à quatre
          lettres à effacer, ce qui nous permet de passer à l'état $4$.
    \item La touche $\touche{\retour^4 bb}$ nous permet d'écrire le $bb$ restant, ce qui 
          nous mène à l'état $0$ où on peut accepter le mot.
\end{itemize}

Nous allons maintenant formaliser un peu tout cela. Pour commencer, nous
énonçons quelques définitions utiles pour la suite.

\begin{Definition}[Action sur les entiers]
    Soient $K$ un clavier de $\RK$ et $t = \retour^kw \in K$. On étend
    l'opération $\cdot$ aux entiers naturels en posant pour tout
    $n \in \Nat$,
    \[
        n \cdot t = \max{0, n - k} + \taille{w}.
    \]
    On étend cette définition à $K^*$ en posant
    $n \cdot \epsilon = n$ et $n \cdot (\tau t) = (n \cdot \tau) \cdot t$.
\end{Definition}
\begin{Remarque}
    L'intérêt de cette définition réside dans le fait que
    pour tous $u \in A^*$ et $\tau \in K^*$,
    on a $\taille{u \cdot \tau} = \taille{u} \cdot \tau$.
\end{Remarque}
\begin{Remarque}
    Les notations $\act{\tau}$, $\actEff{\tau}$ et $\cdotEff$
    sont introduites de manière analogue sur les entiers
    naturels.
\end{Remarque}

Nous établissons maintenant l'existence de la décomposition à l'aide 
de plusieurs lemmes. Pour commencer, considérons deux mots $u_0$ et $v$.
S'il existe $\tau$ permettant de passer de $u_0$ à $vu$ de manière
efficiente pour un certain $u$ (représentant un surplus de lettres), 
alors l'action de $\tau$ peut être décomposée.

\begin{Lemme}
\label{decompositionSansFlecheacteff}%
Soient $K$ un clavier de $\RK$ et $u_0, v \in A^*$. Soit 
$\ell \in \nats$. Il existe $\tau \in K^+$ et $u \in A^*$ tels que $\size{u} = \ell$ et $u_0 \actEff{\tau} vu$ si et seulement s'il existe un entier $k > 0$, $v_0, \ldots , v_k \in A^*, x_0, \ldots, x_k \in A^*$, $0 \leq s_1, \ldots, s_k \leq \normeInf{K}$ et $\tau_0,\tau_1, \ldots, \tau_k \in K^*$ tels que :

\begin{enumerate}
    \item $u_0 = v_0x_0$ et $v = v_0 \cdots v_k$.
    \item Pour tout $1 \leq i \leq k$, $\retour^{s_i} v_ix_i \in K$.
    \item En posant $s_{k+1} = \ell$, pour tout $0 \leq i \leq k$, $\size{x_i} \actEff{\tau_i} s_{i+1}$.
\end{enumerate}
\end{Lemme}

\begin{Preuve}
    Les $\retour^{s_i}v_ix_i$ correspondent aux touches qui écrivent une partie
    de $v$, et les $\tau_i$ correspondent aux touches qui vont modifier $x_i$ en $x_i'$
    de taille $s_{i + 1}$ (qui pourront alors être effacés par 
    $\retour^{s_{i + 1}}v_{i + 1}x_{i + 1}$).
    
    \begin{description}
        \item[$\impliedby$] On pose $\tau = \tau_0 t_1 \tau_1 \cdots t_k\tau_k$. On peut facilement vérifier que $u_0 \actEff{\tau} vu$ avec pour un $u$ tel que $\size{u}=\ell$.
        \item[$\implies$] Soit $\ell \in \nats$, supposons qu'il existe $u \in A^*$ et $\tau \in K^*$ tels que $\size{u} = \ell$ et $u_0 \actEff{\tau} vu$.
        $\tau$ est de la forme $t_1 \ldots t_n$, avec $t_i = \retour^{r_i}w_i$.
        
        Nous voulons alors trouver les touches « clés », celles qui écriront les morceaux de
        $v$. Pour tout $i \in \intervE[1; n]$, on pose 
        \begin{gather*}
            u_i = u_0 \cdot (t_1 \ldots t_i)\\
            y_i = u_0 \cdot (t_1 \ldots t_{i - 1}) \retour^{r_i}
        \end{gather*}
        Ainsi, $u_i$ est le mot obtenu après avoir appuyé sur les $i$ premières touches
        de $\tau$, $y_i$ est ce même mot sauf que l'on n'a pas écrit $w_i$ (on a juste
        appliqué les retours de la dernière touche). En particulier, $y_i$ est un préfixe
        de $u_{i - 1}$ et $u_i = y_iw_i$.
        
        On pose également
        \[
            j = \min\ensC{i}{\forall i' > i, \text{$v$ est un préfixe de $y_{i'}$}}
        \]
        avec $j = 0$ si $v$ est un préfixe de tous les $y_i$. On a alors que 
        $j$ est l'indice de la touche après laquelle le préfixe $v$ du mot
        final $vu$ est écrit et n'est plus affecté. Nous procédons par induction sur $v$.
        
        Si $j = 0$ (ce qui correspond également au cas de base de l'induction),
        alors $v$ est un préfixe de tous les $y_i$ et donc de tous les $u_i$.
        En posant $v_0=v$ et $x_0$ tel que $u_0 = v_0x_0$, on a que $x_0 \actEff{\tau} u$. 
        On pose alors $k = 0$ et $\tau_0 = \tau$ et on obtient le résultat.
        
        Sinon, si $j > 0$, on a $u_j = v z$ (si $j < n$ alors $v$ est un préfixe de $y_{j+1}$ 
        et donc de $u_j$, et si $j = n$ alors $u_j = vu$) avec $z \actEff{t_{j+1} \ldots t_{n}} u$.
        Donc, $t_j$ est en fait la dernière touche clé de la décomposition que nous recherchons
        et $t_{j + 1} \ldots t_{n}$ est la dernière séquence de touches, celle qui transforme
        $u_j = vz$ en $vu$ sans toucher à $v$. 
        
        Il existe $u'$ de taille $r_j$ telle que $u_{j - 1} = y_ju'$. Par minimalité de $j$, 
        $v$ n'est pas un préfixe de $y_j$, mais puisqu'ils sont
        tous deux préfixes de $u_j$, alors $y_j$ est un préfixe strict de $v$.
        
        Nous pouvons alors appliquer l'hypothèse d'induction avec $v' = y_j$.
        On a $u_0 \actEff{t_1...t_{j-1}} v' u'$ avec $\size{u'} = r_j$,  donc il existe 
        $k' > 0$, $v_0 \in A^*$, $v_1, \ldots, v_{k'} \in A^*, x_0, \ldots, x_{k'} \in A^*$, 
        $0 \leq s_1, \ldots, s_{k'} \leq \normeInf{K}$ et $\tau_0,\tau_1, \ldots, \tau_{k'} \in K^*$ tels que :
        \begin{enumerate}
            \item $u_0 = v_0x_0$ et $v' = v_0\cdots v_{k'}$.
            \item Pour tout $1 \leq i \leq k'$, $\retour^{s_i} v_ix_i \in K$.
            \item En posant $s_{k'+1} = \size{u'} = r_j$, pour tout $0 \leq i \leq k$, $\size{x_i} \actEff{\tau_i} s_{i+1}$.
        \end{enumerate}
        On obtient le début de la décomposition recherchée. On pose alors
        \begin{gather*}
            k = k' + 1\\
            \tau_k = t_{j+1} \ldots t_n\\
            v_k \text{ tel que } v = v'v_k\\
            s_k = \size{u'} = r_j\\
            s_{k+1} = \ell\\
            x_k = z 
        \end{gather*}
        Et on vérifie bien les trois conditions du lemme.
    \end{description}
\end{Preuve}
Maintenant, nous affinons un peu ce lemme. Nous voulons pouvoir écrire n'importe quel
mot $v$ à partir de $\epsilon$. L'idée est que la première touche qu'on applique
n'agit pas forcément de manière efficiente, mais les suivantes oui. En effet, si
on écrit $v$ à partir de $t_1 \ldots t_n$ et que $t_j$ (pour $j > 1$) 
n'agit pas de manière efficiente sur $\epsilon \cdot t_1 \cdots t_{j - 1}$, 
alors ce n'est pas la peine d'appliquer $t_1\ldots t_{j - 1}$.

\begin{Lemme}[Décomposition monotone d'une exécution de $\RK$]
\label{decompositionsansFLeche}%
Soit $K$ un clavier de $\RK$ et soit $v \in A^*$. Il existe 
$\tau \in K^+$ tel que $\epsilon \act{\tau} v$ si et seulement
s'il existe un entier $k > 0$, $v_1, \ldots , v_k \in A^*$, 
$x_1, \ldots, x_k \in A^*$, $0 \leq s_1, \ldots, s_k \leq \normeInf{K}$ et $\tau_1, \ldots, \tau_k \in K^*$ tels que :

\begin{enumerate}
    \item $v = v_1\cdots v_k$.
    \item Pour tout $1 \leq i \leq k$, $\retour^{s_i} v_ix_i \in K$.
    \item En posant $s_{k+1} = 0$, pour tout $1 \leq i \leq k$, $\size{x_i} \actEff{\tau_i} s_{i+1}$.
\end{enumerate}
En particulier, on note que $(\epsilon \cdot t_1) \actEff{\tau'} v$, où
$\tau' = t_2 \tau_2 \ldots t_n \tau_n$ et $t_1 = \retour^{s_1} v_1 x_1$.
\end{Lemme}

\begin{Preuve}
\begin{description}
    \item[$\impliedby$] On pose $\tau = \retour^{s_1} v_1x_1 \tau_1 \cdots \retour^{s_k} v_kx_k \tau_k$ et on a le résultat.
    \item[$\implies$] $\tau$ est de la forme $t_1\ldots t_n$ avec $t_i = \retour^{r_i} w_i$.
    On pose $u_0 = \epsilon$ et pour tout $i \in \intervE[1; n]$,
    \begin{gather*}
        u_i = \epsilon \cdot (t_1 \ldots t_i)\\
        y_i = \epsilon \cdot (t_1 \ldots t_{i - 1}) \retour^{r_i}
    \end{gather*}
    Soit maintenant $j$ l'indice maximal tel que $y_j = \epsilon$. 
    Alors $\epsilon \cdot (t_j \cdots t_n) = v$. De plus,
    pour tout $i > j$, $y_i \neq \epsilon$, ce qui veut dire que 
    l'action des $t_i$ pour $i > j$ est efficiente. 
    Donc $(\epsilon \cdot t_j) \actEff{t_{j + 1} \ldots t_n} v$.
    
    En appliquant le \myref{decompositionSansFlecheacteff} avec $\ell = 0$,
    on obtient le résultat.
\end{description}
\end{Preuve}

Et nous avons montré l'existence de la décomposition voulue.
On garde une petite partie du résultat dans le lemme qui suit.

\begin{Lemme}
\label{oneWastefulKey}%
Soit $v \in A^*$, soit $K \subseteq \retour^*A^*$ un clavier. Il existe 
$\tau \in K^+$ tel que $\epsilon \act{\tau} v$ si et seulement s'il existe 
$t \in K$ et $\tau' \in K^*$ tels que $(\epsilon \cdot t) \actEff{\tau'} v$.
\end{Lemme}

Pour construire l'automate que nous recherchons, il nous faut calculer 
les différents types de transition. Les transitions correspondants à 
un bout de décomposition sont faciles à calculer : pour toute touche
$t = \retour^iw$ et pour toute décomposition $w = vx$, on a une transition
de $i$ à $\size{w}$ étiquetée par $w$.

Les $\epsilon$-transitions sont un peu plus compliquées à calculer. Il 
nous faut déterminer, pour chaque $0 \leq n_1, n_2 \leq \normeInf{K}$, 
s'il existe $\tau$ qui permet de passer de $n_1$ à $n_2$ (donc de 
transformer les $x_i$ en $x_i'$).

C'est maintenant cette transformation qui nous intéresse. 
Nous allons montrer que nous pouvons calculer en temps polynomial s'il existe 
une suite de touches qui transforme, de manière efficiente un mot de taille 
$n_1$ en un mot de taille $n_2$. Notons qu'étant donné que les $x_i$ et $x_i'$
de la décomposition sont écrits et effacés par une seule touche, leurs longueurs 
ne dépassent pas $\normeInf{K}$, ainsi il nous suffit bien de nous restreindre
aux entiers inférieurs à $\normeInf{K}$.

Dans les lemmes qui suivent, nous commençons par donner des conditions 
qui permettent de passer de $x$ à $x'$ sous certaines conditions.
Le but est de distinguer plusieurs cas.

Nous allons confondre $x$ et $x'$ avec leur taille respective (ce qui nous
intéresse, c'est la taille de ces mots et pas leurs lettres).

\begin{Lemme}
\label{Frobenius}%
Soit $K$ un clavier de $\RK$. On note $E = \ensC{\abs{w} - r}{\retour^r w \in K}$
et $p = \PGCD{E}$. Soient $x$ et $x'$ deux entiers tels que $0 \leq x, x' \leq \normeInf{K}$.
S'il existe $\retour^{r_1}w_1, \retour^{r_2}w_2 \in K$ tels que
\begin{itemize}
    \item $\size{w_1} - r_1 < 0 < \size{w_2} - r_2$ ;
    \item $r_2 \leq x$ et $\size{w_1} \leq x'$;
    \item $p$ divise $x' - x$,
\end{itemize}
alors il existe une suite $\tau = t_1 \ldots t_n \in K^*$ telle que
$x \actEff{\tau} x'$.
\end{Lemme}

\begin{Preuve}
    $E$ correspond aux nombres de lettres écrits par chaque touche, négatif si la
    touche efface plus qu'elle n'écrit). On commence par séparer $E$ suivant ce
    critère. On pose
    \[
        E_+ = \ensC{i}{i \in E \text{ et } i > 0} \text{ et }
        E_- = \ensC{i}{i \in E \text{ et } i > 0}
    \]
    et $p^+$ et $p^-$ leurs PGCDs respectifs,
    \[
        p^+ = \PGCD{E_+} \text{ et } p^- = \PGCD{E_-}. 
    \]
    On a $\PGCD{p^+, p^-} = p$, donc, par
    le théorème de Bézout, il existe $\alpha, \beta \in \N$ tel
    que $\alpha p^+ - \beta p^- = p$.
    
    On considère alors $M$ un multiple de $p^+$ et de $p^-$ supérieur à $\normeInf{K}$
    et
    \[
        D = x' + (M^2 + M)(r_1 - \size{w_1}) - x - M(\size{w_2} - r_2).
    \] 
    $D$ est positif car $(M^2 + M)(r_1 - \size{w_1}) \geq M^2 + M \geq M\norme{K} + \norme{K}
    \geq M(\size{w_2} - r_2) + x$. De plus, $D$ est un multiple de $p$, donc il existe 
    $c \in \N$ tel que $D = cp$.
    
    On note alors $A = c\alpha p^+ + M^2$ et $B = c \beta p^- + M^2$. Ce sont 
    respectivement des multiples de $p^+$ et de $p^-$ et on a de plus 
    $A - B = cp = D$.
    
    Ainsi, $A \geq M^2 \geq \normeInf{K}^2 \geq \max{E_+}^2$ et de même, 
    $B \geq \max{E_-}^2$, donc par le théorème de Schur, \cite{alfonsin2005diophantine},
    il existe des coefficients $\suite[i]{a_i}[E_+]$ et 
    $\suite[i]{b_i}[E_-]$ tels que
    \[
        \sum_{i \in E_+} ia_i = A \text{ et }
        \sum_{i \in E_-} ib_i = B.
    \]
    
    Nous avons maintenant de quoi construire la suite de touches qui nous permettra
    de passer de $x$ à $x'$ de manière efficiente.
    \begin{itemize}
        \item On part de $x$. On applique $M$ fois la touche $\retour^{r_2}w_2$.
        L'exécution est bien efficiente ($r_2 \leq x$ et la touche écrit plus de lettres
        qu'elle n'en efface) et elle mène à $x + M(\size{w_2} - r_2)$.
        \item Pour tout $i \in E$, on choisit une touche $t_i = \retour^{r_i} w_i$
        telle que $i = \size{w_i} - r_i$. Pour tout $i \in E_+$, on applique $a_i$ 
        fois la touche $t_i$ (dans un ordre arbitraire). L'exécution est efficiente
        car on commence avec plus de $M$ lettres et que les touches appliquées 
        écrivent plus de lettres qu'elles n'en effacent. Et on obtient la 
        configuration $x + M(\size{w_2} - r_2) + A$.
        \item Pour tout $i \in E_-$, on applique $b_i$ fois la touche $t_i$
        (dans un ordre arbitraire). Cela mène à la configuration
        \[
            x + M(\size{w_2} - r_2) + A - B = x + M(\size{w_2} - r_2) + D
                                            = x' + (M^2 + M)(r_1 - \size{w_1}).
        \]
        La configuration finale est supérieure à $M$ et les touches appliquées écrivent 
        moins de lettres qu'elles n'en effacent, donc l'exécution est efficiente.
        \item On termine en appliquant $M + M^2$ fois la touche $\retour^{r_1}w_1$
        ce qui nous donne bien la configuration $x'$. La configuration finale est 
        supérieure à $r_1$ et la touche efface plus de lettres qu'elle n'en
        écrit, donc l'exécution est là encore efficiente.
    \end{itemize}
    Et on obtient bien une suite d'opérations $\tau \in K^*$ telle que
    $x \actEff{\tau} x'$.
\end{Preuve}

Après une condition nécessaire, nous donnons des condition nécessaires
et suffisantes dans d'autres cas particuliers.
Pour commencer, si toutes les touches négatives écrivent strictement plus que $x'$, 
alors on ne peut utiliser que des touches positives, et en particulier $x \leq x'$.

\begin{Lemme}
\label{nogoodnegative}%
Soit $K$ un clavier de $\RK$. Soient $x$ et $x'$ deux entiers tels que 
$0 \leq x, x' \leq \normeInf{K}$. On suppose que pour tout 
$\retour^{r} w \in K$ tel que $\size{w} < r$, $\size{w} > x'$. Alors les 
propositions suivantes sont équivalentes.
\begin{itemize}
    \item Il existe $\tau \in K^*$ tel que $x \actEff{\tau} x'$.
    \item Il existe $\tau = t_1 \ldots t_n \in K^*$ tel que $x \actEff{\tau} x'$
          et pour tout $i$, $x \leq x \cdot (t_1 \ldots t_i) \leq x'$.
\end{itemize}
\end{Lemme}

\begin{Preuve}
\begin{description}
    \item[$\impliedby$] Oui.
    \item[$\implies$] Supposons qu'il existe $\tau = t_1 \ldots t_n$ tel que
    $x \actEff{\tau} x'$. Raisonnons par l'absurde et supposons qu'il existe 
    $i \in \intervE[1; n]$ tel que $t_i$ soit de la forme $\retour^r w$
    avec $\size{w} < r$ (et donc, par hypothèse, $\size{w} > x'$). On
    considère le $i$ maximal vérifiant cette propriété.
    
    Comme $t_{i + 1}, \ldots, t_n$ écrivent plus de lettres qu'elles n'en effacent,
    alors $x \cdot (t_1 \ldots t_i) \leq x' < \size{w}$. Mais $x \cdot (t_1 \ldots t_i)$
    est le mot obtenu après avoir appliqué en dernière touche $t_i$, donc
    $x \cdot (t_1 \ldots t_i) \geq \size{w}$, ce qui est absurde !
    
    Donc, toutes les touches de $\tau$ ajoutent au moins autant de lettres qu'elles
    n'en effacent. La suite $n \cdot (t_1 \ldots t_i)$ est donc croissante
    ce qui nous donne l'encadrement voulu,
    \[
        \forall i \in \intervE[1; n], x \leq x \cdot (t_1 \ldots t_i) \leq x'.
    \]
\end{description}
\end{Preuve}

On fait maintenant le cas symétrique, c'est-à-dire le cas où on ne peut appliquer
aucune touche positive. Le résultat et la preuve sont donc sensiblement semblables.

\begin{Lemme}
\label{nogoodpositive}%
Soit $K$ un clavier de $\RK$. Soient $x$ et $x'$ deux entiers tels que 
$0 \leq x, x' \leq \normeInf{K}$. On suppose que pour tout 
$\retour^{r} w \in K$ tel que $\size{w} > r$, $r > x$. Alors les 
propositions suivantes sont équivalentes.
\begin{itemize}
    \item Il existe $\tau \in K^*$ tel que $x \actEff{\tau} x'$.
    \item Il existe $\tau = t_1 \ldots t_n \in K^*$ tel que $x \actEff{\tau} x'$
          et pour tout $i$, $x' \leq x \cdot (t_1 \ldots t_i) \leq x$.
\end{itemize}
\end{Lemme}

\begin{Preuve}
\begin{description}
    \item[$\impliedby$] Oui.
    \item[$\implies$] Supposons qu'il existe $\tau = t_1 \ldots t_n$ tel que
    $n \actEff{\tau} m$. Raisonnons par l'absurde et supposons qu'il existe 
    $i \in \intervE[1; n]$ tel que $t_i$ soit de la forme $\retour^r w$
    avec $\size{w} > r$ (et donc, par hypothèse, $r > x$). On
    considère le $i$ minimal vérifiant cette propriété.
    
    Comme $t_1, \ldots, t_{i - 1}$ effacent plus de lettres qu'elles n'en écrivent,
    alors $x \cdot (t_1 \ldots t_i) \leq x < r$. Mais $t_i$ efface $r$ lettres,
    ce qui contredit l'efficience de l'exécution !
    
    Donc, toutes les touches de $\tau$ effacent au moins autant de lettres qu'elles
    n'en effacent. La suite $x \cdot (t_1 \ldots t_i)$ est donc décroissante
    ce qui nous donne l'encadrement voulu,
    \[
        \forall i \in \intervE[1; n], x' \leq x \cdot (t_1 \ldots t_i) \leq x.
    \]
\end{description}
\end{Preuve}

Avec ces différents cas, nous pouvons déterminer en temps polynomial
s'il est possible de passer de $x$ à $x'$ de manière efficiente.

\begin{Lemme}
\label{Polynomialeff}%
Le problème suivant est décidable en temps polynomial.
\[
\begin{cases}
    \text{\textsc{Donnée}}:   & 
    \text{$K$ un clavier de $\RK$ et $0 \leq x, x' \leq \normeInf{K}$} \\
    \text{\textsc{Question}}: & 
    \text{Existe-t-il $\tau = t_1 \ldots t_n \in K^*$ tel que $x \actEff{\tau} x'$ ?}
\end{cases}
\]
\end{Lemme}

\begin{Preuve}
    Soient $K$ un clavier de $\RK$ et $0 \leq x, x' \leq \normeInf{K}$. On procède
    par cas.
\begin{itemize}
    \item Si pour tout $\retour^r w \in K$ tel que $\size{w} - r < 0$, $\size{w} > x'$, alors
    on construit un graphe dont les sommets sont les entiers de $0$ à $\normeInf{K}$ et tel
    qu'il y ait une transition de $s$ à $s'$ si et seulement si il existe $t \in K$ telle que 
    $s \cdot t = s'$. D'après le \myref{nogoodnegative}, il existe $\tau$ tel que 
    $x \actEff{\tau} x'$ si et seulement si il y a un chemin de $x$ à $x'$ dans ce graphe.
    \item Si pour tout $\retour^r w \in K$ tel que $0 < \size{w} - r$, $r > x$, alors
    on construit le même graphe que pour le cas précédent. D'après le \myref{nogoodpositive},
    il existe $\tau$ tel que $x \actEff{\tau} x'$ si et seulement si il y a un chemin de 
    $x$ à $x'$ dans ce graphe.
    \item S'il existe $\retour^{r_1} {w_1}, \retour^{r_2} {w_2} \in K$ tels que 
    $\size{w_1} - r_1 < 0 <\size{w_2} - r_2$, $r_2 \leq x$, $\size{w_1} \leq x'$ et 
    $p$ divise $x' - x$, alors, par le \myref{Frobenius}, il existe $\tau$ tel que 
    $x \actEff{\tau} x'$.
    \item Sinon, $p$ ne divise pas $x - x'$ : on montre par une récurrence immédiate
    sur $\tau$ que pour tout $\tau \in K^*$, $x \cdot \tau \neq x'$.
\end{itemize}
\end{Preuve}

Nous avons maintenant tous les résultats nécessaires pour construire 
un automate reconnaissant un langage de $\RK$ en suivant l'intuition donnée plus haut.

Nous nous autorisons dans la définition suivante à étiqueter les transitions avec des mots de $A^*$ (de longueur au plus $\normeInf{K}$) et pas seulement avec des lettres. 
Ceci permet une présentation plus claire et il est facile de tirer de cette construction un automate étiqueté par des lettres.

\begin{Definition}
\label{defautoRK}%
Soit $K$ un clavier de $\RK$, on définit $\Auto{K}$ l'automate fini non-déterministe 
dont les états sont les éléments de $\set{\Init} \cup \intervE[0; \normeInf{K}]$.

L'unique état initial est $\Init$, l'unique état final est $0$.

L'ensemble des transitions est 
$\Delta = \Delta_{\Init} \cup \Delta_{\text{touche}} \cup \Delta_{\tau}$ avec

\begin{align*}
    \Delta_{\Init}         &= \ensC{\Init \act{v} n}
                           {\exists m, w, \retour^m vw \in K \land \size{w} = n}\\
    \Delta_{\text{touche}} &= \ensC{n_1 \act{v} n_2}
                           {\exists w, \retour^{n_1} vw \in K \land \size{w} = n_2}\\
    \Delta_{\tau}          &= \ensC{n_1 \act{\epsilon} n_2}
                           {\exists \tau \in K^*, n_1 \actEff{\tau} n_2}
\end{align*}
\end{Definition}

\begin{Lemme}
\label{AutoEqSansFleche}%
Soit $K$ un clavier de \RK, $\Auto{K}$ et $K$ reconnaissent le même langage.
\end{Lemme}

\begin{Preuve}
Soit $v \in A^*$. Supposons que $v \in \langage{K}$, par le \myref{decompositionsansFLeche}, il existe $k>0$, $v_1, \ldots , v_k \in A^*$, 
$x_1, \ldots, x_k \in A^*$, $0 \leq s_1, \ldots, s_k \leq \normeInf{K}$ et $\tau_1, \ldots, \tau_k \in K^*$ tels que :

\begin{enumerate}
    \item $v = v_1 \cdots v_k$.
    \item Pour tout $1 \leq i \leq k$, $\retour^{s_i} v_i x_i \in K$.
    \item En posant $s_{k + 1} = 0$, pour tout $1 \leq i \leq k$, $\size{x_i} \actEff{\tau_i} s_{i + 1}$.
\end{enumerate}

On peut alors aisément constater que l'exécution $\Init \act{v_1} \size{x_1} \act{\epsilon} s_2 \act{v_2} \size{x_2} \cdots \act{v_k} \size{x_k} \act{\epsilon} 0$ accepte $v$.

À présent supposons que $v \in \langage{\Auto{K}}$, il existe une exécution acceptant $v$.

En remarquant que :
\begin{itemize}
    \item pour tout état $n \in \intervE[0; \normeInf{K}]$ il existe une transition 
    $n \act{\epsilon} n$ ;
    \item pour toutes transitions $n_1 \act{\epsilon} n_2, n_2 \act{\epsilon} n_3$, il existe une transition $n_1 \act{\epsilon} n_3$.
\end{itemize}
on voit que l'on peut supposer cette exécution de la forme

\[
    \Init \act{v_1} n_1 \act{\epsilon} n'_2 
          \act{v_2} n_2 \act{\epsilon} \cdots 
          \act{v_k} n_k \act{\epsilon} n'_{k + 1} = 0.
\]
La définition de l'automate nous permet alors de poser
\begin{itemize}
    \item pour tout $1 \leq i \leq k$, $\tau_i \in K^*$ tel que $n_i \actEff{\tau_i} n'_i$ ;
    \item pour tout $1 \leq i \leq k$, $x_i \in A^*$ et $n'_1$ tels que 
          $\retour^{n'_i}v_ix_i \in K$ et $\size{x_i} = n_i$ ;
    \item $s_i = n'_i$ pour tout $1 \leq i \leq k+1$.
\end{itemize} 
Et on obtient que $v \in \langage{K}$ en appliquant le \myref{decompositionsansFLeche}.
\end{Preuve}

\begin{Lemme} \label{PolyAutoSansfleche}%
    Soit $K$ un clavier de $\REK$, on peut construire $\Auto{K}$ en temps polynomial 
    en la taille de $K$.
\end{Lemme}

\begin{Preuve}
D'après le \myref{Polynomialeff}, pour tous $0 \leq n_1, n_2 \leq \normeInf{K}$, on 
peut déterminer en temps polynomial en $\normeInf{K}$ s'il existe $\tau \in K^*$ 
tel que $n_1 \actEff{\tau} n_2$.

En conséquence, l'automate $\Auto{K}$ peut être construit en temps polynomial.
\end{Preuve}

\begin{Theoreme}\label{thm-rk-rat}%
    Soit $K \in \RK$. Alors $\langage{K}$ est rationnel et on peut
    construire en temps polynomial un automate $\automate{K}$ non
    déterministe reconnaissant $\langage{K}$.
\end{Theoreme}

\begin{Preuve}
    Soit $K$ un clavier de $\RK$. En appliquant le \myref{lem-caracterisation_rek} 
    et le \myref{thm-rek_in_rat}, nous pouvons construire un clavier $K'$ en forme 
    normale, équivalent à $K$ et de taille polynomiale en celle de $K$, puis un 
    automate $\Auto{K'}$ de taille polynomiale en celle de $K'$ (et donc en celle
    de $K$), le tout en temps polynomial. 
    
    De plus, par \myref{AutoEqSansFleche}, $\langage{\Auto{K'}} = \langage{K'}$,
    ce qui donne le résultat.
\end{Preuve}

Nous allons à présent étendre ces résultats à $\REK$.
L'idée est assez naturelle, une exécution d'un clavier (T, F) de $\REK$, c'est
une exécution du clavier $T$ de $\RK$ suivie de l'application d'une touche de $F$.

\begin{restatable}[$\REK$ est inclus dans \Rat{}]{Theoreme}
                 {ThmREKRat}%
    \label{thm-rek_in_rat}%
Soit $K = (T, F) \in \REK$. Alors $\langage{K}$ est rationnel et on peut
construire en temps polynomial un automate $\automate{K}$ non
déterministe reconnaissant $\langage{K}$.
\end{restatable}

\begin{Preuve}
Soit $K \eqDef (T, F)$ un clavier de $\REK$. On pose $L \eqDef \langage{K}$ et 
$L_T = \langage{T}$ où $T$ est vu comme un clavier de $\RK$.

Par \myref{AutoEqSansFleche} et \myref{PolyAutoSansfleche}, 
$L_T$ est rationnel et on peut construire en temps polynomial un 
automate le reconnaissant. De plus, on remarque que 
\[
    L = \bigcup_{f \in F} L_f
\]
où $L_f \eqDef \ensC{w \cdot f}{w \in L_T}$.

Soit $f$ une touche finale de la forme $\retour^k u \entree$. 
Alors $L_f = (L_T/A^k)u + u$ où 
\[
    L_T/A^k = \ensC{w}{\exists v \in A^k, wv \in L_T}
\]
est le quotient de $L_T$ par $A^k$ et est donc rationnel.

L'union, la concaténation et le quotient par $A^k$ sont calculables en temps
polynomial, ce qui nous donne le résultat.
\end{Preuve}

\begin{Remarque}
Une construction explicite de l'automate est également possible, par
exemple en se basant sur l'automate obtenu pour le clavier $T$ de $\RK$.
\begin{itemize}
    \item L'état $0$ n'est plus final et on rajoute un état $\Fin$ qui lui est final.
    \item Pour tous état $i$ et $t_f = \retour^i w \in F$, on rajoute 
          une transition de $i$ à $\Fin$ étiquetée par $w$. Ça correspond
          à lire $vx$ (où $\size{x} = i$) avec $T$, puis appliquer $t_f$.
    \item Pour tout $t_f = \retour^r w \in F$, on rajoute une transition
          de $\Init$ à $\Fin$ étiquetée par $w$.
\end{itemize}
\end{Remarque}

%% file: 4Langages/2_GREK.tex
\section{Les claviers sans flèche droite}

Dans cette section, nous allons montrer que les langages de $\GREK$ sont
algébrique. L'idée est d'utiliser le fait que rien de ce qui est écrit à 
droite du curseur ne peut être affecté par une touche sans flèche droite.

On construit un automate à pile devinant une suite de touches et maintenant 
l'invariant « Après avoir exécuté une suite de touches $t_1\cdots t_n$, 
en posant $u$ le contenu de la pile et $v$ le mot lu par l'automate, on a $\config{\epsilon}{\epsilon} \cdot t_1 \cdots t_n = \config{u}{\miroir{v}}$ ».

Les opérations élémentaires de clavier sont faciles à traduire en transitions 
de manière à respecter cet invariant. Écrire une lettre revient à l'empiler, 
effectuer un retour arrière revient à supprimer la tête de pile, et exécuter 
une flèche gauche revient à dépiler la tête de pile et lire la lettre correspondante.

Après avoir deviné la suite de touches, l'automate dépile tous les éléments de sa pile
en les lisant. On a alors lu le miroir du mot obtenu en exécutant cette suite de touches.
Il suffit alors de construire un automate à pile reconnaissant le miroir du précédent.

\begin{Definition}
Soit $K = (T,F)$ un clavier de $\GREK$. On définit l'automate à pile $\AutoMiroir{K}$ comme suit.
\begin{itemize}
    \item Son ensemble d'états est $\Pref(T \cup F) \cup \set{\Fin}$.
    \item Son alphabet d'entrée est $A$, son alphabet de pile est $A \cup \set{\bot}$, $\bot$ étant le symbole de fond de pile.
    \item $\epsilon$ est le seul état initial.
    \item $\Fin$ est le seul état final. L'automate accepte uniquement dans l'état $\Fin$ avec une pile vide.
    \item L'ensemble des transitions est
    $\Delta = \Delta_A \cup \Delta_{\gauche} \cup \Delta_{\retour} \cup \Delta_{\text{boucle}} \cup \Delta_{\Fin}$
    avec
\begin{align*}
    \Delta_{A} =
        &\set{t \transpile{\epsilon}{\depile a}{\rien} ta \mid ta \in \Pref(T \cup F), a \in A}\\
   \Delta_{\gauche} =
        &\set{t \transpile{a}{\rien}{\empile a} t\gauche \mid t\gauche \in \Pref(T \cup F), a \in A} \cup \\
        &\set{t \transpile{\epsilon}{\depile \bot}{\empile \bot} t\gauche \mid t\gauche \in \Pref(T \cup F)}\\
    \Delta_{\retour} =
        &\set{t \transpile{\epsilon}{\rien}{\empile a} t\retour \mid t\retour \in \Pref(T \cup F), a \in A} \cup \\
        &\set{t \transpile{\epsilon}{\depile \bot}{\empile \bot} t\retour \mid t\retour \in \Pref(T \cup F)}\\
    \Delta_{\text{boucle}} =
        &\set{t \transpile{\epsilon}{\rien}{\rien} \epsilon \mid t \in T}\\
    \Delta_{\Fin} =
        &\set{t \transpile{\epsilon}{\rien}{\rien} \Fin \mid t \in F} \cup\\
        &\set{\Fin \transpile{a}{\rien}{\empile a} \Fin \mid a \in A} \cup\\
        &\set{\Fin \transpile{\epsilon}{\rien}{\empile \bot} \Fin}
\end{align*}
\end{itemize}
\end{Definition}

On utilise la notation $\etatpile{s}{u}$ pour la configuration de l'automate 
$\AutoMiroir{K}$ dans l'état $s$ avec $u$ comme contenu de pile (de bas en haut).

\begin{Lemme}\label{AutoGREK}%
    Pour tout clavier $K = (T,F)$ de $\GREK$, 
    $\langage{K} = \miroir{\langage{\AutoMiroir{K}}}$.
\end{Lemme}

\begin{Preuve}
Soient $u,v \in A^*$, soit $t = \sigma_1\cdots \sigma_k \in  T$.

Montrons que pour tout $1 \leq i \leq k$, il existe une exécution de $\AutoMiroir{K}$ lisant $\miroir{v'}$ de $\etatpile{\epsilon}{\bot u}$ dans la pile à $\etatpile{\sigma_1 \cdots \sigma_i}{\bot u'}$ si et seulement si $\config{u}{v} \cdot \sigma_1 \cdots \sigma_i = \config{u'}{v'v}$ 

On procède par induction sur $i$. Si $i =0$, alors la propriété est trivialement vraie.
Si $i>0$, commençons par observer que la seule transition vers $\sigma_1 \cdots \sigma_i$ est depuis $\sigma_1 \cdots \sigma_{i-1}$.
Alors par hypothèse d'induction en appliquant les transitions jusqu'à $\sigma_1 \cdots \sigma_{i-1}$ on lit un mot $\miroir{v''}$ et on obtient un contenu de pile $u''$ tels que $\config{u}{v} \cdot \sigma_1 \cdots \sigma_{k-1} = \config{u''}{v''v}$.

On distingue quatre cas :

\begin{itemize}
    \item Si $\sigma_i = a \in A$, alors $\config{u}{v} \cdot \sigma_1 \cdots \sigma_{i} = \config{u''a}{v''v}$. De plus en appliquant la transition de $\sigma_1 \cdots \sigma_{i-1}$ à $\sigma_1 \cdots \sigma_{i}$ on obtient un contenu de pile $u''a$ et on ne lit pas de lettre.

    \item Si $\sigma_i = \gauche$ ou $\retour$ et $u'' = \epsilon$, alors $\config{u}{v} \cdot \sigma_1 \cdots \sigma_{i} = \config{u''}{v''v}$. De plus en appliquant la transition de $\sigma_1 \cdots \sigma_{i-1}$ à $\sigma_1 \cdots \sigma_{i}$ on obtient un contenu de pile $u''$ et on ne lit pas de lettre.

    \item Si $\sigma_i = \gauche$ et $u'' = wa$, $w \in A^*,a \in A$ alors $\config{u}{v} \cdot \sigma_1 \cdots \sigma_{i} = \config{w}{av''v}$. De plus en appliquant la transition de $\sigma_1 \cdots \sigma_{i-1}$ à $\sigma_1 \cdots \sigma_{i}$ on obtient un contenu de pile $w$ et on lit la lettre $a$.

    \item Si $\sigma_i = \retour$ et $u'' = wa$, $w \in A^*,a \in A$, alors $\config{u}{v} \cdot \sigma_1 \cdots \sigma_{k} = \config{w}{v''v}$. De plus en appliquant la transition de $\sigma_1 \cdots \sigma_{i-1}$ à $\sigma_1 \cdots \sigma_{i}$ on obtient un contenu de pile $w$ et on ne lit pas de lettre.
\end{itemize}

On obtient le résultat.

De plus toutes les transitions d'un $t \in T$ vers $\epsilon$  sont des $\epsilon$-transitions sans effet sur la pile.

On en conclut que pour tous $u, v, u', v' \in A^*$, il existe une exécution de $\etatpile{\epsilon}{\bot u}$ à $\etatpile{\epsilon}{\bot u'}$ lisant $\miroir{v'}$ si et seulement si il existe $t \in T$ telle que $\config{u}{v}\cdot t = \config{u'}{v'v}$.

De même il existe une exécution de $\etatpile{\epsilon}{\bot u}$ à $\etatpile{\Fin}{\bot u'}$ lisant $\miroir{v'}$ si et seulement si il existe $t \in F$ telle que $\config{u}{v}\cdot t = \config{u'}{v'v}$.

Une induction immédiate généralise ceci à tout $\tau \in T^*$.

De plus l'automate accepte sur pile vide, et en $\Fin$ l'automate dépile et lit toute la pile puis accepte, donc un mot $w$ est accepté si et seulement si il existe $u,v$ tels que $w=\miroir{v}\miroir{u}$  et une exécution de l'automate lisant $\miroir{v}$ de $\etatpile{\epsilon}{\bot}$ à $\etatpile{\Fin}{u\bot}$.

Donc un mot $w$ est accepté si et seulement s'il existe $u,v \in A^*$ et $t_1, \ldots t_n \in T, t' \in F$ tels que $w = \miroir{v}\miroir{u}$ et $\config{u}{v} = \config{\epsilon}{\epsilon} \cdot t_1\cdots t_n t'$, donc si et seulement si $\miroir{w} \in \langage{K}$.

On a bien $\langage{\AutoMiroir{K}} = \miroir{\langage{K}}$.
\end{Preuve}

\begin{Theoreme} \label{thm-grek_dans_alg}%
Soit $L$ un langage de $\GREK$. Alors $L$ est algébrique et on peut construire en temps polynomial un automate à pile non déterministe reconnaissant $L$.
\end{Theoreme}

\begin{Remarque}
    Si l'on se restreint à des claviers de $\REK$, on se rend compte
    que l'automate à pile construit ne contient que des 
    $\epsilon$-transitions. Le mot lu est donc
    entièrement stocké sur la pile. On en déduit que le langage
    d'un clavier de $\REK$ correspond, à peu de choses près, au
    langage de pile de l'automate à pile qui le reconnaît.
    
    Le langage de pile d'un automate à pile étant rationnel,
    cette observation peut mener à une autre preuve que les 
    claviers de $\REK$ sont rationnels.
\end{Remarque}

%% file: 4Langages/3_FEK.tex
\section{Les claviers sans retour}

\begin{Theoreme} \label{th-fek_contextuel}%
    Tout langage de $\FEK$ est contextuel.
\end{Theoreme}

\begin{Preuve}
    Étant donné un clavier $K = (T, F)$ de $\FEK$, il est facile de construire 
    $\automate{K}$, un automate linéairement borné qui reconnaît $\langage{\FEK}$.
    Étant donné une entrée $w$, l'automate procède ainsi.
    \begin{itemize}
        \item Il divise le ruban en deux parties : une pour mémoriser l'entrée
        (donc de taille linéaire) et une pour simuler une exécution de $K$
        (le \myref{th-sumo} permet de la garder de taille linéaire).
        \item Il devine une séquence de touches de $T$ suivie par une touche
              de $F$ et applique leur effet à la volée. Après l'application
              de chaque touche, l'automate rejette si le mot sur la deuxième
              partie du ruban a une taille supérieure à $\size{w}$.
        \item Après avoir deviné et appliqué la touche de $F$, il compare
              le mot obtenu à $w$ et accepte si et seulement si ils sont
              égaux.
    \end{itemize}
    Et on a bien $\langage{\automate{K}} = \langage{K}$.
    
    Notons que $\automate{K}$ est de taille polynomiale en celle de $K$.
\end{Preuve}

Nous fournissons également une preuve où nous construisons une grammaire
croissante pour $K$.

\begin{Preuve}
Rappelons que les langages contextuels sont équivalents aux langages de grammaires 
croissantes (avec mot vide autorisé) \cite{carton2008}, c'est-à-dire des grammaires 
où pour toute production $\alpha \to \beta$ on a $\alpha \leq \beta$. La preuve 
consiste à construire une grammaire croissante $\mathcal{G}$ qui reconnaît
$\fraiche L\fraiche$, pour $L$ un langage de $\FEK$ et $\fraiche \not \in A$.

Soit $K = (T, F)$ un clavier de $\FEK$. On définit sur l'alphabet $A \cup \{\fraiche\}$ où $\fraiche \not\in A$,
\[
    \mathcal{G} = \left(
        \{|\}, 
        A, 
        \set{ S \to \fraiche | \fraiche} \cup P \cup P_{\epsilon} \cup P_F, 
        S
    \right)
\]
une grammaire croissante. Les $\fraiche$ représentent le début et la fin du mot. Dans cette construction, ils ne sont pas enlevés à la toute fin. 

On prend $P_{\epsilon} = \set{S \to \fraiche \fraiche}$ s'il existe une touche acceptante 
$t_F \in F$ qui ne contient pas de symbole de $A$, et $P_{\epsilon} = \emptyset$ sinon. 
Il reste à déterminer $P$ et $P_F$.

L'idée est de définir, pour chaque touche $t$, une règle qui déplace le curseur $|$ et 
ajoute des symboles de $A$, en définissant l'action de chaque lettre de $t$. Pour 
construire l'action du mot entier, on unifie les actions de ses lettres.

On se munit de variables de mot $\alpha, \beta, \ldots \in \mathcal{V}$ (infini dénombrable).
Soit $U$ l'union des ensembles de couples suivants :

\begin{align*}
    U_a &= \set{ (\alpha | \beta, \alpha a | \beta )} \text{ pour tout $a \in A$}\\
    U_{\gauche} &= \set{ (\fraiche | \beta, \fraiche | \beta) } \cup
                   \ensCtq{ \alpha a | \beta, \alpha | a \beta)}{a \in A} \\
    U_{\droite} &= \set{ (\alpha | \fraiche, \alpha | \fraiche) } \cup
                   \ensCtq{ \alpha | a \beta, \alpha a | \beta)}{a \in A}.
\end{align*}

Pour deux couples $q_1 = (u_1, v_1)$ et $q_2 = (u_2, v_2) \in U$ (alpha-renommés différemment), 
on définit l'unification de $q_1 \sqcup q_2$ par $\text{échec}$ si l'unification de $v_1$ et $u_2$ 
est $\text{échec}$, et sinon par $(u_1 \sigma, v_2 \sigma)$, où $\sigma$ est la substitution 
telle que $v_1 \sigma = u_1 \sigma$.

On pose $U_{\epsilon} = \emptyset$ pour toute touche $t = \sigma_1 \ldots \sigma_n$, 
$U_t = (...(U_{\sigma_1} \star U_{\sigma_2}) \star ... ) \star U_{\sigma_n}$, où la composition\footnote{En fait, $\star$ est associative.} $\star$ est définie par

\[
    U_{\sigma} \star U_{\sigma'} = \ensC{ q \sqcup q' \neq \text{échec}}
                                        {q \in U_\sigma, q' \in U_{\sigma'}}.
\]

À un ensemble $X$ de couples, on peut maintenant associer l'ensemble des productions 
$P(X) = \ensC{u\sigma_{\epsilon} \to v\sigma_{\epsilon}}{(u,v) \in X }$, où 
$\sigma_{\epsilon}$ est la substitution qui à toute variable de mot associe 
$\epsilon$. On définit alors $P = \bigcup_{t \in T} P(U_t)$.

Les couples des $U_s$ pour $s \in A \cup \{\gauche, \droite\}$ sont croissants.
De plus, il est facile de voir que l'unification conserve cette croissance, 
puisque les substitutions la conserve (car chaque variable est présente dans 
la composante de gauche si et seulement si elle est présente dans la composante 
de droite du couple). Ainsi, les productions de $P$ sont bien toutes croissantes.

Pour $P_F$, créé à partir des touches acceptantes, il faut s'assurer de ne pas 
avoir de production où effacer le curseur $|$ vient faire décroître la production.
L'idée est donc de concaténer la production avec des productions strictement 
croissantes, d'où $P_F = \bigcup_{t_F \in F} P(U'_{t_F})$ où

\[
    U'_{t_F} =
        \bigcup_{
            \substack{t \in T \\ \exists a \in A, a \in t}
        }
        U_t \star (U_{t_F} \star \{ (\alpha | \beta, \alpha \beta) \}).
\]

Notons que la taille de $P$ est bornée par $1 + \bigl(\size{T} + \size{F}\bigr) \size{A}^{2 \normeInf{K}}$.

On voit bien que $\langage{\mathcal{G}} = \fraiche L \fraiche$.

\end{Preuve}

%% file: 4Langages/complexite.tex
\section{Quelques résultats de décidabilité}

Les résultats que nous avons démontré jusqu'ici nous permettent d'obtenir
quelques résultats sur les problèmes de mot et d'universalité de certaines
classes de clavier. Nous les exposons ici, ainsi que d'autres résultats
de décidabilité.

\subsection{Le problème du mot}

Pour commencer les différentes caractérisations obtenues nous donnent des
bornes de complexité pour le problème du mot dans la plupart
des classes de clavier.

\begin{Theoreme}
    Le problème du mot est décidable en temps polynomial
    sur les claviers $\REK$.
\end{Theoreme}
\begin{Preuve}
    Par le \myref{thm-rek_in_rat}.
\end{Preuve}

\begin{Theoreme}
    Le problème du mot est décidable en temps polynomial
    sur les claviers $\GREK$.
\end{Theoreme}
\begin{Preuve}
    Par le \myref{thm-grek_dans_alg}.
\end{Preuve}

Le \myref{th-fek_contextuel} nous permet de montrer
que le problème du mot est dans PSPACE. Pour améliorer
cette borne, nous allons montrer que si un mot de taille 
$n$ est reconnu par un clavier de $\FEK$, alors il existe
une exécution de taille polynomiale qui mène à ce mot.

\begin{Lemme} \label{lem-borne_touches_fek}%
    Soit $K = (T, F)$ un clavier de $\FEK$ et $w \in \langage{K}$.
    Alors, il existe une exécution $\tau = t_1 \cdots t_n \in T^*F$
    telle que $n \leq \size{w}^2 + 1$ et 
    $\config{\epsilon}{\epsilon} \cdot \tau = \config{u}{v}$.
\end{Lemme}
\begin{Preuve}
    Soit $u, v \in A^*$ et $\tau = t_1 \cdots t_n \in T^*F$ telle que $w = uv$ et 
    $\config{\epsilon}{\epsilon} \cdot \tau = \config{u}{v}$. Nous allons montrer
    que si $n > \size{w}^>2 + 1$, alors il existe une exécution plus courte qui
    mène à $\config{u}{v}$. Supposons donc $n > \size{w}^2 + 1$. Pour tout 
    $0 \leq i \leq n$, on pose
    \[
        \config{u_i}{v_i} = \config{\epsilon}{\epsilon} \cdot t_1 \cdots t_i
        \quad \text{ et } \quad 
        \ell_i = \size{u_iv_i}.
    \]
    Par le \myref{th-sumo}, la séquence $(\ell_0, \ldots, \ell_n)$ est croissante. Puisque 
    $n > \size{w}^2 + 1$ et $\ell_n = \size{w}$, alors il existe $0 \leq i \leq n - \size{w}$
    telle que $\ell_i = \ell_{i + 1} = \cdots = \ell_{i+\size{w}}$ et, toujours par le
    \myref{th-sumo}, $u_iv_i = u_{i+1}v_{i+1} = \cdots = u_{i+\size{w}}v_{i+\size{w}}$. On a
    donc $\size{w}$ configurations de suite avec le même mot. On distingue alors deux cas.
    \begin{itemize}
        \item Si $u_iv_i = w$ alors l'exécution $t_1 \cdots t_i$ écrit $w$.
        \item Sinon, il y a $\size{w} > \ell_i$ configurations de même taille $\ell_i$
              et avec le même mot, donc deux de ces configurations sont identiques : 
              il existe $i \leq j < k \leq i + \size{w}$ telle que  
              $\config{u_j}{v_j} = \config{u_k}{v_k}$. L'exécution 
              $t_1 \cdots t_j t_{k+1} \cdots t_n$ écrit alors $w$.
    \end{itemize}
    Et on a le résultat.
\end{Preuve}
\begin{Theoreme}
    Le problème du mot sur les claviers de $\FEK$ est dans NP.
\end{Theoreme}
\begin{Preuve}
   Par le \myref{lem-borne_touches_fek}, il nous suffit de deviner une suite de touches
   de taille polynomiale, de l'exécuter, et de vérifier qu'elle mène bien au mot.
\end{Preuve}

\subsection{Universalité}

De même, nous tirons la plupart de nos résultats sur le problème d'universalité de nos différentes caractérisations.

\begin{Theoreme}
    L'universalité est décidable en temps polynomial sur
    les claviers de $\REK$.
\end{Theoreme}
\begin{Preuve}
    Par le \myref{thm-rek_in_rat}.
\end{Preuve}

Et une étude plus fine de certaines sous-classes de $\REK$
nous donne des résultats plus précis.

\begin{Theoreme}
    L'universalité est décidable en temps polynomial sur les claviers de $\MK$.
\end{Theoreme}
\begin{Preuve}
    Soit $K \subseteq A^*$ un clavier. Le langage de $K$ est $K^+$,
    donc $K$ est universel si et seulement si $K^+ = A^*$. Montrons
    à présent que $K^+ = A^*$ si et seulement si $A\cup \set{\epsilon} \subseteq K$.
    \begin{itemize}
        \item Supposons que $A \cup \set{\epsilon} \subseteq K$. Alors, clairement
              $A^* = (A\cup \epsilon)^+ \subseteq K^+$.
        \item Réciproquement, supposons que $A \cup \set{\epsilon} \nsubseteq K$.
              Alors soit $\epsilon \notin K$, auquel cas $\epsilon \notin K^+$,
              soit il existe $a \in A$ tel que $a \notin K$, auquel cas
              $a \notin K^+$. Dans les deux cas $A^* \neq K^+$.
    \end{itemize}
    On peut donc tester l'universalité en regardant si
    $A \cup \set{\epsilon} \subseteq K$, ce qui est possible en temps polynomial.
\end{Preuve}

Nous avons également un résultat plus précis pour $\RK$
grâce au lemme qui suit.

\begin{Lemme} \label{lem-rek-ex-eq}%
    Soit $K$ un clavier de $\RK$ de norme strictement positif. 
    Alors, pour tout $\tau \in K^*$, il existe $\tau' \in K^*$ 
    de forme normale $\retour^{k'}w'$ avec $k' < \normeInf{K}$ tel que 
    $\epsilon \cdot \tau = \epsilon \cdot \tau'$.
\end{Lemme}
\begin{Preuve}
    On montre le résultat par induction sur $\tau$.
    Le résultat est vraie pour une suite de touches vide (on garde
    la même suite). Soit $\tau \in K^*$. On suppose qu'il existe 
    $\tau' \in K^*$ de forme normale $\retour^{k'} w'$ avec 
    $k' < \normeInf{K}$ tel que 
    $\epsilon \cdot \tau = \epsilon \cdot \tau'$. Soit $t \in K$.
    On regarde la suite $t \tau$.
    
    Il existe $k$ et $w$ tel que $t$ est équivalente à 
    $\retour^k w$. Soit $n = \size{w} - k'$.
    \begin{itemize}
        \item Si $n > 0$, alors $t\tau$ est équivalente à
        $\retour^{k} w[1, n]w'$ et $t \tau$ convient (en particulier,
        $n > 0$ implique $\size{w} > 0$ et donc $k < \norme{K}$).
        \item Sinon, $\epsilon \cdot t \tau = w \cdot \tau = w \cdot \tau' =
              w' = \epsilon \cdot \tau$ et $\tau'$ convient.
    \end{itemize}
    Ce qui montre le résultat.
    
    Remarquons également que le lemme est une conséquence triviale
    de la décomposition des exécutions de $\RK$ et plus 
    particulièrement du \myref{oneWastefulKey}.
\end{Preuve}

\begin{Theoreme}
    L'universalité sur les claviers de $\RK$ est dans coNP.
\end{Theoreme}

\begin{Preuve}
    Soit $K$ un clavier de $\RK$ de norme strictement positive. 
    Nous allons montrer que pour reconnaître un mot $w$ de taille 
    $\size{w} > \normeInf{K}$, il suffit de reconnaître tous les 
    mots de taille strictement inférieure à $\size{w}$. 
    
    Soit donc $w$ tel que $\size{w} > \normeInf{K} \geq 1$. On 
    suppose que pour tout $w'$ de taille $\size{w'} < \size{w}$,
    on a $w' \in \langage{K}$. Il existe $a \in A$ et $w_2 \in A^*$ 
    tel que $w = aw_2$.
    
    Par hypothèse, $w_2 \in L$, donc il existe $\tau_2 \in K^+$ tel que 
    $\epsilon \cdot \tau_2 = w_2$. Le \myref{lem-rek-ex-eq} nous permet
    de supposer que $\tau_2$ est équivalente à $\retour^k w_2$ avec
    $k < \normeInf{K}$.
    
    De plus, par hypothèse, $a^{k + 1}$ est reconnu par $K$ (car 
    $k + 1 < \normeInf{K} + 1 \leq \size{w}$). Ainsi, il 
    existe $\tau_1 \in K^+$ tel que $\epsilon \cdot \tau_1 = a^{k + 1}$.
    On a alors $\epsilon \cdot \tau_1 \tau_2 = aw_2$.
    
    Ce résultat nous permet de montrer par induction que si $K$ 
    reconnaît tous les mots de taille inférieure ou égale à
    $\normeInf{K}$, alors $K$ est universel.
    
    Pour montrer qu'un clavier $K$ n'est pas universel, on peut
    alors accepter si $\normeInf{K}$ est nul, et sinon deviner un mot 
    $w$ de taille $\size{w} \leq \normeInf{K}$
    et vérifier (en temps polynomial) que $w \not \in \langage{K}$.
\end{Preuve}

\subsection{Vacuité de l'intersection de \GK}

Finalement, nous présentons un résultat intéressant sur 
les claviers de $\GK$. Nous nous plaçons sur l'alphabet 
$A = A' \cup \set{\fraiche}$ et nous posons $f \colon {A'}^* \to A^*$
la fonction qui ajoute un $\fraiche$ à la droite de chaque lettre,
c'est-à-dire que si $u = a_1 \ldots a_n$, alors
$f(u) = a_1 \fraiche a_2 \fraiche \ldots a_n \fraiche$.

De plus, nous posons $K_{\pal} = \ensC{aa \gauche}{a \in A} \cup \set{\epsilon}$,
donc $\langage{K_{\pal}} = \ensC{w\miroir{w}}{w \in A^*}$ est 
le langage des palindromes pairs.

\begin{Theoreme}
    Le problème suivant est indécidable.
    \[
    \problemeDecision*{Palindrome}
                      {Un clavier $K$ de $\GK$}
                      {$\langage{K} \cap \langage{K_{\pal}} = \emptyset$ ?}
    \]
\end{Theoreme}

\begin{Preuve}
Nous allons réduire le problème de correspondance de Post. Considérons
$(u_i,v_i)_{i \in \intervE[1,n]}$ une instance de PCP. Pour tout $i \in \intervE[1, n]$, 
on pose  $t_i = f\left(u_i\right)\miroir{f\left(v_i\right)} \gauche^{2\taille{v_i}}$ et on considère le clavier 
\[
    K = \ensC{t_i}{i \in \intervE[1,n]}.
\]
Montrons que $\langage{K} \cap \langage{K_{\pal}} \neq \emptyset$ si et seulement si
$(u_i,v_i)_{i \in \intervE[1,n]} \in \PCP$. Une récurrence sur $k \in \N$ permet de montrer
que pour toute suite d'indices $i_1, \ldots, i_k \in \intervE[1,n]$, 
\[
    \config{\epsilon}{\epsilon} \cdot (t_{i_1} \ldots t_{i_k}) = 
    \config{f\left(u_{i_1}\right) \cdots f\left(u_{i_k}\right)}{\miroir{f\left(v_{i_k}\right)} \cdots \miroir{f\left(v_{i_1}\right)}}.
\]
De plus, on note que si $k > 0$, alors 
$f\left(u_{i_1}\right) \cdots f\left(u_{i_k}\right) \miroir{f\left(v_{i_k}\right)} \cdots \miroir{f\left(v_{i_1}\right)}$ a
exactement un facteur $\fraiche\fraiche$ avec la dernière lettre de $f\left(u_{i_k}\right)$
et la première lettre de $\miroir{f\left(v_{i_k}\right)}$. Ainsi, ce mot est un palindrome
si et seulement si $\miroir{f\left(v_{i_k}\right)} \cdots \miroir{f\left(v_{i_1}\right)}$  est le miroir
de $f\left(u_{i_1}\right) \cdots f\left(u_{i_k}\right)$ donc si et seulement si
\[
    f\left(u_{i_1}\right) \cdots f\left(u_{i_k}\right) =  f\left(v_{i_1}\right) \cdots f\left(v_{i_k}\right).
\]
Il nous suffit maintenant de montrer que 
$f\left(u_{i_1}\right) \cdots f\left(u_{i_k}\right) = f\left(v_{i_1}\right) \cdots f\left(v_{i_k}\right)$ si et seulement si 
$u_{i_1} \cdots u_{i_k} = v_{i_1} \cdots v_{i_k}$. Et c'est bien le cas.

Donc $K$ contient un palindrome de taille paire si et seulement s'il
existe $k > 0$ et $i_1 \ldots i_k \in \intervE[1, n]$ tel que 
$u_{i_1} \cdots u_{i_k} = v_{i_1} \cdots v_{i_k}$.
\end{Preuve}

\begin{Corollaire}
    Le problème suivant est indécidable.
    \[
    \problemeDecision*{Vacuité de l'intersection}
                      {$K_1$ et $K_2$ dans $\GK$}
                      {$\langage{K_1} \cap \langage{K_2} = \emptyset$ ?}
    \]
\end{Corollaire}

\begin{Remarque}
    Cela signifie notamment que $\GK$ capture une part raisonnablement bonne
    de l'expressivité des algébriques, puisque ce problème est décidable 
    pour deux langages rationnels.
\end{Remarque}

%% file: 5ContreExemples/ce.tex
\chapter{La structure de \FREK}

Dans ce chapitre, nous allons établir plus précisément la structure de $\FREK$
en comparant les différentes classes entre elles.

Pour commencer, nous montrons que chaque flèche ajoute vraiment un peu plus d'expressivité.

\input{5ContreExemples/gk_not_in_rek}
\input{5ContreExemples/fk_not_in_grek}

\input{5ContreExemples/rk_not_in_fek_3}

\input{5ContreExemples/ek_not_in_fk}
\input{5ContreExemples/ek_not_in_grk}
\input{5ContreExemples/gek_not_in_frk_21}

\input{5ContreExemples/chomsky}

\section*{La hiérarchie des claviers}

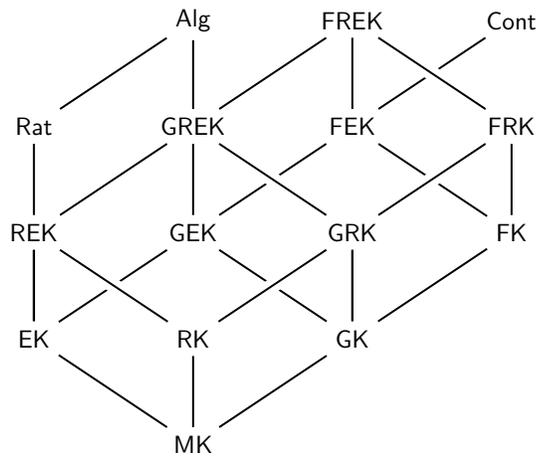
\begin{figure}[H]
\centering
    \begin{tikzpicture}
        \node (Alg)   at (0, 1.5) {$\Alg$};
        \node (GREK) at (0, 0) {$\GREK$};
        \node (Rat) at (-2.25, 0) {$\Rat$};
        \node (REK)  at (-2.25, -1.5) {$\REK$};
        \node (GEK)  at (0, -1.5) {$\GEK$};
        \node (GRK)  at (2.25, -1.5) {$\GRK$};
        \node (EK)  at  (-2.25, -3) {$\EK$};
        \node (RK)  at  (0, -3) {$\RK$};
        \node (GK)  at  (2.25, -3) {$\GK$};
        \node (MK)  at (0, -4.5) {$\MK$};
        \node (FK)  at  (4.5, -1.5) {$\FK$};
        \node (FEK)  at (2.25, 0) {$\FEK$};
        \node (FRK)  at (4.5, 0) {$\FRK$};
        \node (FREK)  at (2.25, 1.5) {$\FREK$};
        \node (Cont)  at (4.5, 1.5) {$\Cont$};
        \draw[thick, >=latex] (MK) -- (EK) -- (REK) -- (GREK) -- (GRK) -- (GK) --
        (MK) -- (RK);
        \draw[thick, >=latex] (GEK) -- (GK) -- (GRK) -- (RK) -- (REK) --
        (EK) -- (GEK) -- (GREK) -- (Alg);
        \draw[thick, >=latex] (REK) -- (Rat);
        \draw[thick, >=latex] (GK) -- (FK) -- (FRK) -- (FREK) -- (GREK);
        \draw[thick, >=latex] (GRK) -- (FRK) -- (FREK) -- (FEK);
        \draw[thick, >=latex] (GEK) -- (FEK) -- (FK);
        \draw[thick, >=latex] (FEK) -- (Cont);
        \draw[thick, >=latex] (Rat) -- (Alg);
        \draw[preaction={draw=white,line width=6pt}, thick, >=latex] (REK) -- (RK) -- (GRK);
        \draw[preaction={draw=white,line width=6pt}, thick, >=latex] (GREK) -- (GRK) -- (FRK);
        \draw[preaction={draw=white,line width=6pt}, thick, >=latex] (FREK) -- (FRK);
    \end{tikzpicture}
    \caption{Hiérarchie des langages}
    \label{fig-hierarchie}
\end{figure}

Finalement, on obtient la \myref{fig-hierarchie} qui résume les inclusions
des classes étudiées et établit une hiérarchie des claviers.
Néanmoins, deux questions sont pour l'instant sans réponse : a-t-on
$\REK \subset \FRK$ ? $\EK \subset \FRK$ ?

%% file: 5ContreExemples/gk_not_in_rek.tex
\section{\GK{} n'est pas inclus dans \REK}

Pour montrer que $\GK$ n'est pas dans $\REK$ il suffit
d'exhiber un langage de $\GK$ algébrique. Par exemple,
les palindromes \emph{de longueur paire} sont facilement
faisables dans $\GK$. Ici, nous allons nous intéresser
au langage $L$ des mots de Dyck (les mots bien
parenthésés).

\begin{Proposition}
    Le langage $L$ n'est pas dans $\REK$.
\end{Proposition}
\begin{Preuve}
    $L$ n'est pas rationnel, et le \myref{thm-rek_in_rat}
    permet de conclure.
\end{Preuve}

\begin{Proposition} \label{ex-dyck-gk}%
    Soit $K = \set{ \touche{()\gauche}, \touche{\gauche}}$.
    Alors $\langage{K} = L$.
\end{Proposition}

\begin{Preuve}
    Les mots engendrés sont clairement dans $L$, donc $\langage{K} \subset L$.

    On montre la réciproque par récurrence sur la longueur de $u \in L$.
    Montrons ainsi qu'il existe une suite de touche $\tau$ telle que
    telle que pour tout $x, y$, $\config{x}{y} \cdot \tau = \config{x}{uy}$.
    \begin{itemize}
        \item Pour $u = \epsilon$, on prend $\tau = \epsilon$.
        \item Soit $u$ bien parenthésé de taille $2(n + 1)$. Il existe $v$ et $w$
              de tailles respectives $2k$ et $2(n - k)$ tel que $u = (v)w$. Par
              hypothèse de récurrence, il existe $\tau_v$  et $\tau_w$ tel que
              $\config{x}{y} \cdot \tau_w = \config{x}{wy}$ et
              $\config{x(}{)wy} \cdot \tau_v = \config{x(}{v)wy}$.
              En posant $\tau = \tau_w \touche{() \gauche} \tau_v
              \touche{\gauche}$, on obtient bien
              $\config{x}{y} \cdot \tau = \config{x}{uy}$.
    \end{itemize}
    On a alors le résultat pour tout $u$ non vide en appliquant le résultat obtenu
    pour $x = y = \epsilon$. Pour le mot vide, on a $\epsilon \cdot \gauche = \epsilon$
    (on ne peut pas prendre $\tau = \epsilon$, car une exécution contient au moins
    une touche).
\end{Preuve}

\begin{Corollaire}
    $\RK$, $\EK$ et $\REK$ ne contiennent aucune classe contenant le symbole $\gauche$.
\end{Corollaire}

%% file: 5ContreExemples/fk_not_in_grek.tex
\section{\FK{} n'est pas inclus dans \GREK}

Soit $L = \ensC{ab^{n+1}a^n}{n \in \N}$.

\begin{Proposition}
    Soit $K = \set{\touche{a \gauche^2\droite b}}$. Alors, $\langage{K} = L$.
\end{Proposition}

\begin{Preuve}
    En appliquant $\touche{a \gauche^2\droite b}$ sur $\config{\epsilon}{\epsilon}$ on obtient 
    $\config{ab}{\epsilon}$, et en l'appliquant sur une configuration de la forme $\config{ub}{v}$ 
    on obtient $\config{ubb}{av}$. Le mot obtenu en l'appliquant $n$ fois sur 
    $\config{\epsilon}{\epsilon}$ est donc $ab^{n+1}a^n$, ce clavier produit donc bien le 
    langage $\ensC{ab^{n+1}a^n}{n \in \N}$.
\end{Preuve}

Nous allons maintenant montrer que ce clavier n'est pas dans $\GREK$.

\begin{Lemme} \label{lem-ak}%
    Soit $K = (T, F)$ reconnaissant $L$. Alors, pour tout $\tau \in T^*$,
    $\config{\epsilon}{\epsilon} \cdot \tau$ est de la forme $\config{u}{a^k}$. 
\end{Lemme}

\begin{Preuve}
Puisque $abba \in L$, il existe $\tau \in T^*$ et $t_f \in F$ telle que 
$\config{\epsilon}{\epsilon} \cdot \tau t_f = \config{x}{y}$ avec $xy = abba$.

Soit $\config{u}{v}$ avec $v$ contenant un $b$. Alors, par le \myref{thm-fondamental_grek},
$\config{u}{v} \cdot \tau t_f$ est de la forme $\config{u'x}{v'v}$ avec $y$ un sous-mot
de $v'$. Par suite, $abba$ est un sous mot de $u'xv'v$, ce qui est impossible 
car $u'xv'v \in L$. 
\end{Preuve}

\begin{Proposition}
    Le langage $L$ n'est pas dans $\GREK$.
\end{Proposition}
\begin{Preuve}
    Supposons que $K = (T, F)$ de $\GREK$ reconnaisse $L$.
    Soient $\tau \in T^*$ et $t_f \in F$. 
    On pose 
    \[
        \config{x}{y} = \config{\epsilon}{\epsilon} \cdot t_f 
        \quad \text{ et } \quad
        \config{u}{v} = \config{\epsilon}{\epsilon} \cdot \tau.
    \]
    Il existe $n \in N$ tel que $xy = ab^{n + 1}a^n$.
    Par le \myref{thm-fondamental_grek}, $\config{\epsilon}{\epsilon} \cdot \tau t_f$
    est de la forme $\config{u'x}{v'v}$ avec $y$ un sous-mot de $v'$.
    
    Puisque $ab^{n + 1}a^n$ est un sous mot de $xv'$ et que $u'xv'v \in L$,
    alors $u'$ est vide (sinon $u'xv'v$ commence par au moins deux $a$). 
    
    Par le \myref{lem-encadrement_taille}, 
    $\size{v'v} - \size{v} \leq \size{t} \leq \normeInf{K}$,
    c'est-à-dire $\size{v'} \leq \normeInf{K}$. De plus, toujours par le 
    \myref{lem-encadrement_taille}, $\size{x} \leq \normeInf{K}$.
    
    Mais, par le \myref{lem-ak}, $v$ ne contient que des $a$. Alors,
    tous les $b$ de $u'xv'v$ sont dans $x$ et $v'$ ce qui borne 
    le nombre de $b$ qu'il est possible d'obtenir et mène donc
    à une contradiction.
\end{Preuve}

\begin{Corollaire}
    $\EK$, $\RK$, $\GK$, $\GEK$, $\GRK$, $\REK$ et $\GREK$ ne
    contiennent aucune classe contenant le symbole $\droite$.
\end{Corollaire}

%% file: 5ContreExemples/rk_not_in_fek_3.tex
\section{\RK{} n'est pas inclus dans \FEK}

\label{sec-rk_not_in_fek}

On se place sur un alphabet $A = \set{a, b, \dollar, \diese}$.

On considère le langage $L$ engendré par le clavier de $\RK$
$\set{\touche{\retour a \diese \dollar}, \touche{\retour\retour b \diese \dollar\dollar}}$.
Nous allons montrer que ce langage n'est pas dans $\FEK$.

Dans toute la suite nous noterons $t_a = \retour a \diese \dollar$ et 
$t_b = \retour \retour b \diese\dollar\dollar$.

\begin{restatable}{Lemme}{LemFormeLC} \label{lem-forme-lc}%
Soit $x \colon \intervE[1; n] \mapsto \set{a, b}$. On a 
\[ 
    \config{\epsilon}{\epsilon} \cdot t_{x_1} \cdots t_{x_n} 
    = x_1 w_{x_1x_2} x_2 w_{x_2x_3} x_3 \cdots w_{x_{n - 1} x_n} x_n v_{x_n}
\]
avec $w_{aa} = w_{bb} = \diese$, $w_{ab} = \epsilon$, $w_{ba} = \diese\dollar$, $v_a = \diese\dollar$ et $v_b = \diese\dollar\dollar$.

En particulier, pour tous $u_1, u_2, u_3 \in A^*$ si $u_1 b u_2 a u_3 \in L$, 
alors $u_2$ contient un $\dollar$. 
\end{restatable}

\begin{Preuve}
Nous procédons par récurrence sur $n$. 
Pour $n = 1$ la propriété est claire.

Soit $n > 1$, soit $x \colon \interv[1; n] \mapsto \set{a,b}$. On pose $x' = x_1 \cdots x_{n - 1}$ et
on suppose 
\[
    \config{\epsilon}{\epsilon}\cdot t_{x_1} \cdots t_{x_{n - 1}} 
    = x_1 w_{x_1x_2}  \cdots x_{n - 1} v_{x_{n - 1}}.
\]
On distingue alors quatre cas.

\begin{itemize}
    \item Si $x_{n - 1} = x_n = a$, alors $v_{x_{n - 1}} = \diese\dollar$ et donc
        \[
            \config{\epsilon}{\epsilon}\cdot t_{x_1}\cdots t_{x_n} 
            = x_1 w_{x_1x_2}  \cdots x_{n - 1} \diese a \diese\dollar.
        \]
    
    \item Si $x_{n - 1} = x_n = b$, alors $v_{x_{n - 1}} = \diese\dollar\dollar$ et donc 
        \[
            \config{\epsilon}{\epsilon}\cdot t_{x_1}\cdots t_{x_n} 
            = x_1 w_{x_1x_2}  \cdots x_{n - 1} \diese b \diese\dollar\dollar.
        \]
    
    \item Si $x_{n - 1} = a$ et $x_n = b$, alors $v_{x_{n - 1}} = \diese\dollar$ et donc 
        \[
            \config{\epsilon}{\epsilon}\cdot t_{x_1}\cdots t_{x_n} 
            = x_1 w_{x_1x_2}  \cdots x_{n - 1} b \diese\dollar\dollar.
        \]
    
    \item $x_{n - 1} = b$ et $x_n = a$, alors $v_{x_{n - 1}} = \diese\dollar\dollar$ et donc 
        \[
            \config{\epsilon}{\epsilon}\cdot t_{x_1}\cdots t_{x_n}
            = x_1 w_{x_1x_2}  \cdots x_{n - 1} \diese \dollar a \diese\dollar.
        \]
\end{itemize}
Dans tous les cas la propriété est vérifiée.
\end{Preuve}

\begin{Proposition}
    Le langage $L$ n'est pas dans $\FEK$.
\end{Proposition}

\begin{Preuve}
Supposons qu'il existe un clavier $K = (T,F)$ de $\FEK$ reconnaissant $L$.
Ici, nous allons noter les touches de $K$ avec un « $k$ » pour ne pas
les confondre avec les touches du clavier de $\RK$.

Puisque $\diese \dollar \in K$, il existe $\tau_f \in T^*F$ qui produit
ce mot.

De plus, $w = ({a\diese})^{3\normeInf{K}}a\diese \dollar \in L$ (obtenu en appliquant
$3\normeInf{K}$ fois la touche $t_a$). Donc il existe une exécution de $K$ 
produisant ce mot. Par le \myref{lem-non-eff}, comme $\size{k}_a \leq \normeInf{K}$ 
pour tout $k \in T \cup F$, cette exécution contient au moins quatre touches
contenant $a$, donc au moins trois dans $T$ (car il y a une seule touche finale 
par exécution). 

Comme $w$ contient un seul $\dollar$, une de ces touches, $k_a$, contient un $a$, 
mais pas de $\dollar$ ni de $b$.

De même, on montre en partant de $({b\diese})^{3\normeInf{K}}b\diese \dollar \dollar$
qu'il existe $k_b \in T$ qui contient un $b$, mais pas de $\dollar$ ni de $a$.

Soit $\tau = k_a k_b$. On note 
\[
    \config{u}{v} = \config{\epsilon}{\epsilon} \cdot \tau
    \quad \text{ et } \quad
    \config{u'}{v'} = \config{u}{v} \cdot \tau_f
\]
On a que $u'v'$ contient un seul $\dollar$. De ce fait, et puisque $u'v' \in L$, 
$u'v'$ est de la forme $wa\diese\dollar$ avec $w$ qui ne contient pas de $\dollar$.

Puisque $wa$ contient au moins un $b$, $wa$ est de la forme $u_1bu_2a$, donc par le 
\myref{lem-forme-lc}, $u_2$ contient un $\dollar$, ce qui est impossible. 
En conclusion, $L \not\in \FEK$. 
\end{Preuve}

\begin{Corollaire}
    $\EK$, $\GEK$ et $\FEK$ ne contiennent aucune classe contenant le
    symbole $\retour$.
\begin{gather*}
    \RK \not \subset \EK      \qquad   \RK \not \subset \GK     \qquad \RK \not\subset \FK\\
    \RK \not \subset \GEK     \qquad   \RK \not \subset \FEK\\
    \GRK \not \subset \GEK    \qquad   \FRK \not\subset \FEK\\
    \GREK \not \subset \GEK   \qquad  \FREK \not \subset \FEK.
\end{gather*}
\end{Corollaire}

%% file: 5ContreExemples/ek_not_in_fk.tex
\section{\EK{} n'est pas inclus dans \FK}

L'étude de l'apport de l'entrée est un peu plus compliquée. En effet, nous
n'avons pas de preuve que $\EK$ n'est pas inclus dans $\FRK$.

Nous allons montrer que les langages finis non vide
ne sont pas dans $\FK$.

\begin{Proposition} \label{prop-finite_not_in_fk}%
Si $L$ est fini et dans $\FK$ alors $L \subset \set{\epsilon}$.
\end{Proposition}

\begin{Preuve}
Nous raisonnons par contraposée. S'il existe $w$ non vide dans $L$,
alors $w$ contient une lettre $a$. Le \myref{lem-iter-lettre} nous
donne l'existence de $\suite{w_n}$ telle que $\forall n \in \N$,
$\taille{w_n}_a \geqslant n$.

Donc $L$ est infini.
\end{Preuve}

\begin{Corollaire}
    $\FK$ et $\GK$ ne contiennent aucune classe contenant le symbole $\entree$.
\end{Corollaire}

\begin{Remarque} \label{rem-finite_rk_not_fk}%
    Les langages finis sont également dans $\RK$. En posant,
    pour $L$ un langage fini, $M = \max\limits_{w \in L}{\taille{w}}$ et 
    \[
        K = \ensC{\retour^M w}{ w \in L},
    \]
    alors $K$ reconnaît le langage $L$.
    
    On obtient une nouvelle preuve (plus simple) que $\FK$ et $\GK$ ne 
    contiennent aucune classe contenant le symbole $\retour$, résultat
    déjà établi en \cref{sec-rk_not_in_fek}.
\end{Remarque}

%% file: 5ContreExemples/ek_not_in_grk.tex
\section{\EK{} n'est pas inclus dans \GRK} \label{sec_ek_not_in_grk}%

On considère le langage $L \eqDef (a^2)^*(b+b^2)$. 

\begin{Proposition}
    Soit $K = \set{\touche{aa}, \touche{b \entree}, \touche{bb\entree}}$.
    Alors $\langage{K} = L$.
\end{Proposition}

Nous montrons maintenant que $L$ n'est pas dans $\GRK$. On commence par 
montrer que si $K$ de $\GRK$ reconnaît $L$, alors le curseur reste
toujours tout à gauche du mot, et donc $K$ est un clavier de $\RK$.

\begin{Lemme}
    Si $K$ de $\GRK$ reconnaît $L$, alors $K$ est équivalent à un 
    clavier de $\RK$.
\end{Lemme}

\begin{Preuve}
    Puisque $b^2 \in L$, il existe une suite de touches $\tau_{b^2}$ de $K$ 
    telle que $\config{\epsilon}{\epsilon} \cdot \tau_{b^2} = \config{x}{y}$ 
    avec $xy = b^2$. est reconnu par $\tau$. 
    
    Soit $\tau \in K^*$. On pose $\config{u}{v} = \config{\epsilon}{\epsilon} \cdot \tau$.
    Par le \myref{thm-fondamental_grek}, 
    $\config{\epsilon}{\epsilon} \cdot \tau \tau_{b^2}$ est de la forme
    $\config{u'x}{v'v}$ avec $y$ un sous-mot de $v'$. Ainsi $xv'$
    contient deux $b$, et puisque $u'xv'v \in L$, on obtient que
    $v = \epsilon$. 
    
    Supposons alors une touche $t$ contenant la flèche gauche. On a 
    $t$ de la forme $t_1 \gauche t_2$. Il existe $\tau$ telle que 
    $\config{\epsilon}{\epsilon} \cdot \tau = \config{a^{2M}b}{\epsilon}$
    avec $M = \normeInf{K}$.
    On a $\config{a^{2M}}{\epsilon} \cdot t_1 = \config{u}{v}$
    avec $u$ non vide donc de la forme $u'a$.
    Alors $\config{a^{2M}}{\epsilon} \cdot t = \config{u'}{av} \cdot t_2$
    et le $a$ restera toujours à droite du curseur donc la seconde
    composante de la configuration n'est pas vide.
    
    On en déduit qu'aucune touche de $K$ ne contient $\gauche$, et $K$
    est donc un clavier de $\RK$.
\end{Preuve}

Il nous suffit alors de montrer que $L$ n'est pas dans $\RK$.

\begin{Proposition} \label{prop-ek_not_in_grk}%
    Le langage $L$ n'est pas dans $\RK$.
\end{Proposition}

\begin{Preuve}
Supposons qu'il existe un clavier $K$ de $\RK$ qui reconnaît $L$. Le 
\myref{lem-caracterisation_rek} nous permet de supposer qu'il est sous forme normale.

Puisque $b^2 \in L$, alors il existe une suite de touches $\tau$ telle que
$\epsilon \cdot \tau = b^2$. La suite $\tau$ est alors, par mise sous
forme normale, équivalent à $\retour^kb^2$ avec $k \in \N$. Nous
raisonnons alors suivant la valeur de $k$.
\begin{itemize}
    \item Si $k = 0$, alors $\tau \sim b^2$ ; on a alors
          $\epsilon \cdot \tau \cdot \tau = b^2 \cdot \tau = b^4 \in L$.
    \item Si $k = 1$, alors $\tau \sim \retour b^2$ ; on a alors
          $\epsilon \cdot \tau \cdot \tau = b^2 \cdot \tau = b^3 \in L$.
    \item Si $k > 1$ et $\bm{k}$ \textbf{pair} ; on a $a^{2k}b \in L$, et donc
          $a^{2k}b \cdot \tau = a^{k + 1}b^2 \in L$.
    \item Si $k > 1$ et $\bm{k}$ \textbf{impair} ; on a $a^{2k}b^2 \in L$, et donc
          $a^{2k}b^2 \cdot \tau = a^{k + 2}b^2 \in L$.
\end{itemize}
Dans tous les cas, on aboutit à une contradiction, ce qui permet de
conclure la preuve.
\end{Preuve}

\begin{Corollaire}
    $\GK$, $\RK$ et $\GRK$ ne contiennent aucune classe contenant le symbole $\entree$.
\end{Corollaire}

Nous remarquons de plus que ce langage est dans $\FRK$. Il ne
permet donc pas de montrer que $\EK$ n'est pas inclus dans
$\FRK$ (résultat que nous ne connaissons pas encore).

\begin{Proposition}
    Soit $K = \set{\textcolor{color2}{t_1}, \textcolor{color3}{t_2}}$, avec 
    $\textcolor{color2}{t_1 = b\droite \droite \gauche \gauche \retour}$ et 
    $\textcolor{color3}{t_2 =b\droite \droite \gauche \gauche \retour aa\gauche \gauche}$.
    Alors, $\langage{K} = L$.
\end{Proposition}

\begin{Preuve}
    Pour montrer l'inclusion de $\langage{K}$ dans $(a^2)^*(b+b^2)$, on 
    considère les ensembles de configurations représentés sur le schéma suivant. 
    Les transitions rouges (flèches pleines) désignent l'application de $t_1$, 
    les bleues (pointillés) celle de $t_1$. La notation $\config{\epsilon}{L}$ 
    désigne l'ensemble de configuration $\ensC{\config{\epsilon}{w}}{w \in L}$. 
    On vérifie facilement que depuis une configuration de chacun de ces langages, 
    en appliquant la touche $t_i$ on obtient une configuration du langage atteint
    par la transition $t_i$ sur le schéma. Les mots correspondant aux configurations 
    de tous ces langages étant dans $(a^2)^*(b+b^2)$, on obtient l'inclusion.
    
    \begin{center}
        \begin{tikzpicture}
            \node (e) {$\config{\epsilon}{\epsilon}$};
            \node (a) [below left = 0.5cm and 0.25cm of e] {$\config{\epsilon}{b}$};
            \node (bbaa) [below right = 1cm and 0.15cm of a] {$\config{\epsilon}{(aa)^+bb}$};
            \node (bba) [below right = 0.5cm and 0.25cm of e] {$\config{\epsilon}{(aa)^+b}$};
            \node (aa) [below left = 0.5cm and 0.15cm of a]{$\config{\epsilon}{bb}$};
            \draw[->,thick, >=latex, color2, dashed] (e) to (a);
            \draw[->,thick, >=latex, color3] (e) to (bba);
            \draw[->,thick, >=latex, color2, dashed, bend right] (a) to (aa);
            \draw[->,thick, >=latex, color3, bend left] (a) to (bbaa);
            \draw[->,thick, >=latex, color2, dashed, loop right] (bbaa) to (bbaa);
            \draw[->,thick, >=latex, color3, loop below] (bbaa) to (bbaa);
            \draw[->,thick, >=latex, color2, dashed, loop right] (bba) to (bba);
            \draw[->,thick, >=latex, color3, loop below] (bba) to (bba);
            \draw[->,thick, >=latex, color3] (aa) to (bbaa);
            \draw[->,thick, >=latex, color2, dashed, loop left] (aa) to (aa);
        \end{tikzpicture}
    \end{center}
    Pour l'inclusion inverse, il nous suffit d'observer que pour tout $k \in \nats$,
    \begin{gather*}
    \config{\epsilon}{\epsilon} \act{t_2^k} \config{\epsilon}{b^{2k+2}a}\\ 
    \config{\epsilon}{\epsilon} \act{t_1t_2^k} \config{\epsilon}{b^{2k}aa}\\
    \config{\epsilon}{\epsilon} \act{t_1} \config{\epsilon}{aa} \act{t_1} \config{\epsilon}{aa}.
    \end{gather*}
\end{Preuve}

%% file: 5ContreExemples/gek_not_in_frk_21.tex
\section{\GEK{} n'est pas inclus dans \FRK}

Et finalement, à défaut de montrer que $\EK$ n'est pas inclus
dans $\FRK$, nous allons montrer que $\GEK$ n'est pas inclus
dans $\FRK$. En particulier, cela montrera que tous les 
claviers ne sont pas automatiques et que l'entrée ajoute
bien de l'expressivité !

Nous nous plaçons sur l'alphabet $A = \set{a, b, c}$ et nous considérons
$L = L_1 \cup L_2$ avec 
$L_1 = \ensC{w c \miroir{w}}{w \in \set{a,b}^*}$ et
$L_2 = \ensC{w cc \miroir{w}}{w \in \set{a,b}^*}$.

\begin{Proposition}
    Soit $K = \set{\touche{aa}\gauche, \touche{bb}\gauche, \touche{c}\entree, \touche{cc}\entree}$. 
    Alors $\langage{K} = L$. 
\end{Proposition}
\begin{Preuve}
    Notons $t_a = \touche{aa\gauche}$ et $t_b = \touche{bb\gauche}$. Une induction
    montre que $\langage{\set{t_a, t_b}}$ est le langage des palindromes et que 
    les configurations sont de la forme $\config{w}{\miroir{w}}$. Et les touches
    finales de $K$ permettent d'affirmer que $\langage{K} = L$.
\end{Preuve}

Nous allons montrer que $L \notin \FRK$. L'intuition est la suivante. 
Supposons avoir un clavier $K$ de $\FRK$ pour $L$. Puisque nous voulons reconnaître 
à peu de choses près des palindromes, il nous faut rester près du centre du mot 
(pour pouvoir rajouter des $a$ ou des $b$ de chaque côté du $c$ ou du $cc$).

Avec des mots très grands, cela reviendra à être très loin du bord et à
appliquer les touches loin des bords. En particulier, on obtient qu'il y
a une touche $t$ qui assure un $a$ loin des bords. Et cette touche 
assure même deux $a$ loin des bords (en l'appliquant, il faut pouvoir 
rester dans le langage).

On peut alors regarder le comportement de cette touche sur $b^ncb^n$
et $b^nccb^n$ avec $n$ très grand, de manière à appliquer la touche 
loin des bords. La touche fera alors exactement la même chose dans les
deux cas. En particulier, on remarque ces deux points.
\begin{itemize}
    \item Les $a$ seront à la même distance $d$ dans les deux mots obtenus.
          Donc les deux mots obtenus sont tous les deux de la même forme
          (donc soit ils ont tous les deux un seul $c$, soit ils en ont
          tous les deux deux).
    \item La touche ajoute dans les deux cas un nombre de lettres $\delta$,
          et donc si $b^ncb^n$ devient un mot de taille paire, alors
          $b^nccb^n$ devient un mot de taille impaire et vice-versa.
          Cela implique que les deux mots obtenus n'ont pas le même nombre
          de $c$.
\end{itemize}

On commence par montrer que la distance des $a$ est la même
si on applique une même touche à deux grandes configurations
qui ne contiennent pas de $a$.

\begin{Lemme} \label{lem-distance_des_a}%
    Soient $t \in T$ une touche et $u, v, u', v' \in A^*$ des mots
    ne contenant pas de $a$ et de longueurs supérieures ou égales 
    à $\size{t}$. Alors
    \[
        \distance_a\left( \config{u}{v} \cdot t   \right) =
        \distance_a\left( \config{u'}{v'} \cdot t \right).
    \]
\end{Lemme}
\begin{Preuve}
    Par le \myref{prop-ecriture_efficiente}, $t$ écrit un $a$ en position $i$ 
    depuis $\config{u'}{v'}$ si et seulement si $t$ écrit un $a$ en position $i$ 
    depuis $\config{u}{v}$.
    
    Et puisque $\config{u}{v}$ ne contient pas de $a$, alors, par le \myref{prop-ecriture}, 
    $(\config{u}{v} \cdot t)_i = a$ si et seulement si $t$ écrit un $a$ en position $i$
    depuis $\config{u}{v}$. De même, $(\config{u'}{v'} \cdot t)_i = a$ si et seulement si 
    $t$ écrit un $a$ en position $i$ depuis $\config{u'}{v'}$.
    
    Donc $(\config{u}{v} \cdot t)_i = a$ si et seulement si  $(\config{u'}{v'} \cdot t)_i = a$,
    ce qui montre le résultat.
\end{Preuve}

Nous pouvons maintenant revenir à la preuve de notre résultat. On suppose
avoir $K$ qui reconnaît $L$.

On commence par prouver que le curseur reste proche du centre du mot.
\begin{Lemme}\label{resteaumilieu}%
Soit $\config{u}{v}$ atteint par une exécution de $K$ et $t \in K$ tel que, en notant 
$\config{u'}{v'} = \config{u}{v} \cdot t$, on a $uv \neq u'v'$.
Alors $\abs{\size{u}-\size{v}\enlarger} \leq 6\normeInf{K}$
et $\abs{\size{u'}-\size{v'} \enlarger} \leq 8\normeInf{K}$.
\end{Lemme}
\begin{Preuve}
Supposons $\abs{\size{u}-\size{v}\enlarger} > 6\normeInf{K} \geq 6\size{t}$. 
Alors, soit $\size{u}-\size{v} > 6\size{t}$, soit $\size{v} - \size{u} > 6\size{t}$.
On suppose être dans la première situation, l'autre cas est similaire. On a donc
\[
    \size{u} = \frac{\size{u}}{2} + \frac{\size{u}}{2} > 
              \frac{\size{u}}{2} + \frac{\size{v} + 6\size{t}}{2} 
\]
D'où $\size{u} \geq \frac{\size{uv}}{2} + 3\size{t}$.

Comme $uv \in L$, $uv$ est de la forme $wc\miroir{w}$ ou $wcc\miroir{w}$ 
pour un certain $w \in \set{a,b}^*$. 

On obtient du \myref{lem-localite} qu'il existe un préfixe commun $u_p$ à $u$ et 
$u'$ tel que 
\[
    \size{u_p} \geq \size{u} - \size{t} 
               \geq \frac{\size{uv}}{2} + 2\size{t} 
               \geq \frac{\size{uv}}{2} + 2,
\]
la dernière inégalité étant vraie puisque $t$ est non vide (car $uv' \neq uv$). 

Donc $u_p$ est de la forme $wcy$ (et $uv = wc\miroir{w}$) ou $wccy$ (et $uv = wcc\miroir{w}$) 
avec $y \in \set{a,b}^+$. Comme $u'v'$ est dans $L$, dans les deux cas
$uv=u'v'$, ce qui contredit nos hypothèses.

En conclusion, on a bien $\abs{\size{u}-\size{v}\enlarger} \leq 6 \normeInf{K}$.

Le \myref{lem-encadrement_taille} nous donne 
$\abs{\size{u} - \size{u'}\enlarger} \leq \size{t}$ et 
$\abs{\size{v} - \size{v'}\enlarger} \leq \size{t}$. De ce fait,
\[
    \abs{\size{u'}-\size{v'} \enlarger} 
    \leq \abs{\size{u}-\size{v} \enlarger} + \abs{\size{u} - \size{u'} \enlarger} 
                                           + \abs{\size{v} - \size{v'} \enlarger} 
    \leq 6\normeInf{K} + 2 \size{t} \leq 8 \normeInf{K}.
\]
\end{Preuve}

\begin{Lemme}\label{existetouchea}%
Si $K \in \FRK$ reconnaît $L$, alors $K$ contient une touche qui assure un $a$
loin des bords.
\end{Lemme}

\begin{Preuve}
Soit $w =  a^{6\normeInf{K}}cca^{6\normeInf{K}}$. Il existe $t_1 \cdots t_n \in T^*$
tel que $\config{\epsilon}{\epsilon} \cdot t_1 \cdots t_n = \config{u}{v}$ avec $uv = w$. 
Soit $i$ minimal tel que le nombre de $a$ dans 
$\config{u_i}{v_i} = \config{\epsilon}{\epsilon} \cdot t_1 \cdots t_i$ soit maximal. On pose 
\[
    \config{u_{i-1}}{v_{i-1}} = \config{\epsilon}{\epsilon} \cdot t_1 \cdots t_{i-1}. 
\]
En particulier $\config{u_i}{v_i}$ contient au moins $12\normeInf{K}$ $a$. 
Le \myref{lem-encadrement_taille} donne alors
\begin{equation}
    \label{eq_name-todo}
    \size{u_{i-1}v_{i-1}} \geq \size{u_iv_i} - \size{t_i} 
                          \geq 12\normeInf{K} - \normeInf{K}
                          \geq 11 \normeInf{K}.
\end{equation}

De plus, par le \myref{resteaumilieu}, $\abs{\size{u_{i-1}} - \size{v_{i-1}} \enlarger} \leq 6 \normeInf{K}$, donc, par inégalité triangulaire,
\[
    \size{u_{i-1}v_{i-1}} \leq  \size{u_{i-1}} + \bigl(\size{u_{i-1}} + 6 \normeInf{K}\bigr).
\]
En supposant $\size{u_{i-1}} < \normeInf{K}$, on aurait 
$\size{u_{i-1}v_{i-1}} \leq 8 \normeInf{K}$, ce qui contredit 
\eqref{eq_name-todo}. Donc 
$\size{u_{i-1}} \geq \normeInf{K}$.

De même, on montre que $\size{v_{i-1}} \geq \normeInf{K}$. Ainsi $u_{i-1}$ et $v_{i-1}$ 
sont de longueurs supérieures ou égales à $\normeInf{K}$, et 
le \myref{cor-assure_a_loin_bord} nous donne que $t_i$ assure un $a$ loin des bords, car, 
par  minimalité de $i$, $\config{u_{i-1}}{v_{i-1}}$ contient strictement moins de $a$
que $\config{u_i}{v_i}$.
\end{Preuve}

\begin{Lemme}
Le langage $L$ n'est pas reconnu par un clavier de $\FRK$.
\end{Lemme}

\begin{Preuve}
    Supposons qu'il existe un clavier $K$ de $\FRK$ reconnaissant $L$.
    Alors, il existe des suites de touches minimales $\tau$ et $\tau'$ qui
    mènent respectivement aux configurations $\config{u}{v}$ et $\config{u'}{v'}$
    avec $uv = b^{5\normeInf{K}} c b^{5\normeInf{K}}$
    et $u'v' = b^{5\normeInf{K}} cc b^{5\normeInf{K}}$.
    
    Notons $\config{x}{y}$ l'avant-dernière configuration correspondant à l'exécution
    de $\tau$. Par minimalité de $\tau$, $xy \neq uv$ et le \myref{resteaumilieu} donnent 
    alors $\abs{\size{u} - \size{v} \enlarger} \leq 8\normeInf{K}$. 

    De même, on montre que $\abs{\size{u'} - \size{v'} \enlarger} \leq 8 \normeInf{K}$.  
    
    En particulier, comme $\size{uv}, \size{u'v'} > 10 \normeInf{K}$ on a $\size{u}, \size{v}, \size{u'}, \size{v'} \geq \normeInf{K}$.
    
    Par le \myref{existetouchea}, il existe $t \in T$ telle que $t$ assure un $a$ loin des bords.
    On pose alors
    \[
        \config{u_a}{v_a} = \config{\epsilon}{\epsilon}\cdot \tau t 
        \quad \text{et} \quad
        \config{u'_a}{v'_a} = \config{\epsilon}{\epsilon} \cdot \tau' t.
    \] 
    Par le \myref{lem-distance_des_a}, on a 
    $\distance_a(u_av_a) = \distance_a(u_a' v_a')$. On pose 
    \[
        d = \distance_a(u_av_a)
        \quad \text{et} \quad
        \delta = \sum_{x \in A} \size{t}_x - \size{t}_{\retour}.
    \]
    On a $\size{u} \geq \size{t}$, $\size{v} \geq \size{t}$ et $t$ assure un
    $a$ loin des bords, donc $u_av_a$ contient au moins un $a$ et donc en 
    contient au moins deux car $u_av_a \in L$. On en déduit que $d \in \Nat$.
    
    De plus, puisque $u_av_a$ et $u_a'v_a'$ sont dans $L$, alors si $d$ est pair
    ils sont tous les deux  dans $L_2$ et sinon tous les deux dans 
    $L_1$.
    
    Mais si $\delta$ est pair alors par le \myref{lem-egalite_taille_loin_bords},
    \[
    \begin{cases*}
        \size{u_av_a} = \size{uv} +\delta = 10\normeInf{K} + 1 + \delta 
                            & est impair\\
        \size{u_a'v_a'} = \size{u'v'} +\delta = 10\normeInf{K} + 2 + \delta 
                            & est impair
    \end{cases*}
    \]
    donc $u_av_a \in L_1$ et $u_a'v_a' \in L_2$,
    ce qui contredit le fait qu'ils sont dans le même $L_j$.

    De même si $\delta$ est impair alors $u_av_a \in L_2$ et 
    $u_a'v_a' \in L_1$, ce qui contredit le fait qu'ils sont 
    dans le même $L_j$.

    On obtient une contradiction, donc $L$ n'est pas reconnu par 
    un clavier de $\FRK$.
\end{Preuve}

\begin{Corollaire}
    $\RK$, $\FK$ et $\FRK$ ne contiennent aucune classe contenant les symboles $\gauche$ et $\entree$.
\end{Corollaire}

%% file: 5ContreExemples/chomsky.tex
\section{Comparaison à la hiérarchie de \bsc{Chomsky}}

Nous comparons également les différentes classes de langages à celles
de la hiérarchie de \bsc{Chomsky}.

\subsection{\Rat{} n'est pas inclus dans \FEK}
Le résultat tombe trivialement des résultats établis précédemment.

\begin{Theoreme}
    $\Rat$ n'est pas inclus dans $\FEK$.
\end{Theoreme}
\begin{Preuve}
    $\REK$ n'est pas inclus $\FEK$ et $\REK$ est inclus dans $\Rat$.
\end{Preuve}

\begin{Corollaire}
    $\Rat$ n'est inclus dans aucune sous-classe de $\FEK$.
\end{Corollaire}

\subsection{\Rat{} n'est pas inclus dans \GREK} \label{subsec-rat_not_in_grek}
    On pose $L = a^* + b^*$.

\begin{Lemme}
    Le langage $L$ n'est pas dans $\GREK$.
\end{Lemme}
\begin{Preuve}
    Supposons qu'il existe $K = (T, F)$ de $\GREK$ qui reconnaisse
    $L$. Il existe $\tau_a \in T^*$ et $f_a \in F$ tel que
    $\config{\epsilon}{\epsilon} \cdot \tau_a f_a = \config{u_a}{v_a}$
    avec $u_av_a = a$.
    
    Il existe $\tau_b \in T^*$ et $f_b \in F$ tel que
    $\config{\epsilon}{\epsilon} \cdot \tau_b f_b = \config{u_b}{v_b}$ avec
    $u_bv_b = b^{1 + \normeInf{K}(\size{\tau_a} + 2)}$.
    
    Par le \myref{lem-localite}, en appliquant $\tau_b$ à 
    $\config{\epsilon}{\epsilon}$, on obtient un mot $w$ 
    contenant un nombre de $b$ supérieur ou égal à 
    $1 + \normeInf{K}(\size{\tau_a} + 1)$.
    
    Regardons alors, le mot $w$ obtenu avec 
    $\config{\epsilon}{\epsilon} \cdot \tau_b \tau_a f_a$. Par le 
    \myref{thm-fondamental_grek}, ce mot contient au moins un $a$.
    Mais, en appliquant $\tau_a f_a$ on peut supprimer au plus
    $\normeInf{K}(\size{\tau_a} + 1)$ lettres de $w$, donc
    le mot obtenu contient au moins un $b$.
    On obtient donc un mot avec des $a$ et des $b$, impossible, d'où
    le résultat.
\end{Preuve}

\begin{Corollaire}
    $\Rat$ n'est inclus dans aucune sous-classe de $\GREK$.
\end{Corollaire}

\begin{Remarque}
    On montre facilement que $L$ n'est pas non plus dans $\FEK$.
\end{Remarque}
\begin{Preuve}
    Supposons qu'il existe $K = (T, F)$ de $\FEK$ qui reconnaisse
    $L$. Puisque $a^{\normeInf{K} + 1}$ et $b^{\normeInf{K} + 1}$
    sont dans $L$, alors il existe $t_a$ et $t_b$ dans $T$ qui
    écrivent respectivement un $a$ et un $b$. Puisqu'on ne peut
    pas effacer dans $\FEK$, alors pour toute toute touche finale
    $f$, $\config{\epsilon}{\epsilon} \cdot t_a t_b f$ contient
    un $a$ et un $b$, d'où le résultat.
\end{Preuve}

\begin{Remarque}
    Ce langage ne semble pas non plus être dans $\FREK$ et est
    un bon candidat pour montrer que $\Rat$ n'est pas dans $\FREK$.
\end{Remarque}

%% file: 6Stabilite/stabilite.tex
\chapter{Les propriétés de stabilité}

Et finalement, nous étudions quelques propriétés de stabilité des
différentes classes de clavier.

\input{6Stabilite/morphisme}
\input{6Stabilite/miroir}
\input{6Stabilite/intersection}
\input{6Stabilite/union}
\input{6Stabilite/concatenation}

%% file: 6Stabilite/morphisme.tex
\section{Stabilité par morphisme}

La stabilité des langages de clavier par morphisme bien choisi est un résultat 
relativement évident. Par exemple, si l'on a un morphisme lettre à lettre, il 
est évident qu'il suffit de remplacer chaque lettre de chaque touche du clavier
par son image par le morphisme.

\begin{Definition}
    Soit $g \colon A^* \to B^*$ un morphisme lettre à lettre. On définit le prolongement de $g$,
    noté également $g$, de $(A \cup S)^*$ dans $(B \cup S)^*$
    en posant $g(\retour) = \retour$, $g(\gauche) = \gauche$ 
    et $g(\droite) = \droite$.
\end{Definition}

\begin{Theoreme}
    Soient $L$ reconnu par un clavier $K = (T, F)$ et $g \colon A^* \to B^*$ un morphisme 
    lettre à lettre. Alors, $g(L) = \ensC{g(w)}{w \in L}$ est reconnu par le clavier
    \[
        g(K) = \bigl(\ensC{g(t)}{t \in T}, \ensC{g(t)}{t \in F}\bigr).
    \]
    En particulier, chaque classe de clavier est stable par morphisme puisque les symboles
    spéciaux de $K$ et de $g(K)$ sont les mêmes.
\end{Theoreme}
\begin{Preuve}
    Soit $\tau \in (T \cup F)^*$. On montre par induction sur que 
    $\config{\epsilon}{\epsilon} \cdot \tau = \config{u}{v}$ si et 
    seulement si $\config{\epsilon}{\epsilon} \cdot g(\tau) = \config{g(u)}{g(v)}$.
    Pour cela, il suffit de le montrer pour les opérations élémentaires.
\end{Preuve}

%% file: 6Stabilite/miroir.tex
\section{Stabilité par miroir}

La stabilité par miroir est celle qui, après celle par morphisme lettre
à lettre,
semble la plus probable. Néanmoins, si on peut facilement inverser
les flèches, ce n'est pas le cas du retour, car il ne fait rien
sur une configuration de la forme $\config{\epsilon}{u}$.

Avec les deux flèches et sans le retour, tout se passe effectivement
très bien.

\begin{Theoreme} \label{th-fek_stable_miroir}%
    $\MK$, $\FK$ et $\FEK$ et sont stables par miroir.
\end{Theoreme}
\begin{Preuve}
Pour $\MK$ on constate simplement qu'étant donné $K$ un clavier de $\MK$, on a $\miroir{K}^* = \miroir{K^*}$. En conséquence, le clavier $\miroir{K}$ de $\MK$ reconnaît le miroir de $\langage{K}$.

Pour $\FEK$ et $\FK$ on pose $\overline{\cdot}$ le morphisme de mots généré par
\begin{align*}
    \overline{\droite} = \gauche \qquad
    \overline{\gauche} = \droite \qquad
    \overline{a} = a \gauche  \text{ si } a \in A
\end{align*}
Si $\config{u}{v}$ est une configuration, on note $\miroir{\config{u}{v}} = \config{\miroir{v}}{\miroir{u}}$. On peut montrer par une récurrence élémentaire  sur $n$ que pour toute suite 
d'opérations élémentaires $\sigma_1 \ldots \sigma_n$, 
\[
c_0 \act{\sigma_1...\sigma_n} c_n \iff 
\miroir{c_0} \act{\overline{\sigma_1}...\overline{\sigma_n}} \miroir{c_n}.
\]

Soit $L$ un langage de $\FEK$ et $K = (T, F)$ un clavier le reconnaissant. 
Soit $\overline{K} = (\overline{T}, \overline{F})$ où le morphisme est 
appliqué respectivement sur chacune des touches. Pour tout mot $w$, l'exécution
$\config{\epsilon}{\epsilon} \act{t_1\ldots t_n} w$ peut se renverser en une exécution
$\config{\epsilon}{\epsilon} \act{\overline{t_1} \ldots \overline{t_n}} \miroir{w}$ 
de $\overline{K}$ et vice-versa. Donc $w \in L$ si et seulement si $\miroir{w} \in \langage{\overline{K}}$,
d'où $\langage{\overline{K}} = \miroir{L}$. De plus, 
$\overline{K}$ est un clavier de $\FEK$, et est automatique si et seulement si $K$ 
l'est aussi, ce qui montre le résultat pour $\FK$.
\end{Preuve}

\begin{Theoreme}
    $\REK$, $\RK$ et $\EK$ ne sont pas stables par miroir.
\end{Theoreme}
\begin{Preuve}
    Considérons $L = b^*a$ et $\miroir{L} = ab^*$. On a $b^*a \in \EK$
    (avec le clavier $\set{\touche{b}, \touche{a\entree}}$) et $b^*a \in \RK$, reconnus
    par le clavier $\set{\touche{\retour a}, \touche{\retour b a}}$ (en effet,
    $b^na \cdot \touche{\retour a} = b^n a$, $b^na \cdot \touche{\retour ba} = b^{n + 1}a$,
    $\epsilon \cdot \touche{\retour a} = a$ et $\epsilon \cdot \touche{\retour ba} = ba$).
    
    Montrons maintenant que $\miroir{L} \not \in \REK$. Supposons que $K = (T, F)$ un clavier
    de $\REK$ reconnaisse $\miroir{L}$. Puisque $w = ab^{\normeInf{K}} \in \miroir{L}$, il
    existe $(\tau, t_f) \in T^* \times F$ qui mène à $w$. En particulier, $t_f$ n'a écrit
    que des $b$ et est donc équivalente $\retour^k a^n$ pour $k$ et $n$ deux entiers.
    
    En appliquant $t_f$ seule, on obtient que $a^n \in \miroir{L}$. Absurde, donc
    $\miroir{L} \not \in \REK$.
    
    On a alors $L$ dans $\EK$, dans $\RK$ et dans $\REK$, et $\miroir{L} \not\in \REK$,
    ce qui prouve le résultat.
\end{Preuve}

\begin{Theoreme}
    $\GRK$ n'est pas stable par miroir.
\end{Theoreme}
\begin{Preuve}
    Soit $L = a^*(b + b^2)$. Nous avons montré en \myref{prop-ek_not_in_grk} 
    que $L \not\in \GRK$. Mais on montre que $\miroir{L} = (b + b^2)a^*$
    est reconnu par 
    \[
        K = \set{
            \touche{\retour^2 a\gauche b},
            \touche{\retour^2 a\gauche bb},
            \touche{\retour^2 b},
            \touche{\retour^2 bb}
        }.
    \]
    L'invariant conservé est que la configuration est
    $\config{\epsilon}{\epsilon}$, de la forme
    $\config{b}{a^n}$ ou de la forme $\config{bb}{a^n}$.
    En particulier, les touches $\touche{\retour^2 a\gauche b}$ et
    $\touche{\retour^2 a\gauche bb}$ permettent de rajouter 
    exactement un $a$ à la configuration.
    
    La \cref{fig-miroir-grk} nous montre comment agissent
    les touches sur les différents types de configuration.
    
    \begin{figure}
        \centering
        \begin{tikzpicture}
            \node (e) {$\config{\epsilon}{\epsilon}$};
            \node (b) [below left = 1.5cm and 0.3cm of e] {$\config{b}{a^n}$};
            \node (bb) [below right = 1.5cm and 0.3cm of e] {$\config{bb}{a^n}$};
            \draw[->, thick, >=latex, bend right, color4] (e) to (b);
            \draw[->, thick, >=latex, bend left, color4] (bb) to (b);
            \draw[->, thick, >=latex, loop left, color4] (b) to (b);
            
            \draw[->, thick, >=latex, dotted, color3] (e) to (b);
            \draw[->, thick, >=latex, dotted, bend left = 10, color3] (bb) to (b);
            \draw[->, thick, >=latex, dotted, loop below, color3] (b) to (b);
            
            \draw[->, thick, >=latex, dashed, bend left, color2] (e) to (bb);
            \draw[->, thick, >=latex, dashed, bend left, color2] (b) to (bb);
            \draw[->, thick, >=latex, dashed, loop right, color2] (bb) to (bb);
            
            \draw[->, thick, >=latex, loosely dashdotted, color1] (e) to (bb);
            \draw[->, thick, >=latex, loosely dashdotted, bend left = 10, color1] (b) to (bb);
            \draw[->, thick, >=latex, loosely dashdotted, loop below, color1] (bb) to (bb);
        \end{tikzpicture}
        \caption{Le clavier $\set{\touche{\retour^2 a\gauche b},\touche{\retour^2 a\gauche bb},
            \touche{\retour^2 b}, \touche{\retour^2 bb}}$.}
        \label{fig-miroir-grk}
    \end{figure}

    Les transitions pleines correspondent à l'application de 
    {\color{color4}$\touche{\retour^2 a\gauche b}$}, celles en tirets à l'application
    de {\color{color2}$\touche{\retour^2 a\gauche b}$}, celles en pointillés
    à l'application de {\color{color3}$\touche{\retour^2 b}$} et
    celles en mi-pointillés mi-tirets à l'application de
    {\color{color1}$\touche{\retour^2 bb}$}.
    
\end{Preuve}

\begin{Theoreme}
    $\GEK$ et $\GREK$ ne sont pas stables par miroir.
\end{Theoreme}
\begin{Preuve}
    Soit $K = (\set{\touche{a\gauche\gauche b}}, \set{\touche{\gauche c}})$ un 
    clavier de $\GEK$ et $L = \langage{K}$. On montre que $L = c + cb(ba)^*a$. En 
    effet,
    \begin{align*}
        \config{\epsilon}{\epsilon} \cdot \touche{a\gauche\gauche b} &= \config{b}{a}  &  
        \config{\epsilon}{\epsilon} \cdot \touche{\gauche c}  &=  \config{c}{\epsilon}\\
        \config{b}{(ba)^na} \cdot \touche{a\gauche\gauche b} &= \config{b}{(ba)^{n + 1}a} &
        \config{b}{(ba)^na} \cdot \touche{\gauche c} &= \config{c}{b(ba)^{n + 1}a}
    \end{align*}
    Ainsi, $\miroir{L} = c + a(ab)^*bc$. Supposons avoir $K' = (T, F)$ qui reconnaisse
    $\miroir{L}$. Toute touche finale $f$ écrit un $c$ (car en l'appliquant seule, on doit 
    obtenir un mot de $\miroir{L}$), et le \myref{lem-insensibilite-position} nous
    assure qu'elle écrit ce $c$ depuis tout contexte.
    
    Si une touche $t$ de $T$ contient le symbole $\gauche$, alors en l'appliquant à 
    $\config{u}{v}$ tel que $\size{uv} > \normeInf{K}$ obtenu à partir de 
    $\tau \in T$ (une telle configuration existe bien, car dans le cas contraire, 
    les mots de $\miroir{L}$ sont de taille bornée par $2\normeInf{K}$), on 
    obtient une configuration $\config{x}{y}$ avec $y$ non vide. 
    
    En appliquant une touche finale à $\config{x}{y}$, le \myref{thm-fondamental_grek}
    nous assure qu'on obtient un mot $w$ contenant un $c$ qui n'est pas en dernière
    position (car toute touche finale écrit un $c$). Absurde, donc $T$ ne contient
    pas le symbole $\gauche$ et est un clavier de $\RK$.
    
    Il y a une suite de touches $\tau \in T^*$ qui écrit un mot $w_1$ de la forme
    $a(ab)^{\normeInf{K}}x_1$. Par le \myref{lem-forme_normale_touche_rek}, 
    $\tau$ est équivalente à $\retour^k w$. De plus, il existe une suite de 
    touches $\tau' \in T^*$ qui écrit un mot $w_2$ tel que $\size{w_2} > k$. 
    
    Alors $\config{\epsilon}{\epsilon} \cdot \tau \tau' = \config{w}{\epsilon}$
    avec $w$ de la forme $w'a(ab)^{\normeInf{K}}x_1$ où $w'$ est non vide (
    par le \myref{lem-localite}).
    
    Soit maintenant $f \in F$. Alors, par le \myref{lem-localite},
    $\config{w}{\epsilon} \cdot f$ mène à un mot de $\miroir{L}$
    ayant pour préfixe $w'aab$. Impossible, car le facteur $aa$ n'apparaît 
    qu'en tout début de mot pour un mot de $\miroir{L}$.
\end{Preuve}

\begin{Theoreme}
    $\GK$ n'est pas stable par miroir.
\end{Theoreme}
\begin{Preuve}
    Soit $K = \set{\touche{a\gauche\gauche bc}}$ un clavier de $\GK$
    et $L = \langage{K}$. Pour commencer, on montre que 
    $L = \ensC{b^nc(ca)^{n - 1}a}{n \in \N^*}$. En effet,
    \[
        \config{\epsilon}{\epsilon} \cdot \touche{a\gauche\gauche bc}
        = \config{bc}{a} 
        \quad \text{ et } \quad
        \config{b^nc}{(ca)^{n - 1}a} \cdot \touche{a\gauche\gauche bc} 
        = \config{b^{n + 1}c}{(ca)^na}. 
    \]
    Ainsi, il nous suffit de montrer que $\miroir{L} = \ensC{a(ac)^{n - 1}cb^n}{n \in \N^*}$
    n'est pas reconnaissable par un clavier de $\GK$. Supposons avoir $K$ de $\GK$ 
    de langage $\miroir{L}$.
    Toute touche de $K$ écrit autant de $a$ que de $b$ et de $c$, et le 
    \myref{lem-insensibilite-position} nous assure qu'elle les écrit dans toutes
    les configurations. Soit $t$ qui écrit des lettres. 
    
    Il existe $\tau \in K^*$ qui mène à $\config{u}{v}$ avec $uv = acb$.
    On pose $\config{u'}{v'} = \config{u}{v} \cdot \tau$. Alors $u'v'$ 
    contient au moins deux $a$ et au moins deux $c$.
    
    Supposons que $c \in v$, alors on obtient par le \myref{thm-fondamental_grek}
    que $cb$ est un suffixe de $u'v'$. Absurde, car $u'v$ contient au moins deux $b$.
    
    Donc $c \in u$ et $v$ ne contient pas de $c$. Le même raisonnement montre que 
    $v'$ ne contient pas de $c$. En particulier, pour écrire $u'v'$ à partir de 
    $\config{u}{v}$, on n'est jamais passé à gauche du $c$ déjà présent dans $uv$ 
    et on obtient que $ac$ est un préfixe de $u'v'$. Absurde, car $u'v'$ contient 
    au moins deux $a$.
\end{Preuve}

%% file: 6Stabilite/intersection.tex
\section{Stabilité par intersection}

Il semble assez clair que les classes de langages de clavier
ne sont pas stables par intersection. Nous allons montrer ce résultat en trois
fois. Une première fois pour $\EK$ et $\MK$, puis pour les claviers sans
flèches, et enfin pour les claviers avec flèches.

\subsection{\EK{} et \MK{}}

\begin{Theoreme}
    $\MK$ et $\EK$ ne sont pas stables par intersection.
\end{Theoreme}
\begin{Preuve}
    On considère $L = \ensC{bw \enlarger}{w \in (a+b)^*, \size{w} \text{ est impair et } aa \notin \Fact{w}}$. On a
    \[
        L = (ab + bb + ba)^* \cap (ba + b)^*.
    \]
    On a bien $(ab + bb + ba)^*$ et $(ba + b)^*$ dans $\MK$. Nous allons montrer que $L$
    n'est pas dans $\EK$.
    
    Supposons qu'il existe $K = (T, F)$, un clavier de $\EK$ de langage $L$. 
    On a $L = T^*F$, et puisque tous les mots de $L$ sont de taille paire, 
    alors tous les mots de $T$ et de $F$, donc toutes les touches de $T$ et de $F$
    sont de longueurs paires.
    
    Puisque $w = (ba)^{\normeInf{K}} \in L$, alors il existe $t_1 \ldots t_n f \in T^*F$
    tel que $t_1 \ldots t_n f = w$ et $n > 0$. Puisque $\size{f}$ est paire, 
    alors $t_n$ se termine par un $a$ (on peut supposer que $T$ ne contient pas
    la touche vide).
    
    De même, $w' = bb(ab)^{\normeInf{K}} \in L$, donc il existe $t_1' \ldots t_m' f' \in T^*F$
    tel que $t_1' \ldots t_m' f' = w'$ et $m > 0$. Puisque $\size{f'}$ est paire,
    alors $f'$ commence par un $a$.
    
    Mais $t_nf' \in L$ et contient un facteur $aa$, d'où une contradiction. 
\end{Preuve}

\subsection{\RK{} et \REK{}}

Pour montrer que $\RK$ et $\REK$ ne sont pas stables par intersection, 
nous allons utiliser $L_1$ du clavier $K_1 = \set{\touche{\retour a \diese \dollar},
\touche{\retour\retour b \diese \dollar\dollar}}$ vue en \cref{sec-rk_not_in_fek}. 
Nous rappelons ici le \myref{lem-forme-lc} (avec les notations 
$t_a = \touche{\retour a \diese \dollar}$ et 
$t_b = \touche{\retour\retour b \diese \dollar\dollar}$).

\LemFormeLC*

\begin{Theoreme}
    $\RK$ et $\REK$ ne sont pas stables par intersection.
\end{Theoreme}
\begin{Preuve}
    On pose $L_2$ le langage $(a + b + \diese)^+ \dollar^2$ reconnu
    par le clavier 
    \[
        K_2 = \set{\touche{\retour^2a\dollar^2}, \touche{\retour^2b\dollar^2}, \touche{\retour^2\diese\dollar^2}}.
    \]
    Par le \myref{lem-forme-lc}, on a que l'intersection $L$ de $L_1$ et $L_2$ est
    $\bigl[(a\diese)^*a + \epsilon\bigl] (b\diese)^+\dollar^2$. Nous allons
    montrer que ce langage n'est pas dans $\REK$.
    
    Supposons avoir $K = (T, F)$ un clavier de $\REK$ de langage $L$. Nous
    pouvons supposer que $K$ est sous forme normale.
    
    Puisque $L$ contient $(a\diese)^{2\normeInf{K} + 1}b\diese \dollar^2$, il 
    existe $\tau_a \in T^*$ qui écrit $(a\diese)^{2\normeInf{K} + 1}w$ pour un 
    certain $w$.
    
    De même, puisque $L$ contient $(b\diese)^{(\size{\tau} + 1)\normeInf{K} + 1}\dollar^2$,
    alors il existe $\tau_b \in T^*$ qui écrit $(b\diese)^{(\size{\tau} + 1)\normeInf{K} + 1}w'$
    pour un certain $w'$.
    
    En appliquant $\tau_b\tau_a$, le \myref{lem-localite} nous assure qu'on obtient
    un mot qui commence par un $b$. Mais le nombre de $a$ de ce mot est au moins 
    $\normeInf{K} + 1$. Ainsi, en lui appliquant une touche finale $f$, on obtient
    que $L$ contient un mot qui commence par un $b$ (toujours par le \myref{lem-localite}), 
    et qui contient au moins un $a$ ($f$ efface au plus $\normeInf{K}$ lettres).
    Contradiction, un tel mot ne peut pas être dans $L$.
\end{Preuve}

\subsection{Au-dessus de \GK{}}

Et finalement, on s'intéresse aux classes restantes.

\begin{Theoreme}
    Aucune classe contenant $\gauche$ n'est stable par intersection.
\end{Theoreme}

\begin{Preuve}
    On s'intéresse à $L = \ensC{a^nb^nc^n}{n \in \N}$. On a que $L$
    est l'intersection des quatre langages suivants.
    \begin{align*}
        \langage{\set{a, b, c\gauche}} &= (a + b)^*c^* 
        & \langage{\set{ab, ba, c, \gauche}} &= \ensC{w \in \set{a, b, c}^* \enlarger}{\size{w}_a = \size{w}_b}\\
        \langage{\set{a, b\gauche, c\gauche}} &= a^*(b + c)^* 
        & \langage{\set{bc, cb, a, \gauche}} &= \ensC{w \in \set{a, b, c}^* \enlarger}{\size{w}_c = \size{w}_b}
    \end{align*}
    Nous allons montrer que $L$ n'est pas dans $\FREK$. Supposons 
    qu'il existe $K = (T, F)$, un clavier de langage $L$.
    
    On pose $M = 6\normeInf{K} + 1$. Soit $\tau f \in T^*F$ une 
    exécution de taille minimale menant à accepter le mot 
    $w = a^Mb^Mc^M$. On pose $\config{u}{v}$ la configuration atteinte.
    Puisque $w > \normeInf{K}$, alors $\tau$ est non vide, donc il
    existe $t \in T$ et $\tau' \in T^*$ tel que $\tau = \tau't$.
    
    On a soit $a^Mb^{3\normeInf{K} + 1}$ préfixe de $u$, soit
    $b^{3\normeInf{K} + 1}c^M$ suffixe de $v$. Nous supposons
    être dans le premier cas, l'autre cas est similaire.
    Nous posons alors $\config{u'}{v'} = \config{\epsilon}{\epsilon} \cdot \tau'$
    et $\config{x}{y} =  \config{\epsilon}{\epsilon} \cdot \tau' f$.
    
    Par le \myref{lem-localite}, on a $a^Mb^{\normeInf{K} + 1}$ préfixe de $u'$,
    puis $a^Mb$ est un préfixe de $x$. Mais puisque $xy \in L$, on a alors forcément
    $xy = a^Mb^Mc^M = w$. Ceci contredit la minimalité de $\tau f$ et nous donne
    alors le résultat.
\end{Preuve}

%% file: 6Stabilite/union.tex
\section{Stabilité par union}

    \begin{Theoreme}
        $\MK$, $\EK$, $\RK$, $\GK$, $\FK$, $\GRK$, $\GEK$, $\FEK$ et $\GREK$
        ne sont pas stables par union.
    \end{Theoreme}
    \begin{Preuve}
        Nous avons montré en sous-section \ref{subsec-rat_not_in_grek} que
        $a^* + b^*$ n'est pas dans $\GREK$ ni dans $\FEK$. Or $a^*$ et $b^*$
        sont dans $\MK$.
    \end{Preuve}

    \subsection{\FRK{} et \FREK{} ne sont pas stables par union}

    On pose $L_a = \ensC{a^nca^n}{n \in \N}$ et $L_b = \ensC{b^ncb^n}{n \in \N}$.
    Nous allons montrer que $L = L_a \cup L_b$ n'est pas dans $\FREK$.

    Notons qu'ils sont bien dans $\FRK$, reconnus respectivement par les claviers
    $K_a = \set{\retour c, \retour aca \gauche}$ et
    $K_b = \set{\retour c, \retour bcb \gauche}$.

    Nous supposons donc qu'il existe un clavier $K = (T, F)$ de $\FREK$ qui
    reconnaît $L$.

    Pour commencer, nous allons montrer que toute touche de $K$ contient au
    moins autant de $\retour$ que de lettres. 
    Le lemme qui suit est un préliminaire à ce résultat.

    \begin{Lemme} \label{lem-union-frek-center}%
        Soit $M \in \N$. Alors, il existe $\tau_a \in T^*$ telle
        que $\config{\epsilon}{\epsilon} \cdot \tau_a = \config{u_a}{v_a}$ avec
        $\size{u_a} > M$ et $a^{\normeInf{K}+1}$ préfixe
        de $u_a$. De même, il existe $\tau_b \in T^*$ telle que
        $\config{\epsilon}{\epsilon} \cdot \tau_b = \config{u_b}{v_b}$ avec
        $\size{u_b} > M$ et $b^{\normeInf{K}+1}$ préfixe de $u_b$.
    \end{Lemme}

    \begin{Preuve}
        Puisque $w = a^{M + 4 \normeInf{K}} c a^{M + 4\normeInf{K}} \in L$, alors
        il existe $\tau \in T^*$ et $f \in F$ tels que
        $\config{\epsilon}{\epsilon} \cdot \tau f = \config{x}{y}$ avec $w = xy$.

        Posons $\config{u}{v} = \config{\epsilon}{\epsilon} \cdot \tau$. Nous
        pouvons supposer que $\tau$ est de taille minimale. Clairement, $\tau$ est
        non vide donc est de la forme $\tau' t$. Nous posons alors
        $\config{u'}{v'} = \config{\epsilon}{\epsilon} \cdot \tau'$.

        Supposons $\size{u} \leq M$. Alors, par le \myref{lem-encadrement_taille},
        $\size{x} \leq M + \normeInf{K}$, donc $\size{y} \geq M + 7\normeInf{K}+1$.
        On en déduit que $a^{3\normeInf{K}}ca^{M + 4\normeInf{K}}$ est un suffixe
        de $y$, et par le \myref{lem-localite},
        $a^{2\normeInf{K}}c a^{M + 4\normeInf{K}}$ est un suffixe de $v'$.

        Toujours par \myref{lem-localite}, $c a^{M + 4\normeInf{K}}$ est un suffixe
        du mot obtenu avec $\config{\epsilon}{\epsilon}\cdot \tau' f$. Puisque ce mot
        est dans $L$, il s'agit de $w$, ce qui contredit la minimalité de $\tau$.

        Donc $\size{u} \geq M$. Une autre application du \myref{lem-localite} nous
        donne que $a^{\normeInf{K}+1}$ est un préfixe de $u$. On peut alors prendre
        $\tau_a = \tau$.

        Une preuve similaire nous donne l'existence de $\tau_b$.
    \end{Preuve}

    \begin{Lemme} \label{lem-union-frek-retours-lettres}%
        Soit $t \in T$. Alors $t$ contient au moins autant de $\retour$ que
        de lettres, \ie{}
        $\size{t}_{\retour} \geq \size{t}_a + \size{t}_b + \size{t}_c$.
    \end{Lemme}

    \begin{Preuve}
        Soit $M = 3(\normeInf{K} + 4)$ et soient $\tau_a, \tau_b$, $u_a, u_b$ et $v_a, v_b$
        donnés par le \myref{lem-union-frek-center} avec la constante $M^2$. Supposons qu'il
        existe $t \in T$ tel que $\size{t}_{\retour} < \size{t}_a + \size{t}_b + \size{t}_c$.
        On considère $\tau = t^M$. Alors $\tau$ écrit de manière effective $\normeInf{K}+2$
        fois $a$, $b$ ou $c$.

        Si ce sont des $c$ ou des $b$, alors on pose $\config{u}{v} = \config{u_a}{v_a} \cdot \tau$.
        Par le \myref{lem-localite}, $a^{\normeInf{K} + 1}$ est un préfixe de
        $u$. En effet, $\tau$ contient au plus $M$ symboles et
        \[
        \size{u_a} >
        M^2 \geq
        3(\normeInf{K}+4)\normeInf{K} + \normeInf{K} + 1.
        \]
        Mais le nombre de $b$ (ou de $c$) de $uv$ est alors supérieur ou égal à
        $\normeInf{K} + 2$, donc en appliquant une touche finale à $\config{u}{v}$,
        on obtient un mot qui commence par un $a$ et qui contient deux $b$ ou deux
        $c$. Impossible puisque ce mot doit appartenir à $L$.

        Si ce sont des $a$, on fait une preuve similaire en posant cette fois
        $\config{u}{v} = \config{u_b}{v_b} \cdot \tau$.

        Donc aucune touche n'écrit plus de lettres qu'elle n'en efface.
    \end{Preuve}

    En particulier, ce lemme nous apporte deux choses.
    \begin{enumerate}
        \item La longueur du mot ne peut augmenter que si l'on est près du bord
              gauche (au moins un $\retour$ ne doit pas avoir d'effet).
        \item $\normeInf{K} \geq 2$.  
    \end{enumerate}

    On montre maintenant que le nombre de lettres de la configuration ne peut pas trop
    augmenter lorsqu'on est trop loin du centre. En effet, si par exemple l'on rajoute 
    beaucoup de $b$ en étant trop à droite du centre du mot, le mot sera trop déséquilibré 
    et  appliquer une touche finale ne mènera pas à un palindrome.

    \begin{Lemme} \label{lem-union-frek-bord-gauche}%
        Soient $\tau_0 \in T^*$ et $t_1, \ldots, t_n \in T$.
        Pour tout $i \in \intervE[0, n]$, on pose
        \[
        \config{u_i}{v_i} = \config{\epsilon}{\epsilon}
        \cdot \tau_0 t_1\cdots t_i.
        \]
        Si $\size{u_nv_n} \geq \size{u_0v_0} + 4\normeInf{K} + 1$, alors il existe $i \in \intervE[0, n]$ tel que
        \[
        \abs{\enlarger\size{u_i} - \size{v_i}} \leq 8 \normeInf{K} + 2 \leq 9 \normeInf{K}.
        \]
    \end{Lemme}

    \begin{Preuve}
        Supposons que $\size{u_nv_n} \geq \size{u_0v_0} + 4\normeInf{K} + 1$ et que pour tout $i \in \intervE[0, n]$,
        \[
        \abs{\enlarger\size{u_i} - \size{v_i}} \geq 8 \normeInf{K} + 2.
        \]
        Puisque $\abs{\enlarger\size{u_{i + 1}} - \size{u_i}} \leq \normeInf{K}$
        et $\abs{\enlarger\size{u_{i + 1}} - \size{u_i}} \leq \normeInf{K}$, alors 
        $\size{u_i} - \size{v_i}$ et $\size{u_{i + 1}} - \size{v_{i + 1}}$ ont le 
        même signe. En particulier, ce signe est alors le même pour tout $i$. 
        Nous supposons ce signe positif, l'autre cas étant similaire.

        Soit $f \in F$. Pour tout $i$, on pose $\config{u^f_i}{v_i^f} = \config{u_i}{v_i}\cdot f$. Par le \myref{lem-encadrement_taille},
        on a
        \[
        \size{u_i^f} - \size{v_i^f} \geq 6\normeInf{K} + 2.
        \]
        Puisque  $u_i^fv_i^f \in L$, nous avons alors $u_i^f$ de la forme
        $a^mca^p$ ou $b^mcb^p$ avec $p \geq 3\normeInf{K} + 1$. Nous supposons
        être dans le premier cas, l'autre cas étant similaire.

        Par le \myref{lem-localite}, $a^mca^{2\normeInf{K} + 1}$ est un préfixe
        de $u_i$, donc $a^mca^{\normeInf{K} + 1}$ est un préfixe de
        $u_{i + 1}$ et donc $a^mca$ est un préfixe de $u_{i + 1}^f$. Puisque
        $u_{i + 1}^f v_{i + 1}^f \in L$, cela impose
        $u_{i + 1}^f v_{i + 1}^f = a^mca^m$.

        Par suite, pour tout $i$, on a $u_i^fv_i^f = a^mca^m$. En particulier,
        on a $u_n^fv_n^f = u_0^fv_0^f$, et par le \myref{lem-encadrement_taille},
        \[
        \abs{\enlarger \size{u_nv_n} - \size{u_0v_0}} \leq 4\normeInf{K},
        \]
        ce qui contredit les hypothèses.
    \end{Preuve}

    Nous avons montré que le nombre de lettres ne pouvait augmenter que lorsqu'on était
    près du bord gauche du mot, et que nous ne pouvions pas trop augmenter le nombre de
    lettres du mot en étant loin de son centre.

    Ainsi, pour écrire un très long mot $a^Mca^M$, il nous faut faire des allers-retours
    entre le bord gauche et le centre du mot. Nous isolons alors un moment de l'exécution
    au cours duquel le curseur est loin du centre et du bord, va au bord gauche puis
    retourne loin du bord et du centre après avoir rajouté des lettres.

    On obtient alors deux configurations avec un nombre différents de lettres. Mais
    l'application d'une touche finale $f$ a le même comportement dans les deux cas
    (car nous sommes loin des bords). Donc les deux mots obtenus ne peuvent pas être
    les mêmes (on est parti de mots de taille différente).

    Cependant, comme nous étions toujours entre le bord gauche et le centre, mais loin
    du centre, un suffixe $ca^M$ est commun aux deux mots (l'application de $f$
    n'a pas pu le modifier car trop loin), et les deux mots doivent être égaux !

    C'est ceci que nous allons formaliser dans les lemmes qui suivent. 
    
    Nous posons $M = 30 \normeInf{K}$, $\tau = t_1 \ldots t_n f \in T^*F$ une 
    exécution menant à $\config{u}{v}$ avec $\config{u}{v} = a^Mca^M$ et pour tout 
    $i \in \intervE[0, n]$
    \[
        \config{u_i}{v_i} = \config{\epsilon}{\epsilon} \cdot t_1 \cdots t_i
        \text{ et }
        w_i = u_iv_i.
    \]
    De plus, pour tout $i \in \intervE[1; n]$, nous définissons les prédicats 
    $B$ (pour près du bord) et $C$ (pour près du centre)  par 
    \begin{align*}
        B(i) &\text{ si et seulement si } 
            \size{u_i} \leq \normeInf{K}\\
        C(i) &\text{ si et seulement si } 
            \abs{\enlarger\size{u_i} - \size{v_i}} \leq 9 \normeInf{K}
    \end{align*}

On va découper l'exécution en notant des moments clés : un $a_i$ correspond à la première
fois qu'on est près du bord, puis entre un $a_i$ et un $b_i$ on ne va jamais près du centre.
Un $b_i$ correspond à la dernière fois qu'on est près du bord avant d'aller près du centre.
Un $c_i$ correspond à la première fois qu'on est près du centre après s'être éloigné du bord
(après $b_i$). Et un $d_i$ correspond à la dernière fois qu'on est près du centre avant 
de retourner près du bord.

\begin{Lemme} \label{lem-union-sequence}%
    Il existe $m \geq 2$, et 
    $a_0 \leq b_0 < c_0 \leq d_0 < \ldots < c_{m - 1} \leq d_{m - 1} < a_m \leq b_m$ tels que 
    \begin{itemize}
        \item pour tout $i \geq a_0$, $\size{w_i} \geq 11 \normeInf{K}$ ;
        \item pour tout $0 \leq i \leq m$, $B(a_i)$ et $B(b_i)$ sont vrais ;
        \item pour tout $0 \leq i < m$, $C(c_i)$ et $C(d_i)$ sont vrais ;
        \item pour tout $0 \leq i \leq m$, pour tout $a_i \leq j \leq b_i$, $C(j)$ est faux ;
        \item pour tout $0 \leq i < m$, pour tout $c_i \leq j \leq d_i$, $B(j)$ est faux ;
        \item pour tout $0 \leq i < m$, pour tout $b_i < j < c_i$, $B(j)$ et $C(j)$ sont faux ;
        \item pour tout $0 \leq i < m$, pour tout $d_i < j < a_{i + 1}$, $B(j)$ et $C(j)$ sont faux ;
        \item $\size{w_{b_0 + 1}} < \size{w_{a_m}}$.
    \end{itemize}
\end{Lemme}

\begin{Preuve}
Soit $j_1$ le plus petit entier tel que pour tout $i \geq j_1$, 
$\size{w_i} \geq 11 \normeInf{K}$. En particulier, on a 
$11 \normeInf{K} \leq \size{w_{j_1}} \leq 12 \normeInf{K}$ et, par 
le \myref{lem-union-frek-retours-lettres}
$\size{u_{j_1}} \leq \normeInf{K}$ (donc $B(j_1)$ est vrai).

Il existe $j_2 > j_1$ tel que $\size{w_{j_2}} \geq 27 \normeInf{K}$.
Là, encore, on peut supposer, grâce au \myref{lem-union-frek-retours-lettres} que 
$\size{u_{j_2}} \leq \normeInf{K}$ (donc $B(j_2)$ est vrai).

On pose $a_0 = j_1$ et on construit la séquence inductivement. Supposons avoir
$a_i$. On a deux cas.
\begin{itemize}
    \item Si pour tout $a_i < k < j_2$, $C(k)$ est faux, alors on pose $m = i$
          et $b_m = j_2$.
    \item Sinon, il existe $a_i < k < j_2$ tel que $C(k)$ est vrai. On pose alors
    \begin{align*}
        c_i = \min \ensC{k}{a_i < k < j_2 \text{ et $C(k)$ est vrai}} \\
        b_i = \max \ensC{k}{a_i \leq k < c_i \text{ et $B(k)$ est vrai}}.
    \end{align*}
\end{itemize}
Notons que $b_i$ existe bien car $B(b_i)$ est vrai. De plus,
par construction, on a que pour tout $a_i \leq j \leq b_i$, $C(j)$ est faux
et que pour tout $b_i < j < c_i$, $B(j)$ et $C(j)$ sont faux). 
    
Et si on a $c_i$, on fait de même. On pose
\begin{align*}
    a_{i + 1} = \min \ensC{k}{c_i < k \leq j_2 \text{ et } B(k) }\\
    d_i = \max \ensC{k}{c_i \leq k < a_{i + 1} \text{ et } C(k) }.
\end{align*}
Notons que $b_{i + 1}$ existe bien car $B(j_2)$ est vrai et que $c_i'$ 
existe bien car $C(c_i)$ est vrai. Et on a bien que pour tout 
$c_i \leq j \leq d_i$, $B(j)$ est faux
et que pour tout $d_i < j < a_{i + 1}$, $B(j)$ et $C(j)$ sont faux).

Le \myref{lem-union-frek-retours-lettres} nous assure que 
$\size{w_{b_0}} \leq 17 \normeInf{K}$ et 
$\size{w_{a_m}} \geq 22 \normeInf{K}$. En appliquant le 
\myref{lem-encadrement_taille}, on obtient alors 
\[
    \size{w_{b_0 + 1}} \leq 18 \normeInf{K} < \size{w_{a_m}}.
\]

Il nous reste à vérifier, que $m \geq 2$.

Soit $j_3 > j_1$ le plus petit entier tel que $\size{w_{j_3}} \geq 18 \normeInf{K}$.
En particulier, on a $18 \normeInf{K} \leq \size{w_{j_3}} \leq 19 \normeInf{K}$ et 
le \myref{lem-union-frek-retours-lettres} nous permet de supposer que 
$\size{u_{j_3}} \leq \normeInf{K}$. De plus, on a $j_3 < j_2$.

Le \myref{lem-union-frek-retours-lettres} nous assure alors qu'il existe deux indices 
$i_1$ et $i_2$ tels que $j_1 < i_1 < j_3 < i_2 < j_2$ et $C(i_1)$ 
et $C(i_2)$ sont vrais. On en déduit que $m \geq 2$ 
(il y a au moins deux-allers retours).
\end{Preuve}

Ce lemme nous assure qu'il existe des passages centre-bord-centre. En trouvant 
deux indices $i$ et $j$ loin du bord dans un de ces passages avec 
$\size{u_iv_i} \neq \size{u_jv_j}$, et on pourra conclure.

\begin{Lemme}
    Le langage $L$ n'est pas dans $\FREK$.
\end{Lemme}
\begin{Preuve}
    On considère $a_0 \leq b_0 < c_0 \leq d_0 < \ldots < c_{m - 1} \leq d_{m - 1} < a_m \leq b_m$
    tels que définis par le \myref{lem-union-sequence}.
    Le \myref{lem-union-frek-retours-lettres} nous assure que lors de 
    tout passage loin du bord, la taille du mot ne peut que décroître,
    c'est-à-dire que pour tout $0 < i < m$,
    \begin{equation} \label{eq-utile}
        \size{w_{b_i + 1}} \geq \size{w_{b_i + 2}} \cdots 
        \geq \size{w_{a_{i + 1} - 1}} \geq \size{w_{a_{i + 1}}}.
    \end{equation}
    
    Supposons que pour tout $i$, pour tout $k_1$, $k_2$ tels que
    $d_i \leq k_1 < a_{i + 1}$ et $b_{i + 1} < k_2 \leq c_{i + 1}$, on a
    $\size{w_{k_1}} = \size{w_{k_2}}$. 
    
    En particulier, on obtient grâce à l'\cref{eq-utile} que pour tout 
    $k_1, k_2$ de la séquence qui ne sont pas entre un $a_i$ et un $b_i$ 
    (donc loin des bords), on a $\size{w_{k_1}} = \size{w_{k_2}}$. Alors, 
    toujours grâce à l'\cref{eq-utile},
    $\size{w_{a_m}} \geq \size{w_{a_m - 1}}$ et donc
    $\size{w_{a_m}} \leq \size{w_{b_0 + 1}}$, ce qui contredit le 
    \myref{lem-union-sequence}.
    
    Don, il existe $i$, $d_i < k_1 < a_{i + 1}$ et 
    $b_{i + 1} < k_2 < c_{i + 1}$ tels que 
    $\size{w_{k_1}} \neq \size{w_{k_2}}$, c'est-à-dire qu'il existe
    un aller-retour durant lequel la taille du mot varie.
    
    On note $\config{x_1}{y_1} = \config{u_{k_1}}{v_{k_1}} \cdot t$
    et $\config{x_2}{y_2} = \config{u_{k_2}}{v_{k_2}} \cdot t$.
    On suppose $x_1y_1 = a^Nca^N$ avec $N \in \N$ (le cas où le mot
    contient des $b$ est similaire).
    
    Notons que pour tous $k_1 \leq i \leq k_2$, on a 
    $\size{v_i} - \size{u_i} > 9 \normeInf{K}$.
    De plus, pour tous $k_1 \leq i, j \leq k_2$, on a 
    $\abs{\enlarger\size{w_i} - \size{w_j}} < 5 \normeInf{K}$,
    par le \myref{lem-union-frek-bord-gauche}.

    Soit $f \in F$. On note $\config{x_1}{y_1} = \config{u_{k_1}}{v_{k_1}} \cdot f$.
    On a $x_1y_1 \in L$. (On va supposer $x_1y_1 = a^Nca^N$, le cas avec des $b$ est
    On a,  par le \myref{lem-encadrement_taille} que
    $2N \leq \size{w_{k_1}} \leq 2N + \normeInf{K}$. 
    
    Et puisque, $\size{v_{k_1}} - \size{u_{k_1}} > 9 \normeInf{K}$,
    on a, que $\size{v_{k_1}} > N + 4 \normeInf{K}$ et que 
    $\size{u_{k_1}} < N - 5\normeInf{K}$. Et le \myref{lem-localite}
    nous donne que $a^{3\normeInf{K}}ca^N$ est un suffixe de $v_{k_1}$.
    
    Nous allons montrer que pour tout $k_1 \leq i \leq k_2$,
    on a $\size{v_i} \geq N + 2\normeInf{K}$. En
    effet, si on avait $\size{v_i} < N + 2 \normeInf{K}$,
    alors on aurait $\size{u_i} < N - 7 \normeInf{K}$,
    et donc $\size{w_i} < 2N - 5\normeInf{K} < \size{w_{k_1}} - 5\normeInf{K}$.
    Absurde.
    
    Ainsi, pour tout $k_1 \leq i \leq k_2$, on a 
    $\size{v_i} \geq N + 2\normeInf{K}$. En particulier,
    une récurrence et le \myref{lem-localite} nous permet d'avoir
    que $v_{k_1}[\size{v_{k-1}} - N + \normeInf{K}]$ est un suffixe
    de tous les $v_i$ et donc $a^{\normeInf{K}}ca^N$ est un
    suffixe de tous les $v_i$.
    
    Notons alors $\config{x_2}{y_2} = \config{u_{k_2}}{v_{k_2}} \cdot f$.
    Par le \myref{lem-localite}, $ca^N$ est un suffixe de $y_2$, et
    donc $x_2y_2 = a^Nca^N = x_1y_1$.
    
    Mais, puisque dans les deux configurations $\config{u_{k_1}}{v_{k_1}}$
    et $\config{u_{k_2}}{v_{k_2}}$, la touche $t$ agit de manière 
    efficiente (on est loin des bords), alors
    \[
        \size{x_2y_2} - \size{w_{k_2}} = \size{x_1y_1} - \size{w_{k_1}}. 
    \]
    Absurde, car $\size{w_{k_1}} \neq \size{w_{k_2}}$ !
    
    D'où le résultat.
\end{Preuve}

%% file: 6Stabilite/concatenation.tex
\section{Stabilité par concaténation}

De même, nous allons montrer qu'aucune classe de clavier n'est
stable par concaténation. On s'intéresse d'abord
aux classes de clavier sans flèches, et ensuite à celles qui 
restent.

\subsection{Les claviers sans flèches}

Soit $L = a^*c^*$. Nous allons montrer par l'absurde que
$L$ n'est pas dans $\REK$. On possède une suite de touches 
$\tau$ qui écrit un $a$ et valide. Mais on possède également 
une suite de touches $\tau'$ qui écrit un nombre aussi grand 
de $c$ que souhaité, sans valider. On conclut en regardant 
$\epsilon \cdot \tau' \cdot \tau'$.

\begin{Proposition} \label{prop-rat_not_in_rek}%
    Le langage $L$ n'est pas dans $\REK$.
\end{Proposition}
\begin{Preuve}
Raisonnons par l'absurde et supposons qu'il existe un clavier $K$ de $\REK$
reconnaissant $L$. Nous posons $M = \normeInf{K}$.

Puisque $a \in L$, il existe une suite de touches transientes $t_1\ldots t_i$
et une touche finale $t_f$ telles que $\epsilon \cdot t_1 \ldots t_it_f = a$.
La suite $t_1 \ldots t_it_f$ est alors, par mise sous forme normale,
équivalente à $\tau = \retour^ka\entree$.

De même, puisque $c^{k + M + 1} \in L$, il existe il existe une suite de touches
transientes $t'_1\ldots t'_j$ et une touche finale $t'_f$ telles que
$\epsilon \cdot t'_1 \ldots t'_jt'_f = c^{k + M + 1}$.
Puisque $\taille{t_f} \leq M$, le \myref{lem-localite} nous permet de
dire que $u = \epsilon \cdot t'_1 \ldots t'_j$ admet
$c^{k + 1}$ comme préfixe.

Mais on a alors $w \eqDef \epsilon \cdot (t'_1\ldots t'_j) \cdot \tau \in L$,
c'est-à-dire que $u[1, \taille{u} - k]a \in L$.

Puisque $u$ admet $c^{k + 1}$ comme préfixe, $u[1, \taille{u} - k]$
admet $c$ comme préfixe et donc $w$ est de la forme $cva$ (avec $v \in A^*$)
ce qui est impossible.
\end{Preuve}

\begin{Theoreme}
    $\MK$, $\RK$, $\EK$ et $\REK$ ne sont pas stables par concaténation.
\end{Theoreme}
\begin{Preuve}
    $L$ n'est pas dans $\REK$ or $a^*$ et $b^*$ sont dans $\MK$.
\end{Preuve}

\subsection{Les autres classes}

\begin{Theoreme}
    $\GRK$ n'est pas stable par concaténation.
\end{Theoreme}
\begin{Preuve}
    Soit $L = (a^2)^*(b + b^2)$. Nous avons montré en \cref{sec_ek_not_in_grk}
    que $L \not\in \GRK$. Mais $(a^2)^*$ est reconnu par $\set{\touche{aa}}$
    et $(b + b^2)$ par $\set{\touche{\retour^2 b}, \touche{\retour^2 b^2}}$.
\end{Preuve}

\begin{Theoreme}
    $\GK$ et $\FK$ ne sont pas stables par concaténation.
\end{Theoreme}
\begin{Preuve}
    Soit $L = a^nb^nc^md^m$. Le langage $a^nb^n$ est reconnu par $\set{\touche{ab\gauche}}$.
    Supposons $L$ reconnu par $K$. Il existe une touche $t_1$ de $K$ qui écrit un $a$ (et
    alors elle écrit aussi un $b$). De même, il existe une touche $t_2$ de $K$ qui écrit un
    $c$ et elle écrit aussi un $d$.
    Soient $M = 2\normeInf{K}$ et $\tau$ qui mène à $\config{u}{v}$ avec $uv = a^Mb^Mc^Md^M$.
    On a alors deux cas.
    \begin{itemize}
        \item Si $u \leq 2M$, en appliquant $t_2$ à $\config{u}{v}$, on obtient un mot 
        $w'$ tel que $\size{w'}_d > M$ et qui a pour suffixe $cd^M$ (par le
        \myref{lem-localite}).
        \item Sinon, on applique cette fois $t_1$ à $\config{u}{v}$ et on obtient un mot 
        $w'$ tel que $\size{w'}_a > M$ et qui a pour préfixe $a^Mb$ (par le
        \myref{lem-localite}).
    \end{itemize}
    Absurde dans les deux cas, d'où le résultat.
\end{Preuve}

Pour montrer que les autres classes de clavier ne sont pas stables par concaténation,
on s'intéresse à $L = \ensC{a^nca^na^mca^m}{n, m \in \N}$, concaténation de 
$L_1 = \ensC{a^nca^n}{n \in \N}$ avec lui-même. Commençons par montrer que ce dernier
langage est dans $\FRK$ et dans $\GEK$. Il est bien dans $\GEK$, reconnu par
$\left(\enlarger\set{\touche{ab\gauche}}, \set{\touche{c}}\right)$.

\begin{Lemme}
    Le langage $L_1$ est dans $\FRK$.
\end{Lemme}
\begin{Preuve}
    On pose $K = \set{\touche{\retour c}, 
    \touche{\retour aca \gauche}}$. Alors $\langage{K} = L_1$.
    En effet,
    \begin{align*}
        \config{\epsilon}{\epsilon} \cdot \touche{\retour c} 
                                    &= \config{c}{\epsilon} &
        \config{a^nc}{a^n} \cdot \touche{\retour c} 
                                    &= \config{a^nc}{a^n}\\
        \config{\epsilon}{\epsilon} \cdot \touche{\retour aca \gauche}
                                    &= \config{ac}{a} &
         \config{a^nc}{a^n} \cdot \touche{\retour aca \gauche}
                                    &= \config{a^{n + 1}c}{a^{n + 1}}
    \end{align*}
\end{Preuve}

Nous allons montrer que $L$ n'est pas dans $\FREK$. Supposons avoir
$K = (T, F)$ de langage $L$. Nous allons montrer qu'il existe $t \in T$
qui écrit ou supprime des lettres loin des bords.

\begin{Lemme} \label{lem-concatenation_frek_helper}%
    Il existe $t \in T$ tel que $\size{t}_{\retour} \neq \size{t}_{A}$.
\end{Lemme}
\begin{Preuve}
On pose $M = 7\normeInf{K}$ et $w = a^Mca^M a^Mca^M$. Soit 
$\tau f \in T^*f$ qui reconnaisse $w$.
Par le \myref{lem-encadrement_taille}, $\tau >2$ et donc $\tau = \tau' t$
avec $t \in T$.
On pose 
\[
    \config{u}{v} = \config{\epsilon}{\epsilon} \cdot \tau f \text{ et }
    \config{x}{y} \config{\epsilon}{\epsilon} \cdot \tau'.
\]
On commence par montrer qu'avant d'appliquer $tf$, on est loin des bords et du
centre.

Supposons $\size{x} < 2\normeInf{K}$ (le cas $\size{y} < 2\normeInf{K}$ est
similaire). Alors, en appliquant le \myref{lem-encadrement_taille}, on obtient 
que  $\size{u} < 4\normeInf{K}$. On observe alors que
$a^{3\normeInf{K}}ca^Ma^Mca^M$ est un  suffixe de $v$ et le \myref{lem-localite}
donne que $a^{\normeInf{K}}ca^Ma^Mca^M$ est un suffixe de $y$.  

En appliquant  $f$ à $\config{x}{y}$, on obtient (toujours avec le \myref{lem-localite}) un mot $w'$ de suffixe $ca^Ma^Mca^M$. Puisque 
$w' \in L$, alors $w' = w$ et $\tau f$ n'est pas minimal.

Supposons maintenant que $\size{t}_{\retour} = \size{t}_{A}$. On note
$\config{u'}{v'} = \config{x}{y} \cdot f$. Puisque $\size{x} > 2\normeInf{K}$ 
et $\size{y} > 2 \normeInf{K}$, on obtient, par le
\myref{lem-egalite_taille_loin_bords}, que $\size{uv} = \size{u'v'}$.

Supposons maintenant $\size{u} > \size{v}$ (l'autre cas est similaire).
Alors $a^Mca^M$ est un préfixe de $u$. Par le \myref{lem-localite},
$a^Mca^{5\normeInf{K}}$ est un préfixe de $x$, et encore avec le
\myref{lem-localite}, on obtient que $a^Mca^{3\normeInf{K}}$ est un
préfixe de $u'$. 

Donc $u'v'$ est de la forme $a^Mca^Ma^Nca^N$ avec $N \in \N$. Mais
puisqu'on a $\size{uv} = \size{u'v}$, on en déduit que $uv = u'v'$,
ce qui est impossible par minimalité de $\tau f$.

Donc il existe $t$ tel que $\size{t}_{\retour} \neq \size{t}_{A}$.
\end{Preuve}

\begin{Lemme}
    Le langage $L$ n'est pas dans $\FREK$.
\end{Lemme}
\begin{Preuve}
    On pose $M = 10\normeInf{K}$. Soit $\tau = t_1 \ldots t_n \in T^*F$ qui mène à 
    $\config{u}{v}$ avec $\config{u}{v} = a^Mca^Ma^Mca^M$. Pour tout 
    $i \in \intervE[0; n]$, on pose
    \[
        \config{u_i}{v_i} = \config{\epsilon}{\epsilon} \cdot t_1 \cdots t_i.
    \]
    Supposons $\size{u} > \size{v}$ (l'autre cas est similaire).
    Soit $i$ le premier indice tel que pour tout $k \geq i$, $\size{v_k} > M + 4\normeInf{K}$
    (puisque $\size{v_n} = \size{u} > 2M$, alors, par le \myref{lem-localite}, $i < n$).
    En particulier, on note que
    \[
        M + 4\normeInf{K} < \size{v_i} \leq M + 5\normeInf{K}
    \]
    et le \myref{lem-localite} avec une récurrence nous donne que
    $a^{3\normeInf{K}}ca^M$ est un suffixe de $v_k$ pour tout $k \in \intervE[i; n]$
    et donc de $v_i$.
    
    Soit $f \in F$. Posons $\config{u'}{v'} = \config{u_i}{v_i} \cdot f$. 
    Alors 
    \begin{itemize}
        \item $u'v' \in L$ ;
        \item le \myref{lem-localite} assure que $a^{2\normeInf{K}}ca^M$ est un suffixe de $v'$.
        \item le \myref{lem-encadrement_taille} assure que $\size{v'} \leq M + 6\normeInf{K}$.
    \end{itemize}
    Donc $u'v'$ est de la forme $a^Nca^Na^Mca^M$ et $a^Nca^Na^{4\normeInf{K}}$ est un suffixe 
    de $u'$. On en déduit, par le \myref{lem-localite}, que 
    $a^Nca^Na^{3\normeInf{K}}$ est un préfixe de $u_i$.
    
    On a alors que  $a^{3\normeInf{K}}ca^M$ est un suffixe de $v_i$ et que
    $a^Nca^Na^{3\normeInf{K}}$ est un préfixe de $u_i$. Soit $t$ une touche telle que 
    $\size{t}_{\retour} \neq \size{t}_A$ 
    (conformément au \myref{lem-concatenation_frek_helper}). Posons 
    \[
        \config{x}{y} = \config{u_i}{v_i} \cdot tf \text{ et }
        \config{x'}{y'} = \config{u_i}{v_i} \cdot ttf 
    \]
    Par le \myref{lem-localite}, $a^Nc$ est un préfixe de $x$ et de $x'$ et $ca^M$ est
    un suffixe de $y$ et de $y'$. Puisque $xy \in L$ et $x'y' \in L$, on en 
    déduit $xy = xy$. Mais en appliquant le \myref{lem-egalite_taille_loin_bords},
    on obtient
    \begin{align*}
        \size{xy} &= \size{u_iv_i} - \size{t}_{\retour} + \size{t}_A  - \size{f}_{\retour} + \size{f}_A\\
        \size{x'y'} &= \size{u_iv_i} - 2\size{t}_{\retour} + 2\size{t}_A  - \size{f}_{\retour} + \size{f}_A
    \end{align*}
    Il est alors impossible d'avoir $\size{xy} = \size{x'y'}$ puisque 
    $\size{t}_{\retour} \neq \size{t}_A$.
\end{Preuve}    

\begin{Theoreme}
    $\GEK$, $\FEK$, $\GREK$, $\FRK$ et $\FREK$ ne sont pas stables par concaténation.
\end{Theoreme}
\begin{Preuve}
    Les langages $a^nca^n$ et $b^ncb^n$ sont dans $\GEK$ et dans $\FRK$, mais
    ne sont pas dans $\FREK$.
\end{Preuve}

Notons qu'il est très simple de montrer que $L$ n'est pas dans $\FEK$ ou dans $\GREK$.

\begin{Lemme}
    Le langage $L$ n'est pas dans $\FEK$.
\end{Lemme}
\begin{Preuve}
    On applique le \myref{lem-distanciation_minimale} avec la lettre $c$.
\end{Preuve}

\begin{Lemme}
    Le langage $L$ n'est pas dans $\GREK$.
\end{Lemme}
\begin{Preuve}
    On suppose $L$ reconnu par $K = (T, F)$. Soit $f \in F$. Alors $f$ écrit deux 
    $c$ (car en appliquant $t$ on obtient un mot de $L$). En particulier, le 
    \myref{lem-insensibilite-position} nous assure que $f$ écrit ces deux $c$ 
    depuis n'importe quelle configuration. On en déduit que tout mot reconnu
    par $K$ contient deux $c$ à une distance au plus $\normeInf{K}$. Absurde,
    d'où le résultat.
\end{Preuve}

%% file: 6Conclusion/Conclusion.tex
\addchap{Conclusion}

Les claviers paraissent à première vue extrêmement erratiques dans leur expressivité : ils reconnaissent aisément les mots de Dyck ou les palindromes, mais certains langages rationnels semblent leur résister. Ils renversent notre vision usuelle des machines, car ne peuvent ni lire ni avoir d'état, mais peuvent effacer et se déplacer à l'aveugle dans les mots. Une simple paire de touches de quelques caractères peut mener à un langage à première vue abominable et les opérations usuelles sur les langages n'offrent guère de propriétés de clôture.

En nous intéressant aux différentes opérations élémentaires sur les claviers, nous avons pu voir que chacune apportait une grande expressivité et que la hiérarchie des langages de clavier était stricte et contrôlée par les classes usuelles de langages (rationnels, algébriques, contextuels). Cette étude permet également de mieux comprendre les opérations qu'admettent ces différentes classes : si les automates finis suffisent en l'absence de flèches, il faut une pile pour gérer les déplacements gauche et une grammaire contextuelle pour les déplacements quelconques.

Les claviers n'ayant pas fini de nous livrer leurs secrets, de nombreuses questions restent en suspens.
\begin{itemize}
    \item A-t-on $\EK \subset \FRK$ ? (dernière question de la hiérarchie)
    \item $\FREK$ est-il inclus dans les algébriques ? Nous pensons que non, après regardé le clavier $\set{\touche{a\droite\droite},\touche{b\gauche\gauche}}$ qui semble avoir un comportement très erratique.
    \item $\FREK$ est-il inclus dans une classe de langages connue ? Nous pensons que les langages de clavier sont reconnaissables par des machines de \bsc{Turing} en espace linéaire. Il suffirait pour cela de majorer le nombre de lettres à effacer pour obtenir un mot.
    \item $\FREK$ permet-il d'exprimer tous les langages rationnels ? Nous pensons que non et cherchons à montrer que $a^*b^*c^*$ ou $a^*+b^*+c^*$ ne sont pas de clavier.
    \item Caractériser la décidabilité et la complexité du problème du mot et de l'universalité
    pour toutes les classes.
    \item Conclure sur les propriétés de clôture des différentes classes de clavier.
    \item Étudier d'autres extensions, sur lesquelles nous ne nous sommes pas penchés pour le moment : claviers de Büchi (avec une sémantique à définir proprement), claviers avec états, etc.
\end{itemize}

%% file: Annexes/annexes.tex
\appendix

\chapter{Preuves}

\input{Annexes/1_generalites_annexes}
\input{Annexes/5_touche_annexes}
\input{Annexes/2_sans_fleches_annexes}
\input{Annexes/3_sans_fleche_droite_annexes}
\input{Annexes/4_sans_retour_annexes}

%% file: Annexes/1_generalites_annexes.tex
\section{Généralités}
    
\subsection[]{Preuve du \myref{lem-localite}} \label{proof-localite}

\lemLocalite*

\begin{Preuve}
    Pour tout entier $i$, nous allons noter $w_p(u, i) = u[1, \taille{u} - i]$ et
    $w_s(v, i) = v[i+ 1, \taille{v}]$.
    
    On montre le résultat pour les opérations élémentaires et une récurrence nous
    donnera le résultat. Soit $\sigma$ une opération élémentaire. On note 
    $\config{u'}{v'} = \config{u}{v} \cdot \sigma$. On a plusieurs cas.
    \begin{itemize}
        \item Si $\sigma = a \in A$, alors $u' = ua$ et
              $v' = v$, et on a bien le résultat.
        \item Si $\sigma = \retour$, on a $v' = v$. Pour montrer
              que $u[1, \size{u} - 1]$ et $u'[1, \size{u} - 1]$ sont préfixes de 
              $u'$ et de $u$, on distingue deux cas.
              \begin{itemize}
                    \item Si $u$ est vide, alors $u'$ est vide et on a
                         $w_p(u, 1)$ et $w_p(u', 1)$ vides.
                    \item Sinon, $u$ est de la forme $w_p(u, 1)c$
                          et donc $u' = w_p(u, 1)$.
              \end{itemize}
              Dans les deux cas, on a le résultat.
        \item Si $\sigma = \gauche$, on distingue deux cas.
              \begin{itemize}
                  \item Si $u$ est vide, alors $u' = u$ et $v' = v$.
                    \item Sinon, $u$ est de la forme $w_p(u, 1)c$
                          donc $u' = w_p(u, 1)$ et $v' = c v$.
              \end{itemize}
              Dans les deux cas, on a le résultat.
        \item Si $\sigma = \droite$, on distingue deux cas.
              \begin{itemize}
                  \item Si $v$ est vide, alors $u' = u$ et $v' = v$.
                  \item Sinon, $v$ est de la forme $cw_s(v, 1)$
                          donc $u' = u_i c$ et $v' = w_s(v, 1)$.
              \end{itemize}
              Dans les deux cas, on a le résultat.
    \end{itemize}
    
    Faisons maintenant la récurrence. Soit $t = \sigma_1 \ldots \sigma_n$. 
    On note $\config{u_0}{v_0} = \config{u}{v}$ et pour tout entier 
    $0 \leq i \leq n$, on note 
    $\config{u_i}{v_i} = \config{u_0}{v_0} \cdot \sigma_1 \cdots \sigma_i$
    (en particulier, pour $0 < i \leq n$, $\config{u_i}{v_i} =
    \config{u_{i - 1}}{v_{i - 1}} \cdot \sigma_i$).
    
    Nous allons montrer par récurrence sur $i$ que pour tout
    $0 \leq i \leq n$, $w_p(u, i)$ et $w_p(u_i, i)$ sont des préfixes de $u_i$ et de $u$ et que $w_s(v, i)$ et 
    $w_s(v_i, i)$ sont des suffixes de $v_i$ et de $v$. Le 
    résultat est vrai pour $i = 0$ ($u_0 = u$ et $v_0 = v$).
    
    Supposons le résultat vrai à l'étape $i$. Ainsi,
    $w_p(u, i)$ et $w_p(u_i, i)$ sont des préfixes de $u_i$ 
    et de $u$ et que $w_s(v, i)$ et $w_s(v_i, i)$.
    Or le résultat préliminaire nous donne que
    $w_p(u_i, 1)$ et $w_p(u_{i + 1}, 1)$ sont des préfixes
    de $u_{i + 1}$ et de $u_i$ ce qui nous donne le résultat
    pour les préfixes. Et on a la même chose pour les suffixes.
\end{Preuve}

\subsection[]{Preuve du \myref{lem-efficience_loin_bords}} \label{proof-efficience_loin_bords}

\lemEfficienceLoinDesBords*

\begin{Preuve}
Procédons par récurrence sur $n$. Si $n=0$, alors la propriété est triviale.

Supposons le résultat vrai pour un entier $n$. Soient
$t = \sigma_1 \ldots \sigma_{n + 1}$ une suite d'opérations élémentaires
$\config{u}{v}$ une configuration telle que
$n + 1 \leq \min{\taille{u}, \taille{v}}$. On note $t'$ la suite d'opérations
élémentaires $\sigma_1 \ldots \sigma_n$. Par hypothèse de récurrence, il existe
$\config{u_n}{v_n}$ telle que $\config{u}{v} \actEff{t'} \config{u_n}{v_n}$.

Le \myref{lem-localite} nous donne de plus que $u_n$ et $v_n$ contiennent au moins
un caractère. Ainsi $u_n$ est de la forme $u'a$ et $v_n$ est de la forme $bv'$
avec $a, b \in A$. On a
\begin{align*}
    \config{u_n}{v_n} &\actEff{c} \config{u_n c}{v_n} \text{ si $c \in A$}.\\
    \config{u_n}{v_n} &\actEff{\retour} \config{u'}{v_n}.\\
    \config{u_n}{v_n} &\actEff{\gauche} \config{u'}{av_n}.\\
    \config{u_n}{v_n} &\actEff{\droite} \config{u_nb}{v'}.
\end{align*}
Ainsi, peu importe la valeur de $\sigma_{n + 1}$, il existe une configuration
$\config{u_{n + 1}}{v_{n + 1}}$ telle que $\config{u_n}{v_n} \actEff{\sigma_{n + 1}}
\config{u_{n + 1}}{v_{n + 1}}$ et donc telle que
$\config{u}{v} \actEff{t} \config{u_{n + 1}}{v_{n + 1}}$.

Et donc on a bien le résultat.
\end{Preuve}

\subsection[]{Preuve du \myref{lem-egalite_taille_loin_bords}}
\label{proof-egalite_taille_loin_bords}

\lemEgaliteTailleLoinDesBords*

\begin{Preuve}
Procédons par induction sur $n$. Si $n=0$ alors la propriété est triviale.

Supposons la propriété vraie pour un entier $n$. Soient
$t = \sigma_1 \ldots \sigma_{n + 1}$ une suite d'opérations élémentaires
et $\config{u}{v}$ une configuration telle que $\size{u} \geq n + 1$ et
$\size{v} \geq n + 1$. On note $t'$ la suite $\sigma_1 \ldots \sigma_n$,
$\config{u_n}{v_n} = \config{u}{v} \cdot t'$ et $\config{u_{n + 1}}{v_{n + 1}} = \config{u}{v} \cdot t$.

Le lemme de localité nous donne que $u_n$ et $v_n$ contiennent au moins
un caractère. Ainsi $u_n$ est de la forme $u'a$ et $v_n$ est de la forme $bv'$, avec $a,b \in A$.
On a
\[
\left\{
\begin{aligned}
    u_{n + 1} &= u_n c && \; v_{n + 1} = v_n  \enspace &&\text{ si $\sigma_{n + 1} = c \in A$}\\
    u_{n + 1} &= u'    && \; v_{n + 1} = v_n  \enspace &&\text{ si $\sigma_{n + 1} = \retour$}\\
    u_{n + 1} &= u'    && \; v_{n + 1} = av_n \enspace &&\text{ si $\sigma_{n + 1} = \gauche$}\\
    u_{n + 1} &= u_n b && \; v_{n + 1} = v'   \enspace &&\text{ si $\sigma_{n + 1} = \droite$}
\end{aligned}
\right.
\]
L'hypothèse d'induction appliquée à $t'$ et à $\config{u_n}{v_n}$, à savoir
\[
    \size{u_nv_n} = \size{uv} - \size{t'}_{\retour} + \sum_{x \in A} \size{t'}_x
\]
nous permet de conclure et d'obtenir le résultat :
\[
    \size{u_{n + 1}v_{n + 1}} = \size{uv} - \size{t}_{\retour} + \sum_{x \in A} \size{t}_x.
\]
Ceci conclut notre récurrence et nous donne le résultat.
\end{Preuve}

\subsection[]{Preuve de la \myref{prop-croissance_automatique_bornee}}
\label{proof-croissance_automatique_bornee}

\propCroissanceDesAutomatiquesBornee*

\begin{Preuve}
    Si pour tout mot $w \in L$, $\size{w} \leq \normeInf{K}$, la propriété est trivialement vraie.

    Sinon, on considère $w \in L$ tel que $\size{w} > \normeInf{K}$.
    Le \myref{lem-encadrement_taille} nous assure qu'une exécution menant
    à $w$ est au moins de taille 2. En effet, si on avait $w$ en une seule touche $t$, on aurait
    \[
        \size{w} \leq \size{\epsilon} + \sum_{a \in A} \size{t}_a \leq \normeInf{K}.
    \]
    Ainsi, il existe $t_0 \ldots t_{n + 1}$ ($n \in \nats$)
    telle que $\config{\epsilon}{\epsilon} \cdot t_0 \cdots t_{n + 1} = \config{u}{v}$ et $w = uv$. Pour tout $i \in\intervE[0, n + 1]$,
    on pose
    \[
        \left\{
        \begin{aligned}
            \config{u_i}{v_i} &= \config{\epsilon}{\epsilon} \cdot t_1 \cdots t_i\\
            w_i               &= u_iv_i
        \end{aligned}.
        \right.
    \]
    Les $w_i$ sont dans $L$ (car $K$ est automatique) et le \myref{lem-encadrement_taille}
    nous assure que pour tout $i \in \intervE[0, n]$,
    $\abs{\enlarger \size{w_{i + 1}} - \size{w_i}} \leq \normeInf{K}$.
    De plus, $0 \leq \size{w_0} \leq \normeInf{K}$ et $\size{w_n} = \size{w} > \normeInf{K}$, donc il existe
    $k \in \intervE[0, n - 1]$ tel que $w_k \neq w$ et
    \[
        \size{w} - \normeInf{K} \leq \size{w_k} \leq \size{w}.
    \]
    Ainsi, pour tout $w \in L$ de taille strictement supérieure à $\normeInf{K}$,
    il existe $w' \in L$ tel que $w' \neq w$ et
    \[
        \size{w} - \normeInf{K} \leq \size{w'} \leq \size{w},
    \]
    ce qui démontre la proposition.
\end{Preuve}

\subsection[]{Preuve de la \myref{prop-croissance_bornee}}
\label{proof-croissance_bornee}

\propCroissanceBornee*

\begin{Preuve}
Si pour tout $w \in L$, $\taille{w} \leq 3 \normeInf{K}$, alors le résultat
est vrai.

Sinon, on considère $w$ tel que $\taille{w} > 3 \normeInf{K}$. Le
\myref{lem-encadrement_taille} assure qu'on ne peut pas obtenir
$w$ en appuyant sur une seule touche. Ainsi, il existe $n > 0$ et 
$t_1 \ldots t_n f \in T^*F$ qui mène à $w$. Pour tout $i$ 
entre $1$ et $n$, on pose
\[
    \config{u_i}{v_i} = \config{\epsilon}{\epsilon} \cdot t_1 \cdots t_i
\]
et on pose $\config{u_{n + 1}}{v_{n + 1}} = \config{\epsilon}{\epsilon} \cdot t_1 \cdots t_n f$.

Soit $i$ l'indice maximal tel que $\size{u_iv_i} < \size{w} - \normeInf{K}$. Alors, 
on a $\size{u_{i + 1}v_{i + 1}} > \size{u_iv_i}$ et 
$\size{u_{i + 1}v_{i + 1}} \geq \size{w} - \normeInf{K}$. De plus, le
\myref{lem-encadrement_taille} nous donne que  
$\size{u_{i + 1}v_{i + 1}} - \size{u_iv_i} \leq \normeInf{K}$.
On en déduit que 

\[
    \size{w} - 2\normeInf{K} \leq 
    \size{u_iv_i} < \size{w} - \normeInf{K}.
\]

En appliquant une touche finale à $\config{u_i}{v_i}$, on obtient  un mot $w'$ de $L$.
On montre alors avec le \myref{lem-encadrement_taille} que 
\[
    \size{w} - 3\normeInf{K} \leq 
    \size{w'} < \size{w}.
\]
Ainsi, pour tout $w \in L$ de taille strictement supérieure à $3 \normeInf{K}$,
il existe $w' \in L$ tel que $\size{w} - 3 \normeInf{K} \leq \size{w'} < \size{w}$,
ce qui démontre la proposition.
\end{Preuve}

\subsection[]{Preuve du \myref{lem-monotonie}}
\label{proof-monotonie}

\lemMonotonie*

\begin{Preuve}
    On procède par induction sur la touche $t$. Si $t = \epsilon$,
    le résultat est vrai.

    Sinon, $t = t'\sigma$, alors en notant
    $\config{\alpha}{\beta} = \config{u}{v} \cdot t'$ et
    $\config{\alpha'}{\beta'} = \config{u'}{v'} \cdot t'$,
    on a, par l'hypothèse d'induction,
    \[
        \size{\alpha} \leq \size{\alpha'}  \text{ et }
        \size{\beta} \leq \size{\beta'}.
    \]
    Nous analysons les différents cas possibles.
    \begin{itemize}
        \item Si $\sigma = a \in A$ alors $x = \alpha a$, $x' = \alpha' a$,
              $y = \beta$ et $y' = \beta'$ donc les inégalités sont bien
              préservées.
        \item Si $\sigma = \retour$, on a $y = \beta$ et $y' = \beta'$.
              Pour le côté gauche, on distingue trois cas.
              \begin{itemize}
                  \item Si $\alpha' = \epsilon$, alors $\alpha = \epsilon$
                        et donc $x' = x = \epsilon$.
                  \item Si $\alpha = \epsilon$ et $\alpha' \neq \epsilon$,
                        alors $\size{x} = 0 \leq \size{x'}$.
                  \item Si $\alpha \neq \epsilon$ et $\alpha' \neq \epsilon$,
                        alors $\size{x} = \size{\alpha} - 1$ et
                        $\size{x'} = \size{\alpha'} - 1$.
              \end{itemize}
               Les inégalités sont bien préservées.
        \item Si $\sigma = \gauche$, on distingue trois cas.
              \begin{itemize}
                \item Si $\alpha' = \epsilon$, alors $\alpha = \epsilon$
                      et donc $x' = x = \epsilon$, $y = \beta$ et $y' = \beta'$.
                \item Si $\alpha = \epsilon$ et $\alpha' \neq \epsilon$, alors
                      on a $\size{x} = 0 \leq \size{x'}$ du côté droit et
                      $\size{y} = \size{\beta}$ et
                      $\size{y'} = \size{\beta'} + 1$ du côté gauche.
                \item Si $\alpha \neq \epsilon$ et $\alpha' \neq \epsilon$,
                      alors on a $\size{x} = \size{\alpha} - 1$ et
                      $\size{x'} = \size{\alpha'} - 1$ du côté droit et
                      $\size{y} = \size{\beta} + 1$ et
                      $\size{y'} = \size{\beta'} + 1$ du côté gauche.
            \end{itemize}
            Les inégalités sont bien préservées.
        \item Si $\sigma = \droite$, on distingue trois cas.
            \begin{itemize}
                \item Si $\beta' = \epsilon$, alors $\beta = \epsilon$ et donc
                      $y' = y = \epsilon$, $x = \alpha$ et $y' = \alpha'$.
                \item Si $\beta = \epsilon$ et $\beta' \neq \epsilon$, alors on
                      a $\size{y} = 0 \leq \size{y'}$ du côté droit et
                      $\size{x} = \size{\alpha}$ et
                      $\size{x'} = \size{\alpha'} + 1$ du côté gauche.
                \item Si $\beta \neq \epsilon$ et $\beta' \neq \epsilon$,
                      alors on a $\size{y} = \size{\beta} - 1$ et
                      $\size{y'} = \size{\beta'} - 1$ du côté droit et
                      $\size{x} = \size{\alpha} + 1$ et
                      $\size{x'} = \size{\alpha'} + 1$ du côté gauche.
            \end{itemize}
            Et les inégalités sont bien préservées.
    \end{itemize}
    Ceci permet de conclure notre induction et donc notre preuve.
\end{Preuve}

\subsection[]{Preuve du \myref{lem-efficience_correct}}

\label{proof-lem-efficience-correct}

\lemEfficienceCorrect*

\begin{Preuve}
    On montre le résultat par récurrence sur $n$. Le résultat est
    vrai si $n = 0$. Si, $n > 0$ on a 
    \[
        \config{u}{v} \actEff{\sigma_1 \ldots \sigma_n} \config{u_n}{v_n}
        \actEff{\sigma_{n + 1}} \config{u_{n + 1}}{v_{n + 1}}.
    \]
    Par hypothèse d'induction, 
    \[
        \config{xu}{vy} \actEff{\sigma_1 \ldots \sigma_n} \config{xu_n}{v_ny}.
    \]
    On regarde le comportement de $\sigma_{n + 1}$.
    \begin{itemize}
        \item Si $\sigma_{n + 1} = a \in A$, 
              alors $\config{u_{n + 1}}{v_{n + 1}} = \config{u_na}{v_n}$ et
              $\config{xu_n}{v_ny} \actEff{a} \config{xu_na}{v_ny}$.
        \item Si $\sigma_{n + 1} = \retour$, alors, par efficience, $u_n$ n'est
        pas vide et est de la forme $u'a$ et donc 
        $\config{u_{n + 1}}{v_{n + 1}} = \config{u'}{v_n}$ et
            $\config{xu_n}{v_ny} \actEff{\retour} \config{xu'}{v_ny}$.
        \item Si $\sigma_{n + 1} = \gauche$, alors, par efficience, $u_n$ n'est
        pas vide et est de la forme $u'a$ et donc 
        $\config{u_{n + 1}}{v_{n + 1}} = \config{u'}{av_n}$ et
            $\config{xu_n}{v_ny} \actEff{\gauche} \config{xu'}{av_ny}$.
        \item Si $\sigma_{n + 1} = \droite$, alors, par efficience, $v_n$ n'est
        pas vide et est de la forme $av'$ et donc 
        $\config{u_{n + 1}}{v_{n + 1}} = \config{u_na}{v'}$ et
        $\config{xu_n}{v_ny} \actEff{\droite} \config{xu_na}{v'y}$.
    \end{itemize}
    et on obtient le résultat voulu.
\end{Preuve}

%% file: Annexes/5_touche_annexes.tex
\section{Preuve des propriétés sur les touches}

\subsection[]{Preuve de la \myref{prop-ecriture}}
\label{proof-prop-ecriture}

\propEcriture*

\begin{Preuve}
    Pour tout $i \in \intervE[1; n]$, on pose
    \[
    \begin{aligned}
        \config{u_i}{v_i}         &= \config{u}{v} \cdot \sigma_1 \ldots \sigma_i       \qquad&
            \config{x_i'}{y_n'}  &= \fc{u}{v} \cdot \sigma_1 \ldots \sigma_i\\
        \config{u_i'}{v_i'}      &= \config{u}{v} \cdot f_t(\sigma_1 \ldots \sigma_i)  \qquad &
            \config{x_i}{y_i}    &= \fc{u}{v} \cdot f_t(\sigma_1 \ldots \sigma_i)
    \end{aligned}
    \]
    Une induction sur $i$ permet alors de montrer que pour tout 
    $i \in \intervalleEntier{1}{n}$,
    \[
        \size{u_i} = \size{x_i} = \size{u_i'} = \size{x_i'} 
        \quad \text{et} \quad 
        \size{v_i} = \size{y_i} = \size{v_i'} = \size{y_i'}
    \]
    et que pour tout $a \in A$,
    \[
    \begin{aligned}
        \config{u_i}{v_i}_j = a 
        &\text{ ssi } \config{x_i}{y_i}_j = \cEnt{k} \text{ et } \config{u}{v}_k = a  &&\text{ ou }
            \config{x_i}{y_i}_j = \tEnt{k} \text{ et } t_k = a\\ 
        &\text{ ssi } \config{u_i'}{v_i'}_j = a &&\text{ ou } 
            \config{u_i'}{v_i'}_j = \tEnt{k} \text{ et } t_k = a\\ 
        &\text{ ssi } \config{x_i'}{y_i'}_j = \cEnt{k} \text{ et } \config{u}{v}_k = a &&\text{ ou } \config{x_i'}{y_i'}_j = a.
    \end{aligned}
    \]
\end{Preuve}

\subsection[]{Preuve de la \myref{prop-ecriture_efficiente}}
\label{proof-prop-ecriture_efficiente}

\propEcritureEfficiente*

\begin{Preuve}
    Soit $n = \size{u}, m = \size{v}, p = \size{x}$ et $q = \size{y}$
    et $a$ une lettre.
    On pose 
    \[
        \config{u'}{v'} = \config{a^n}{a^m} \cdot f_t(t)
        \quad \text{et} \quad 
        \config{x'}{y'} = \config{a^p}{a^q} \cdot f_t(t).
    \]
    Par la \myref{cor-touche_aveugle} et la \myref{prop-touche_aveugle}, 
    on a que $t$ écrit son $k$\ieme symbole en position $i$ depuis 
    $\config{u}{v}$ (respectivement depuis $\config{x}{y}$) 
    si et seulement si $\config{u'}{v'}_i = \tEnt{k}$ (respectivement 
    $\config{x'}{y'}_i = \tEnt{k}$).
    Posons $N = \max{n, p}$ et $M = \max{m, q}$. Par le
    \myref{lem-efficience_meme_taille}, $t$ agit de manière efficiente
    sur $\config{a^n}{a^m}$ et sur  $\config{a^p}{a^q}$ , donc par le \myref{lem-efficience_correct},
    \[
        \config{a^N}{a^M} \cdot f_t(t) = \config{a^{N - n}u'}{v'a^{M - m}}
        \quad \text{et} \quad
        \config{a^N}{a^M} \cdot f_t(t) = \config{a^{N - p}x'}{y'a^{M - q}}.
    \]
    En particulier, on a d'un côté $\config{a^N}{a^M}_i = \tEnt{k}$ 
    si et seulement si $\config{u'}{v'}_i = \tEnt{k}$ et de l'autre 
    $\config{a^N}{a^M}_i = \tEnt{k}$ si et seulement si 
    $\config{x'}{y'}_i = \tEnt{k}$. D'où le résultat.
\end{Preuve}

%% file: Annexes/2_sans_fleches_annexes.tex
\section{Les claviers sans flèches}

\subsection[]{Preuve du \myref{lem-caracterisation_config_rek}}
\label{proof-caracterisation_config_rek}

\lemCaracterisationConfigRek*

\begin{Preuve}
    On montre le résultat pour les actions élémentaires, et on
    l'obtient sur les touches par une induction immédiate sur
    la longueur de la touche. L'action d'une suite de touches
    $\tau$ étant équivalent à celle de la concaténation des
    touches de $\tau$, on aura le résultat.

    Les actions élémentaires qui sont faisables ici sont
    le retour arrière et l'écriture d'une lettre $a$ et on a
    \begin{align*}
        \config{u}{\epsilon} \cdot a &=  \config{ua}{\epsilon}\\
         \config{u}{\epsilon} \cdot \retour &=
         \begin{cases}
            \config{u}{\epsilon} & \text{si $u = \epsilon$}\\
            \config{v}{\epsilon} & \text{si $u = vb$}
         \end{cases}
    \end{align*}
    ce qui montre le résultat pour les actions élémentaires.
\end{Preuve}

\subsection[]{Preuve du \myref{lem-forme_normale_touche_rek}}
\label{proof-forme_normale_touche_rek}
\lemFormeNormaleToucheREK*

\begin{Preuve}
L'idée est de supprimer les séquences $a\retour$ (avec $a \in A$)de la
touche tant qu'il y en a (car $a\retour$ est équivalent à la séquence
vide).
     
Supposons que $t$ est de la forme $t_1 a \retour t_2$
avec $t_1, t_2 \in S^*$ et $a \in A$. Soit $u$
une configuration. Notons $v = u \cdot t_1$. On a
\begin{align*}
    u \cdot t &=
    \bigl((v \cdot a) \cdot \retour\bigr) \cdot t_2\\
    &= (va \cdot \retour) \cdot t_2\\
    &= v \cdot t_2\\
    &= (u \cdot t_1) \cdot t_2
    = u \cdot t_1t_2.
\end{align*}
Ainsi, si $t$ contient une lettre suivie d'un retour, il existe une touche
équivalente, strictement plus courte, sans cette séquence d'opérations.
Il existe donc une touche $t'$ de même action que $t$ sans aucune séquence d'opérations de la forme $a \retour$, c'est-à-dire de la forme $\retour^k w$.

Nous pouvons bien trouver la touche correspondante en temps polynomial. En
effet, nous parcourons la touche $t$ jusqu'à trouver une lettre suivie d'un
retour. S'il n'y en a pas, on retourne la touche, sinon on supprime cette
séquence et on réitère l'opération.

À chaque étape, la taille de la touche considérée diminue d'au moins 1, et donc
il y a au plus $\size{t}$ étapes avec, à l'étape $k$, le parcours d'une touche
de taille inférieure ou égale à $\size{t} - k$. On majore donc notre nombre
d'opérations par
\[
    \sum_{k = 0}^{\size{t}} \size{t} - k = O(\size{t}^2).
\]

\end{Preuve}

%% file: Annexes/3_sans_fleche_droite_annexes.tex
\section{Les claviers sans flèche droite}

\subsection[]{Preuve du \myref{thm-fondamental_grek}}
\label{proof-fondamental_grek}
\thmFondamentalGREK*

\begin{Preuve}
Procédons par induction sur $n$. Si $n=0$ alors la propriété est trivialement vraie.

Si $n > 0$, posons 
$\config{x_{n-1}}{y_{n-1}} = \config{\epsilon}{\epsilon} \cdot \sigma_1 \cdots \sigma_{n-1}$. 
Soit $\config{u}{v}$ une configuration. 
On pose $c_n = \config{u}{v} \cdot \sigma_1 \cdots \sigma_n$. 
Par hypothèse d'induction, il existe $u_{n-1}$ et $v_{n-1}$ tels que 
\[
    c_{n-1} = \config{u_{n-1}x_{n-1}}{v_{n-1}v}
\]
avec $y_{n-1}$ un sous-mot de $v_{n-1}$ et $u_{n-1}$ un préfixe de $u$.

On distingue alors plusieurs cas.
\begin{itemize}
    \item Si $\sigma_n = a \in A$, alors $x_n = x_{n-1} a$ et $y_n = y_{n-1}$. On 
          pose $u_n = u_{n-1}$ et $v_n = v_{n-1}$.
    
    \item Si $\sigma_n = \retour$ et $x_{n-1} = \epsilon$, alors $x_n = \epsilon$ 
    et $y_n = y_{n-1}$. On pose $v_n = v_{n - 1}$ et
    \[
        u_n = \begin{cases*}
            u_{n-1}' & si $u_{n-1} = u_{n-1}'a$ ($a \in A$)\\
            \epsilon & sinon
        \end{cases*}.
    \]
    
     \item Si $\sigma_n = \retour$ et $x_{n-1} = x_{n-1}'a$ avec $a \in A$, 
           alors $x_n = x_{n - 1}'$ et $y_n = y_{n - 1}$. On pose 
           $u_n = u_{n - 1}$ et $v_n = v_{n - 1}$.

    \item Si $\sigma_n = \gauche$ et $x_{n-1} = \epsilon$, alors 
          $x_n = \epsilon$ et $y_n = y_{n + 1}$. On distingue deux cas.
    \begin{itemize}
        \item Si $u_{n-1} = \epsilon$, on pose $u_n = \epsilon$ et $v_n = v_{n-1}$.
        \item Si $u_{n-1} = u_{n-1}' a$ avec $a \in A$, on pose $u_n = u_{n - 1}'$
              et $v_n = a v_{n-1}$.
    \end{itemize}
    
    \item Si $\sigma_n = \gauche$ et $x_{n-1} = x_{n-1}' a$ avec $a \in A$, alors
          $x_n = x_{n - 1}'$ et $y_n = a y_{n - 1}$. On pose $u_n = u_{n-1}$ et 
          $v_n = a v_{n-1}$.
\end{itemize}
Dans tous les cas, on a bien $c_n = \config{u_n x_n}{v_n v}$ avec $y_n$ sous-mot de $v_n$ 
et $u_n$ préfixe de $u$.

L'induction est démontrée.
\end{Preuve}

\subsection[]{Preuve du \myref{lem-insensibilite-position}}
\label{proof-insensibilite-position}
\lemInsensibilitePosition*

\begin{Preuve}
    Soit $\config{x}{y}$ une configuration. La \myref{prop-touche_aveugle} permet de 
    supposer qu'elle ne contient pas de $a$ et la \myref{rem-extension-touche-ecriture}, 
    permet de supposer que $t_1 = a$ et que $t$ écrit son premier symbole depuis 
    $\config{u}{v}$. On a que $t$ est de la forme $\sigma_1 \ldots \sigma_n$ et pour 
    tout $i$ entre $1$ et $n$, on pose 
    \[
        \config{u_i}{v_i} = \config{u}{v} \cdot \sigma_1 \ldots \sigma_i
        \qquad \text{ et } \qquad
        \config{x_i}{y_i} = \config{x}{y} \cdot \sigma_1 \ldots \sigma_i.
    \]
    Soit $i$ le premier indice tel que $u_i$ ou $x_i$ est vide, et $i = n$
    si un tel entier n'existe pas (notons que $i > 1$ car $u_1$ et $x_1$
    contiennent au moins un $a$ écrit par $\sigma_1$). 
    Puisque $t$ écrit son premier symbole depuis $\config{u}{v}$, alors 
    $\sigma_1 \ldots \sigma_i$ écrit un $a$ depuis $\config{u}{v}$ à une
    certaine position $j$. 
    
    De plus, $\sigma_1 \ldots \sigma_i$ agit de manière efficiente sur
    $\config{u}{v}$ et sur $\config{x}{y}$, donc par la 
    \myref{prop-ecriture_efficiente}, $t$ écrit un $a$ depuis $\config{x}{y}$
    à la position $j$. Puisque $u_i$ ou $x_i$ est vide, alors $j > 0$ et on a 
    $y_i[j] = v_i[j] = a$. Et ce $a$ ne peut pas être effacé par 
    $\sigma_{i + 1} \ldots \sigma_n$ car il est à droite du curseur.
    
    Ainsi, $\config{x_n}{y_n}$ contient un $a$, hors $\config{x}{y}$ n'en 
    contenait pas, ce qui montre le résultat, par la
    \myref{prop-alternative_ecrire_a}.
\end{Preuve}

%% file: Annexes/4_sans_retour_annexes.tex
\section{Les claviers sans retour}

\subsection[]{Preuve du \myref{lem-iter-lettre}}
\label{proof-iter-lettre}
\lemIterationLettre*

\begin{Preuve}
    Soit $K$ un clavier de $\FK$ qui reconnaît $L$. Il existe une
    suite de touches $\tau = t_1 \ldots t_k$ telle que
    $\config{\epsilon}{\epsilon} \cdot \tau = \config{u}{v}$
    avec $uv = w$.

    Puisque $a$ est une lettre de $w$, alors il existe $i \in \intervE[1, k]$
    telle que $\size{t_i}_a \geq 1$. On pose alors $u_0 = v_0 = \epsilon$,
    et pour tout $n \in \N$, $\config{u_{n + 1}}{v_{n + 1}} = \config{u_n}{v_n} \cdot t_i$ et $w_n = u_nv_n$.

    Une récurrence sur $n$ montre que $\size{w_n}_a \geq n$. Le résultat est
    vrai pour $n = 0$ et s'il est vrai pour $n \in \N$, alors
    le \myref{lem-non-eff} et l'hypothèse de récurrence permettent de
    conclure :
    \[
        \size{w_{n + 1}}_a = \size{w_n}_a + \size{t_i}_a \geq
        \size{w_n}_a + 1 \geq n + 1.
    \]
    Et on a le résultat.
\end{Preuve}

\subsection[]{Preuve du \myref{lem-iter-lettre-fek}}
\label{proof-iter-lettre-fek}

\lemIterationLettreFEK*

\begin{Preuve}
    Soit $w$ tel que $\size{w}_a > \normeInf{K}$. Le \myref{lem-encadrement_taille}
    nous donne qu'une exécution acceptant $w$ est au moins de longueur 2
    (car $\size{w} > \normeInf{K}$).

    Soit $\tau = t_0 \ldots t_{n + 1}$ une suite de touches menant à accepter $w$.
    On note $w'$ le mot obtenu en appuyant sur $\tau' = t_0 \ldots t_n$.
    Puisque $\size{w}_a > \normeInf{K}$, alors, par le \myref{lem-non-eff},
    \[
        \size{w'}_a = \size{w}_a - \size{t_{n+1}}_a \geq \size{w}_a - \size{t_{n+1}}
        \geq \size{w}_a - \normeInf{K} > 0.
    \]
    On en déduit qu'il existe $t$ transiente tel que $t_a > 0$.
    Pour tout $k \in \N$, on pose alors $w_k = u_kv_k$ où
    \[
        \config{u_k}{v_k} = \config{\epsilon}{\epsilon} \cdot
        \underbrace{t \cdots t}_{k \text{ fois }}
        \cdot t_{n + 1}.
    \]
    Une récurrence immédiate, à l'aide du \myref{lem-non-eff}, montre que $\size{w_k}_a \geq k$.
\end{Preuve}